\documentclass[onecolumn,prd,floatfix,showpacs,preprintnumbers,nofootinbib,%
superscriptaddress]{revtex4-2}
\usepackage{mathrsfs}
\usepackage{amssymb}	
\usepackage{amsmath}
\usepackage{graphicx}
\usepackage{subdepth}
\usepackage{xcolor}
\definecolor{lcolor}{rgb}{0.5,0,0}
\definecolor{citcolor}{rgb}{0,0.3,0.0}

\usepackage[breaklinks,colorlinks,urlcolor=blue,citecolor=citcolor,linkcolor=lcolor]{hyperref}

\usepackage{tikz}

\makeatletter
\def\l@subsubsection#1#2{}
\makeatother

\newcommand{\Pt}{{\mathbf{P}}}

\newcommand{\rt}{{\mathbf{r}}}
\newcommand{\xt}{{\mathbf{x}}}
\newcommand{\bt}{{\mathbf{b}}}
\newcommand{\yt}{{\mathbf{y}}}

\newcommand{\qt}{{\mathbf{q}}}
\newcommand{\kt}{{\mathbf{k}}}
\newcommand{\ktpzero}{{\mathbf{k}_{0'}}}

\newcommand{\Rt}{{\mathbf{R}}}

\newcommand{\Kt}{\mathbf{K}}
\newcommand{\Ht}{\mathbf{H}}
\newcommand{\Lt}{\mathbf{L}}

\newcommand{\ed}{\mathrm{ED}}

\newcommand{\epst}{\boldsymbol{\varepsilon}}

\newcommand{\kaz}{\kappa_{z}}
\newcommand{\kac}{\kappa_{\chi}}

\newcommand{\kvec}{{\hat{k}}}

\newcommand{\qvec}{{\hat{q}}}

\newcommand{\lo}{{\textnormal{LO}}}
\newcommand{\nlo}{{\textnormal{NLO}}}

\newcommand{\epsl}{{\varepsilon\!\!\!/}}

\newcommand{\ksl}{{k\!\!\!/}}

\newcommand{\dk}{{\widetilde{\mathrm{d} k}}}

\newcommand{\dkpzero}{{\widetilde{\mathrm{d} k_{0'}}}}
\newcommand{\dkppzero}{{\widetilde{\mathrm{d} k_{0''}}}}
\newcommand{\dkpone}{{\widetilde{\mathrm{d} k_{1'}}}}

\newcommand{\ud}{\, \mathrm{d}}

\newcommand{\nc}{{N_\mathrm{c}}}

\newcommand{\cf}{C_\mathrm{F}}

\newcommand{\nr}[1]{(\ref{#1})}

\newcommand{\as}{\alpha_{\mathrm{s}}}

\newcommand{\fig}{Fig.~}
\newcommand{\figs}{Figs.~}
\newcommand{\eq}{Eq.~}

\newcommand{\eqs}{Eqs.~}

\newcounter{diag}
\newcommand{\namediag}[1]{\refstepcounter{diag} \thediag \label{#1}}

\renewcommand{\thediag}{\textnormal{(\alph{diag})}}

\begin{document}

\author{G. Beuf}
\affiliation{
Theoretical Physics Division, National Centre for Nuclear Research, Pasteura 7, Warsaw 02-093, Poland}

\author{T. Lappi}
\affiliation{
Department of Physics, %
 P.O. Box 35, 40014 University of Jyv\"askyl\"a, Finland}
\affiliation{
Helsinki Institute of Physics, P.O. Box 64, 00014 University of Helsinki,
Finland}

\author{R. Paatelainen}
\affiliation{
Helsinki Institute of Physics, P.O. Box 64, 00014 University of Helsinki,
Finland}
\affiliation{
Department of Physics, %
P.O. Box 64, 00014 University of Helsinki, 
Finland
}

\title{Massive quarks in NLO dipole factorization for DIS: Transverse photon}
\preprint{HIP-2021-47/TH}


\begin{abstract}
We calculate the  light cone wave functions for the QCD Fock components in a transverse virtual photon necessary for applications at next-to-leading order in the QCD coupling, including quark masses. We present a detailed calculation of both the one loop wave function for the quark-antiquark Fock component and the tree level wave function for the quark-antiquark-gluon Fock component. The quark masses are renormalized in the pole mass scheme, satisfying constraints from the requirement of Lorentz invariance. In particular the quark Pauli form factor at NLO is recovered from the on-shell limit of the quark-antiquark Fock component. We use our result to calculate the next-to-leading order correction to the high energy deep inelastic scattering (DIS) transverse structure function on a dense target in the dipole factorization framework. Together with our earlier result for longitudinal photons, this completes the calculation of the total deep inelastic scattering cross section in the dipole picture with massive quarks at next-to-leading order, enabling a comparison with experimental data.
\end{abstract}

\maketitle
\tableofcontents

\section{Introduction}
\addtocontents{toc}{\protect\setcounter{tocdepth}{0}}

In the limit of high scattering energies, Quantum Chromodynamics (QCD) is believed to exhibit the phenomenon of gluon saturation. This means that  partial wave scattering amplitudes can become of the order of unity, i.e. sensitive to unitarity requirements, even for processes at weak coupling (transverse) momentum scales. Understanding the behavior of QCD in this limit has been the object of much attention both experimentally and theoretically. In addition to high energy hadronic and nuclear collision experiments, this small-$x$ regime of QCD can be  probed in high energy deep inelastic scattering experiments, at the HERA collider and in a future Electron-Ion collider (EIC)~\cite{Accardi:2012qut,AbdulKhalek:2021gbh}. 

A convenient theoretical tool to understand the behavior of QCD in this limit is provided by the Color Glass Condensate (CGC) effective theory description~\cite{Iancu:2003xm,Weigert:2005us,Gelis:2010nm}. Here one starts from the fact that in the small-$x$ limit the dominant degrees of freedom of the scattering are gluonic states in the Fock state of the high energy target hadron or nucleus. One then formulates the high energy scattering process as an eikonal scattering off a classical color field~\cite{Bjorken:1970ah}. The advantage of this eikonal approximation is that one works in transverse coordinate space, where the unitarity of the scattering matrix is manifest. The relevant physical degrees of freedom describing the dense gluonic target are eikonal scattering amplitudes of a dilute probe in the target gluon field, light-like Wilson lines. The Wilson lines resum nonlinear interactions involving any number of gluons in the target, and are thus very well adapted for describing the nonlinear physics of gluon saturation. For the deep inelastic scattering (DIS) process at high energy the eikonal scattering approach leads to the \emph{dipole picture}~\cite{Nikolaev:1990ja,Nikolaev:1991et,Mueller:1993rr,Mueller:1994jq,Mueller:1994gb}, where one factorizes the perturbative partonic structure of the virtual photon from the eikonal scattering of the partonic states from the possibly dense target.

In recent years the eikonal scattering picture of CGC has  been systematically pushed to next-to-leading order (NLO) accuracy for several different processes where a dilute projectile interacts with the dense color field of the target. The evolution of the Wilson lines with~$x$ (known as the Balitsky-Kovchegov equation~\cite{Balitsky:1995ub,Kovchegov:1999ua,Kovchegov:1999yj} for the two-point function or as the JIMWLK equation~\cite{Mueller:2001uk} more generally) is now known to NLO  including the resummations of collinear logarithms needed to stabilize the calculation~\cite{Balitsky:2008zza,Balitsky:2013fea,Kovner:2013ona,Balitsky:2014mca,Beuf:2014uia,Lappi:2015fma,Iancu:2015vea,Iancu:2015joa,Albacete:2015xza,Lappi:2016fmu,Lublinsky:2016meo,Ducloue:2019ezk,Dai:2022imf} and even to NNLO in $\mathcal{N}=4$ super-Yang-Mills theory~\cite{Caron-Huot:2016tzz}.  There have been several calculations of single~\cite{Altinoluk:2011qy,JalilianMarian:2011dt,Chirilli:2012jd,Stasto:2013cha,Kang:2014lha,Altinoluk:2014eka,Altinoluk:2015vax} and double~\cite{Ayala:2016lhd,Iancu:2018hwa,Iancu:2020mos} inclusive parton production at forward rapidity in high energy proton-nucleus collisions.  The inclusive DIS cross section for massless quarks has been calculated~\cite{Balitsky:2010ze,Beuf:2011xd,Balitsky:2012bs,Beuf:2016wdz,Beuf:2017bpd,Ducloue:2017ftk,Hanninen:2017ddy} and the first descriptions of experimental data have recently become available~\cite{Beuf:2020dxl}.
Still in the context of DIS, exclusive scattering processes have been studied at NLO in several papers~\cite{Boussarie:2014lxa,Boussarie:2016bkq,Boussarie:2016ogo,Escobedo:2019bxn,Mantysaari:2021ryb,Mantysaari:2022bsp}, as well as inclusive dijet production~\cite{Caucal:2021ent}.

The formalism of \emph{light cone perturbation theory} (LCPT)~\cite{Kogut:1969xa,Bjorken:1970ah,Lepage:1980fj,Brodsky:1997de} provides a calculational and conceptional tool to  develop the picture of high energy scattering in a systematical perturbative expansion. We use LCPT to formulate a weak coupling Fock-space expansion of the quantum state of the probe, expressed in terms of  \emph{light cone wave functions} (LCWF)'s . When Fourier transformed to transverse coordinate space, the LCWF's are naturally combined with the Wilson lines describing the target to obtain scattering cross sections. 

In this work we address the DIS process with massive quarks at one loop order in QCD perturbation theory. This calculation is highly relevant both for its phenomenological applications and for the development of the LCPT framework. 
On the phenomenological side, consistently describing HERA results for $F_2$ and $F_2^c$ simultaneously has been challenging in leading order fits with BK (or JIMWLK) evolution~\cite{Albacete:2010sy,Mantysaari:2018zdd}. For understanding data from HERA and the EIC in terms of high energy QCD and it will be important to include $F_2^c$ in the description at NLO accuracy. This paper, taken together with our earlier results for longitudinal photons~\cite{Beuf:2021qqa}, provides the full expressions needed to calculate the total DIS cross section including quark masses. 

On the more theoretical level, there are not many LCPT calculations at higher orders in perturbation theory. While the LCPT formulation provides a more physical explicit picture of the scattering process, the technology for loop calculations has been less developed than in covariant theory. Both the Hamiltonian formalism and the light cone gauge break explicit Lorentz-invariance, which can make intermediate expressions more cumbersome. Also renormalization has been less well understood in LCPT than in covariant perturbation theory, both because of the small number of loop calculations, and the complications related to Lorentz-invariance and the gauge choice. In this paper we will be faced with the problem of quark mass renormalization in LCPT.  To our knowledge, our calculation here is the first practical LCPT calculation of a physical observable at NLO where not just the divergent part of the mass counterterm is extracted~\cite{Mustaki:1990im,Harindranath:1993de,Zhang:1993dd,Zhang:1993is}, but one-loop mass renormalization is carried out in full in the pole mass scheme (see also discussion in \cite{Burkardt:1991tj}), and all the finite leftover NLO terms are calculated with full mass dependence. 

This paper is a part of a  series of papers on extending the CGC (or more generally dipole picture) calculations of small-$x$ DIS   at NLO accuracy to massive quarks using Light Cone Perturbation Theory. In our first paper \cite{Beuf:2021qqa} we calculated the photon wave function and the total DIS cross section for longitudinally polarized photons. The final result for the   wave function, and the cross sections for both polarizations, are presented in an accompanying shorter paper~\cite{Beuf:2021srj}, but all the details of the calculation for the transverse photon will be described here.  
Since much of the calculation is relatively similar to the previous calculations with massless quarks~\cite{Beuf:2016wdz,Beuf:2017bpd,Hanninen:2017ddy} and to the longitudinal polarization case~\cite{Beuf:2021qqa},  we will be relatively brief with the introduction in this paper and move straight to the point. We refer the reader to these earlier papers for more background on the physics and the calculational methods.

In addition to the longitudinal photon case being algebraically simpler than our present calculation, the calculation in Ref.~\cite{Beuf:2021qqa} also did not encounter the full complexity of quark mass renormalization in LCPT in the same way as in this paper (although the issue is discussed in Ref.~\cite{Beuf:2021qqa}). We will in  this paper need to fully renormalize the quark mass in both the propagator correction (so called ``kinetic mass'') and vertex correction (``vertex mass''). However, we will treat the issue in a concise way in this paper, ensuring that we get the correct result that maintains Lorentz-invariance of the quark form factor at one loop. We will return to a more detailed discussion of mass renormalization, regularization and the related issue of the self-induced inertias \cite{Pauli:1985pv}  in a separate future paper.

The paper is structured as follows. We will first, in Sec.~\ref{sec:nlointro}, write down the expressions for the total NLO DIS cross section, defining our notations and normalization for the LCWF's. We will then briefly describe the calculation of the leading order cross section in Sec.~\ref{sec:LOsection}. At NLO, before moving to the individual diagrams, we will first discuss the overall spinor structure, kinematics  and mass renormalization procedure of the wavefunctions in Sec.~\ref{sec:spinstucture}.  We then calculate the loop diagrams in Sec.~\ref{sec:oneloopdiags} and combine them to get the momentum space mass renormalized $\gamma^*\to q\bar{q}$ wavefunction in Sec.~\ref{sec:momspacefull}. This is then transformed to mixed longitudinal momentum and transverse coordinate space in Sec.~\ref{sec:NLOmixedspace}. We then write down and Fourier-transform the tree-level $\gamma^* \to q \bar{q}g$ wavefunctions in Sec.~\ref{sec:qqbargwf}. In Sec.~\ref{sec:xsdetails} we input the wave functions into the expressions for the cross section and effectuate the cancellation of the remaining UV divergences between the contributions of the $q\bar{q}$ and $q\bar{q}g$ Fock states, to arrive at our result for the total DIS cross section which is summarized in Sec.~\ref{sec:xsfinal}. We provide brief conclusions and an outlook for the future in Sec.~\ref{sec:conc}. Several more technical parts of the calculation are presented in the Appendices.

\section{Dipole factorization for DIS: Cross section at NLO}
\label{sec:nlointro}

The DIS cross section can be expressed in terms of the cross section of a virtual photon scattering on the hadronic target. Following the discussion presented in \cite{Beuf:2021qqa}, the total NLO cross section for a virtual transverse photon scattering from a classical gluon field takes the form
\begin{equation}
\label{eq:crosssectionformula}
\sigma^{\gamma^{\ast}}_{T} = 2\nc\widetilde{\sum_{q\bar{q}~ \text{F. states}}} \frac{1}{2q^+}
\left| \widetilde{\psi}^{\gamma^{\ast}_{T}\rightarrow q\bar{q}} \right|^2 \mathrm{Re}[1-\mathcal{S}_{01}] +  2\nc\cf\widetilde{\sum_{q\bar{q}g~ \text{F. states}}} \frac{1}{2q^+}
\left| \widetilde{\psi}^{\gamma^{\ast}_{T}\rightarrow q\bar{q}g} \right|^2 \mathrm{Re}[1-\mathcal{S}_{012}] + \mathcal{O}(\alpha_{em}\alpha_s^2),
\end{equation}
where the color factor  $\cf = (\nc^2-1)/(2\nc)$ and  $\nc$ is the number of colors. The phase space sums over the mixed space quark-antiquark ($q\bar q$) and quark-antiquark-gluon ($q\bar q g$) Fock states are given by: 
\begin{equation}
\label{eq:defphasesp}
\begin{split}
\widetilde{\sum_{q\bar{q}~ \text{F. states}}} & = 
\left (\prod_{i=0}^{1}  \int_{0}^{\infty} \frac{\ud k^+_i \Theta(k^+_i)}{2k^+_i(2\pi)}\right )
2\pi\delta\left(q^+ - \sum_{j=0}^{1}k^+_j\right) \left (\prod_{k=0}^{1}\int \ud^{D-2}\xt_k\right ),
\\
\widetilde{\sum_{q\bar{q}g~ \text{F. states}}} & = 
\left (\prod_{i=0}^{2}\int_{0}^{\infty} \frac{\ud k^+_i \Theta(k^+_i)}{2k^+_i(2\pi)}\right )
2\pi\delta\left(q^+ - \sum_{j=0}^{2}k^+_j\right)\left (\prod_{k=0}^{2}\int \ud^{D-2}\xt_k\right ),
\end{split}
\end{equation}
where $\xt_0$ is the transverse coordinate of the quark, $\xt_1$ that of the antiquark and $\xt_2$ that of the gluon. 
The reduced LCWF's $\widetilde{\psi}^{\gamma^{\ast}_{\rm T}\rightarrow q\bar{q}}$ and $\widetilde{\psi}^{\gamma^{\ast}_{\rm T}\rightarrow q\bar{q}g}$  (see the discussion in~\cite{Beuf:2021qqa}) are independent of the photon transverse momentum $\qt$ and cannot depend on the absolute transverse positions of the Fock state partons, just on their differences. Finally, the quark-antiquark $(\mathcal{S}_{01})$ and quark-antiquark-gluon $(\mathcal{S}_{012})$ amplitudes are defined as 
\begin{equation}
\begin{split}
\mathcal{S}_{01} &= 
\frac{1}{\nc} \mathrm{Tr} \left(U_F(\xt_0)U^{\dagger}_{F}(\xt_1) \right),
\\
\mathcal{S}_{012} &= 
\frac{1}{\nc\cf}\mathrm{Tr} \left(t^{b}U_F(\xt_0)t^{a}U^{\dagger}_{F}(\xt_1) \right) U_{A}(\xt_2)_{ba}, \\
\end{split}
\end{equation}
for the quark-antiquark $(\mathcal{S}_{01})$ and quark-antiquark-gluon $(\mathcal{S}_{012})$ amplitudes.  Here the fundamental (F) and the adjoint (A) Wilson lines are defined as light-like path ordered exponentials for a classical gluon target
\begin{equation}
\begin{split}
U_{F}(\xt) & = \mathcal{P}\exp\biggl [-ig\int \ud x^{+} t^{a} A^{-}_{a}(x^{+},0,\xt) \biggr ],\\
U_{A}(\xt) & = \mathcal{P}\exp\biggl [-ig\int \ud x^{+} T^{a} A^{-}_{a}(x^{+},0,\xt) \biggr ],\\
\end{split}
\end{equation}
where $t^a$ and $T^a$ are the generators of the fundamental and adjoint representations, respectively.

What we will do in this paper is to calculate the LCWF $\widetilde{\psi}^{\gamma^{\ast}_{\rm T}\rightarrow q\bar{q}}$  to one loop order (up to $\as$) and the gluon emission LCWF $\widetilde{\psi}^{\gamma^{\ast}_{\rm T}\rightarrow q\bar{q}g}$ to tree level, including quark masses. We will then insert the result into \eq\eqref{eq:crosssectionformula} and subtract and add a term to make explicit a cancellation of a transverse  UV divergence between the two contributions.

\section{Leading order wave function}
\label{sec:LOsection}

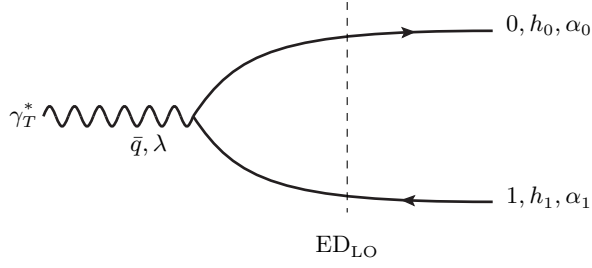
\begin{figure}[tb!]
\centerline{
\includegraphics[width=6.4cm]{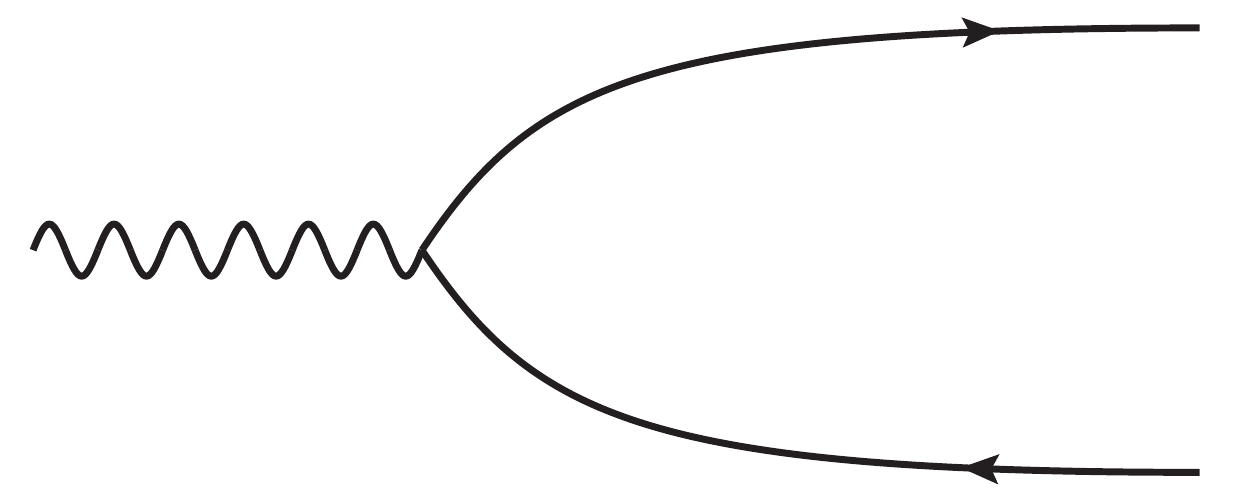}
\begin{tikzpicture}[overlay]
\draw [dashed] (-2.3,2.8) -- (-2.3,0);
\node[anchor=north] at (-2.3cm,-0.2cm) {$\ed_{\lo}$};
\node[anchor=west] at (-0.3,2.5) {$0,h_0,\alpha_0$};
\node[anchor=west] at (-0.3,0.2) {$1,h_1,\alpha_1$};
\node[anchor=west] at (-5.3,0.9) {$\bar q, \lambda$};
\node[anchor=east] at (-6.3,1.3) {$\gamma^{\ast}_{T}$};
 \end{tikzpicture}
}
\rule{0pt}{1ex}
\caption{Time ordered light cone diagram (momenta flows from left to right) contributing to the transverse virtual photon wave function at leading order. Here, the quark (antiquark) helicity and color indices are denoted as $h_0 (h_1)$ and $\alpha_0 (\alpha_1)$. In the vertex the light cone spatial three-momentum $\qvec \equiv (q^+,\qt)=  \kvec_0+\kvec_1$ is conserved, i.e. $ q^+ =  k^+_0 +  k^+_1$ and $\qt = \kt_0+\kt_1$.
}
\label{fig:lovertex}
 \end{figure}

As in the longitudinal case \cite{Beuf:2021qqa}, let us  first write down the leading order LCWF contribution to the transverse photon splitting into a massive quark-antiquark dipole. 
Following the notation shown in \fig\ref{fig:lovertex}, the leading order $\gamma^{\ast}_T\rightarrow q\bar{q}$ LCWF can be written as 
\begin{equation}
\label{eq:LOlcwf}
\Psi^{\gamma^{\ast}_T\rightarrow q\bar{q}}_{\lo}  = \frac{\delta_{\alpha_0\alpha_1}}{\ed_{\lo}}V^{\gamma^{\ast}_T\rightarrow q\bar{q}}_{h_0;h_1}.
\end{equation}
Note that, as in Ref.~\cite{Beuf:2021qqa}, we do not include the delta function for overall 3-momentum conservation in the momentum space LCWF. For the transverse momentum it is included in the definition of the Fourier-transform from momentum to coordinate space, and for the longitudinal momentum in the phase space integration measure~\nr{eq:defphasesp}.
The leading order light cone energy denominator $\ed_{\lo}$ simplifies to
\begin{equation}
\label{eq:LOED}
\begin{split}
\ed_{\lo} & = -\frac{q^+}{2k^{+}_0k^{+}_1}\left[\Pt^2 + \overline{Q}^2 + m^2 \right]\\
& = \frac{1}{(-2q^+)z(1-z)}\left [\Pt^2 + \kappa_z^2 \right].
\end{split}
\end{equation}
Here we have introduced the notations
\begin{align}
\label{eq:PtdefandQ}
\Pt   &= \kt_0 - z\qt = -\kt_1 +  (1-z)\qt = (1-z)\kt_0 -z \kt_1, \\
\overline{Q}^2  &= z(1-z)Q^2,\\
\kappa_z &= \sqrt{z(1-z)Q^2 + m^2},
\end{align}
with the momentum fraction $z = k_0^+/q^+$ and  $1-z = k_1^+/q^+$ with $z \in [0,1]$. The interpretation of $\Pt$ is that it is the relative transverse momentum of the quark-antiquark pair on the light cone, where the plus-momentum plays the role of a mass in a 2-dimensional nonrelativistic system. After Fourier-transforming to transverse coordinate space, $\kappa_z$ will be the typical inverse size of the dipole.

The transverse photon splitting vertex into massive quark-antiquark ($q\bar{q}$) pair is given by   
\begin{equation}
\label{eq:LOvertex}
V^{\gamma^{\ast}_T\rightarrow q\bar{q}}_{h_0;h_1}= +ee_f \bar{u}(0)\epsl_{\lambda}(q) v(1),
\end{equation}
where the parameter $e_f$ is the fractional
charge of quark flavor $f$, $e$ is the QED coupling constant, and the compact notation for the spinors $\bar u(0) = \bar u(k_0,h_0)$ and $v(1) = v(k_1,h_1)$ has been introduced. The transverse momentum dependence in the massive spinors $u$ and $v$ and in the photon polarization vector can be extracted out by using the decomposition derived in \cite{Beuf:2021qqa}. For the leading order quark spinor structure in  \eq\nr{eq:LOvertex}, this decomposition yields
\begin{equation}
\label{eq:LOvertexdecomp}
\bar{u}(0)\epsl_{\lambda}(q) v(1) = \left (\frac{q^+}{2k_0^+k_1^+} \right )
\left\{ 
\left[ \left(\frac{k_0^+-k_1^+}{q^+} \right) \delta^{ij}_{(D_s)} \bar{u}(0)\gamma^{+}v(1) + \frac{1}{2}\bar{u}(0)\gamma^{+}[\gamma^i, \gamma^j]v(1)  
\right] \Pt^{i} - m\bar{u}(0)\gamma^{+}\gamma^{j}v(1)  
\right\} \epst^{j}_{\lambda}.
\end{equation}
Indeed, in each term in the right-hand side of Eq.~\eqref{eq:LOvertexdecomp}, the $\gamma^+$ Dirac matrix projects out the components that depend on the transverse momenta (and masses) from the spinors $u$ and $v$, so that only the components  depending only on the light cone momenta and on the light cone helicities (sometimes called \emph{good components}) survive (see e.g.~\cite{Beuf:2016wdz}).

For the DIS cross section, we need to Fourier transform the momentum space expression of the LCWF, supplemented with the delta function for overall transverse momentum conservation (see \cite{Beuf:2021qqa}), into mixed space. 
The reduced wave function $\widetilde{\psi}^{\gamma^{\ast}_T\rightarrow q\bar{q}}_{\lo}$ is extracted from the two-dimensional Fourier-transform of the wavefunction  $\widetilde{\Psi}^{\gamma^{\ast}_T\rightarrow q\bar{q}}_{\lo}$ by separating an overall kinematical factor related to the total transverse momentum, and a color factor, as:
\begin{equation}
\label{eq:LOmixspace}
\widetilde{\Psi}^{\gamma^{\ast}_T\rightarrow q\bar{q}}_{\lo} = \delta_{\alpha_0\alpha_1}e^{i(\qt/q^{+})\cdot \left (k^+_0\xt_0 + k^+_1\xt_1\right )}\widetilde{\psi}^{\gamma^{\ast}_T\rightarrow q\bar{q}}_{\lo}.
\end{equation}
Thus the reduced LCWF in mixed space is given by the following expression
\begin{equation}
\label{eq:psireducedLO}
\widetilde{\psi}^{\gamma^{\ast}_T\rightarrow q\bar{q}}_{\lo} = -\frac{ee_f}{2\pi}\Biggl \{ \biggl [\left (\frac{k_0^+-k_1^+}{q^+} \right )\delta^{ij}_{(D_s)} \bar{u}(0)\gamma^{+}v(1)  + \frac{1}{2}\bar{u}(0)\gamma^{+}[\gamma^i, \gamma^j]v(1)  \biggr ] \mathcal{F}[\Pt^{i}] - m\bar{u}(0)\gamma^{+}\gamma^{j}v(1) \mathcal{F}[1] \Biggr \} \epst^{j}_{\lambda}.
\end{equation}
Here we have defined a Fourier transform operator $\mathcal{F}$ that corresponds to the Fourier-transform of its argument, \emph{divided by the leading order energy denominator} and an overall factor, as
\begin{equation}
\label{eq:FToperator}
\begin{split}
\mathcal{F}[1] & = 2\pi\int \frac{\ud^{D-2}\Pt}{(2\pi)^{D-2}} \frac{e^{i\Pt \cdot \xt_{01}}}{\Pt^2 + \kappa_z^2},\\
\mathcal{F}[\Pt^i] & = 2\pi\int \frac{\ud^{D-2}\Pt}{(2\pi)^{D-2}} \frac{e^{i\Pt \cdot \xt_{01}} \ \Pt^i }{\Pt^2 + \kappa_z^2}.\\
\end{split}
\end{equation}
We have also introduced the compact notation $\xt_{01} = \xt_0 - \xt_1$ for the difference of transverse coordinates $\xt_0$ and $\xt_1$. The shorthand notation $\mathcal{F}$ will be very convenient in the NLO calculation. The evaluation of such Fourier transforms is discussed in Appendix~\ref{app:FTSqqbarcase}.

It is then straightforward to perform the remaining $(D-2)$-dimensional Fourier integrals in \eq\nr{eq:FToperator}. All in all, the final result for the reduced leading order LCWF in mixed space can be written in the following form  
\begin{equation}
\label{eq:LOlcwfmixfact}
\begin{split}
\widetilde{\psi}^{\gamma^{\ast}_T\rightarrow q\bar{q}}_{\lo} =  \frac{-ee_f}{2\pi}&\left ( \frac{\kappa_z}{2\pi\vert \xt_{01}\vert }\right )^{\frac{D}{2}-2}\Biggl \{ \left[\left (\frac{k_0^+-k_1^+}{q^+} \right )\delta^{ij}_{(D_s)} \bar{u}(0)\gamma^{+}v(1)  + \frac{1}{2}\bar{u}(0)\gamma^{+}[\gamma^i, \gamma^j]v(1)  \right]  \frac{i\xt_{01}^i}{\vert \xt_{01}\vert} \kappa_z K_{\frac{D}{2} - 1}\left (\kappa_z\vert \xt_{01}\vert\right ) \\
&  - m\bar{u}(0)\gamma^{+}\gamma^{j}v(1) K_{\frac{D}{2} -2}\left (\kappa_z \vert \xt_{01}\vert\right ) \Biggr \} \epst^{j}_{\lambda},
\end{split}
\end{equation}
where the function $K_{\nu}(z)$ is the modified Bessel function of the second kind. Setting $D=4$, it is easy to check that one recovers the conventional result for the wavefunction~\cite{Nikolaev:1990ja}.

\section{Structure of the NLO corrections}
\label{sec:spinstucture}

\subsection{Spinor structure and mass renormalization}
\label{sec:spinstuctureandmrenorm}

Before proceeding to calculate these contributions, let us first discuss the general Lorentz and gauge invariance constraints for the wave functions, and its relation to mass renormalization. As discussed already in our previous paper~\cite{Beuf:2021qqa}, the quark mass appears in two separate terms in the QCD light front Hamiltonian. The mass in the free fermion term that determines the relation between light cone energy $k^-$ and light cone momentum $\kt^2+m^2$ is referred to as the ``kinetic mass''. On the other hand, as 
can be explicitly  seen from the decomposition \eqref{eq:LOvertexdecomp}, also a part of the interaction term of the Hamiltonian is linearly proportional to the mass, which we call the ``vertex mass''. Lorentz-invariance at the level of the Lagrangian guarantees that the two masses stay equal. However, the regularization procedure of transverse dimensional regularization and a longitudinal cutoff, which has proven convenient in loop calculations~\cite{Beuf:2016wdz,Beuf:2017bpd,Hanninen:2017ddy}, breaks Lorentz-invariance. One is left with two options. One is to modify the regularization procedure to take this into account. It turns out that such a regularization procedure in fact can be found, by only a slight extension of the one that has been used so far, affecting only the treatment of the ``self-induced inertia'' or ``seagull'' diagrams that vanish in our earlier scheme ~\cite{Pauli:1985pv,Tang:1991rc,Brodsky:1997de}. We will discuss this option in more detail in a future work. The other option is to accept that the two masses will become different after loop contributions, and must be renormalized by separate renormalization conditions at each order in perturbation theory. We have verified explicitly that, at least at one loop, both procedures lead to exactly equivalent results. 

What happens in a practical calculation is the following. As discussed in the case of the longitudinal polarization~\cite{Beuf:2021qqa}, the kinetic mass renormalization is determined by the ``quark propagator correction diagrams''. Here the renormalization condition in the pole mass scheme is determined by the requirement that the leading divergence resulting from the energy denominator approaching zero in the on-shell limit is absorbed into mass renormalization. This procedure, discussed explicitly in~\cite{Beuf:2021qqa}, is identical in the case of the transverse polarization, since the propagator corrections completely factorize from the photon vertex.
If the propagator correction diagrams are evaluated with a Lorentz-invariance conserving regularization scheme, the ensuing counterterm can directly be used to renormalize the vertex mass. If, on the other hand, one uses the straightforward $k^+$-cutoff of ~\cite{Beuf:2016wdz,Beuf:2017bpd,Hanninen:2017ddy}, one needs a separate renormalization condition for the vertex mass, to restore Lorentz-invariance. In the case of the longitudinal photon  polarization, the leading order vertex does not have a light cone helicity flip term proportional to the mass. There is thus no vertex mass parameter to renormalize and, and vertex mass renormalization is not needed.

Since the difficult issues with mass renormalization are related to the interplay of the regularization method and Lorentz-invariance, it is useful to start by checking the constraints of Lorentz-invariance on the wavefunction. 
In the transverse photon case, the NLO correction to the initial-state LCWF for $\gamma^{\ast}_T\rightarrow q\bar{q}$ must be proportional to the polarization vector of the photon. It can also only depend on one transverse momentum vector $\Pt$, and involve matrix elements of different Dirac matrix structures between the quark and antiquark spinors, which can be related to each other by Dirac matrix commutation relations, and by using the Dirac equation. Based on these facts it can be seen that the LCWF can be written as a linear combination of four independent spinor structures.
In fact, for different stages of the calculation it is convenient to use different bases for these spinor structures.
We start by introducing a basis  that is convenient for the loop calculations and  first choose the independent structures as:
\begin{equation}
    \bar u(0) \epsl_{\lambda}(q)v(1) \ ,\quad 
    (\Pt \cdot \epst_{\lambda}) \bar u(0)\gamma^+ v(1) \ ,\quad 
    \frac{(\Pt \cdot \epst_{\lambda})}{\Pt^2}\Pt^j \bar u(0)\gamma^+\gamma^j v(1) \ ,\quad 
    \bar u(0)\gamma^+ \epsl_{\lambda}(q) v(1).
\end{equation}
In terms of these four structures we can write
the wave function up to  NLO in terms of four form factors 
$\mathcal{V}^T, \mathcal{N}^T,  \mathcal{S}^T$ and $\mathcal{M}^T$
which are defined, extracting some constants for future convenience, by
\begin{equation}
\label{NLOformfactors}
\begin{split}
\Psi^{\gamma^{\ast}_T\rightarrow q\bar{q}}_{\rm LO} 
+\Psi^{\gamma^{\ast}_T\rightarrow q\bar{q}}_{\rm NLO} 
& = \delta_{\alpha_0\alpha_1}\frac{ee_f}{\ed_{\lo}} \Biggl \{\bar u(0) \epsl_{\lambda}(q)v(1)\left [1 + \left (\frac{\alpha_s\cf}{2\pi}\right )\mathcal{V}^T\right ] + \frac{q^+}{2k^+_0k^+_1}(\Pt \cdot \epst_{\lambda}) \bar u(0)\gamma^+ v(1) \left (\frac{\alpha_s\cf}{2\pi}\right )\mathcal{N}^T\\
& +  \frac{q^+}{2k^+_0k^+_1} \frac{(\Pt \cdot \epst_{\lambda})}{\Pt^2}\Pt^j m \bar u(0)\gamma^+\gamma^j v(1) \left (\frac{\alpha_s\cf}{2\pi}\right )\mathcal{S}^T + \frac{q^+}{2k^+_0k^+_1}m \bar u(0)\gamma^+ \epsl_{\lambda}(q) v(1)\left (\frac{\alpha_s\cf}{2\pi}\right )\mathcal{M}^T
\Biggr \}. 
\end{split}
\end{equation}
Here the form factor $\mathcal{V}^T$ multiplies the leading-order-like structure, which has both light cone helicity nonflip contributions, and flip contributions proportional to the quark mass. Out of the new structures that only appear at one-loop order, $\mathcal{N}^T$ is a nonflip contribution  and $\mathcal{S}^T$ and $\mathcal{M}^T$ are light cone helicity flip contributions. 

 In general, the parametrization of the $\gamma q \bar{q}$ vertex function based on Lorentz and gauge invariances, relevant for the $\gamma_T^*\rightarrow q\bar{q}$ amplitude, can be written as
\begin{align}
\Gamma^{\mu}(q)
= & F_D(q^2/m^2) \gamma^{\mu} + F_P(q^2/m^2) \frac{q_{\nu}}{2m} i \sigma^{\mu\nu},
\label{1PI_vertex_generic}
\end{align}
where $F_D$ and $F_P$ are the Dirac and Pauli form factors, respectively. 
As explained in Appendix~\ref{sec:Pauliform}, one can show that the form factors ${\cal V}^{T}$, ${\cal N}^{T}$, ${\cal S}^{T}$ and ${\cal M}^{T}$ have to obey the following constraints:
\begin{align}
-\left (\frac{\alpha_s \cf}{2\pi}\right ) \left.\frac{m^2}{\Pt^2}\; {\cal S}^{T}\right|_{\Pt^2= -\overline{Q}^2 -m^2}& = F_P(q^2/m^2)
\label{FP_ST_correspondence}
\\
-\left (\frac{\alpha_s \cf}{2\pi}\right ) \left.\frac{1}{(2z\!-\!1)}\; {\cal N}^{T}\right|_{\Pt^2= -\overline{Q}^2 -m^2}& = F_P(q^2/m^2)
\label{FP_NT_correspondence}
\\
\left (\frac{\alpha_s \cf}{2\pi}\right ) \left.{\cal V}^{T}\right|_{\Pt^2= -\overline{Q}^2 -m^2} & = -1 +F_D(q^2/m^2) +F_P(q^2/m^2)
\label{FD_VT_correspondence} 
\\
\left. {\cal M}^{T}\right|_{\Pt^2= -\overline{Q}^2 -m^2}& = 0
\label{MT_condition}
\, .
\end{align}
All of these constraints are evaluated at what we call the \emph{on-shell point} $Q^2= (\Pt^2+m^2)/(z(1-z))$. The on-shell  point corresponds to the kinematical configuration of a time-like virtual photon decaying to a quark-antiquark pair with four-momentum conservation. For the DIS process the virtual photon is spacelike, and the on-shell point is outside of the physical region. We have, however, analytical expressions for the form factors, and they can easily be extended up to the on-shell point. The on-shell point is, also, the renormalization point in the on-shell, i.e. pole mass,  quark mass renormalization scheme.

Depending on the details of the regularization procedure, we can use these four constraints in two ways. If the regularization procedure preserves Lorentz-invariance, all four conditions serve as cross-checks of the result of the loop calculation. If, on the other hand, Lorentz-invariance needs to be restored, one of the constraint equations becomes a renormalization condition for the vertex mass. To see explicitly how this happens, let us  see how a vertex mass counterterm appears in the wave function~\eqref{NLOformfactors}. For the purpose of mass renormalization, it is convenient to change from the basis in \eqref{NLOformfactors} to a different one, where the longitudinal component of the polarization vector is eliminated using $\varepsilon^+(q)=0, q\cdot \varepsilon(q)=0$. This leads to the following structure
\begin{equation}
\label{eq:NLOformfactorsv2}
\begin{split}
\Psi^{\gamma^{\ast}_T\rightarrow q\bar{q}}_{\rm LO}+\Psi^{\gamma^{\ast}_T\rightarrow q\bar{q}}_{\rm NLO} & = \delta_{\alpha_0\alpha_1}\frac{ee_f}{\ed_{\lo}} \frac{q^+}{2k_0^+k_1^+}\Biggl \{
\Biggl[
\left (\frac{k_0^+-k_1^+}{q^+} \right ) \left [1 + \left (\frac{\alpha_s\cf}{2\pi}\right ) \mathcal{V}^T  \right ] + \left (\frac{\alpha_s\cf}{2\pi}\right )\mathcal{N}^T
\Biggr]
\Pt \cdot \epst_\lambda
\bar u(0)\gamma^+ v(1) 
\\ &
+
\frac{1}{2} \Biggl[1 + \left (\frac{\alpha_s\cf}{2\pi}\right ) \mathcal{V}^T  \Biggr ]
\Pt^{i}\epst^j_\lambda
\bar{u}(0)\gamma^{+}[\gamma^i, \gamma^j]v(1) 
\\ &
+
m
\Biggl[
\left (\frac{\alpha_s\cf}{2\pi}\right )\mathcal{S}^T
\Biggr]
 \left (\frac{\Pt^i\Pt^j}{\Pt^2}\right)
\epst^i_\lambda 
\bar u(0)\gamma^+ \gamma^j  v(1)
\\ &
-
m \Biggl[1 +
 \left (\frac{\alpha_s\cf}{2\pi}\right )\biggl [\mathcal{M}^T + \mathcal{V}^T  \biggr ]
\Biggr]
\epst^i_\lambda
\bar u(0)\gamma^+ \gamma^i  v(1)
\Biggr\},
\end{split}
\end{equation}
where clearly the first two lines are light cone helicity conserving, and the second two flip terms. 

Following a bare perturbation theory approach, the mass appearing in the unrenormalized wave function \eqref{NLOformfactors} is in fact the bare mass $m_0$. One then replaces $m_0 =m-\delta m$ with $\delta m \sim \as$, and inserts this relation into the equation. In a renormalized perturbation approach, on the other hand, one works all the time with the physical quark mass $m$, and inserts an additional 3-point vertex mass counterterm in the Hamiltonian. In both cases, it is obvious that the mass counterterm at order $\as$ is associated with the last line of \eqref{eq:NLOformfactorsv2}, which is the only one where the mass appears at LO, and gives a contribution of the form
\begin{equation}
\Psi^{\gamma^{\ast}_T\rightarrow q\bar{q}}_{\delta m}
=
 \delta_{\alpha_0\alpha_1}\frac{ee_f}{\ed_{\lo}} \frac{q^+}{2k_0^+k_1^+}\Biggl \{
\delta m
\ 
\epst^i_\lambda
\bar u(0)\gamma^+ \gamma^i  v(1)
\Biggr\}.
\end{equation}
Since the form factor  $\mathcal{M}^T$ is the only one that is \emph{only} associated with the same spinor structure, it is clear that the vertex mass renormalization will affect only the form factor $\mathcal{M}^T$. Thus we can separate $\mathcal{M}^T$ into the loop contribution $\mathcal{M}^T_{\text{loop}}(\Pt,z,Q^2)$, which depends on the kinematics, and the constant counterterm
\begin{equation}
\label{eq:MTrenorm}
\mathcal{M}^T = \mathcal{M}^T_{\text{loop}}(\Pt,z,Q^2) + \mathcal{M}_{\text{c.t.}}^T.
\end{equation}
It is now clear how to use one of the form factor constraints, namely the last one~\eqref{MT_condition}, as a mass renormalization condition. One simply determines the counterterm contribution to 
$\mathcal{M}^T$ by requiring~\eqref{MT_condition} to be satisfied, i.e.
\begin{equation}\label{eq:Mcalcondition}
\mathcal{M}_{\text{c.t.}}^T = - \left. \mathcal{M}^T_{\text{loop}}(\Pt,z,Q^2)\right|_{\Pt^2= -\overline{Q}^2 -m^2}.
\end{equation}
This also points to a consistency condition that the loop calculation must, and will, satisfy. Namely, when $\Pt^2$ is set to  the on-shell value $\Pt^2= -\overline{Q}^2 -m^2$ the form factor $\mathcal{M}^T_{\text{loop}}(\Pt,z,Q^2)$ must no longer depend on $z$ or $Q^2$. Also, if a Lorentz-invariance preserving regularization has been used,  \eqref{eq:Mcalcondition} must be automatically satisfied with the mass counterterm extracted from the kinetic mass renormalization condition. Thus we have  established the effect of the vertex mass renormalization. Once the loop contribution $\mathcal{M}^T_{\text{loop}}(\Pt,z,Q^2)$ is calculated, we will simply subtract from it its value at the on-shell point, corresponding to the vertex mass counterterm.
The additional check of our result obtained from comparing the first three constraint equations \nr{FP_ST_correspondence}, \nr{FP_NT_correspondence} and \nr{FD_VT_correspondence} to the well known result for the Pauli form factor at one loop in QCD (identical to the QED result~\cite{Schwinger:1951nm,Peskin:1995ev} up to the replacement $\alpha_s\, C_F \leftrightarrow \alpha_{em}\, e_f^2$) is done in detail in Appendix~\ref{sec:Pauliform}.

In practical calculations of loop diagrams, we will want to use a symmetry under exchanging the quark and the antiquark to restore some contributions, without calculating the corresponding diagrams explicitly. The relevant symmetry here is that the light cone wave function should stay invariant if, for a fixed photon light front  helicity, one exchanges both the momenta and the helicities of the quark and the antiquark. 
In terms of the invariant momenta the quark and antiquark transverse momenta are given by $\kt_0=\Pt+z\qt$ and $\kt_1=-\Pt + (1-z)\qt$. Thus the exchange of three-momenta $\kvec_0 \leftrightarrow \kvec_1$ is achieved  by  the substitution $z \mapsto 1-z$ and $\Pt \mapsto -\Pt$. The spinor matrix elements 
$\bar u(0)\gamma^+ v(1),$ $\bar{u}(0)\gamma^{+}[\gamma^i, \gamma^j]v(1)$  and $\bar u(0)\gamma^+ \gamma^i  v(1)$  
in \eq\eqref{eq:NLOformfactorsv2} are independent of transverse momenta, and symmetric under the exchange $z \mapsto 1-z$.
Under exchanging the light cone helicities $h_0 \leftrightarrow h_1$, the first light cone helicity conserving matrix element  $\bar u(0)\gamma^+ v(1)\sim \delta_{h_0,-h_1}$ and the light cone helicity flip one
$\bar u(0)\gamma^+ \gamma^i  v(1)\sim \delta_{h_0,h_1}$ are symmetric, and the second heliclity conserving one $\bar{u}(0)\gamma^{+}[\gamma^i, \gamma^j]v(1) \sim h_0 \delta_{h_0,-h_1}$ antisymmetric. The scalar form factors only depend on the square of the transverse momentum $\Pt$, but have a nontrivial dependence on the momentum fraction $z$. 
Requiring the invariance of the wave function under the simultaneous exchange $\kvec_0 \leftrightarrow \kvec_1, h_0 \leftrightarrow h_1$
 we can then deduce the following symmetrization requirements for the scalar form factors:
 \begin{eqnarray}
 \label{eq:Vsymmetry}
 \mathcal{V}^T (1-z) &=&  \mathcal{V}^T(z)
\\
 \label{eq:Nsymmetry}
 \mathcal{N}^T (1-z) &=&  -\mathcal{N}^T(z)
\\
 \label{eq:Ssymmetry}
 \mathcal{S}^T (1-z) &=&  \mathcal{S}^T(z)
\\
 \label{eq:Msymmetry}
 \mathcal{M}^T (1-z) &=&  \mathcal{M}^T(z).
 \end{eqnarray}
We will use these conditions to deduce the correct sign for the contributions of diagrams related to each other by the exchange of the quark and antiquark.

Before moving on, let us note that for the purposes of Fourier-transforming the wavefunction to mixed space (which is done after mass renormalization) and for calculating the total DIS cross section, it is convenient to change the basis yet again. In fact, the Fourier-transform of both the $\mathcal{S}^T$ contribution and the $\mathcal{M}^T + \mathcal{V}^T $ one are  more easily calculated if one splits the symmetric tensor $\Pt^i\Pt^j$ into its trace and a traceless part. This additional reorganization changes the decomposition \eqref{eq:NLOformfactorsv2} into the form
\begin{equation}
\label{eq:NLOformfactorsv3}
\begin{split}
\Psi^{\gamma^{\ast}_T\rightarrow q\bar{q}}_{\rm LO}+\Psi^{\gamma^{\ast}_T\rightarrow q\bar{q}}_{\rm NLO} & = \delta_{\alpha_0\alpha_1}\frac{ee_f}{\ed_{\lo}} \frac{q^+}{2k_0^+k_1^+}\Biggl \{
\Biggl[
\left (\frac{k_0^+-k_1^+}{q^+} \right ) \left [1 + \left (\frac{\alpha_s\cf}{2\pi}\right ) \mathcal{V}^T  \right ] + \left (\frac{\alpha_s\cf}{2\pi}\right )\mathcal{N}^T
\Biggr]
\Pt \cdot \epst_\lambda
\bar u(0)\gamma^+ v(1) 
\\ &
+
\frac{1}{2} \Biggl[1 + \left (\frac{\alpha_s\cf}{2\pi}\right ) \mathcal{V}^T  \Biggr ]
\Pt^{i}\epst^j_\lambda
\bar{u}(0)\gamma^{+}[\gamma^i, \gamma^j]v(1) 
\\ &
+
m
\Biggl[ 
\left (\frac{\alpha_s\cf}{2\pi}\right )\mathcal{S}^T
\Biggr]
\left (\frac{\Pt^i\Pt^j}{\Pt^2} -\frac{\delta^{ij}}{2}\right)
\epst^i_\lambda 
\bar u(0)\gamma^+ \gamma^j  v(1)
\\ &
-
m \Biggl[1 +
 \left (\frac{\alpha_s\cf}{2\pi}\right )\biggl [\mathcal{M}^T + \mathcal{V}^T  -\frac{\mathcal{S}^T}{2}\biggr ]
\Biggr]
\epst^i_\lambda
\bar u(0)\gamma^+ \gamma^i  v(1)
\Biggr\}.
\end{split}
\end{equation}
It will also turn out that the expression for the combination $\mathcal{M}^T + \mathcal{V}^T  -\mathcal{S}^T/2$ will be simpler than the terms separately.
 The spinor matrix elements of this basis $\bar u(0)\gamma^+ v(1),$  $ \bar{u}(0)\gamma^{+}[\gamma^i, \gamma^j]v(1) $ and $\bar u(0)\gamma^+ \gamma^i  v(1)$ do not depend on the transverse momentum $\Pt$ and can be factorized from the Fourier-transform.
The result of the Fourier-transform of this decomposition is the one already given in the accompanying shorter paper \cite{Beuf:2021srj}.


\subsection{Diagrams, energy denominators and kinematics}

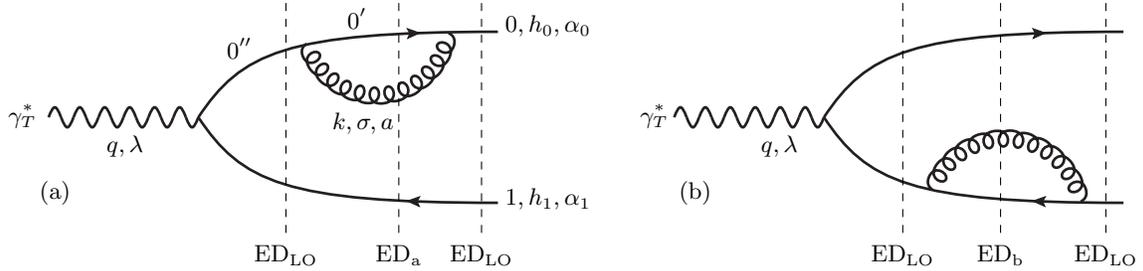
\begin{figure}[tbh!]
\centerline{
\includegraphics[width=6.4cm]{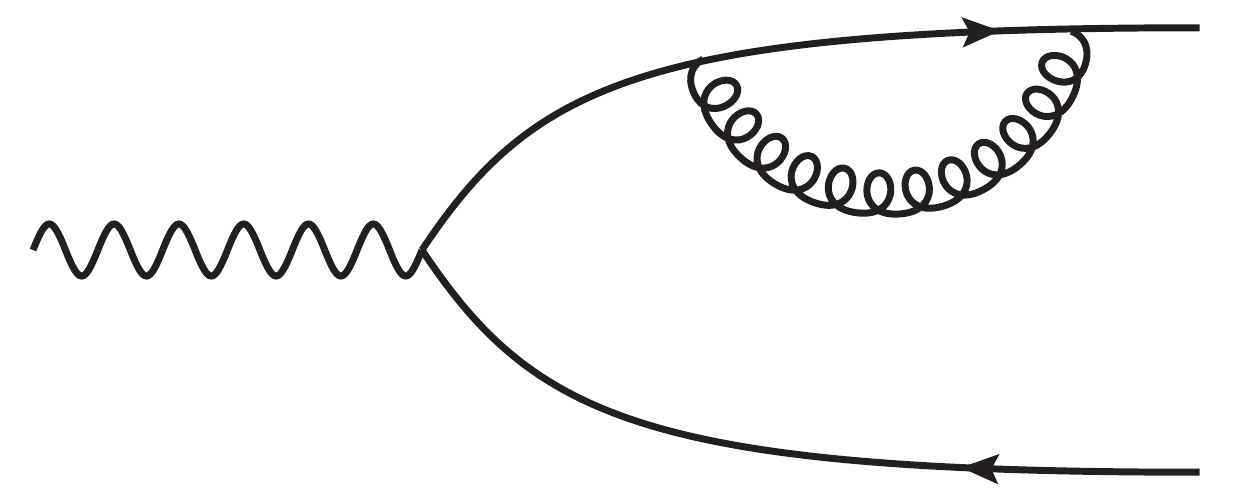}
\hspace*{1.7cm}
\includegraphics[width=6.4cm]{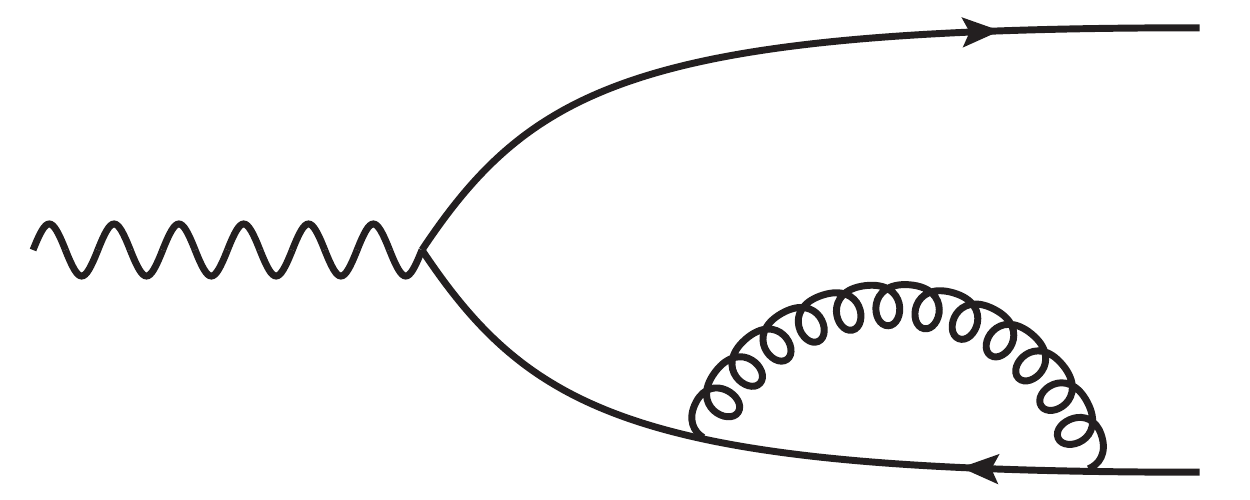}
\begin{tikzpicture}[overlay]
\draw [dashed] (-11.5,2.8) -- (-11.5,-0.3);
\node[anchor=north] at (-11.5cm,-0.3cm) {$\ed_{\lo}$};
\draw [dashed] (-10.0,2.8) -- (-10.0,-0.3);
\node[anchor=north] at (-10.0cm,-0.3cm) {$\ed_{\rm a}$};
\draw [dashed] (-8.9,2.8) -- (-8.9,-0.3);
\node[anchor=north] at (-8.9cm,-0.3cm) {$\ed_{\lo}$};
\node[anchor=east] at (-14.7,1.3) {$\gamma^{\ast}_{T}$};
\node[anchor=west] at (-14.0,0.9) {$q, \lambda $};
\node[anchor=west] at (-8.7,2.5) {$ 0,h_0,\alpha_0$};
\node[anchor=west] at (-8.7,0.2) {$1,h_1,\alpha_1$};
\node[anchor=west] at (-11.0,1.2) {$k,\sigma,a$};
\node[anchor=west] at (-10.8,2.6) {$0'$};
\node[anchor=west] at (-12.4,2.2) {$0''$};
\draw [dashed] (-3.3,2.8) -- (-3.3,-0.3);
\node[anchor=north] at (-3.3cm,-0.3cm) {$\ed_{\lo}$};
\draw [dashed] (-2.0,2.8) -- (-2.0,-0.3);
\node[anchor=north] at (-2.0cm,-0.3cm) {$\ed_{\rm b}$};
\draw [dashed] (-0.6,2.8) -- (-0.6,-0.3);
\node[anchor=north] at (-0.6cm,-0.3cm) {$\ed_{\lo}$};
\node[anchor=west] at (-5.3,0.9) {$q, \lambda $};
\node[anchor=east] at (-6.3,1.3) {$\gamma^{\ast}_{T}$};
\node[anchor=south west] at (-15.0cm,0cm) {\namediag{diag:oneloopSEUPT}};
\node[anchor=south west] at (-6.5cm,0cm) {\namediag{diag:oneloopSEDOWNT}};
 \end{tikzpicture}
}
\rule{0pt}{1ex}
\caption{Time ordered one-loop quark self-energy diagrams \ref{diag:oneloopSEUPT} and \ref{diag:oneloopSEDOWNT} contributing to the transverse virtual photon LCWF at NLO. Diagram \ref{diag:oneloopSEUPT}: Imposing  plus and transverse momentum conservation at each vertex gives: $\qvec = \kvec_{0''} + \kvec_1$, $\kvec_{0''} = \kvec_{0'} + \kvec$ and $\kvec_{0'}+ \kvec = \kvec_0$.} 
\label{fig:selfenergyT}
 \end{figure}

\begin{figure}[tbh!]
\centerline{
\includegraphics[width=6.4cm]{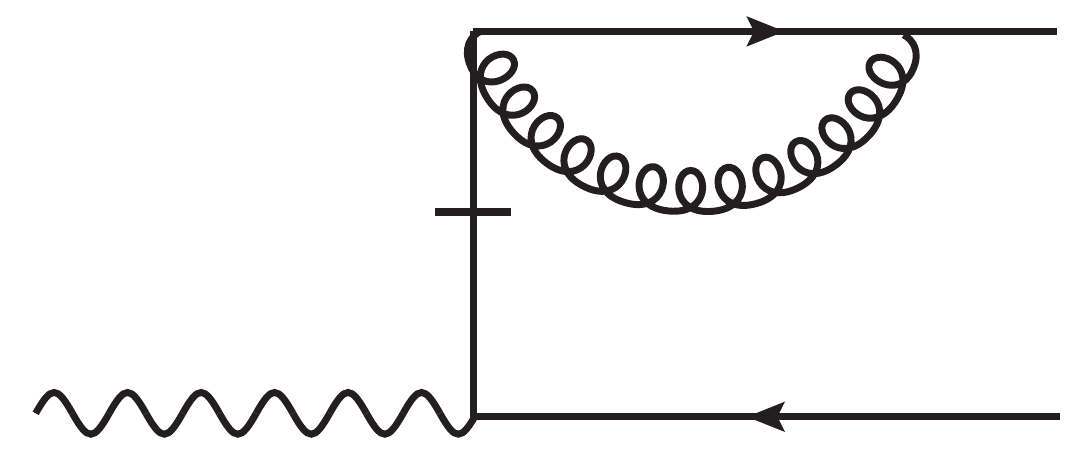}
\hspace*{1.7cm}
\includegraphics[width=6.4cm]{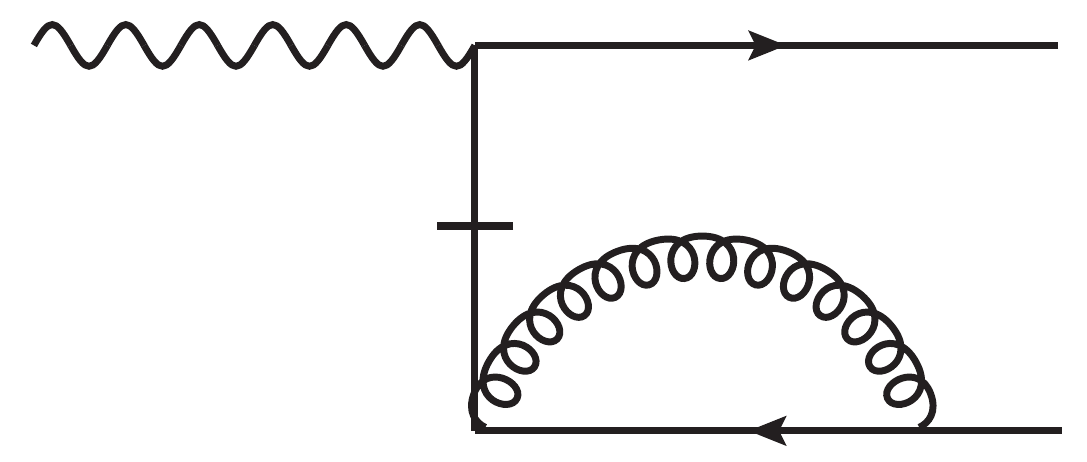}
\begin{tikzpicture}[overlay]
\draw [dashed] (-10.6,2.8) -- (-10.6,-0.3);
\node[anchor=north] at (-10.6cm,-0.3cm) {$\ed_{\rm a}$};
\draw [dashed] (-8.9,2.8) -- (-8.9,-0.3);
\node[anchor=north] at (-8.9cm,-0.3cm) {$\ed_{\lo}$};
\node[anchor=east] at (-14.7,0.4) {$\gamma^{\ast}_{T}$};
\node[anchor=west] at (-14.0,-0.1) {$q, \lambda $};
\node[anchor=west] at (-8.6,2.5) {$0,h_0,\alpha_0$};
\node[anchor=west] at (-8.6,0.2) {$1,h_1,\alpha_1$};
\node[anchor=west] at (-11.6,1.2) {$k,\sigma,a$};
\node[anchor=west] at (-11.3,2.8) {$0'$};
\draw [dashed] (-2.3,2.8) -- (-2.3,-0.3);
\node[anchor=north] at (-2.3cm,-0.3cm) {$\ed_{\rm b}$};
\draw [dashed] (-0.6,2.8) -- (-0.6,-0.3);
\node[anchor=north] at (-0.6cm,-0.3cm) {$\ed_{\lo}$};
\node[anchor=west] at (-5.3,2.1) {$q, \lambda$};
\node[anchor=east] at (-6.3,2.4) {$\gamma^{\ast}_{T}$};
\node[anchor=south west] at (-15.0cm,1.4cm) {\namediag{diag:oneloopSEUPTinst}};
\node[anchor=south west] at (-6.5cm,.4cm) {\namediag{diag:oneloopSEDOWNTinst}};
 \end{tikzpicture}
}
\rule{0pt}{1ex}
\caption{Time ordered one-loop instantaneous self-energy diagrams \ref{diag:oneloopSEUPTinst} and \ref{diag:oneloopSEDOWNTinst} contributing to the transverse virtual photon LCWF at NLO. Diagram \ref{diag:oneloopSEUPTinst}: Imposing  plus and transverse momentum conservation at each vertex gives: $\qvec = \kvec_{0} + \kvec_1$ and $\kvec_{0'} + \kvec = \kvec_0$.} 
\label{fig:selfenergyTinst}
 \end{figure}

\begin{figure}[tbh!]
\centerline{
\includegraphics[width=6.4cm]{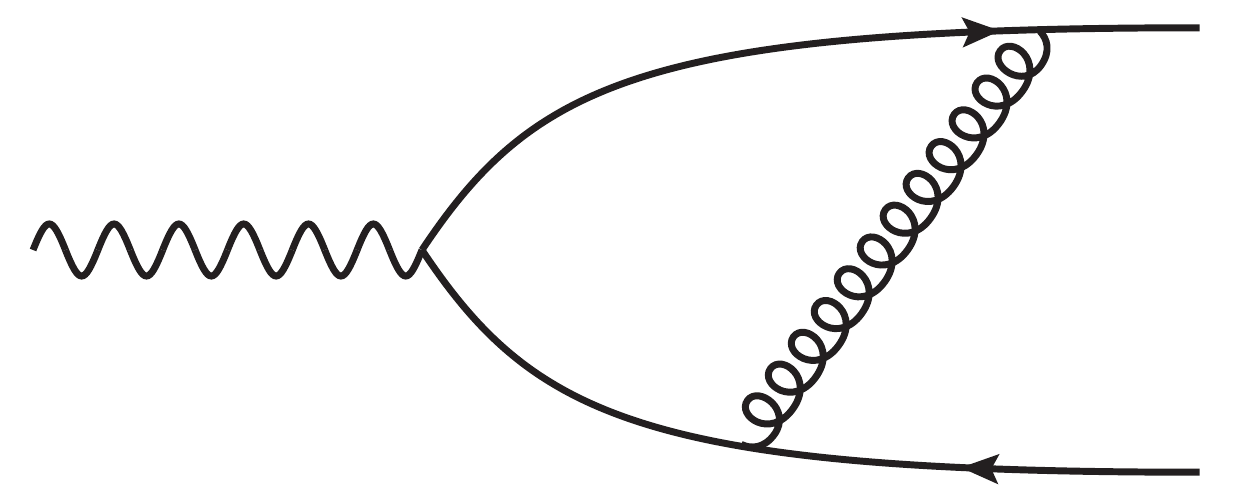}
\hspace*{1.7cm}
\includegraphics[width=6.4cm]{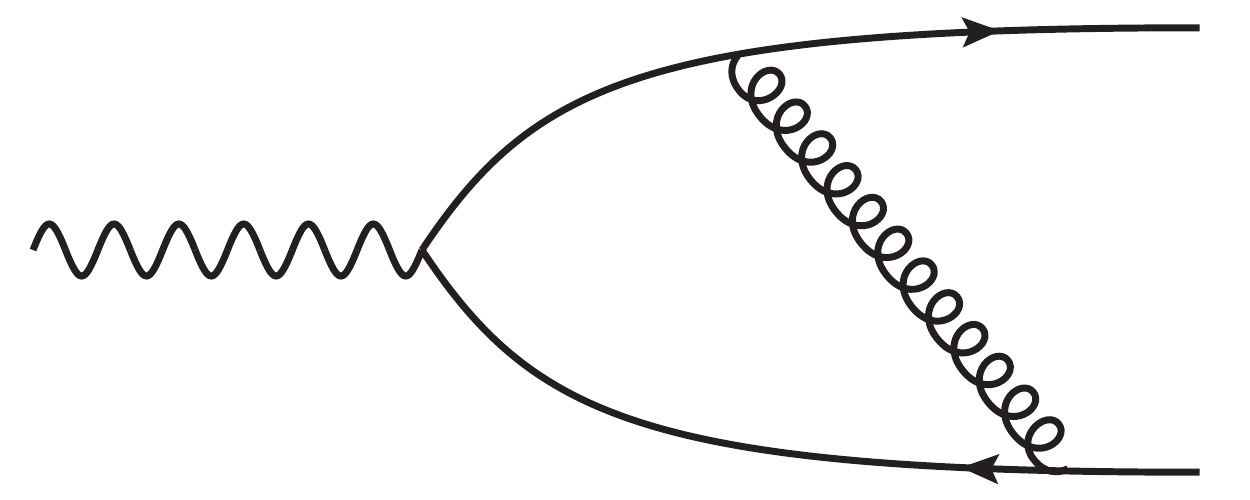}
\begin{tikzpicture}[overlay]
\draw [dashed] (-11.7,2.8) -- (-11.7,-0.3);
\node[anchor=north] at (-11.7cm,-0.3cm) {$\ed_{\rm v}$};
\draw [dashed] (-10.0,2.8) -- (-10.0,-0.3);
\node[anchor=north] at (-10.0cm,-0.3cm) {$\ed_{\rm a}$};
\draw [dashed] (-8.9,2.8) -- (-8.9,-0.3);
\node[anchor=north] at (-8.9cm,-0.3cm) {$\ed_{\lo}$};
\node[anchor=east] at (-14.7,1.3) {$\gamma^{\ast}_{T}$};
\node[anchor=west] at (-14.0,0.9) {$q, \lambda$};
\node[anchor=west] at (-8.7,2.5) {$0,h_0,\alpha_0$};
\node[anchor=west] at (-8.7,0.2) {$1,h_1,\alpha_1$};
\node[anchor=west] at (-10.0,1.2) {$k,\sigma,a$};
\node[anchor=east] at (-11.9,2.4) {$0'$};
\node[anchor=east] at (-11.9,0.3) {$1'$};
\draw [dashed] (-3.3,2.8) -- (-3.3,-0.3);
\node[anchor=north] at (-3.3cm,-0.3cm) {$\ed_{\rm v}$};
\draw [dashed] (-2.0,2.8) -- (-2.0,-0.3);
\node[anchor=north] at (-2.0cm,-0.3cm) {$\ed_{\rm b}$};
\draw [dashed] (-0.6,2.8) -- (-0.6,-0.3);
\node[anchor=north] at (-0.6cm,-0.3cm) {$\ed_{\lo}$};
\node[anchor=west] at (-5.3,0.9) {$q, \lambda$};
\node[anchor=east] at (-6.3,1.3) {$\gamma^{\ast}_{T}$};
\node[anchor=south west] at (-15.0cm,0cm) {\namediag{diag:vertexqbaremT}};
\node[anchor=south west] at (-6.5cm,0cm) {\namediag{diag:vertexqemT}};
 \end{tikzpicture}
}
\rule{0pt}{1ex}
\caption{Time ordered one-loop vertex diagram \ref{diag:vertexqbaremT} and \ref{diag:vertexqemT} and contributing to the transverse virtual photon LCWF at NLO. Diagram \ref{diag:vertexqbaremT}: Imposing plus and transverse momentum conservation at each vertex gives: $\qvec = \kvec_{0'} + \kvec_{1'} $, $\kvec_{0'}  + \kvec = \kvec_{0} $ and $\kvec_{1'}  = \kvec + \kvec_{1} $.}
\label{fig:vertexT}
\end{figure}
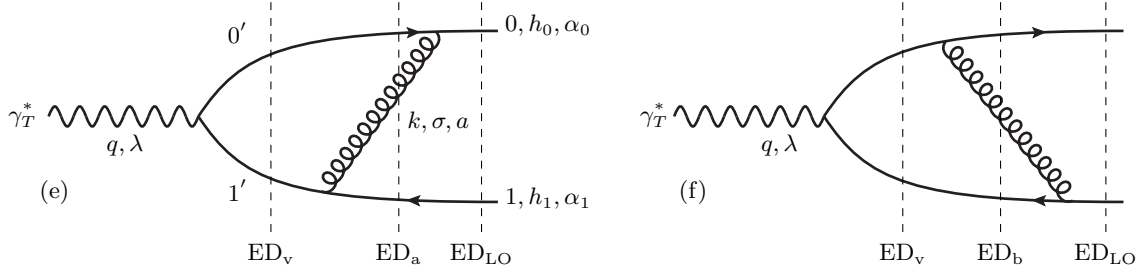

\begin{figure}[tbh!]
\centerline{
\includegraphics[width=6.4cm]{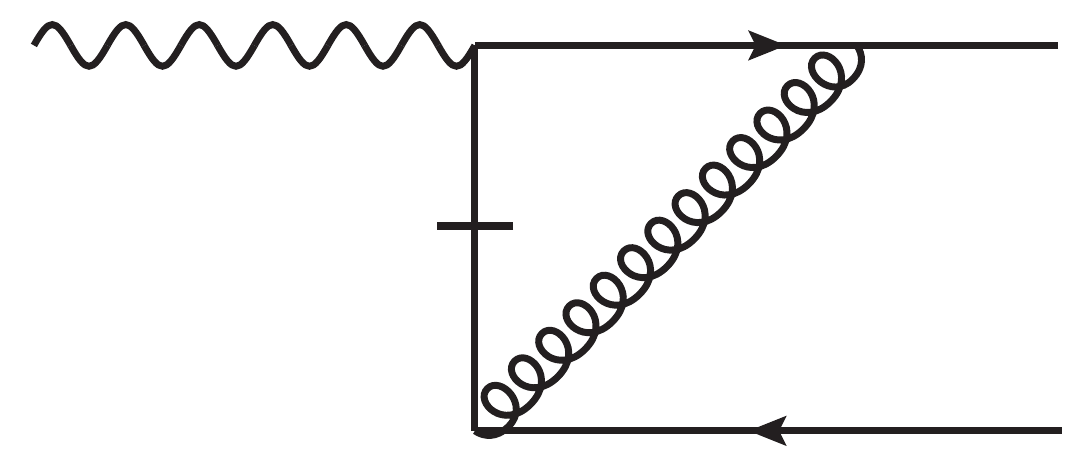}
\hspace*{1.7cm}
\includegraphics[width=6.4cm]{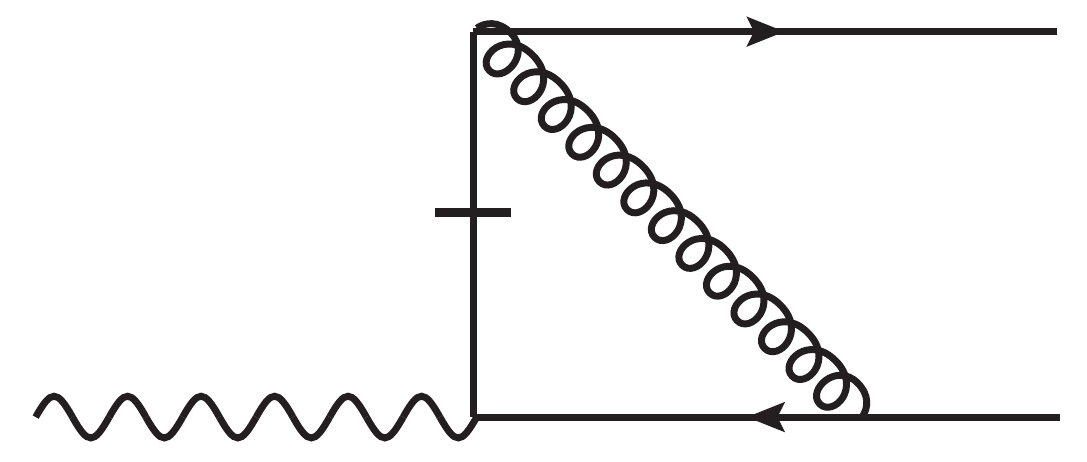}
\begin{tikzpicture}[overlay]
\draw [dashed] (-11.2,2.8) -- (-11.2,-0.3);
\node[anchor=north] at (-11.1cm,-0.3cm) {$\ed_{\rm a}$};
\draw [dashed] (-8.9,2.8) -- (-8.9,-0.3);
\node[anchor=north] at (-8.9cm,-0.3cm) {$\ed_{\lo}$};
\node[anchor=east] at (-14.7,2.5) {$\gamma^{\ast}_{T}$};
\node[anchor=west] at (-14.0,2.1) {$q, \lambda $};
\node[anchor=west] at (-8.6,2.5) {$0,h_0,\alpha_0$};
\node[anchor=west] at (-8.6,0.2) {$1,h_1,\alpha_1$};
\node[anchor=west] at (-10.6,1.2) {$k,\sigma,a$};
\node[anchor=west] at (-11.1,2.7) {$0'$};
\draw [dashed] (-2.9,2.8) -- (-2.9,-0.3);
\node[anchor=north] at (-2.9cm,-0.3cm) {$\ed_{\rm b}$};
\draw [dashed] (-0.6,2.8) -- (-0.6,-0.3);
\node[anchor=north] at (-0.6cm,-0.3cm) {$\ed_{\lo}$};
\node[anchor=west] at (-5.3,-0.1) {$q, \lambda $};
\node[anchor=east] at (-6.3,0.4) {$\gamma^{\ast}_{T}$};
\node[anchor=south west] at (-15.0cm,0.4cm) {\namediag{diag:oneloopvertexinst1}};
\node[anchor=south west] at (-6.5cm,1.4cm) {\namediag{diag:oneloopvertexinst2}};
 \end{tikzpicture}
}
\rule{0pt}{1ex}
\caption{Time ordered one-loop instantaneous vertex diagrams \ref{diag:oneloopvertexinst1} and \ref{diag:oneloopvertexinst2} contributing to the transverse virtual photon LCWF at NLO. Diagram \ref{diag:oneloopvertexinst1}: Imposing plus and transverse momentum conservation at each vertex gives: $\qvec = \kvec_{0} + \kvec_1$, $\kvec_{0'} + \kvec = \kvec_{0}$.} 
\label{fig:instvertex}
 \end{figure}

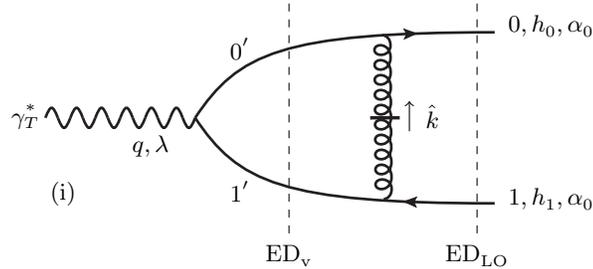
\begin{figure}[tbh!]
\centerline{
\includegraphics[width=6.4cm]{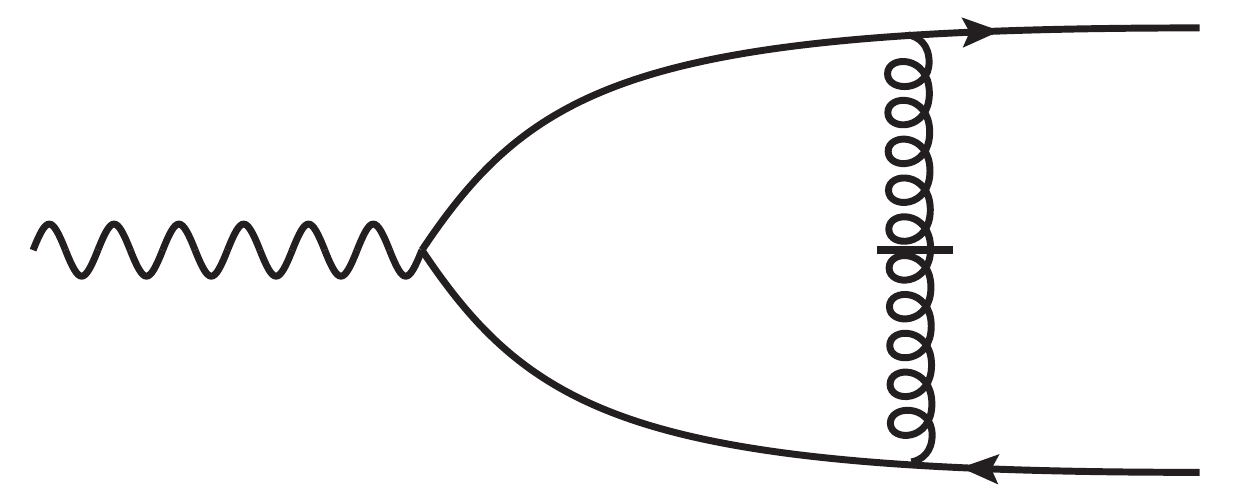}
\begin{tikzpicture}[overlay]
\draw [dashed] (-3.1,2.8) -- (-3.1,-0.3);
\node[anchor=north] at (-3.1cm,-0.3cm) {$\ed_{\rm v}$};
\draw [dashed] (-0.6,2.8) -- (-0.6,-0.3);
\node[anchor=north] at (-0.6cm,-0.3cm) {$\ed_{\lo}$};
\node[anchor=west] at (-0.3,2.5) {$0,h_0,\alpha_0$};
\node[anchor=west] at (-4.0,2.2) {$0'$};
\node[anchor=west] at (-4.0,0.3) {$1'$};
\node[anchor=west] at (-0.3,0.2) {$1,h_1,\alpha_0$};
\node[anchor=west] at (-5.3,0.9) {$q, \lambda $};
\node[anchor=west] at (-1.4,1.3) {$\kvec$};
\draw[->] (-1.5,1.1)--(-1.5,1.5);
\node[anchor=east] at (-6.3,1.3) {$\gamma^{\ast}_{T}$};
\node[anchor=south west] at (-6.5cm,0cm) {\namediag{diag:gluoninst}};
 \end{tikzpicture}
}
\rule{0pt}{1ex}
\caption{Time ordered one-loop instantaneous diagram \ref{diag:gluoninst} contributing to the transverse virtual photon LCWF at NLO. We only calculate this in the kinematics $k_0^+>k_{0'}^+,$ the rest is restored by symmetry,  Imposing plus and transverse momentum conservation gives: $\qvec = \kvec_{0'} + \kvec_{1'} $ and $\kvec_{0'} + \kvec = \kvec_{0}$, where $\kvec$ is the momentum flowing up to the quark }
\label{fig:gluoninst}
\end{figure}

The LCPT diagrams relevant for the calculation of the $\gamma^{\ast}_T\rightarrow q\bar{q}$ LCWF at NLO with massive quarks are shown in \figs\ref{fig:selfenergyT}, \ref{fig:selfenergyTinst}, \ref{fig:vertexT}, \ref{fig:instvertex} and \ref{fig:gluoninst}. There are four propagator correction diagrams, \ref{diag:oneloopSEUPT} and \ref{diag:oneloopSEDOWNT} in \fig\ref{fig:selfenergyT} and \ref{diag:oneloopSEUPTinst} and \ref{diag:oneloopSEDOWNTinst} in \fig\ref{fig:selfenergyTinst}, with a gluon loop attached to the quark or the antiquark. Then, in \figs\ref{fig:vertexT} and \ref{fig:instvertex}, there are four vertex correction diagrams \ref{diag:vertexqbaremT}, \ref{diag:vertexqemT}, \ref{diag:oneloopvertexinst1} and \ref{diag:oneloopvertexinst2}, corresponding to two different kinematical possibilities, with longitudinal momentum (which is always positive) flowing either up from the antiquark to the quark or vice versa. Finally, in \fig\ref{fig:gluoninst} there is an instantaneous gluon exchange \ref{diag:gluoninst} between the quark and the antiquark, in which the gluon momentum can either flow upwards into the quark or downwards into the antiquark. It is convenient to split up the contribution from this diagram into terms that are combined with  one or the other of the diagrams in \fig\ref{fig:vertexT} according to the direction of the momentum flow. Due to the symmetry of the kinematics by exchange of the quark and the antiquark between the two classes of graphs, only the calculation of half of the diagrams is necessary; in this case we will calculate the ones labeled \ref{diag:oneloopSEUPT},  \ref{diag:oneloopSEUPTinst}, \ref{diag:vertexqbaremT}, \ref{diag:oneloopvertexinst1} and the part of \ref{diag:gluoninst} where the momentum  flows to the quark as in \ref{diag:vertexqbaremT}.
 
The energy denominators appearing in the diagrams are the same as in the longitudinal photon case \cite{Beuf:2021qqa}. Following the notation introduced in \figs\ref{fig:selfenergyT} and \ref{fig:vertexT}, and imposing the plus and the transverse momentum conservation we find
\begin{equation}
\label{eq:EDa}
\begin{split}
\ed_{\rm a} &= q^{-} - \biggl [k^{-}_{0'} + k^ {-} + k^{-}_1\biggr ] = \frac{\qt^2 -Q^2}{2q^+} - \Biggl [\frac{\kt^2_{0'} + m^2}{2k^{+}_{0'}} + \frac{\kt^2}{2k^+} + \frac{\kt^2_1+m^2}{2k^{+}_1}\biggr ]\\
& = \frac{-k^+_0}{2k^+(k^+_0-k^+)}\Biggl [\left (\kt - \frac{k^+}{k^+_0}\kt_0\right )^2 + \frac{k^+(k^+_0-k^+)q^+}{(k^+_0)^2k^+_1}\Biggl \{\Pt^2 + \overline{Q}^2 + \frac{k^+_0(q^+-k^+)}{q^+(k^+_0-k^+)}m^2\Biggr \}\Biggr ].
\end{split}
\end{equation}
and
\begin{equation}
\label{eq:EDv}
\begin{split}
\ed_{\rm v}  &= q^{-} - \biggl [k^{-}_{0'} + k^{-}_{1'}\biggr ] = \frac{\qt^2 -Q^2}{2q^+} - \biggl [\frac{\kt^2_{0'} + m^2}{2k^{+}_{0'}} + \frac{\kt^2_{1'} + m^2}{2k^{+}_{1'}} \biggr ]\\ 
& = \frac{-q^+}{2(k^+_0-k^+)(k^++k^+_1)}\Biggl [\left (\left (\kt - \frac{k^+}{k^+_0}\kt_0\right ) + \Lt\right )^2 + \frac{(k^+_0-k^+)(k^++k^+_1)}{k^+_0k^+_1}\overline{Q}^2 + m^2 \Biggr ],
\end{split}
\end{equation}
where $\Lt = -(k^+_0-k^+)\Pt/k^+_0$. It is also convenient to introduce the change of variables $(\kt,k^+)\mapsto (\Kt,\xi)$ by parameterizing the transverse momentum integration in
the loop by the relative momentum $\Kt$ of the gluon with respect to the quark after the loop as 
\begin{equation}
\label{eq:KTdef}
\Kt = \kt - \frac{k^+}{k^{+}_0}\kt_0 = \kt - \xi\kt_0 \quad \text{with} \quad k^+ = \xi zq^+.
\end{equation}
In these notations the energy denominators in \eqs\nr{eq:EDa} and \nr{eq:EDv} can be cast into the following form
\begin{equation}
\label{eq:EDdef}
\begin{split}
\ed_{\rm a} & =  \frac{1}{(-2q^+)z\xi(1-\xi)}\biggl [\Kt^2 + \Delta_1 \biggr ],\\
\ed_{\rm v} & = \frac{1}{(-2q^+)z(1-\xi)(1-z(1-\xi))}\biggl [\left (\Kt + \Lt\right )^2 + \Delta_2 \biggr ],\\
\end{split}
\end{equation}
where the momentum variable $\Lt$ and the coefficients $\Delta_1, \Delta_2$ are given by:
\begin{equation}
\label{eq:delta1and2}
\begin{split}
\Lt & = -(1-\xi)\Pt,\\
\Delta_1 & = \frac{\xi(1-\xi)}{(1-z)}\biggl (\Pt^2 + \overline{Q}^2 + m^2\biggr ) + \xi^2m^2,\\
\Delta_2 & = (1-\xi)\left (1 + \frac{\xi z}{(1-z)}\right )\overline{Q}^2  + m^2.
\end{split}
\end{equation}

In the following subsections, we present the detailed computation of the NLO form factors corresponding to the one-loop self-energy diagrams, one-loop instantaneous diagrams and one-loop vertex diagrams.   

\section{Calculation of the loop diagrams}
\label{sec:oneloopdiags}

\subsection{One-loop quark self-energy contribution} 
\label{selfenergycontribution}

In this section we compute the contribution to the $\gamma^{\ast}_T\rightarrow q\bar{q}$ LCWF in \eq\nr{NLOformfactors} from the massive quark self-energy diagrams \ref{diag:oneloopSEUPT} and \ref{diag:oneloopSEDOWNT} shown in \fig\ref{fig:selfenergyT}. Applying the diagrammatic LCPT rules formulated in momentum space yields the expression
\begin{equation}
\label{eq:diagaTLCWF}
\begin{split}
\Psi^{\gamma^{\ast}_T\rightarrow q\bar{q}}_{\ref{diag:oneloopSEUPT}} & =  \int \dk \int \dkpzero \int \dkppzero (2\pi)^{D-1}\delta^{(D-1)}(\bar k_{0'} + \bar k - \bar k_0)(2\pi)^{D-1}\delta^{(D-1)}(\bar k_{0''} - \bar k_{0'} -  \bar k)\frac{N^T_{\ref{diag:oneloopSEUPT}}}{(\ed_{\lo})^2\ed_{\rm a}}\\
& = \frac{1}{16\pi}\int_{0}^{k^+_0}\frac{\ud k^+}{k^+k^+_0(k^+_0-k^+)}\int\frac{\ud^{D-2}\kt}{(2\pi)^{D-2}}\frac{N^T_{\ref{diag:oneloopSEUPT}}}{(\ed_{\lo})^2\ed_{\rm a}},
\end{split}
\end{equation}
where the energy denominators are written down in \eqs\nr{eq:LOED} and \nr{eq:EDa}, and the spinor structure coming from the vertices (note that the summation is implicit over the internal helicities, gluon polarization and color) is given by the numerator
\begin{equation}
\label{eq:numa}
N^T_{\ref{diag:oneloopSEUPT}} = +ee_fg^2t^{a}_{\alpha_0\alpha_{0'}}t^{a}_{\alpha_{0'}\alpha_1} \biggl [\bar{u}(0)\epsl_{\sigma}(k)u(0')\biggr ]\biggl [\bar{u}(0')\epsl^{\ast}_{\sigma}(k)u(0'')\biggr ]\biggl [\bar{u}(0'')\epsl_{\lambda}(q)v(1)\biggr ].
\end{equation}
By making the change of variables $(\kt,k^+)\mapsto (\Kt,\xi)$ (see \eq\nr{eq:KTdef}) and regulating the small $k^+\rightarrow 0$ (or $\xi \rightarrow 0)$ divergences by an explicit cutoff $k^+>\alpha q^+$ (or $\xi > \alpha/z$) with the dimensionless parameter $\alpha>0$, we can simplify the expression in \eq\nr{eq:diagaTLCWF} to    
\begin{equation}
\label{eq:lcwfaint}
\begin{split}
\Psi^{\gamma^{\ast}_T\rightarrow q\bar{q}}_{\ref{diag:oneloopSEUPT}} & = \frac{-1}{8\pi q^+(\ed_{\lo})^2}\int_{\alpha/z}^{1}\frac{\ud \xi}{z}\int\frac{\ud^{D-2}\Kt}{(2\pi)^{D-2}}\frac{N^T_{\ref{diag:oneloopSEUPT}}}{[\Kt^2 + \Delta_1]},
\end{split}
\end{equation}
where the detailed calculation of the numerator in \eq\nr{eq:numa}, performed in Appendix~\ref{app:nums}, gives 
\begin{equation}
\label{eq:numav2}
N^T_{\ref{diag:oneloopSEUPT}} = \frac{2g^2\cf}{\xi^2(1-\xi)}\delta_{\alpha_0\alpha_1}V^{\gamma^{\ast}_T \rightarrow q\bar{q}}_{h_0;h_1}\Biggl \{\biggl [1 + (1-\xi)^2\biggr ]\Kt^2  + m^2\xi^4 + \frac{(D_s-4)}{2}\xi^2\biggl [\Kt^2 + m^2\xi^2\biggr ] \Biggr \}.
\end{equation}
Similarly as in the longitudinal case, we note that from this expression one obtains both the FDH scheme result by taking the limit $D_s \rightarrow 4$, and the CDR one by setting $D_s = D$.  

Following the same steps as presented in~\cite{Beuf:2021qqa}, we find
\begin{equation}
\label{eq:finalURSEa}
\begin{split}
\Psi^{\gamma^{\ast}_T\rightarrow q\bar{q}}_{\ref{diag:oneloopSEUPT}}  = \Psi^{\gamma^{\ast}_T\rightarrow q\bar{q}}_{\lo}\left (\frac{\alpha_s\cf}{2\pi} \right ) \Biggl \{-\int_{\alpha/z}^{1}\frac{\ud \xi}{\xi}\biggl [1 + (1-\xi)^2\biggr ]\mathcal{A}_0(\Delta_1) &- \frac{2(1-z)m^2}{[\Pt^2 + \overline{Q}^2 + m^2]}\int_{0}^{1}\ud \xi \mathcal{A}_0(\Delta_1)\\
& - \frac{(D_s-4)}{2}\int_{0}^{1}\ud \xi  \xi \mathcal{A}_0(\Delta_1)\Biggr \},
\end{split}
\end{equation}
where the elementary scalar integral $\mathcal{A}_0$ is defined in Ref.~\cite{Beuf:2016wdz} (see Appendices D and E there):
\begin{equation}
\mathcal{A}_0(\Delta_1) = 4\pi(\mu^2)^{2-D/2}\int\frac{\ud^{D-2}\Kt}{(2\pi)^{D-2}}\frac{1}{[\Kt^2 + \Delta_1]}.
\end{equation}
We now perform the renormalization of the kinetic mass precisely as in our earlier calculation for the longitudinal photon~\cite{Beuf:2021qqa} (see Sec. V~C). In the end, this amounts to a mass subtraction that makes the leading singularity at the on shell point $\Pt^2 + \overline{Q}^2+m^2\to 0$, i.e. $\Delta_1^2\to \xi^2m^2$ (see \eq\eqref{eq:delta1and2}) cancel. Note that the denominator $\Pt^2 + \overline{Q}^2+m^2$ in the second term of \eq\eqref{eq:finalURSEa} is precisely the kind of doubling of the energy denominator in 
$\Psi^{\gamma^{\ast}_T\rightarrow q\bar{q}}_{\lo}$ that results from a Taylor series development in powers of the mass counterterm. Thus the effect of mass renormalization is to replace  $\mathcal{A}_0(\Delta_1)$ in the second term of \eq\eqref{eq:finalURSEa} by $\mathcal{A}_0(\Delta_1)-\mathcal{A}_0(\xi^2m^2)$. 
With some abuse of notation we will continue denoting the mass-renormalized wavefunction with the same notation $ \Psi^{\gamma^{\ast}_T\rightarrow q\bar{q}}_{\ref{diag:oneloopSEUPT}} $.
Together with the explicit expression~\cite{Beuf:2016wdz} 
\begin{equation}
\label{eq:A0exp}
\mathcal{A}_0(\Delta_1) = \Gamma\left (2 - \frac{D}{2}\right ) \biggl [\frac{\Delta_1}{4\pi\mu^2}\biggr ]^{\frac{D}{2}-2} =  \frac{(4\pi)^{2-\frac{D}{2}}}{(2-\frac{D}{2})}\Gamma\left (3-\frac{D}{2}\right ) - \log\left (\frac{\Delta_1}{\mu^2}\right ) + \mathcal{O}(D-4),
\end{equation}
we can now evaluate the wavefunction as
\begin{equation}
\label{eq:finalRESEafinal}
\begin{split}
\Psi^{\gamma^{\ast}_T\rightarrow q\bar{q}}_{\ref{diag:oneloopSEUPT}}  = \Psi^{\gamma^{\ast}_T\rightarrow q\bar{q}}_{\lo}\left (\frac{\alpha_s\cf}{2\pi} \right )\mathcal{V}^{T}_{\ref{diag:oneloopSEUPT}},
\end{split}
\end{equation}
where the form factor $\mathcal{V}^{T}_{\ref{diag:oneloopSEUPT}}$ is given by 
\begin{equation}
\label{eq:Va}
\begin{split}
\mathcal{V}^{T}_{\ref{diag:oneloopSEUPT}} = \biggl [\frac{3}{2} &+ 2\log\left (\frac{\alpha}{z}\right )\biggr ]\Biggl \{ \frac{(4\pi)^{2-\frac{D}{2}}}{(2-\frac{D}{2})}\Gamma\left (3-\frac{D}{2}\right )  + \log\left (\frac{(1-z)\mu^2}{\overline{Q}^2 + m^2}\right ) - \log\left (\frac{\Pt^2 + \overline{Q}^2 + m^2}{\overline{Q}^2 + m^2}\right )\Biggr \}\\
&- \log^2\left (\frac{\alpha}{z}\right ) -\frac{\pi^2}{3} + 3 + \frac{1}{2}\frac{(D_s-4)}{(D-4)} + \mathcal{I}_{\mathcal{V}_{\ref{diag:oneloopSEUPT}}} + O(D-4).
\end{split}
\end{equation}
The factor $(D_s-4)/2(D-4)$ is the regularization scheme dependent coefficient coming from the following integral
\begin{equation}
\label{eq:Kscheme}
- \frac{(D_s-4)}{2}\int_{0}^{1}\ud \xi \, \xi A_0(\Delta_1) = \frac{1}{2}\frac{(D_s-4)}{(D-4)} + O(D-4),
\end{equation}
and we have defined the function $\mathcal{I}_{\mathcal{V}_{\ref{diag:oneloopSEUPT}}}$ as
\begin{equation}\label{eq:defIVa}
\mathcal{I}_{\mathcal{V}_{\ref{diag:oneloopSEUPT}}} = (\Pt^2 + \overline{Q}^2 + m^2)\int_{0}^{1}\frac{\ud \xi}{\xi} \left (-\frac{2\log(\xi)}{(1-\xi)} + \frac{(1+\xi)}{2}\right )\Biggl \{\frac{1}{\Pt^2 + \overline{Q}^2 + m^2}- \frac{1}{\Pt^2 + \overline{Q}^2 + m^2 + \frac{(1-z)\xi}{(1-\xi)}m^2}\Biggr \}.
\end{equation}
Note that this integral over $\xi \in [0,1]$ is finite. 

The mass renormalized LCWF for the quark self-energy diagram in diagram \ref{diag:oneloopSEDOWNT} shown in  \fig\ref{fig:selfenergyT} can be now easily obtained by using the symmetry between the diagrams \ref{diag:oneloopSEUPT} and \ref{diag:oneloopSEDOWNT}. As discussed in Sec.~\ref{sec:spinstuctureandmrenorm}, this implies that the contribution to the form factor $\mathcal{V}^{T}$ from the antiquark propagator correction diagram~\ref{diag:oneloopSEDOWNT}  is obtained  from the diagram~\ref{diag:oneloopSEUPT} by the substitution
$z \mapsto 1-z$. Using this symmetry we get
\begin{equation}
\label{eq:finalRESEbfinal}
\begin{split}
\Psi^{\gamma^{\ast}_T\rightarrow q\bar{q}}_{\ref{diag:oneloopSEDOWNT}}  = \Psi^{\gamma^{\ast}_T\rightarrow q\bar{q}}_{\lo}\left (\frac{\alpha_s\cf}{2\pi} \right )\mathcal{V}^{T}_{\ref{diag:oneloopSEDOWNT}},
\end{split}
\end{equation}
where from \eq\eqref{eq:Va} one obtains
\begin{equation}
\label{eq:Vb}
\begin{split}
\mathcal{V}^{T}_{\ref{diag:oneloopSEDOWNT}} = \biggl [\frac{3}{2} & + 2\log\left (\frac{\alpha}{1-z}\right )\biggr ]\Biggl \{ \frac{(4\pi)^{2-\frac{D}{2}}}{(2-\frac{D}{2})}\Gamma\left (3-\frac{D}{2}\right )   + \log\left (\frac{z \mu^2}{\overline{Q}^2 + m^2}\right )  - \log\left (\frac{\Pt^2 + \overline{Q}^2 + m^2}{\overline{Q}^2 + m^2}\right )\Biggr \}\\
& - \log^2\left (\frac{\alpha}{1-z}\right ) -\frac{\pi^2}{3} + 3 + \frac{1}{2}\frac{(D_s-4)}{(D-4)}  + \mathcal{I}_{\mathcal{V}_{\ref{diag:oneloopSEDOWNT}}} + O(D-4)
\end{split}
\end{equation}
with 
\begin{equation}
\mathcal{I}_{\mathcal{V}_{\ref{diag:oneloopSEDOWNT}}} = (\Pt^2 + \overline{Q}^2 + m^2)\int_{0}^{1}\frac{\ud \xi}{\xi} \left (-\frac{2\log(\xi)}{(1-\xi)} + \frac{(1+\xi)}{2}\right )\Biggl \{\frac{1}{\Pt^2 + \overline{Q}^2 + m^2}- \frac{1}{\Pt^2 + \overline{Q}^2 + m^2 + \frac{z\xi}{(1-\xi)}m^2}\Biggr \}.
\end{equation}

Summing the expressions in \eqs\nr{eq:finalRESEafinal} and \nr{eq:finalRESEbfinal}, we obtain for the full contribution of the one-loop quark self-energy to the $\gamma^{\ast}_{T}\rightarrow q\bar q$ LCWF the result 
\begin{equation}
\label{eq:finalSE}
\begin{split}
\Psi^{\gamma^{\ast}_T\rightarrow q\bar{q}}_{\ref{diag:oneloopSEUPT} + \ref{diag:oneloopSEDOWNT}}  = \Psi^{\gamma^{\ast}_T\rightarrow q\bar{q}}_{\lo}\left (\frac{\alpha_s\cf}{2\pi} \right )\mathcal{V}^{T}_{\ref{diag:oneloopSEUPT} + \ref{diag:oneloopSEDOWNT}},
\end{split}
\end{equation}
where the NLO form factor $\mathcal{V}^{T}_{\ref{diag:oneloopSEUPT} + \ref{diag:oneloopSEDOWNT}}$ can be written as 
\begin{equation}
\label{eq:Vselfenergy}
\begin{split}
\mathcal{V}^{T}_{\ref{diag:oneloopSEUPT} + \ref{diag:oneloopSEDOWNT}}  =   2\biggl [\frac{3}{2}  &+ \log\left (\frac{\alpha}{z}\right ) + \log\left (\frac{\alpha}{1-z}\right )\biggr ]\Biggl \{  \frac{(4\pi)^{2-\frac{D}{2}}}{(2-\frac{D}{2})}\Gamma\left (3-\frac{D}{2}\right )   + \log\left (\frac{\mu^2}{\overline{Q}^2 + m^2}\right )  - \log\left (\frac{\Pt^2 + \overline{Q}^2 + m^2}{\overline{Q}^2 + m^2}\right )\Biggr \} \\
& + \biggl [\log(z) + \log(1-z)  \biggr ]\left (\frac{3}{2} + 2\log(\alpha) \right ) - 4\log(z)\log(1-z) - \log^2\left (\frac{\alpha}{z}\right ) - \log^2\left (\frac{\alpha}{1-z}\right )\\
& -\frac{2\pi^2}{3} + 6 + \frac{(D_s-4)}{(D-4)} + \mathcal{I}_{\mathcal{V}_{\ref{diag:oneloopSEUPT}}} +  \mathcal{I}_{\mathcal{V}_{\ref{diag:oneloopSEDOWNT}}} + O(D-4).
\end{split}
\end{equation}

\subsection{One-loop instantaneous contribution} 
\label{inscorrection}

Let us then calculate the one-loop instantaneous diagrams shown in \figs\ref{fig:selfenergyTinst}, \ref{fig:instvertex} and \ref{fig:gluoninst}. We start by writing down the expression for the diagram  \ref{diag:oneloopSEUPTinst} in \fig\ref{fig:selfenergyTinst}. Applying the diagrammatic LCPT rules formulated in momentum space yields the expression
\begin{equation}
\label{eq:diagc}
\begin{split}
\Psi^{\gamma^{\ast}_T\rightarrow q\bar{q}}_{\ref{diag:oneloopSEUPTinst}} & =  \int \dk \int \dkpzero (2\pi)^{D-1}\delta^{(D-1)}(\hat k_{0'} + \hat k - \hat k_0)\frac{1}{\ed_{\lo}\ed_{\rm a}} \frac{N^T_{\ref{diag:oneloopSEUPTinst}}}{2(k_{0'}^+ + k^+)}\\
& = \frac{1}{16\pi}\int_{0}^{k^+_0}\frac{\ud k^+}{k^+k^+_0(k^+_0-k^+)}\int\frac{\ud^{D-2}\kt}{(2\pi)^{D-2}}\frac{N^T_{\ref{diag:oneloopSEUPTinst}}}{\ed_{\lo}\ed_{\rm a}},
\end{split}
\end{equation}
where the spinor structure coming from the vertices is given by the numerator
\begin{equation}
\label{eq:numc}
N^T_{\ref{diag:oneloopSEUPTinst}} = ee_fg^2t^{a}_{\alpha_0\alpha_{0'}}t^{a}_{\alpha_{0'}\alpha_1} \biggl [\bar{u}(0)\epsl_{\sigma}(k)u(0')\biggr ]\biggl [\bar{u}(0')\epsl^{\ast}_{\sigma}(k) \gamma^+ \epsl_\lambda(q)v(1)\biggr ].
\end{equation}
After making the change of variables $(\kt,k^+)\mapsto (\Kt,\xi)$, we can simplify the expression in \eq\nr{eq:diagc} to    
\begin{equation}
\label{eq:diagc2}
\begin{split}
\Psi^{\gamma^{\ast}_T\rightarrow q\bar{q}}_{\ref{diag:oneloopSEUPTinst}} & = \frac{-1}{8\pi q^+ \ed_{\lo}}\int_{\alpha/z}^{1}\frac{\ud \xi}{z}\int\frac{\ud^{D-2}\Kt}{(2\pi)^{D-2}}\frac{N^T_{\ref{diag:oneloopSEUPTinst}}}{[\Kt^2 + \Delta_1]},
\end{split}
\end{equation}
where the numerator \eq\nr{eq:numc} in these variables simplifies (see Appendix~\ref{app:nums}) to
\begin{equation}
\label{eq:numcv2}
N^T_{\ref{diag:oneloopSEUPTinst}} = -ee_f g^2\cf \delta_{\alpha_0\alpha_1}\Biggl \{\biggl [-\frac{2}{\xi} + (D_s-2) \biggr ]\bar u(0)\gamma^+\gamma^i  \epsl_\lambda(q)v(1) \Kt^i + \xi(D_s-2) m\bar u(0)\gamma^+ \epsl_\lambda(q)v(1) \biggr ] \Biggr \}.
\end{equation}
Noting that the term proportional to $\Kt^i$ vanishes upon integration over $\Kt$, we then obtain
\begin{equation}
\label{eq:diagc3}
\begin{split}
\Psi^{\gamma^{\ast}_T\rightarrow q\bar{q}}_{\ref{diag:oneloopSEUPTinst}} & = +ee_f\frac{g^2\cf}{32\pi^2 q^+ \ed_{\lo}}\delta_{\alpha_0\alpha_1} (D_s-2)m\bar u(0)\gamma^+ \epsl_\lambda(q)v(1) \int_{\alpha/z}^{1}\frac{\ud \xi \xi}{z} A_0(\Delta_1).
\end{split}
\end{equation}

Next, we compute the diagram \ref{diag:oneloopvertexinst1} in \fig\ref{fig:instvertex}. The expression in momentum space reads 
\begin{equation}
\label{eq:diagg}
\begin{split}
\Psi^{\gamma^{\ast}_T\rightarrow q\bar{q}}_{\ref{diag:oneloopvertexinst1} } & =  \int \dk \int \dkpzero (2\pi)^{D-1}\delta^{(D-1)}(\hat k_{0'} + \hat k - \hat k_0)\frac{1}{\ed_{\lo}\ed_{\rm a}} \frac{N^T_{\ref{diag:oneloopvertexinst1} }}{2(k_1^+ + k^+)}\\
& = \frac{1}{16\pi}\int_{0}^{k^+_0}\frac{\ud k^+}{k^+(k^+_0-k^+)(k_1^+ + k^+)}\int\frac{\ud^{D-2}\kt}{(2\pi)^{D-2}}\frac{N^T_{\ref{diag:oneloopvertexinst1} }}{\ed_{\lo}\ed_{\rm a}},
\end{split}
\end{equation}
where the spinor structure coming from the vertices is given by the numerator
\begin{equation}
\label{eq:numg}
N^T_{\ref{diag:oneloopvertexinst1} } = -ee_fg^2t^{a}_{\alpha_0\alpha_{0'}}t^{a}_{\alpha_{0'}\alpha_1} \biggl [\bar{u}(0)\epsl_{\sigma}(k)u(0')\biggr ]\biggl [\bar{u}(0')  \epsl_\lambda(q)\gamma^+\epsl^{\ast}_{\sigma}(k)v(1)\biggr ].
\end{equation}
In the $(\Kt,\xi)$ variables, we can simplify the expression in \eq\nr{eq:diagg} to    
\begin{equation}
\label{eq:diag2}
\begin{split}
\Psi^{\gamma^{\ast}_T\rightarrow q\bar{q}}_{\ref{diag:oneloopvertexinst1} } & = \frac{-1}{8\pi q^+ \ed_{\lo}}\int_{\alpha/z}^{1}\frac{\ud \xi}{[1-z+z\xi]}\int\frac{\ud^{D-2}\Kt}{(2\pi)^{D-2}}\frac{N^T_{\ref{diag:oneloopSEUPTinst}}}{[\Kt^2 + \Delta_1]},
\end{split}
\end{equation}
where the numerator in \eq\nr{eq:numg} (see Appendix~\ref{app:nums}) simplifies to 
\begin{equation}
\label{eq:numgv2}
N^T_{\ref{diag:oneloopvertexinst1} } = -ee_fg^2\cf \delta_{\alpha_0\alpha_1}\Biggl \{\biggl (\frac{2}{\xi} \bar u(0)\gamma^+\epsl_\lambda(q)  \gamma^i v(1)  + (D_s -4) \bar u(0)\gamma^+\gamma^i\epsl_\lambda v(1) \biggr )\Kt^i 
 + \xi(D_s-4)m\bar u(0)\gamma^+\epsl_\lambda(q)v(1)\Biggr \}.
\end{equation}
Again, the term proportional to $\Kt^i$ vanishes, and we are left with the following expression
\begin{equation}
\label{eq:diagg3}
\begin{split}
\Psi^{\gamma^{\ast}_T\rightarrow q\bar{q}}_{\ref{diag:oneloopvertexinst1}} & = +ee_f\frac{g^2\cf}{32\pi^2 q^+ \ed_{\lo}}\delta_{\alpha_0\alpha_1} (D_s-4)m\bar u(0)\gamma^+ \epsl_\lambda(q)v(1) \int_{\alpha/z}^{1}\frac{\ud \xi \xi}{[1-z+z\xi]} A_0(\Delta_1).
\end{split}
\end{equation}

Finally, we focus on the diagram \ref{diag:gluoninst} in \fig\ref{fig:gluoninst}. We calculate this in the kinematics $k_0^+>k_{0'}^+$ and denote this contribution as $\ref{diag:gluoninst}_1$. The corresponding expression in momentum space reads 
\begin{equation}
\label{eq:diagi1}
\begin{split}
\Psi^{\gamma^{\ast}_T\rightarrow q\bar{q}}_{\ref{diag:gluoninst}_1} & = \int \dkpzero  \int \dkpone (2\pi)^{D-1}\delta^{(D-1)}(\hat k_{1'} + \hat k_{0'} - \hat q)\frac{1}{\ed_{\lo}\ed_{\rm v}} \frac{N^T_{\ref{diag:gluoninst}_1 }}{(k_{0'}^+ - k_0^+)^2}\\
& = \frac{1}{8\pi}\int_{0}^{k^+_0}\frac{\ud k_{0'}^+}{k_{0'}^+(q^+ - k_{0'}^+)(k_{0'}^+ - k_0^+)^2}\int\frac{\ud^{D-2}\ktpzero}{(2\pi)^{D-2}}\frac{N^T_{\ref{diag:gluoninst}_1 }}{\ed_{\lo}\ed_{\rm v}},
\end{split}
\end{equation}
where the energy denominator $\ed_{\rm v}$ is defined in \eq\nr{eq:EDv} and the spinor structure coming from the vertices is given by the numerator
\begin{equation}
\label{eq:numi1}
N^T_{\ref{diag:gluoninst}_1} = -ee_fg^2t^{a}_{\alpha_0\alpha_{0'}}t^{a}_{\alpha_{0'}\alpha_1} \biggl [\bar{u}(0)\gamma^+u(0')\biggr ]\biggl [\bar{u}(0')\epsl_\lambda(q) v(1') \biggr ]\biggl [\bar{v}(1')\gamma^+v(1)\biggr ].
\end{equation}
By the change of variables $(\ktpzero,k^+_{0'})\mapsto (\Kt,\xi)$ we can simplify the expression in \eq\nr{eq:diagi1} to
\begin{equation}
\label{eq:diagi12}
\begin{split}
\Psi^{\gamma^{\ast}_T\rightarrow q\bar{q}}_{\ref{diag:gluoninst}_1} & = \frac{-1}{4\pi (q^+)^2 \ed_{\lo}}\int_{\alpha/z}^{1}\frac{\ud \xi }{z\xi^2}\int\frac{\ud^{D-2}\Kt}{(2\pi)^{D-2}}\frac{N^T_{\ref{diag:gluoninst}_1}}{[\left (\Kt + \Lt\right )^2 + \Delta_2]},
\end{split}
\end{equation}
where the numerator in \eq\nr{eq:numi1} (see Appendix~\ref{app:nums}) simplifies to 
\begin{equation}
\label{eq:numi2}
\begin{split}
N^T_{\ref{diag:gluoninst}_1} 
 = -ee_fg^2 \cf \delta_{\alpha_0\alpha_1} \biggl \{\biggl (4q^+[1-z(1-\xi)] \epst_\lambda^i \bar{u}(0)\gamma^+ v(1)  & + 2q^+ \bar{u}(0)\epsl_\lambda(q)\gamma^+\gamma^i v(1)\biggr ) \biggl [\Kt^i + \Lt^i\biggr ]\\
& + 2q^+ m\bar{u}(0)\gamma^+ \epsl_\lambda(q) v(1) \biggr \}.
\end{split}
\end{equation}
Now, since the term proportional to $(\Kt + \Lt)^i$ vanishes by symmetry due to the integration in $\Kt$, we obtain 
\begin{equation}
\label{eq:diagi13}
\begin{split}
\Psi^{\gamma^{\ast}_T\rightarrow q\bar{q}}_{\ref{diag:gluoninst}_1} & = +ee_f\frac{g^2\cf}{8\pi^2 q^+ \ed_{\lo}}\delta_{\alpha_0\alpha_1} m\bar{u}(0)\gamma^+ \epsl_\lambda(q) v(1)\int_{\alpha/z}^{1}\frac{\ud \xi }{z\xi^2}A_0(\Delta_2).
\end{split}
\end{equation}

By comparing the expressions in \eqs\nr{eq:diagc3}, \nr{eq:diagg3} and \nr{eq:diagi13}  to \eq\nr{NLOformfactors}, we observe that the instantaneous diagrams contribute only to the form factor $\mathcal{M}^T$ as: 
\begin{equation}
\label{eq:MTc}
\begin{split}
\mathcal{M}^T_{\ref{diag:oneloopSEUPTinst}} & = \frac{(D_s-2)(1-z)}{2}\int_{0}^{1} \ud \xi \xi A_0(\Delta_1)\\
& = -\frac{(1-z)}{2}\frac{(D_s-4)}{(D-4)} + (1-z)\int_{0}^{1} \ud \xi \xi A_0(\Delta_1) + O(D-4),   
\end{split}
\end{equation}
\begin{equation}
\label{eq:MTg}
\begin{split}
\mathcal{M}^T_{\ref{diag:oneloopvertexinst1}} & = \frac{(D_s-4)(1-z)}{2}\int_{0}^{1} \ud \xi \frac{z\xi}{[1-z+z\xi]} A_0(\Delta_1)\\
& = -\frac{(D_s-4)}{(D-4)}\biggl \{(1-z) + \frac{(1-z)^2}{z}\log(1-z)\biggr \} + O(D-4),
\end{split}
\end{equation}
and
\begin{equation}
\label{eq:MTi1}
\mathcal{M}^T_{\ref{diag:gluoninst}_1}  = 2(1-z)\int_{\alpha/z}^{1} \frac{\ud \xi}{\xi^2} A_0(\Delta_2). 
\end{equation}
To simplify the expressions in \eqs\nr{eq:MTc} and \nr{eq:MTg} we utilized the relation $-\frac{(D-4)}{2}A_0(\Delta_1) = 1 + O(D-4)$.

The corresponding contributions from diagrams \ref{diag:oneloopSEDOWNTinst}, \ref{diag:oneloopvertexinst2} and $\ref{diag:gluoninst}_2$ can be obtained from \eqs\nr{eq:MTc}, \nr{eq:MTg} and \nr{eq:MTi1} by symmetry, as $z \leftrightarrow 1-z$.

\subsection{One-loop vertex contribution} 
\label{vertexcontribution}

The only diagrams left to calculate for the full $\gamma^{\ast}_T\rightarrow q\bar{q}$ LCWF at NLO are the non-trivial vertex correction diagrams \ref{diag:vertexqbaremT} and \ref{diag:vertexqemT} shown in \fig\ref{fig:vertexT}.  These two diagrams give contributions to all four form factors introduced in the decomposition \eq\nr{NLOformfactors}. Since the contributions coming from the diagrams \ref{diag:vertexqbaremT} and \ref{diag:vertexqemT} are symmetric to each other by exchange of the quark and antiquark, we need to explicitly calculate only one of them, for example the diagram \ref{diag:vertexqbaremT}.

For diagram \ref{diag:vertexqbaremT}, the momentum space expression of the one-loop LCWF can be written as
\begin{equation}
\label{eq:diageTLCWF}
\begin{split}
\Psi^{\gamma^{\ast}_T\rightarrow q\bar{q}}_{\ref{diag:vertexqbaremT}} & =  \int \dk \int \dkpzero \int \dkpone (2\pi)^{D-1}\delta^{(D-1)}(\bar k_{0'} + \bar k - \bar k_0)(2\pi)^{D-1}\delta^{(D-1)}(\bar k_{1'} - \bar k -  \bar k_1)\frac{N^T_{\ref{diag:vertexqbaremT}}}{\ed_{\rm v}\ed_{\rm a}\ed_{\lo}}\\
& = \frac{1}{16\pi}\int_{0}^{k^+_0}\frac{\ud k^+}{k^+(k^+_0 - k^+)(k^+ + k^+_1)}\int\frac{\ud^{D-2}\kt}{(2\pi)^{D-2}}\frac{N^T_{\ref{diag:vertexqbaremT}}}{\ed_{\rm v}\ed_{\rm a}\ed_{\lo}},
\end{split}
\end{equation}
where the numerator (again the summation is implicit over the internal helicities, gluon polarization and color) is given by 
\begin{equation}
\label{eq:nume}
N^T_{\ref{diag:vertexqbaremT}} = -ee_fg^2\delta_{\alpha_0\alpha_1}\cf\biggl [\bar{u}(0)\epsl_{\sigma}(k)u(0')\biggr ]\biggl [\bar{u}(0')\epsl_{\lambda}(q)v(1')\biggr ]\biggl [\bar{v}(1')\epsl^{\ast}_{\sigma}(k)v(1)\biggr ].
\end{equation}
Applying again the change of variables $\kt \mapsto \Kt$, we obtain
\begin{equation}
\label{eq:vertexcv2}
\begin{split}
\Psi^{\gamma^{\ast}_T\rightarrow q\bar{q}}_{\ref{diag:vertexqbaremT}} & =  \frac{1}{4\pi q^+}\frac{1}{\ed_{\lo}} \int_{0}^{k^+_0} \ud k^+ \left (1 - \frac{k^+}{k_0^+} \right )
\int\frac{\ud^{D-2}\Kt}{(2\pi)^{D-2}}\frac{N^T_{\ref{diag:vertexqbaremT}}}{[\Kt^2 + \Delta_1][\left (\Kt + \Lt\right )^2 + \Delta_2]}.
\end{split}
\end{equation}

Before starting to calculate this contribution, it is useful to make a small power counting argument for the transverse momentum integrals in relation to the quark masses and light cone helicities. One is performing a $(D-2)$-dimensional integral with a total denominator $\sim \Kt^4$ at large $\Kt$ from the energy denominators. Therefore one can have UV-divergences in the integral if the numerator, coming from a product of three elementary quark-gauge boson vertices, is proportional to at least $\Kt^2$. From the decomposition of the elementary vertex~\nr{eq:LOvertexdecomp} we see that light cone helicity conserving terms are proportional to the momentum $\Kt^i$, whereas light cone helicity flip terms are proportional to one power of the mass  $\sim m$. We will thus have two kinds of UV-divergent contributions, proportional to $m^0$ or to $m^1$. The first ones behave just like the corresponding diagrams in the massless case. The second ones, proportional to the mass, will contribute to the renormalization of the quark (vertex) mass, which appears linearly in the leading order vertex. There will also be 4 kinds of finite contributions, proportional to $m^n,$ $n=0,1,2,3$. The number of powers of mass corresponds to the number of light cone helicity flips following the quark line. Thus $n=0$ and $n=2$ contributions are nonflip contributions for the overall diagram and can interfere with each other. Similarly $n=1$ and $n=3$ have an overall helicity flip between the outgoing quark and antiquark, and can interfere with each other. Only even powers of the mass up to $m^4$ will appear in the NLO cross section. Quark masses regulate the energy denominators, so there are no IR divergences in the transverse integrals. 

From a technical point of view, the calculation of the numerator and the corresponding tensor integrals in \eq\nr{eq:vertexcv2} turns out to be a quite tedious task. Therefore, the details of the intermediate steps are documented in Appendix \ref{app:numc}, keeping the main text as readable as possible. Collecting the results from Appendix \ref{app:numc}, we find the following result
\begin{equation}
\label{eq:vertexeresult}
\begin{split}
\Psi^{\gamma^{\ast}_T\rightarrow q\bar{q}}_{\ref{diag:vertexqbaremT}}  = \frac{ee_f}{\ed_{\lo}}\delta_{\alpha_0\alpha_1}&\left (\frac{\alpha_s\cf}{2\pi}\right )\biggl \{\bar u(0) \epsl_{\lambda}(q)v(1) \mathcal{V}^T_{\ref{diag:vertexqbaremT}} + \left ( \frac{2k^{+}_0k^{+}_1}{q^+} \right ) \Pt \cdot \epst_\lambda \bar u(0)\gamma^{+} v(1)\mathcal{N}^T_{\ref{diag:vertexqbaremT}}  \\
& +  \left ( \frac{2k^{+}_0k^{+}_1}{q^+} \right )\frac{\Pt\cdot \epst_{\lambda}}{\Pt^2}\Pt^j m\bar u(0)\gamma^{+}\gamma^{j}v(1) \mathcal{S}^T_{\ref{diag:vertexqbaremT}}    +  \left ( \frac{2k^{+}_0k^{+}_1}{q^+} \right ) m\bar u(0)\gamma^{+}\epsl_{\lambda}(q)v(1) \mathcal{M}^T_{\ref{diag:vertexqbaremT}}   \biggr \},
\end{split}
\end{equation}
where the form factors can be cast into the form:
\begin{equation}
\label{eq:VT}
\begin{split}
\mathcal{V}^T_{\ref{diag:vertexqbaremT}}   = & -\frac{z}{2}\frac{(D_s-4)}{(D-4)} - z + \int_{\alpha/z}^{1} \ud \xi  \biggl [\frac{1}{\xi} - 2z + z\xi \biggr ]\Biggl \{\mathcal{A}_0(\Delta_2) + \frac{\left (\Pt^2 + \overline{Q}^2 + m^2\right )}{\Pt^2} \mathcal{T}_{-} \Biggr \}\\
& +  \int_{\alpha/z}^{1} \ud \xi \biggl [\frac{2}{\xi} + \frac{z\xi}{(1-z)} \biggr ]\mathcal{T}_{+} + zm^2  \int_{0}^{1} \ud \xi (1-\xi)\xi \Biggl \{\frac{(1-2z)}{(1-z)}\frac{\Pt^j \mathcal{B}^j}{\Pt^2} + \xi \mathcal{B}_0 \Biggr \}  + O(D-4) 
\end{split}
\end{equation}
\begin{equation}
\label{eq:NT}
\begin{split}
\mathcal{N}^T_{\ref{diag:vertexqbaremT}}   =&  -z(1-z)\frac{(D_s-4)}{(D-4)} + z(2z-1) + 2z(1-z)\int_{0}^{1} \ud\xi (\xi-2) \Biggl \{\mathcal{A}_0(\Delta_2) +  \frac{\left (\Pt^2 + \overline{Q}^2 + m^2 \right )}{\Pt^2}\mathcal{T}_{-}\Biggr \}\\
& + 2z \int_{0}^{1} \ud\xi (1-\xi)\biggl [2 + \frac{(2z-1)\xi}{(1-z)} \biggr ]\Biggl \{\frac{(\Pt^2 + m^2)}{\Pt^2}\Pt^j \mathcal{B}^j - (1-\xi)\left (\Pt^2 + \overline{Q}^2 + m^2\right )\mathcal{B}_0 \Biggr \}\\
& - 2zm^2 \int_{0}^{1} \ud\xi (1-\xi) \biggl [1 + 2(1-z)(1-\xi) \biggr ]\Biggl \{\frac{\Pt^j \mathcal{B}^j}{\Pt^2} + \xi \mathcal{B}_0 \Biggr \} + O(D-4) 
\end{split}
\end{equation}
\begin{equation}
\label{eq:ST}
\begin{split}
\mathcal{S}^T_{\ref{diag:vertexqbaremT}}   = -z +  2z\int_{0}^{1} \ud\xi (1-\xi) \Biggl \{-\frac{(\Pt^2 + \overline{Q}^2 + m^2)}{\Pt^2}\mathcal{T}_{-} & + \frac{\xi(1-2z)(\Pt^2 + m^2)}{(1-z)\Pt^2}\Pt^j \mathcal{B}^j - \Pt^j \mathcal{B}^j\\
& - \xi(1-\xi)\Pt^2 \mathcal{B}_0 + \xi^2m^2 \mathcal{B}_0  \Biggr \}  + O(D-4) 
\end{split}
\end{equation}
and 
\begin{equation}
\label{eq:MT}
\begin{split}
\mathcal{M}^T_{\ref{diag:vertexqbaremT}}   = & \frac{(D_s-4)}{(D-4)}\biggl [(1-z) + \frac{(1-z)^2}{z}\log(1-z) \biggr ] + (1-z)\int_{\alpha/z}^{1} \ud \xi  \biggl [-\frac{2}{\xi^2}\mathcal{A}_0(\Delta_2) + \mathcal{A}_0(\Delta_1)\biggr ]\\
& +  \int_{\alpha/z}^{1} \ud \xi \biggl [-\frac{1}{\xi} + z \biggr ]\frac{\left (\Pt^2 + \overline{Q}^2 + m^2 \right )}{\Pt^2}\mathcal{T}_{-} - \frac{z^2}{(1-z)}\left (\Pt^2 + \overline{Q}^2 + m^2 \right ) \int_{0}^{1} \ud \xi (1-\xi)^2 \mathcal{B}_0 + O(D-4).
\end{split}
\end{equation}
In the above expressions the following compact notation has been introduced
\begin{equation}
\label{eq:Tminusandplus}
\begin{split}
\mathcal{T}_{-}  & \equiv  -(1-\xi) \biggl [\Pt^j \mathcal{B}^j + \frac{\xi \Pt^2 \mathcal{B}_0}{(1-z)} \biggr ],\\
\mathcal{T}_{+}  & \equiv +(1-\xi) \biggl [(1-\xi)\left (\Pt^2 + \overline{Q}^2 + m^2\right )\mathcal{B}_0 - \Pt^j \mathcal{B}^j + (1-z)\xi m^2 \mathcal{B}_0 \biggr ],
\end{split} 
\end{equation}
where the elementary integrals $\mathcal{B}_0$ and $\mathcal{B}^j$ are the ones defined in Ref.~\cite{Beuf:2016wdz} (see Appendices D and E there).

Diagram \ref{diag:vertexqemT} is obtained from \ref{diag:vertexqbaremT} by exchanging the quark and the antiquark. Thus,  as discussed in \ref{sec:spinstuctureandmrenorm}, we can obtain its contribution to the form factors with the  substitution $z \to 1-z$ and changing the sign  for $\mathcal{N}^T$, so that the symmetry requirements  
\eqs\eqref{eq:Vsymmetry}, \eqref{eq:Nsymmetry}, \eqref{eq:Ssymmetry} and \eqref{eq:Msymmetry} are recovered.

\section{Full momentum space LCWF}
\label{sec:momspacefull}

\subsection{Form factor $\mathcal{V}^T$}

The full form factor $\mathcal{V}^T$ gets contributions from the self-energy diagrams \ref{diag:oneloopSEUPT} and \ref{diag:oneloopSEDOWNT} and from the non-trivial vertex correction diagrams \ref{diag:vertexqbaremT} and \ref{diag:vertexqemT}. Consequently, by symmetry, one has
\begin{equation}
\label{eq:fullVT}
\mathcal{V}^T = \mathcal{V}^{T}_{\ref{diag:oneloopSEUPT}} + \mathcal{V}^T_{\ref{diag:vertexqbaremT}} + \biggl [z \leftrightarrow 1-z \biggr ],
\end{equation}
where the mass renormalized expression for the $\mathcal{V}^{T}_{\ref{diag:oneloopSEUPT}}$ contribution is given in \eq\nr{eq:Va}. To obtain the full $\mathcal{V}^T$ contribution in momentum space, we shall next evaluate further the individual terms appearing in \eq\nr{eq:VT} for $\mathcal{V}^T_{\ref{diag:vertexqbaremT}}$. We follow the same strategy that was used in Ref.~\cite{Beuf:2021qqa}, and study the UV divergent and UV finite parts separately.

\subsubsection{UV divergent part}

To evaluate the UV divergent part in \eq\nr{eq:VT}, we first remind that the UV divergent function $\mathcal{A}_0(\Delta_2)$ can be written as 
\begin{equation}
\label{eq:A0delta2}
\mathcal{A}_0(\Delta_2) = \frac{(4\pi)^{2-\frac{D}{2}}}{(2-\frac{D}{2})}\Gamma\left (3 - \frac{D}{2}\right ) - \log\left (\frac{\Delta_2}{\mu^2} \right ) + \mathcal{O}(D-4).   
\end{equation}
Before performing the integration over the $\xi$, we rewrite the coefficient $\Delta_2$ as 
\begin{equation}
\label{eq:delta2mod}
\Delta_2 = \frac{z}{(1-z)}\overline{Q}^2\left (\xi_{(+)} - \xi\right )\left (\xi - \xi_{(-)}\right ),
\end{equation}
where the zeroes in $\xi$ are given by
\begin{equation}
\label{eq:xizeros}
\xi_{(\pm)} =   1 - \frac{1}{2z}\left (1 \pm \gamma\right ),\quad  \text{with}\quad \gamma = \sqrt{1 + \frac{4m^2}{Q^2}}.
\end{equation}
Substituting \eq\nr{eq:delta2mod} into \eq\nr{eq:A0delta2} and then performing the remaining $\xi$-integral analytically gives the following result
\begin{equation}
\int_{\alpha/z}^{1} \ud\xi \biggl [\frac{1}{\xi} - 2z + z\xi \biggr ]\mathcal{A}_0(\Delta_2) = -\biggl [\frac{3z}{2} + \log\left (\frac{\alpha}{z} \right ) \biggr ]    \frac{(4\pi)^{2-\frac{D}{2}}}{(2-\frac{D}{2})}\Gamma\left (3 - \frac{D}{2}\right ) + \mathcal{I}_{\xi;1} - z\left (2\, \mathcal{I}_{\xi;2} - \mathcal{I}_{\xi;3} \right ),
\end{equation}
where the functions $\mathcal{I}_{\xi;i}$ for $i=1,2,3$  are independent of transverse momenta $\Pt$ and are given explicitly in Appendix~\ref{app:logintegrals}. Putting everything together, we get 
\begin{equation}
\begin{split}
\int_{\alpha/z}^{1} \ud\xi \biggl [\frac{1}{\xi} - 2z + z\xi \biggr ]& \mathcal{A}_0(\Delta_2)  =  -\biggl [\frac{3z}{2}  + \log\left (\frac{\alpha}{z}\right ) \biggr ]\Biggl \{\frac{(4\pi)^{2-\frac{D}{2}}}{(2-\frac{D}{2})}\Gamma\left (3 - \frac{D}{2}\right ) + \log \left (\frac{\mu^2}{\overline{Q}^2 + m^2} \right )  \Biggr \} - \frac{(1 + 5z)}{2} \\
& + \mathrm{Li}_2\left (\frac{1}{1-\frac{1}{2z}(1-\gamma)} \right ) + \mathrm{Li}_2\left (\frac{1}{1-\frac{1}{2z}(1+\gamma)} \right ) + \left (1+ \frac{1}{2z}\right )\gamma\log \left ( \frac{1 + \gamma}{1 + \gamma - 2z}\right ) \\
& - \frac{1}{2z}\biggl [\left (z + \frac{1}{2}\right )(1-\gamma ) + \frac{m^2}{Q^2} \biggr ]\log \left (\frac{\overline{Q}^2 + m^2}{m^2} \right ).
\end{split}
\end{equation}

\subsubsection{UV finite parts}

Let us then focus on computation of the UV finite part in \eq\nr{eq:VT}, which turns out to be quite tricky. It is possible to directly compute the $\mathcal{B}_0$ and $\mathcal{B}^i$ integrals, but the result would be too complicated for further analytical integration, in particular for the required Fourier transform. Instead, we first Feynman parametrize the denominator appearing in   $\mathcal{B}_0$ and $\mathcal{B}^i$ as
\begin{equation}
\label{eq:feynmanparam}
\begin{split}
\left[ \begin{array}{c}
\mathcal{B}_0  \\
\mathcal{B}^j
\end{array} \right] = \int_{0}^{1} \ud x \left[ \begin{array}{c}
1  \\
-x\Lt^i
\end{array} \right]\frac{1}{[x(1-x)\Lt^2 + (1-x)\Delta_1 + x\Delta_2]},
\end{split}
\end{equation}  
where $\Lt = -(1-\xi)\Pt$ and the denominator can be rewritten as
\begin{equation}
\label{eq:denomfeynman}
\begin{split}
\mathcal{D} \equiv x(1-x)\Lt^2 + (1-x)\Delta_1 + x\Delta_2 & = (1-\xi)\biggl [(1-x)\Pt^2 + \overline{Q}^2 + m^2\biggr ]\biggl [x(1-\xi) + \frac{\xi}{(1-z)}\biggr ]\\
& + \xi m^2\biggl [\xi(1-x) + x\left (1 - \frac{z(1-\xi)}{(1-z)}\right )\biggr ].
\end{split}
\end{equation}
From \eqs\nr{eq:feynmanparam} and \nr{eq:denomfeynman} we then find
that a combination appearing in the definition of $\mathcal{T}_+$ in \eq\eqref{eq:Tminusandplus} can be expressed as
\begin{equation}
\label{eq:vcBidentity}
\begin{split}
-\Pt^i \mathcal{B}^i  + (1-\xi)\left (\Pt^2 + \overline{Q}^2 + m^2 \right )\mathcal{B}_0  & = \int_{0}^{1} \frac{\ud x}{\mathcal{D}} (1-\xi)\left [(1-x)\Pt^2 + \overline{Q}^2 + m^2\right ]\\
& = \frac{\mathcal{I}_{+} }{(1-\xi)}- \xi m^2\int_{0}^{1}\frac{\ud x}{\mathcal{D}}  \frac{ \xi(1-x) + x[1 - \frac{z(1-\xi)}{(1-z)}] }{\left [x(1-\xi) + \frac{\xi}{(1-z)}\right ]},
\end{split}
\end{equation}
where the function $\mathcal{I}_+$ is given by the simple expression 
\begin{equation}
\label{eq:Iplusfunction}
\mathcal{I}_+ \equiv \int_{0}^{1}\ud x \frac{(1-\xi)}{\left [x(1-\xi) + \frac{\xi}{(1-z)}\right ]} = -\log(\xi) + \log\biggl (1 -z(1-\xi)\biggr ). 
\end{equation}
The relation in \eq\nr{eq:vcBidentity} turns out to be very useful in what follows.

The difficulty in this part of the calculation is that we want to perform the divergent integral $\int_{\alpha/z}^{1} \ud \xi/\xi$  analytically. Being able to do this requires further manipulations of the somewhat complicated functions $\mathcal{T}_{\pm}$ in the integrand. However,  the relations in \eqs\nr{eq:feynmanparam} -- \nr{eq:Iplusfunction}  make it possible to split the  $\int_{\alpha/z}^{1} \ud \xi/\xi$ integral into a divergent part that can be integrated analytically, and a remaining more complicated term that is finite in the limit $\alpha \to 0$. 
To this end, we first substitute the expression  \nr{eq:vcBidentity} into the definition of  $\mathcal{T}_+$ in \eq\nr{eq:Tminusandplus} yielding the following  result
\begin{equation}
\label{eq:Tminusandplusmod}
\begin{split}
\mathcal{T}_{+}  & = \mathcal{I}_{+}  - \xi (1-\xi) m^2\int_{0}^{1} \frac{\ud x}{\mathcal{D}}  \frac{ \xi(1-x) + x[1 - \frac{z(1-\xi)}{(1-z)}] }{\left [x(1-\xi) + \frac{\xi}{(1-z)}\right ]} + \xi (1-\xi) (1-z)m^2 \mathcal{B}_0\\
& = \mathcal{I}_{+}  +  \frac{\xi (1-\xi)^2z^2m^2}{(1-z)} \int_{0}^{1} \frac{\ud x}{\mathcal{D}}\frac{x}{\left [x(1-\xi) + \frac{\xi}{(1-z)}\right ]}.
\end{split} 
\end{equation}
Furthermore, we note that one can rewrite the function $\mathcal{T}_{-}$ from \eq\nr{eq:Tminusandplus} as\footnote{This relation follows from the Passarino-Veltman tensor decomposition for $\mathcal{B}^i$.}
\begin{equation}
\label{eq:Tminusmod}
\begin{split}
\mathcal{T}_{-} & =  \mathcal{A}_0(\Delta_1) - \mathcal{A}_0(\Delta_2) - \mathcal{T}_{+} + \frac{z^2\xi(1-\xi)}{(1-z)}m^2\mathcal{B}_0\\
& = -\log\left (\frac{\Delta_1}{\mu^2} \right ) +  \log\left (\frac{\Delta_2}{\mu^2} \right ) - \mathcal{T}_{+} + \frac{z^2\xi(1-\xi)}{(1-z)}m^2\mathcal{B}_0.\\
\end{split}
\end{equation}
Finally, combining the expressions in \eqs\nr{eq:Tminusandplusmod} and \nr{eq:Tminusmod}, we arrive at the following expression
\begin{equation}
\label{eq:Tminusmodv2}
\begin{split}
\mathcal{T}_{-} & =  -\log\left (\frac{\Delta_1}{\mu^2} \right ) +  \log\left (\frac{\Delta_2}{\mu^2} \right ) - \mathcal{I}_+  +  m^2 \int_{0}^{1}  \frac{\ud x}{\mathcal{D}} \, \frac{z^2}{(1-z)^2}\frac{\xi^2(1-\xi)}{\left [x(1-\xi) + \frac{\xi}{(1-z)}\right ]}.\\
\end{split}
\end{equation}

With these intermediate results at hand, we obtain after the direct integration the term proportional to $\mathcal{T}_-$ in the expression for $\mathcal{V}^T_{\ref{diag:vertexqbaremT}}$  given in \eq\eqref{eq:VT}
\begin{equation}
\label{eq:intovertauminus}
\begin{split}
\int_{\alpha/z}^{1} \ud\xi & \biggl [\frac{1}{\xi} - 2z + z\xi \biggr ]\mathcal{T}_{-}  =  -\mathcal{I}_{\xi;1} + z\left (2\,\mathcal{I}_{\xi;2} - \mathcal{I}_{\xi;3} \right ) - \mathcal{J}_{\xi;1} + z\left (2\, \mathcal{J}_{\xi;2} - \mathcal{J}_{\xi;3} \right )\\
& - \Biggl \{\frac{(1-z)}{2} +  \frac{1}{2}\log^2\left (\frac{\alpha}{z} \right ) + \frac{(1 + 2z - 3z^2)}{2z}\log(1-z) - \log(1-z)\log\left (\frac{\alpha}{z}\right ) - \mathrm{Li}_2\left (- \frac{z}{1-z} \right ) \Biggr \}\\
& +  m^2\int_{0}^{1} \ud \xi \int_{0}^{1} \frac{\ud x}{\mathcal{D}} \times \frac{z^2}{(1-z)^2} \frac{\xi(1-\xi)[1 + z\xi(\xi-2)]}{\left [x(1-\xi) + \frac{\xi}{(1-z)}\right ]},
\end{split}
\end{equation}
where the functions $\mathcal{J}_{\xi;i}$ with $i=1,2,3$ which depend explicitly on the transverse momentum $\Pt$ are given in Appendix~\ref{app:logintegrals}. The point of these manipulations is that we have now managed to split the integral~\eqref{eq:intovertauminus} that had a divergence in the momentum fraction regulated by $\alpha$ into, on the one hand, a part where both the momentum fraction and Feynman parameter integrals have been evaluated and, on the other hand, a part proportional to $m^2$ where the integrals remain, but the regulator has been taken to $\alpha=0$.

The contribution~\eqref{eq:intovertauminus} is, as it appears in the expression 
for $\mathcal{V}^T_{\ref{diag:vertexqbaremT}}$  in \eq\eqref{eq:VT}, multiplied with $1/\Pt^2$. However, as can both be seen  in the general decomposition in \eq\nr{NLOformfactors} and deduced from the fact that the original expressions before transverse momentum integrals  never have energy denominators $\sim \Pt^2$, such a transverse IR-divergent behavior is completely spurious and must in fact cancel. Thus we know that in fact the expression~\eqref{eq:intovertauminus} must in fact vanish in the limit $\Pt^2\to 0$.  
Indeed, one can show that if we rewrite $\mathcal{T}_{-} = \mathcal{T}_{-} - \lim_{\Pt^2 \to 0}\mathcal{T}_{-}  + \lim_{\Pt^2 \to 0}\mathcal{T}_{-}$, where $\lim_{\Pt^2 \to 0}\mathcal{T}_{-}$ is obtained from \eq\nr{eq:Tminusmodv2} with  $\Pt^2 \to 0$, we can cast the integral result in \eq\nr{eq:intovertauminus} into the following form \begin{equation}
\label{eq:intovertauminusv2}
\begin{split}
\int_{\alpha/z}^{1} \ud\xi  \biggl [\frac{1}{\xi} - 2z + z\xi \biggr ]&\mathcal{T}_{-}   = \biggl [\frac{3z}{2} + \log\left (\frac{\alpha}{z}\right )\biggr ] \log\left (\frac{\Pt^2 + \overline{Q}^2 + m^2}{\overline{Q}^2 + m^2} \right ) \\
&  -  \Pt^2 m^2 \int_{0}^{1} \ud \xi \biggl ( \frac{ \log(\xi)}{(1-\xi)^2} + \frac{z}{1-\xi} + \frac{z}{2} \biggr )\frac{(1-z)}{\Pt^2 + \overline{Q}^2 +  m^2 + \frac{\xi(1-z)}{(1-\xi)}m^2}\frac{1}{ \overline{Q}^2 +  m^2 + \frac{\xi(1-z)}{(1-\xi)}m^2} \\
&  -  \Pt^2 m^2\int_{0}^{1} \ud \xi \int_{0}^{1} \frac{\ud x}{\mathcal{D}}\frac{(1-x)}{\mathcal{D}_{\Pt^2 \to 0}} \, \frac{z^2}{(1-z)^2} \xi(1-\xi)^2[1 + z\xi(\xi-2)],
\end{split}
\end{equation}
where
\begin{equation}
\mathcal{D}_{\Pt^2 \to 0} = (1-\xi)(\overline{Q}^2 + m^2)\biggl [x(1-\xi) + \frac{\xi}{1-z}\biggr] + \xi m^2 \bigg [\xi + x(1-\xi) - x(1-\xi) \frac{z}{1-z} \biggr ].    
\end{equation}
The ``double'' denominator $\mathcal{D} \mathcal{D}_{\Pt^2 \to 0}$ appears after combining the original expression $\sim 1/\mathcal{D}$ and the subtraction term $\sim 1/ \mathcal{D}_{\Pt^2 \to 0}$ into a common fraction. 
This expression manifestly vanishes at $\Pt^2\to 0$ and thus cancels the factor $1/\Pt^2$ multiplying it in  \eq\eqref{eq:VT},  as it should. As a separate point, we also note that the remaining $\xi$ and $x$ integrals in \eq\nr{eq:intovertauminusv2} are finite and could be done analytically. However, it turns out that they, unlike the parts that diverge as $\alpha\to 0$, need to be kept in an unintegrated form in order to Fourier transform our expressions into mixed space later. The form~\eqref{eq:intovertauminusv2} is the one we then use to obtain the final result.


Finally, for the $\mathcal{T}_+$ term appearing in \eq\nr{eq:VT} we obtain after the direct integration
\begin{equation}
\begin{split}
\int_{\alpha/z}^{1} \ud\xi & \biggl [\frac{2}{\xi}  + \frac{z \xi}{(1-z)} \biggr ]\mathcal{T}_{+}  =  \frac{1}{2} + \frac{(1-z)}{2z}\log(1-z) -  2\log(1-z)\log \left (\frac{\alpha}{z} \right ) +  \log^2\left (\frac{\alpha}{z} \right ) - 2\mathrm{Li}_2\left (- \frac{z}{1-z} \right ) \\
& + m^2\int_{0}^{1} \ud \xi \int_{0}^{1} \frac{\ud x}{\mathcal{D}}   \frac{z^2 }{(1-z)^2} \frac{x (1-\xi)^2 [ 2(1-z) + z\xi^2 ]}{\left [x(1-\xi) + \frac{\xi}{(1-z)}\right ]}.
\end{split}
\end{equation}

\subsubsection{Collecting the UV divergent and finite terms}
\label{sec:variablechange}
Collecting all the UV divergent and UV finite terms together, we can cast the one-loop form factor $\mathcal{V}^T_{\ref{diag:vertexqbaremT}}$ into the following form
\begin{equation}
\label{eq:Vesimp}
\begin{split}
\mathcal{V}^T_{\ref{diag:vertexqbaremT}}  = & -\biggl [\frac{3z}{2} + \log\left (\frac{\alpha}{z}\right )\biggr ]\Biggl \{\frac{(4\pi)^{2-\frac{D}{2}}}{(2-\frac{D}{2})}\Gamma\left (3 - \frac{D}{2}\right ) + \log \left (\frac{\mu^2}{\overline{Q}^2 + m^2} \right ) - \frac{(\Pt^2 + \overline{Q}^2 + m^2)}{\Pt^2}\log\left (\frac{\Pt^2 + \overline{Q}^2 + m^2}{\overline{Q}^2 + m^2} \right ) \Biggr \}\\
& - \frac{7z}{2} - \frac{z}{2}\frac{(D_s-4)}{(D-4)}  - \log(1-z)\biggl [\frac{3}{2} + 2\log \left (\frac{\alpha}{z} \right ) \biggr ] + \log^2\left ( \frac{\alpha}{z} \right ) - 2\mathrm{Li}_2\left (- \frac{z}{1-z} \right )  + \Omega_{\mathcal{V}_{\ref{diag:vertexqbaremT}}} \\
& + \mathrm{Li}_2\left (\frac{1}{1-\frac{1}{2z}(1-\gamma)} \right )  + \mathrm{Li}_2\left (\frac{1}{1-\frac{1}{2z}(1+\gamma)} \right ) + I_{\mathcal{V}_{\ref{diag:vertexqbaremT}}} +  O(D-4),
\end{split}    
\end{equation}
where we have defined the function $\Omega_{\mathcal{V}_{\ref{diag:vertexqbaremT}}}$ as
\begin{equation}
\label{eq:SigmaOmega}
\Omega_{\mathcal{V}_{\ref{diag:vertexqbaremT}}}  = \left (1 + \frac{1}{2z} \right )\biggl [\log(1-z) + \gamma \log \left (\frac{1 + \gamma}{1 + \gamma - 2z} \right ) \biggr ] - \frac{1}{2z}\biggl [ \left (z + \frac{1}{2} \right )(1-\gamma) +  \frac{m^2}{Q^2}\biggr ]\log \left (\frac{\overline{Q}^2 + m^2}{m^2} \right )
\end{equation}
and $I_{\mathcal{V}_{\ref{diag:vertexqbaremT}}}$ is given by the following integral expression
\begin{multline}
\label{eq:VTeintmod}
I_{\mathcal{V}_{\ref{diag:vertexqbaremT}}}  =  -(\Pt^2 + \overline{Q}^2 + m^2)\Biggl  \{ \int_{0}^{1} \ud \xi \biggl ( \frac{ \log(\xi)}{(1-\xi)^2} + \frac{z}{1-\xi} + \frac{z}{2} \biggr )\frac{1}{\Pt^2 + \overline{Q}^2 +  m^2 + \frac{\xi(1-z)}{(1-\xi)}m^2}\frac{(1-z)m^2}{ \overline{Q}^2 +  m^2 + \frac{\xi(1-z)}{(1-\xi)}m^2} \\
  +  m^2\int_{0}^{1} \ud \xi \int_{0}^{1} \frac{\ud x}{\mathcal{D}}\frac{(1-x) }{\mathcal{D}_{\Pt^2 \to 0}}  \frac{z^2 \, \xi(1-\xi)^2}{(1-z)^2}  \left [1 + z\xi(\xi-2) \right ]\Biggr \}\\
 + m^2\int_{0}^{1} \ud \xi \int_{0}^{1} \frac{\ud x}{\mathcal{D}}  \frac{z(1-\xi)}{(1-z)} \Biggl \{\frac{zx(1-\xi)[2 + \frac{z\xi^2}{(1-z)} ]}{\left [x(1-\xi) + \frac{\xi}{(1-z)}\right ]} + \xi^2 (1-z) + x\xi(1-\xi)(1-2z) \Biggr \}.
\end{multline}

The double integrals appearing in \eq\nr{eq:VTeintmod} can be further simplified by performing a set of change of variables. 
Since we will use the same variable change in several parts of our calculation, let us introduce it here for a more general case.
We consider the following double integral of the type appearing e.g. in \eq\nr{eq:VTeintmod}:
\begin{equation}
\label{eq:intoverf}
\int_{0}^{1} \ud \xi \int_{0}^{1} \frac{\ud x}{\mathcal{D}} \,f(\xi,x),
\end{equation}
where the denominator $\mathcal{D}$ is defined in \eq\nr{eq:denomfeynman} and $f$ is some smooth function depending on $\xi$ and $x$. First, introducing the following change of variable
\begin{align}
x\mapsto y= \xi + x(1\!-\!\xi)
\label{cv_x_to_y}\,
\end{align}
yields
\begin{equation}
\label{eq:intoverf2}
\int_{0}^{1} \frac{\ud \xi}{(1\!-\!\xi)}\int_{\xi}^{1} \frac{\ud y}{{\cal D}}\,f(\xi,y),
\end{equation}
where the denominator $\mathcal{D}$ now becomes
\begin{align}
{\cal D} =& \left[y+\frac{z\, \xi}{(1\!-\!z)}\right]\left[(1\!-\!y)(\Pt^2\!+\!m^2)\!+\!(1\!-\!\xi)\overline{Q}^2 \right]
+y^2\, m^2
.  
\label{Denom_in_y}
\end{align}
Next, introducing the following change of variable
\begin{align}
\xi\mapsto \eta= \frac{\xi}{y}
\label{cv_ix_to_eta}
\end{align}
leads to
\begin{align}
\int_{0}^{1} \ud y\, \int_{0}^{1} \frac{\ud \eta}{(1-y\eta)}  \frac{f(\eta,y)}{\left\{\left[1+\frac{z\, \eta}{(1\!-\!z)}\right]\left[(1\!-\!y)(\Pt^2\!+\!m^2)\!+\!(1\!-\!\eta y)\overline{Q}^2 \right]
+y\, m^2\right\}}.
\end{align}
In this expression the denominator is now linear in $y$. Finally, with the change of variable
\begin{align}
\eta\mapsto \chi= z(1\!-\!\eta)= z\left(1\!-\!\frac{\xi}{y}\right)
\label{cv_eta_to_chi},
\end{align}
one obtains
\begin{align}
\frac{1}{z}\int_{0}^{1} \ud y\, \int_{0}^{z} \frac{\ud \chi}{1-y\left (1-\frac{\chi}{z}\right )} \frac{f(\chi,y)}{\left\{
(1\!-\!y)\, \frac{(1\!-\!\chi)}{(1\!-\!z)}\, (\Pt^2\!+\!\overline{Q}^2\!+\!m^2) 
+y\, \left[m^2+\frac{\chi(1\!-\!\chi)}{z(1\!-\!z)}\, \overline{Q}^2\right]
\right\}}.
\end{align}
Interestingly, we will find that  all the double integrals that are written as an our example integral \nr{eq:intoverf}, can be greatly simplified by performing the above chain of three changes of variables.


Now we can further simplify our expression for $I_{\mathcal{V}_{\ref{diag:vertexqbaremT}}}$ in \eq\nr{eq:VTeintmod} by applying  the change of variables to $y$ and $\chi$. After some algebra, we obtain 
\begin{equation}
\label{eq:VTeintmod2}
\begin{split}
&I_{\mathcal{V}_{\ref{diag:vertexqbaremT}}}  =  -(\Pt^2 + \overline{Q}^2 + m^2)\Biggl  \{\int_{0}^{1} \ud \xi \biggl ( \frac{ \log(\xi)}{(1-\xi)^2} + \frac{z}{1-\xi} + \frac{z}{2} \biggr )\frac{m^2}{\left \{\Pt^2 + \overline{Q}^2 + m^2 + \frac{\xi(1-z)}{(1-\xi)}m^2\right \}}\frac{(1-z)}{\left \{\overline{Q}^2 + m^2 + \frac{\xi(1-z)}{(1-\xi)}m^2\right \}}  \\
&  + \int_{0}^{1} \frac{\ud y}{(1-y)} \int_{0}^{z} \frac{\ud \chi}{(1-\chi)^2} \frac{m^2}{\left \{\Pt^2 + \overline{Q}^2 + m^2 + \frac{y}{(1-y)} \frac{(1-z)}{(1-\chi)}\left[m^2+\frac{\chi(1\!-\!\chi)}{z(1\!-\!z)}\, \overline{Q}^2\right] \right \}} \frac{\left (z - \chi\right )\left [1 - 2y\left (z-\chi\right ) + \frac{y^2}{z}\left (z-\chi\right )^2 \right ]}{\left \{\overline{Q}^2 + m^2 + \frac{y}{(1-y)}\frac{(1-z)}{(1-\chi)}\left[m^2+\frac{\chi(1\!-\!\chi)}{z(1\!-\!z)}\, \overline{Q}^2\right]\right \}}\Biggr \}
\\
& + \int_{0}^{1} \frac{\ud y}{(1-y)} \int_{0}^{z} \frac{\ud \chi}{(1-\chi)^2} \frac{m^2}{\left \{\Pt^2 + \overline{Q}^2 + m^2 + \frac{y}{(1-y)} \frac{(1-z)}{(1-\chi)}\left[m^2+\frac{\chi(1\!-\!\chi)}{z(1\!-\!z)}\, \overline{Q}^2\right] \right \}} \frac{(1-z)}{z}\biggl [2z\chi + y^2(z-\chi)(1-2\chi)\biggr ].  
\end{split}
\end{equation}

\subsubsection{The full form factor $\mathcal{V}^T$ in momentum space}

At this stage, it is convenient to combine the results from the propagator correction diagram \ref{diag:oneloopSEUPT} from \eq\nr{eq:Va} and from the vertex correction \ref{diag:vertexqbaremT} that we have just calculated in \eq\nr{eq:Vesimp}. This gives 
\begin{equation}
\label{eq:Vae}
\begin{split}
\mathcal{V}^{T}_{\ref{diag:oneloopSEUPT}}  + \mathcal{V}^T_{\ref{diag:vertexqbaremT}}   = & \biggl [\frac{3(1-z)}{2} + \log\left (\frac{\alpha}{z}\right )\biggr ]\Biggl \{ \frac{(4\pi)^{2-\frac{D}{2}}}{(2-\frac{D}{2})}\Gamma\left (3 - \frac{D}{2}\right ) + \log \left (\frac{\mu^2}{\overline{Q}^2 + m^2} \right )\Biggl \}\\
& +  \Biggl \{-\frac{3}{2} - 2\log\left (\frac{\alpha}{z}\right ) + \frac{(\Pt^2 + \overline{Q}^2 + m^2)}{\Pt^2}\biggl [\frac{3z}{2} +  \log\left (\frac{\alpha}{z}\right )\biggr ]\Biggr \} \log\left (\frac{\Pt^2 + \overline{Q}^2 + m^2}{\overline{Q}^2 + m^2} \right )\\
& + 3 -\frac{7z}{2} + \frac{(1-z)}{2}\frac{(D_s-4)}{(D-4)} - \frac{\pi^2}{3} - 2\mathrm{Li}_2\left (- \frac{z}{1-z} \right ) + \Omega_{\mathcal{V}_{\ref{diag:vertexqbaremT}}}\\
& + \mathrm{Li}_2\left (\frac{1}{1-\frac{1}{2z}(1-\gamma)} \right )  + \mathrm{Li}_2\left (\frac{1}{1-\frac{1}{2z}(1+\gamma)} \right )  + I_{\mathcal{V}_{\ref{diag:vertexqbaremT}}} +  I_{\mathcal{V}_{\ref{diag:oneloopSEUPT}}} + O(D-4).
\end{split}    
\end{equation}
We note that the $\log^2(\alpha/z)$-term and a part of the $\log(\alpha/z)$ term from the vertex corrections canceled against the self-energy contribution, as in the massless case. 

Finally, we are ready to write down the full result for the form factor $\mathcal{V}^{T}$ in \eq\nr{eq:fullVT}. By 
reconstructing the contributions of diagrams \ref{diag:oneloopSEDOWNT}  and  \ref{diag:vertexqemT} 
using the $z \leftrightarrow 1-z$ symmetry, as discussed in Sec. \ref{sec:spinstuctureandmrenorm}, we obtain the full leading order-like  form factor $\mathcal{V}^{T}$ as
\begin{equation}
\label{eq:VTfull}
\begin{split}
\mathcal{V}^{T} = & \biggl [\frac{3}{2} + \log\left (\frac{\alpha}{z}\right ) + \log\left (\frac{\alpha}{1-z}\right )\biggr ]\Biggl \{\frac{(4\pi)^{2-\frac{D}{2}}}{(2-\frac{D}{2})}\Gamma\left (3 - \frac{D}{2}\right ) + \log \left
(\frac{\mu^2}{\overline{Q}^2 + m^2} \right )\\
& + \frac{(\overline{Q}^2 + m^2 - \Pt^2)}{\Pt^2}\log\left (\frac{\Pt^2 + \overline{Q}^2 + m^2}{\overline{Q}^2 + m^2} \right )\Biggl \} + \frac{5}{2} + \frac{1}{2}\frac{(D_s-4)}{(D-4)} - \frac{\pi^2}{3}  + \log^2 \left (\frac{z}{1-z} \right )\\
& +\Omega_{\mathcal{V}} + L + I_{\mathcal{V}} +  O(D-4),
\end{split}    
\end{equation}
where the functions $\Omega_{\mathcal{V}}$ and $I_{\mathcal{V}}$ are defined as
\begin{equation}
\label{eq:OSI}
\begin{split}
 \Omega_{\mathcal{V}} & = \Omega_{\mathcal{V}_{\ref{diag:vertexqbaremT}}}  + \biggl [z \leftrightarrow 1-z \biggr ],  \\   
I_{\mathcal{V}} & = I_{\mathcal{V}_{\ref{diag:vertexqbaremT}}} +    I_{\mathcal{V}_{\ref{diag:oneloopSEUPT}}}  + \biggl [z \leftrightarrow 1-z \biggr ],
\end{split}
\end{equation}
in terms of $\Omega_{\mathcal{V}_{\ref{diag:vertexqbaremT}}} $ from  \eq\eqref{eq:SigmaOmega} and $I_{\mathcal{V}_{\ref{diag:oneloopSEUPT}}}$, $I_{\mathcal{V}_{\ref{diag:vertexqbaremT}}} $ from \eqs \nr{eq:defIVa} and~\nr{eq:VTeintmod2}
and the function $L$, which also appears in the longitudinal photon case \cite{Beuf:2021qqa}, is given by
\begin{equation}
\label{eq:defL}
L = \sum_{\sigma= \pm 1} \Biggl [ \mathrm{Li}_2\left (\frac{1}{1-\frac{1}{2z}(1 +\sigma \gamma)} \right ) + \mathrm{Li}_2\left (\frac{1}{1-\frac{1}{2(1-z)}(1+\sigma\gamma)} \right ) \Biggr ].   
\end{equation}

It is now straightforward to check that in the massless quark limit ($\gamma \to 1$) the functions defined above satisfy
\begin{equation}
L \to \frac{\pi^2}{6} - \frac{1}{2}\log^2\left (\frac{z}{1-z} \right ), \quad \Omega_{\mathcal{V}} \to 0, \quad I_{\mathcal{V}} \to 0,
\end{equation}
and correspondingly the expression in \eq\nr{eq:VTfull} reduces to the one obtained in Refs.~\cite{Beuf:2017bpd,Hanninen:2017ddy}.

\subsection{Form factor $\mathcal{N}^T$}

The form factor $\mathcal{N}^T$ gets contributions only from the two non-trivial vertex correction diagrams \ref{diag:vertexqbaremT} and \ref{diag:vertexqemT}. Again, by using the symmetry discussed in Sec. \ref{sec:spinstuctureandmrenorm}, we have
\begin{equation}
\label{eq:fullNT}
\mathcal{N}^T = \mathcal{N}^{T}_{\ref{diag:vertexqbaremT}} -  \biggl [z \leftrightarrow 1-z \biggr ].
\end{equation}
The expression above is  UV finite and free from the $\xi \to 0$ singularities. Since the calculation follows the same steps as before, we can immediately write down the final result
\begin{equation}
\label{eq:fullNTv2}
\mathcal{N}^T = \Omega_{\mathcal{N}} + I_{\mathcal{N}},
\end{equation}
where the function $\Omega_{\mathcal{N}}$ is given by
\begin{equation}
\label{eq:degOmegaN_1}
\Omega_{\mathcal{N}} = \Omega_{\mathcal{N}_{\ref{diag:vertexqbaremT}}} - \biggl [z \leftrightarrow 1-z \biggr ]
\end{equation}
with
\begin{equation}
\label{eq:degOmegaN}
\Omega_{\mathcal{N}_{\ref{diag:vertexqbaremT}}} =   \frac{1+z-2z^2}{z}\Biggl [\log(1-z) + \gamma \log \left (\frac{1+\gamma}{1+\gamma-2z}  \right ) \Biggr ] + \frac{(1-z)}{z} \Biggl [ \biggl (\frac{1}{2} + z \biggr )(\gamma -1) - \frac{m^2}{Q^2} \Biggr ]\log \left (\frac{\overline{Q}^2 + m^2}{m^2} \right )
\end{equation}
and $I_{\mathcal{N}}$ is given by the following integral expression
\begin{equation}
I_{\mathcal{N}} = I_{\mathcal{N}_{\ref{diag:vertexqbaremT}}} - \biggl [z \leftrightarrow 1-z \biggr ]
\end{equation}
with
\begin{equation}
\label{eq:INev1}
\begin{split}
I_{\mathcal{N}_{\ref{diag:vertexqbaremT}}}  & = (\Pt^2 + \overline{Q}^2 + m^2) \int_{0}^{1}\ud \xi \int_{0}^{1} \frac{\ud x}{\mathcal{D}} \times 2z(1-z) (2-\xi)(1-\xi)\biggl [x(1-\xi) + \frac{\xi}{1-z} \biggr ] \\
& \hspace{-0.6cm} + m^2 \int_{0}^{1}\ud \xi \int_{0}^{1} \frac{\ud x}{\mathcal{D}} \times 2z(1-\xi) \biggl [x(1-\xi) + \xi \biggr ]\Biggl \{\biggl (2 + \frac{(2z-1)\xi}{(1-z)} \biggr )\frac{x(1-\xi) + \xi}{\left [x(1-\xi) + \frac{\xi}{1-z}\right ]} - \biggl [1 + 2(1-z)(1-\xi) \biggr ]\Biggr \}.
\end{split}
\end{equation}
It is straightforward to check that this contribution vanishes in the massless quark limit. This is an agreement with the findings in the massless case in Ref.~\cite{Beuf:2016wdz}.

The double integrals in \eq\nr{eq:INev1} can simplified by performing the change of variables to $(y, \chi)$ introduced in Sec.~\ref{sec:variablechange}. This yields the following result
\begin{equation}
\label{eq:INev2}
\begin{split}
I_{\mathcal{N}_{\ref{diag:vertexqbaremT}}} &=  \frac{2(1-z)}{z}\int_{0}^{1} \frac{\ud y}{(1-y)} \int_{0}^{z} \ud \chi \frac{1}{\left \{\Pt^2 + \overline{Q}^2 + m^2 + \frac{y}{(1-y)} \frac{(1-z)}{(1-\chi)}\left [m^2 + \frac{\chi(1-\chi)}{z(1-z)}\overline{Q}^2 \right ] \right \}}\\
& \hspace{-0.8cm} \times \biggl \{(\Pt^2 + \overline{Q}^2+m^2)\left [y^2 \chi + z(2-y)y \right ] - \frac{y m^2}{(1-\chi)^2} \left [2z^2 \chi (1-y) + \chi (1-2\chi)y + z(1-y-3\chi + 2\chi (1+\chi)y)\right ]\biggr \}.
\end{split}
\end{equation}


\subsection{Form factor $\mathcal{S}^T$}

The form factor $\mathcal{S}^T$ gets contributions only from the two non-trivial vertex correction diagrams \ref{diag:vertexqbaremT} and \ref{diag:vertexqemT}. Again, by using the symmetry discussed in Sec. \ref{sec:spinstuctureandmrenorm}, we have
\begin{equation}
\label{eq:fullST}
\mathcal{S}^T = \mathcal{S}^{T}_{\ref{diag:vertexqbaremT}} +  \biggl [z \leftrightarrow 1-z \biggr ].
\end{equation} 
The fully finite contribution $\mathcal{S}^{T}_{\ref{diag:vertexqbaremT}}$ from \eq\nr{eq:ST} can be easily rewritten into the following form
\begin{equation}
\label{eq:STfinal}
\begin{split}
\mathcal{S}^{T}_{\ref{diag:vertexqbaremT}} =  -z + 2z\int_{0}^{1} \ud \xi (1-\xi) \Biggl \{-\frac{(\Pt^2 + \overline{Q}^2 + m^2)}{\Pt^2}\mathcal{T}_{-} & + \Pt^2 \int_{0}^{1}\frac{\ud x}{\mathcal{D}} (1-\xi)\biggl [x\xi \frac{(1-2z)}{(1-z)} - x - \xi \biggr ]\\
& + m^2 \int_{0}^{1}\frac{\ud x}{\mathcal{D}}  \biggl [x\xi(1-\xi) \frac{(1-2z)}{(1-z)} + \xi^2 \biggr ]\Biggr \},
\end{split}
\end{equation}
with $\mathcal{T}_{-}$ from \eq\nr{eq:Tminusmodv2}.
Again, one can further simplify the expression in \eq\nr{eq:STfinal} by first performing the change of variables to $(y, \chi)$. This yields a very compact expression
\begin{align}
{\cal S}^{T}_{\ref{diag:vertexqbaremT}} = &  -2(1-z) \Pt^2\int_{0}^{1} \ud y \int_{0}^{z} \frac{\ud \chi}{(1-\chi)} \frac{y}{\left\{\Pt^2+\overline{Q}^2+m^2 
+ \frac{y}{(1-y)} \frac{(1-z)}{(1-\chi)}\left[m^2+\frac{\chi(1-\chi)}{z(1-z)} \overline{Q}^2\right]
\right\}}.
\label{ST_result_y_chi}
\end{align}

\subsection{Form factor $\mathcal{M}^T$}

Finally, we deal with the form factor $\mathcal{M}^T$, which also contains the vertex-mass renormalization counter-term   $\mathcal{M}^T_{\text{c.t.}}$. This form factor gets contributions from the instantaneous diagrams \ref{diag:oneloopSEUPTinst},
\ref{diag:oneloopvertexinst1}, \ref{diag:oneloopSEDOWNTinst}, \ref{diag:oneloopvertexinst2},
\ref{diag:gluoninst} and from the vertex correction diagrams  \ref{diag:vertexqbaremT} and \ref{diag:vertexqemT}. Again, by using the symmetry discussed in Sec. \ref{sec:spinstuctureandmrenorm}, we get
\begin{equation}
\label{eq:MTfull}
\mathcal{M}^T = \biggl (\mathcal{M}^T_{\ref{diag:oneloopSEUPTinst}} + \mathcal{M}^T_{\ref{diag:oneloopvertexinst1}} + \mathcal{M}^T_{\ref{diag:gluoninst}_1} + \mathcal{M}^T_{\ref{diag:vertexqbaremT}} +  \biggl [z \leftrightarrow 1-z \biggr ]\biggr ) + \mathcal{M}^T_{\text{ c.t.}}.
\end{equation}
Collecting the expressions obtained in \eqs\nr{eq:MTc}, \nr{eq:MTg}, \nr{eq:MTi1} and \nr{eq:MT} together, we find 
\begin{equation}
\label{eq:MTfull2}
\begin{split}
\mathcal{M}^{T} - \mathcal{M}^T_{\text{c.t.}} = - \frac{(1-z)}{2} \frac{(D_s-4)}{(D-4)} & + \int_{\alpha/z}^{1} \ud \xi \Biggl \{(1-z)(1+\xi)\mathcal{A}_0(\Delta_1) + \biggl [-\frac{1}{\xi} + z \biggr ]\frac{\left (\Pt^2 + \overline{Q}^2 + m^2 \right )}{\Pt^2}\mathcal{T}_{-}\\
& - \frac{z^2}{(1-z)}\left (\Pt^2 + \overline{Q}^2 + m^2 \right ) (1-\xi)^2 \mathcal{B}_0 \Biggr \} +   O(D-4)  + \biggl [z \leftrightarrow 1-z \biggr ],
\end{split}
\end{equation}
with $\mathcal{T}_{-}$ from \eq\nr{eq:Tminusmodv2}.
In the on-shell scheme for mass renormalization, the vertex-mass counterterm $\mathcal{M}^T_{\rm c.t.}$ is determined by the  renormalization condition  $\mathcal{M}^T \to 0$ for $\Pt^2 \to -(\overline{Q}^2 + m^2)$.  Consequently, the expression in  \eq\nr{eq:MTfull2} yields the following result
\begin{equation}
\label{eq:MTct}
\begin{split}
  \mathcal{M}^T_{\text{c.t.}}  &  = \frac{(1-z)}{2} \frac{(D_s-4)}{(D-4)} - (1-z)\int_{0}^{1} \ud \xi (1+\xi)\mathcal{A}_0(\xi^2 m^2)  +   O(D-4)  + \biggl [z \leftrightarrow 1-z \biggr ]\\
& = \frac{1}{2} \frac{(D_s-4)}{(D-4)} - \int_{0}^{1} \ud \xi (1+\xi)\mathcal{A}_0(\xi^2 m^2)  +   O(D-4).
\end{split}
\end{equation}
We discuss the value of the mass counterterm a bit more in Appendix~\ref{app:Mct}, and continue here with our primary objective, the  mass-renormalized LCWF.
The full mass renormalized expression for the form factor $\mathcal{M}^T$ in the general kinematics now reads 
\begin{equation}
\label{eq:MTNLOintstep}
\begin{split}
\mathcal{M}^{T}  = \int_{\alpha/z}^{1} \ud \xi \Biggl \{(1-z)(1+\xi)&\biggl [\mathcal{A}_0(\Delta_1) - \mathcal{A}_0(\xi^2 m^2)\biggr ]  + \biggl [-\frac{1}{\xi} + z \biggr ]\frac{\left (\Pt^2 + \overline{Q}^2 + m^2 \right )}{\Pt^2}\mathcal{T}_{-}\\
& - \frac{z^2}{(1-z)}\left (\Pt^2 + \overline{Q}^2 + m^2 \right ) (1-\xi)^2 \mathcal{B}_0 \Biggr \} +   O(D-4)  + \biggl [z \leftrightarrow 1-z \biggr ].   
\end{split}
\end{equation}
Finally, we note that the first term can be further evaluated by using the following result
\begin{equation}
\begin{split}
\int_{0}^{1} \ud \xi (1+\xi)\biggl [\mathcal{A}_0(\Delta_1) - \mathcal{A}_0(\xi^2 m^2)\biggr ]  & =  (\Pt^2 + \overline{Q}^2 + m^2)\int_{0}^{1}\ud \xi \biggl (-\frac{3}{2(1-\xi)} + \frac{1}{2} \biggr ) \frac{1}{\Pt^2 + \overline{Q}^2 + m^2 + \frac{\xi(1-z)}{(1-\xi)}m^2},
\end{split}
\end{equation}
and that the third term containing $\mathcal{B}_0$  can be rewritten by using the Feynman parametrized form introduced in \eq\nr{eq:feynmanparam}. Combining these final remarks, we obtain the following result
\begin{equation}
\label{eq:MTNLOfinalstep}
\begin{split}
\mathcal{M}^{T}   & =   \Biggl \{\int_{\alpha/z}^{1} \ud \xi \biggl [-\frac{1}{\xi}  + z \biggr ]\frac{\left (\Pt^2 + \overline{Q}^2 + m^2 \right )}{\Pt^2}\mathcal{T}_{-}\\
& + (\Pt^2 + \overline{Q}^2 + m^2)\int_{0}^{1}\ud \xi \biggl (-\frac{3(1-z)}{2(1-\xi)} + \frac{(1-z)}{2} \biggr ) \frac{1}{\Pt^2 + \overline{Q}^2 + m^2 + \frac{\xi(1-z)}{(1-\xi)}m^2}\\
&  -(\Pt^2 + \overline{Q}^2 + m^2)    \int_{0}^{1} \ud \xi \int_{0}^{1} \frac{\ud x}{\mathcal{D}} \frac{z^2}{(1-z)} (1-\xi)^2 \Biggr \} +   O(D-4) \\
& + \biggl [z \leftrightarrow 1-z \biggr ],
\end{split}
\end{equation}
with $\mathcal{T}_{-}$ from \eq\nr{eq:Tminusmodv2}.

\subsection{Form factor $\mathcal{V}^T + \mathcal{M}^T - \mathcal{S}^T/2$}

We are now almost ready to pass on to the next stage of our calculation and Fourier-transform the LCWF to transverse coordinate space. 
As  discussed in Sec. \ref{sec:spinstuctureandmrenorm}, at this stage it is convenient to switch from our original basis of four scalar form factors $\mathcal{V}^T$, $\mathcal{M}^T$, $\mathcal{S}^T$ and $\mathcal{M}^T$ to one where the most rank-2 tensor, most complicated to transform, is traceless. This makes appear, as the coefficient of the ``leading-order-mass-like'' spinor structure, the combination $\mathcal{V}^T + \mathcal{M}^T - \mathcal{S}^T/2$, as seen in \eq\nr{eq:NLOformfactorsv3}. It turns out that  a significant simplification occurs in this particular linear combination. As the last step in evaluating the momentum space LCWF let us now write it down. 
This combination has quite a simple form
\begin{equation}
\label{eq:VMSfull}
\begin{split}
\mathcal{V}^T & + \mathcal{M}^T - \frac{1}{2}\mathcal{S}^T = \biggl [\frac{3}{2} + \log\left ( \frac{\alpha}{z}\right ) +  \log\left ( \frac{\alpha}{1-z}\right ) \biggr ]\Biggl \{\frac{(4\pi)^{2-\frac{D}{2}}}{(2-\frac{D}{2}}\Gamma \left (3 - \frac{D}{2}\right ) + \log \left (\frac{\mu^2}{\overline{Q}^2+ m^2}\right )\\
& - 2\log\left (\frac{\Pt^2 + \overline{Q}^2 + m^2}{\overline{Q}^2 + m^2} \right ) \Biggr \} + \frac{1}{2}\frac{(D_s-4)}{(D-4)} + 3 - \frac{\pi^2}{3}  + \log^2 \left (\frac{z}{1-z} \right ) + L  + \Omega_{\mathcal{V}}  + I_{\mathcal{VMS}} +   O(D-4),
\end{split}
\end{equation}
where $I_{\mathcal{VMS}}$ is given by the following integral expression 
\begin{equation}
\label{eq:IVMSv1}
\begin{split}
I_{\mathcal{VMS}}  & =  (\Pt^2 + \overline{Q}^2 + m^2)\int_{0}^{1}\frac{\ud \xi}{\xi} \left (-\frac{2\log(\xi)}{(1-\xi)} + \frac{(1+\xi)}{2}\right )\Biggl \{\frac{1}{\Pt^2 + \overline{Q}^2 + m^2}- \frac{1}{\Pt^2 + \overline{Q}^2 + m^2 + \frac{\xi(1-z)}{(1-\xi)}m^2}\Biggr \}  \\
& +  (\Pt^2 + \overline{Q}^2 + m^2) \int_{0}^{1}\ud \xi \left (-\frac{3(1-z)}{2(1-\xi)} + \frac{(1-z)}{2}\right )\frac{1}{\Pt^2 + \overline{Q}^2 + m^2 + \frac{\xi(1-z)}{(1-\xi)}m^2} \\
& - (\Pt^2 + \overline{Q}^2 + m^2) \int_{0}^{1} \ud \xi \int_{0}^{1} \frac{\ud x}{\mathcal{D}}  \frac{z^2}{(1-z)} (1-\xi)^2\\
& - (\Pt^2 + \overline{Q}^2 + m^2) \int_{0}^{1} \ud \xi \int_{0}^{1} \frac{\ud x}{\mathcal{D}}  z(1-\xi)^2 \biggl [x\xi \frac{(1-2z)}{(1-z)} - x - \xi \biggr ]\\
& + \int_{0}^{1} \ud \xi \int_{0}^{1} \frac{\ud x}{\mathcal{D}} z(1-\xi)^2 \Biggl \{ (\overline{Q}^2 + m^2)  \biggl [x\xi \frac{(1-2z)}{(1-z)} - x - \xi \biggr ] +    \frac{z m^2}{(1-z)^2} \frac{x[2(1-z) + z\xi^2]}{\left [x(1-\xi) + \frac{\xi}{(1-z)} \right ]} \Biggr \}\\
& + \biggl [z \leftrightarrow 1-z \biggr ].
\end{split}
\end{equation}
We can further simplify the expression in \eq\nr{eq:IVMSv1} by first performing the change of variables to $(y, \chi)$. This yields 
\begin{equation}
\label{eq:IVMSv2}
\begin{split}
I_{\mathcal{VMS}} & =  (\Pt^2 + \overline{Q}^2 + m^2)\int_{0}^{1}\frac{\ud \xi}{\xi} \left (-\frac{2\log(\xi)}{(1-\xi)} + \frac{(1+\xi)}{2}\right )\Biggl \{\frac{1}{\Pt^2 + \overline{Q}^2 + m^2}- \frac{1}{\Pt^2 + \overline{Q}^2 + m^2 + \frac{\xi(1-z)}{(1-\xi)}m^2}\Biggr \}  \\
& +  (\Pt^2 + \overline{Q}^2 + m^2) \int_{0}^{1}\ud \xi \left (-\frac{3(1-z)}{2(1-\xi)} + \frac{(1-z)}{2}\right )\frac{1}{\Pt^2 + \overline{Q}^2 + m^2 + \frac{\xi(1-z)}{(1-\xi)}m^2} \\
&  + (\Pt^2 + \overline{Q}^2 + m^2) \int_{0}^{1} \frac{\ud y}{(1-y)}  \int_{0}^{z} \frac{\ud \chi}{(1-\chi)} \frac{-z + y(1-z) + \frac{y}{z}\left (z - \chi\right )\left [z + \chi y - y(1-z) \right ]}{\left\{
\Pt^2 + \overline{Q}^2 + m^2 
+ \frac{y}{(1-y)} \frac{(1-z)}{(1-\chi)} \left[m^2+\frac{\chi(1 - \chi)}{z(1 - z)} \overline{Q}^2\right]
\right\}}   \\
& + (1-z)\int_{0}^{1} \frac{\ud y}{(1-y)} \int_{0}^{z} \frac{\ud \chi}{(1-\chi)} \frac{-\left (\overline{Q}^2 + m^2 \right )y(1-y) \bigg [1 + \frac{y}{(1-y)}\frac{\chi(1-\chi)}{z(1-z)}\biggr ] +  m^2\frac{ \chi}{(1-\chi)}\left [2 +  \frac{y^2}{z(1-z)}\left (z - \chi\right )^2 \right ]}{\left\{
\Pt^2 + \overline{Q}^2 + m^2 
+ \frac{y}{(1-y)} \frac{(1-z)}{(1-\chi)} \left[m^2+\frac{\chi(1 - \chi)}{z(1 - z)} \overline{Q}^2\right]
\right\}}  \\
& + \biggl [z \leftrightarrow 1-z \biggr ].
\end{split}
\end{equation}

\section{Coordinate space LCWF}
\label{sec:NLOmixedspace}

We now Fourier transform the full NLO result of $\gamma^{\ast}_T \rightarrow q\bar q$ in \eq\nr{eq:NLOformfactorsv3} into mixed space. 
We first factor out the exponential dependence on the center-of-mass coordinate of the dipole and the momentum of the photon as described in \cite{Beuf:2021qqa}. This yields
\begin{equation}
\label{eq:finalNLOmixspace}
\widetilde{\Psi}^{\gamma^{\ast}_T\rightarrow q\bar{q}} = \delta_{\alpha_0\alpha_1}e^{i(\qt/q^{+})\cdot \left (k^+_0\xt_0 + k^+_1\xt_1\right )}\widetilde{\psi}^{\gamma^{\ast}_T\rightarrow q\bar{q}},
\end{equation}
where the reduced LCWF $\widetilde{\psi}^{\gamma^{\ast}_T\rightarrow q\bar{q}}$ up to NLO accuracy in coupling $\alpha_s$ reads
\begin{equation}
\widetilde{\psi}^{\gamma^{\ast}_T\rightarrow q\bar{q}} = \widetilde{\psi}^{\gamma^{\ast}_T\rightarrow q\bar{q}}_{\lo} + \widetilde{\psi}^{\gamma^{\ast}_T\rightarrow q\bar{q}}_{\nlo} + O(e\alpha_s^2).
\end{equation}
Here the leading order result $\widetilde{\psi}^{\gamma^{\ast}_T\rightarrow q\bar{q}}_{\lo}$ is given by \eq\nr{eq:psireducedLO} and the NLO piece Fourier transformed to mixed space reads
\begin{equation}
\label{eq:finalNLOmixspacereducedv2}
\begin{split}
\widetilde{\psi}^{\gamma^{\ast}_T\rightarrow q\bar{q}}_{\nlo}  = -\frac{ee_f}{2\pi} & \left (\frac{\alpha_s\cf}{2\pi}\right )  \Biggl \{ \Biggl [\left (\frac{k_0^+-k_1^+}{q^+} \right )\delta^{ij}_{(D_s)} \bar{u}(0)\gamma^{+}v(1)  + \frac{1}{2}\bar{u}(0)\gamma^{+}[\gamma^i, \gamma^j]v(1)  \Biggr ] \mathcal{F} \left [\Pt^i \mathcal{V}^T \right ]\\
& + \bar u(0)\gamma^+ v(1) \mathcal{F} \left [\Pt^j \mathcal{N}^T \right ] +   m \bar u(0)\gamma^+\gamma^i v(1) \mathcal{F}\left [\left (\frac{\Pt^i \Pt^j}{\Pt^2} - \frac{\delta^{ij}}{2} \right )\mathcal{S}^T \right ]\\
& - m \bar u(0)\gamma^+\gamma^ {j}v(1)\mathcal{F} \left [\mathcal{M}^T + \mathcal{V}^T - \frac{\mathcal{S}^T}{2}\right ]
\Biggr \}\epst^{j}_{\lambda}, 
\end{split}
\end{equation}
where the Fourier operator $\mathcal{F}$ is defined in \eq\nr{eq:FToperator}. We remind the reader that in this basis, our expression for the form factor $\mathcal{S}^T$ is written in a symmetric and traceless form, which turns out to be very convenient for the Fourier transformation into mixed space. In addition, as discussed in previous section, the combination $\mathcal{M}^T + \mathcal{V}^T - \mathcal{S}^T/2$ is quite straightforward to compute both in momentum and mixed space due to the simple analytical structure.

Before computing the Fourier transforms in \eq\nr{eq:finalNLOmixspacereducedv2}, we would like to clarify some important points. Firstly, all the UV finite terms appearing in this expression can be Fourier transformed in four dimensions and only the UV regularization dependent terms (including the $D\rightarrow 4$ pole term) need to be Fourier transformed in $D$ dimensions. Secondly, the terms which are independent of the transverse momenta $\Pt$ trivially factor out from the Fourier transform. Thirdly, the terms exhibit a number of different types of   $\Pt$ dependence. All the corresponding transverse Fourier integrals needed for these terms are given in Appendix~\ref{app:FTSqqbarcase}. 

Putting all these steps together, we first Fourier transform the momentum space expression of $\mathcal{V}^T$ in \eq\nr{eq:VTfull}. The final result can be cast into the following form
\begin{equation}
\label{eq:PiVi}
\begin{split}
    \mathcal{F}\biggl [\Pt^i \mathcal{V}^T \biggr ]  & = \frac{i\xt_{01}^i}{ \vert\xt_{01}\vert} \left (\frac{\kaz}{2\pi\vert \xt_{01}\vert} \right )^{\frac{D}{2}-2}\Biggl \{\biggl [\frac{3}{2} + \log \left ( \frac{\alpha}{z} \right ) + \log \left (\frac{\alpha}{1-z}\right ) \biggr ] \\
    & \times \biggl \{\frac{(4\pi)^{2-\frac{D}{2}}}{(2-\frac{D}{2})}\Gamma\left (3 - \frac{D}{2}\right ) + \log\left (\frac{\vert \xt_{01}\vert^2 \mu^2}{4}\right ) + 2\gamma_E\biggr \} + \frac{1}{2}\frac{(D_s-4)}{(D-4)} \Biggr \} \kaz K_{\frac{D}{2}-1}\left (\vert \xt_{01}\vert \kaz\right )\\
    & + \frac{i\xt_{01}^i}{\vert \xt_{01}\vert} \Biggl \{\biggl [\frac{5}{2} - \frac{\pi^2}{3} + \log^2\left (\frac{z}{1-z} \right ) + \Omega_{\mathcal{V}} + L\biggr ] \kaz K_{1}\left (\vert \xt_{01}\vert \kaz \right )  + \widetilde{I}_{\mathcal{V}} + O(D-4)\Biggr \},
\end{split}
\end{equation}
where the function $\widetilde{I}_{\mathcal{V}}$ is given by
\begin{equation}
\label{eq:IVT}
\begin{split}
  \widetilde{I}_{\mathcal{V}} & = \int_0^1 \frac{\ud \xi}{\xi} \left (\frac{2\log(\xi)}{(1-\xi)} - \frac{(1+\xi)}{2} \right ) \Biggl \{\sqrt{\kaz^2 + \frac{\xi(1-z)}{(1-\xi)}m^2}K_{1}\left (\vert \xt_{01}\vert \sqrt{\kaz^2 + \frac{\xi(1-z)}{(1-\xi)}m^2}\right ) - [\xi \to 0]\Biggr \}\\
 & -\int_0^1 \ud \xi \left (\frac{\log(\xi)}{(1-\xi)^2} + \frac{z}{(1-x)} + \frac{z}{2} \right ) \frac{(1-z)m^2}{\sqrt{\kaz^2 + \frac{\xi(1-z)}{(1-\xi)}m^2}}K_{1}\left (\vert \xt_{01}\vert \sqrt{\kaz^2 + \frac{\xi(1-z)}{(1-\xi)}m^2}\right )\\
& - \int_{0}^{z} \frac{\ud \chi}{(1-\chi)}
\int_{0}^{\infty}\frac{\ud u }{u(u + 1)} \frac{ m^2}{\kac^2} \biggl [2\chi + \left (\frac{u}{1+u}\right )^2 \frac{1}{z}(z-\chi)(1-2\chi) \biggr ]\\
& \hspace{2cm} \times \Biggl \{\sqrt{\kaz^2 + u\frac{(1-z)}{(1-\xi)}\kac^2}K_{1}\left (\vert \xt_{01}\vert \sqrt{\kaz^2 + u\frac{(1-z)}{(1-\xi)}\kac^2}\right ) - [u \to 0] \Biggr \}\\
& -  \int_{0}^{z} \frac{\ud \chi}{(1-\chi)^2}
\int_{0}^{\infty}\frac{\ud u}{(u + 1)} \, \left (z - \chi\right )\left [1 - \frac{2u}{1+u}\left (z - \chi\right ) +  \left (\frac{u}{1+u} \right )^2 \frac{1}{z}\left (z - \chi\right )^2 \right ]\\
& \hspace{2cm}\times \frac{m^2}{\sqrt{\kaz^2 + u\frac{(1-z)}{(1-\chi)}\kac^2}}K_{1}\left (\vert \xt_{01}\vert \sqrt{\kaz^2 + u\frac{(1-z)}{(1-\chi)}\kac^2}\right )\\
& +  \biggl [z \leftrightarrow 1-z \biggr ].
\end{split}
\end{equation}
To obtain \eq\nr{eq:IVT}, we have additionally performed the following
change of variable $y \to u = y/(1-y)$ in \eq\nr{eq:VTeintmod2}. In addition, the above expressions use again  the function $\kappa_v = \sqrt{v(1-v)Q^2 + m^2}$ for the inverse transverse length scale associated with momentum fraction $v \in \{z, \chi\}$. 

Next, we Fourier transform the momentum space expression of $\mathcal{N}^T$ in \eq\nr{eq:fullNT}. The final result can be cast into the following form
\begin{equation}
\label{eq:PiNi}
    \mathcal{F} \biggl [\Pt^j \mathcal{N}^T \biggr ] = \frac{i\xt_{01}^j}{\vert \xt_{01}\vert}  \Biggl \{ \Omega_{\mathcal{N}} \,  \kaz  K_{1}\left (\vert \xt_{01}\vert \kaz \right ) + \widetilde{I}_{\mathcal{N}}  \Biggr \},
\end{equation}
where the function $\widetilde{I}_{\mathcal{N}}$ is given by
\begin{equation}
\label{eq:INT}
\begin{split}
  \widetilde{I}_{\mathcal{N}}   & = \frac{2(1-z)}{z} \int_0^z \ud \chi \int_0^\infty \frac{\ud u}{(u+1)^3} \Biggl \{\left [(2+u)u z + u^2\chi\right ]\sqrt{\kaz^2 + u \frac{(1-z)}{(1-\chi)}\kac^2} \, K_1\left (\vert \xt_{01}\vert  \sqrt{\kaz^2 + u \frac{(1-z)}{(1-\chi)}\kac^2} \right ) 
  \\
&
+ \frac{m^2}{\kac^2} 
\biggl (\frac{z}{1-z} + \frac{\chi}{1 - \chi} [u  - 2z - 2u\chi] \biggr )
\Biggl [\sqrt{\kaz^2 + u \frac{(1-z)}{(1-\chi)}\kac^2} \, K_1\left (\vert \xt_{01}\vert  \sqrt{\kaz^2 + u \frac{(1-z)}{(1-\chi)}\kac^2} \right ) -[u\to 0] \Biggr ]\Biggr \}\\
& -  \biggl [z \leftrightarrow 1-z \biggr ].
\end{split}
\end{equation}
Similarly, Fourier transforming the momentum space expression of $\mathcal{S}^T$ in \eq\nr{eq:fullST} yields
\begin{equation}
\begin{split}
    \mathcal{F} \biggl [\left (\frac{\Pt^i \Pt^j}{\Pt^2} - \frac{\delta^{ij}}{2}\right )\mathcal{S}^T \biggr ] & = \frac{(1 - z)}{2} \left[\frac{\xt_{01}^i\xt_{01}^j}{\vert \xt_{01}\vert^2}- \frac{\delta^{ij}}{2}\right]
\int_{0}^{z} \frac{\ud \chi}{(1 - \chi)}\,
\int_{0}^{\infty}\frac{\ud u}{(u + 1)^2} \,
\vert \xt_{01}\vert  \sqrt{\kaz^2 + u \frac{(1 - z)}{(1 - \chi)} \kac^2}
\\
& \hspace{1cm}
\times 
K_1\left(\vert \xt_{01}\vert \sqrt{\kaz^2 + u \frac{(1 - z)}{(1 - \chi)} \kac^2}\right) +  \biggl [z \leftrightarrow 1-z \biggr ].
\end{split}
\end{equation}

Finally, the result for the Fourier transformed form factor combination $\mathcal{V}^T + \mathcal{M}^T - \mathcal{S}^T/2$ in \eq\nr{eq:VMSfull} reads 
\begin{equation}
\label{eq:VMSscalar}
\begin{split}
& \mathcal{F}\biggl [\mathcal{V}^T + \mathcal{M}^T - \frac{\mathcal{S}^T}{2} \biggr ]  = \left (\frac{\kaz}{2\pi\vert \xt_{01}\vert} \right )^{\frac{D}{2}-2}\Biggl \{\biggl [\frac{3}{2} + \log\left ( \frac{\alpha}{z}\right ) +  \log\left ( \frac{\alpha}{1-z}\right ) \biggr ]\biggl \{\frac{(4\pi)^{2-\frac{D}{2}}}{(2-\frac{D}{2})}\Gamma \left (3 - \frac{D}{2}\right ) + \log \left (\frac{\vert \xt_{01}\vert^2 \mu^2}{4}\right )\\
& + 2\gamma_E \biggr \} +  \frac{1}{2}\frac{(D_s-4)}{(D-4)} \Biggr \}K_{\frac{D}{2}-2}\left (\vert \xt_{01}\vert \kaz \right ) +  \Biggl \{ 3 - \frac{\pi^2}{3}  + \log^2 \left (\frac{z}{1-z} \right )  + \Omega_{\mathcal{V}} + L  \Biggr \}K_{0}\left (\vert \xt_{01}\vert \kaz \right )   + \widetilde{I}_{\mathcal{VMS}} + O(D-4), 
\end{split}    
\end{equation}
where the function $\widetilde{I}_{\mathcal{VMS}}$ is given by
\begin{equation}
\label{eq:IVMST}
\begin{split}
\widetilde{I}_{\mathcal{VMS}}  & = \int_{0}^{1}\frac{\ud \xi}{\xi} \left (\frac{2\log(\xi)}{(1-\xi)} - \frac{(1+\xi)}{2}\right )\Biggl \{ K_{0}\left (\vert \xt_{01}\vert \sqrt{\kaz^2  + \frac{\xi(1-z)}{(1-\xi)}m^2}\right )  -[\xi \to 0] \Biggr \}\\
& + \int_{0}^{1}\ud \xi \left (-\frac{3(1-z)}{2(1-\xi)} + \frac{(1-z)}{2}\right ) K_{0}\left (\vert \xt_{01}\vert \sqrt{\kaz^2  + \frac{\xi(1-z)}{(1-\xi)}m^2}\right )\\
& + \int_{0}^{z} \frac{\ud \chi}{(1-\chi)}\int_{0}^{\infty} \frac{\ud u}{(u+1)^2} \biggl \{ -z - \frac{u}{(1+u)}\frac{(z +u\chi)}{z}(\chi  - (1-z)) \biggr \} 
K_{0}\left (\vert \xt_{01}\vert \sqrt{\kaz^2  + u\frac{(1-z)}{(1-\chi)}\kac^2}\right )\\
& + \int_{0}^{z} \ud \chi \int_{0}^{\infty} \frac{\ud u}{(u+1)^3}  \biggl \{\frac{\kaz^2}{\kac^2} \biggl [1 + u\frac{\chi(1-\chi)}{z(1-z)}\biggr ] - \frac{m^2}{\kac^2} \frac{\chi}{(1-\chi)}\bigg [2\frac{(1+u)^2}{u} +  \frac{u}{z(1-z)}\left (z - \chi\right )^2 \biggr ]\biggr \}\\
&  \hspace{2cm}  \times \Biggl \{K_{0}\left (\vert \xt_{01}\vert \sqrt{\kaz^2  + u\frac{(1-z)}{(1-\chi)}\kac^2}\right ) - [u \to 0]    \Biggr \}\\
& +  \biggl [z \leftrightarrow 1-z \biggr ].
\end{split}
\end{equation}
We have now obtained the full mass-renormalized one-loop mixed space LCWF for $\gamma^*\to q\bar{q}$ with massive quarks, the most important result of our paper. This result, without any details of it derivation, is also shown in the shorter companion paper~\cite{Beuf:2021srj}.

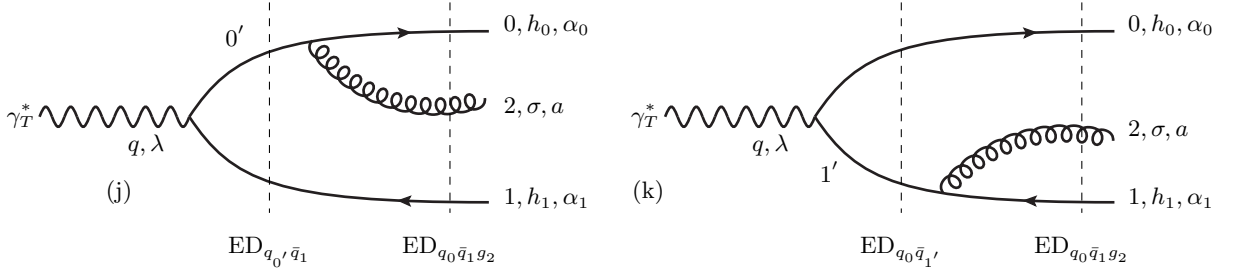
\begin{figure}[tbh!]
\centerline{
\includegraphics[width=6.4cm]{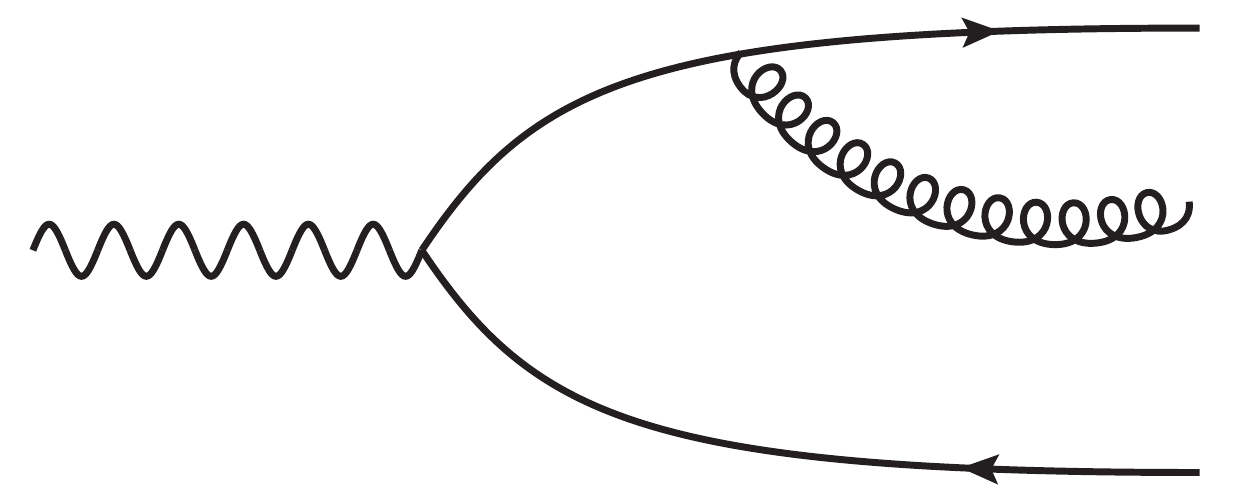}
\hspace*{1.7cm}
\includegraphics[width=6.4cm]{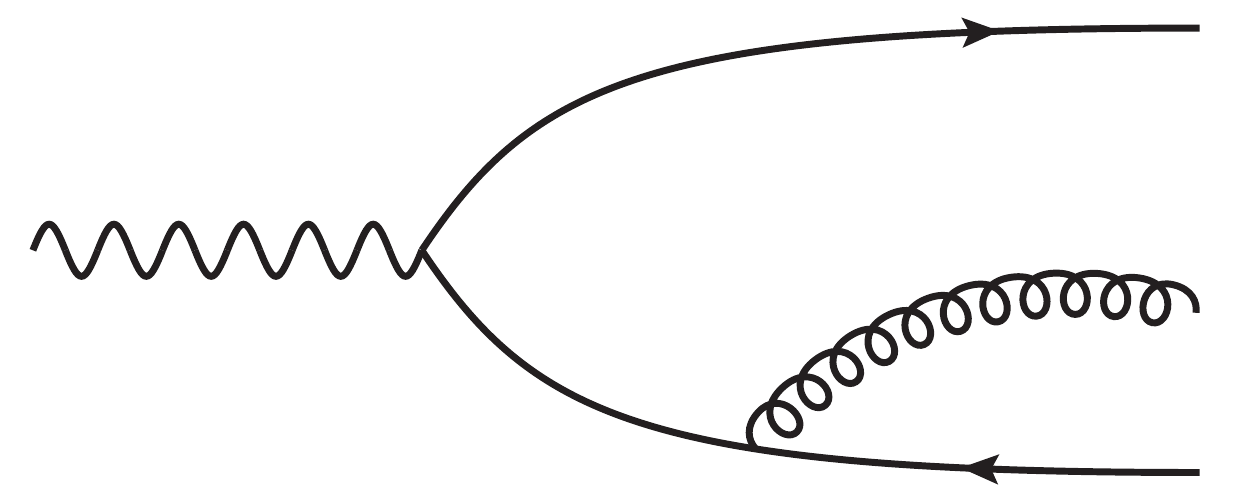}
\begin{tikzpicture}[overlay]
\draw [dashed] (-11.6,2.8) -- (-11.6,0);
\node[anchor=north] at (-11.6cm,-0.2cm) {$\ed_{q_{0'}\bar q_{1}}$};
\draw [dashed] (-9.2,2.8) -- (-9.2,0);
\node[anchor=north] at (-9.2,-0.2) {$\ed_{q_{0}\bar q_{1}g_2}$};
\node[anchor=west] at (-8.6,2.5) {$0,h_0,\alpha_0$};
\node[anchor=west] at (-8.6,1.4) {$2,\sigma,a$};
\node[anchor=west] at (-8.6,0.2) {$1,h_1,\alpha_1$};
\node[anchor=west] at (-13.6,0.9) {$q, \lambda $};
\node[anchor=south west] at (-12.3,2.1) {$0'$};
\node[anchor=east] at (-14.6,1.3) {$\gamma^{\ast}_{T}$};
\draw [dashed] (-3.2,2.8) -- (-3.2,0);
\node[anchor=north] at (-3.2cm,-0.2cm) {$\ed_{q_{0}\bar q_{1'}}$};
\draw [dashed] (-0.8,2.8) -- (-0.8,0);
\node[anchor=north] at (-0.8,-0.2) {$\ed_{q_{0}\bar q_{1}g_2}$};
\node[anchor=west] at (-0.3,2.5) {$0,h_0,\alpha_0$};
\node[anchor=west] at (-0.3,1.1) {$2,\sigma,a$};
\node[anchor=west] at (-0.3,0.2) {$1,h_1,\alpha_1$};
\node[anchor=west] at (-5.3,0.9) {$q, \lambda $};
\node[anchor=south west] at (-4.4,0.3) {$1'$};
\node[anchor=east] at (-6.3,1.3) {$\gamma^{\ast}_{T}$};
\node[anchor=south west] at (-14cm,0cm) {\namediag{diag:qgqbarT}};
\node[anchor=south west] at (-7cm,0cm) {\namediag{diag:qqbargT}};
 \end{tikzpicture}
}
\rule{0pt}{1ex}
\caption{Tree level gluon emission diagrams \ref{diag:qgqbarT} and \ref{diag:qqbargT} contributing to the quark-antiquark-gluon component of the transverse virtual photon wave function at NLO. Imposing  plus and transverse momentum conservation at each vertex gives for diagram \ref{diag:qgqbarT}: $\hat k_{0'} = \hat k_0 + \hat k_2$ and $\hat q = \hat k_0 + \hat k_1 + \hat k_2$. Similarly for diagram \ref{diag:qqbargT}: $\hat k_{1'} = \hat k_1 + \hat k_2$ and $\hat q = \hat k_0 + \hat k_1 + \hat k_2$.}
\label{fig:qgqbarT}
 \end{figure}
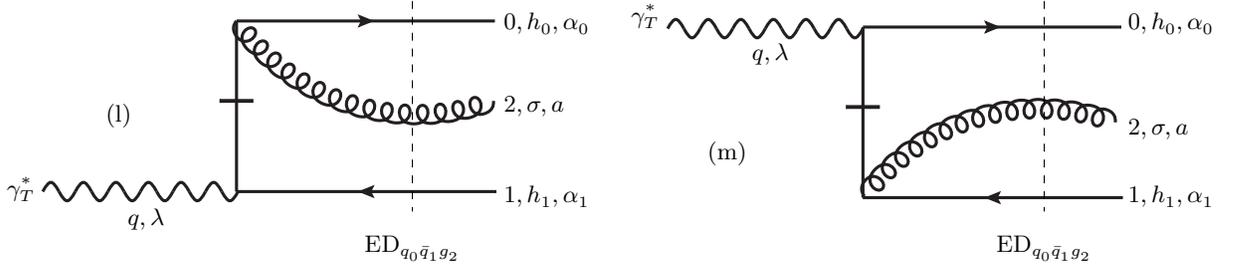
\begin{figure}[tbh!]
\centerline{
\includegraphics[width=6.4cm]{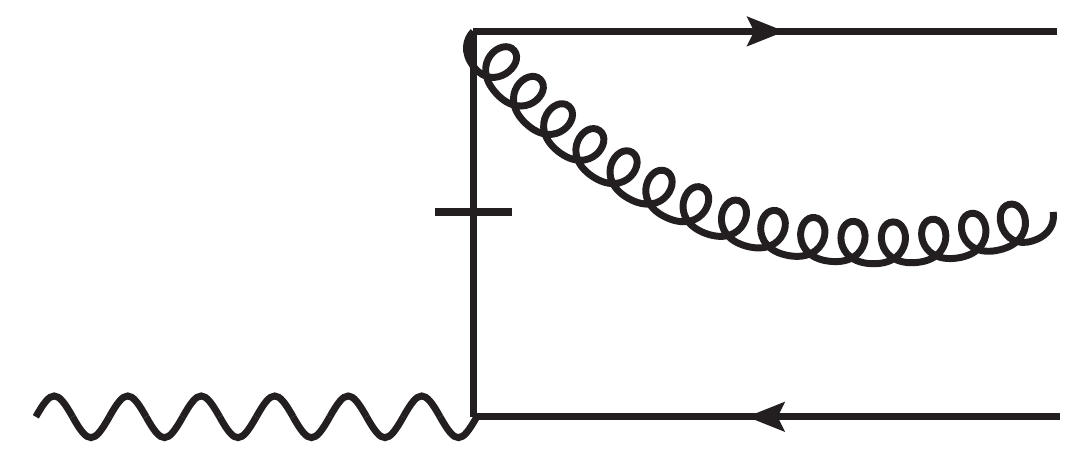}
\hspace*{1.7cm}
\includegraphics[width=6.4cm]{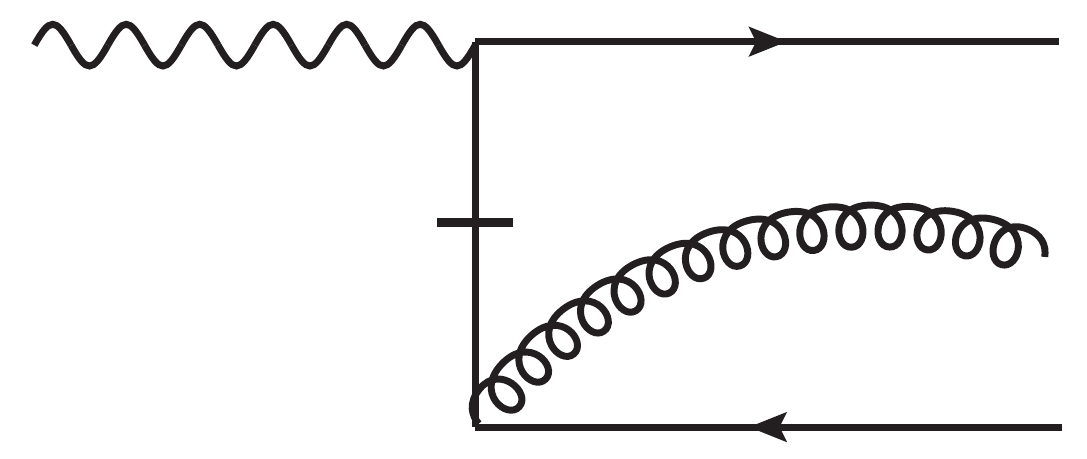}
\begin{tikzpicture}[overlay]
\draw [dashed] (-9.7,2.8) -- (-9.7,0);
\node[anchor=north] at (-9.7,-0.2) {$\ed_{q_{0}\bar q_{1}g_2}$};
\node[anchor=west] at (-8.6,2.5) {$0,h_0,\alpha_0$};
\node[anchor=west] at (-8.6,1.4) {$2,\sigma,a$};
\node[anchor=west] at (-8.6,0.2) {$1,h_1,\alpha_1$};
\node[anchor=west] at (-13.6,-0.1) {$q, \lambda $};
\node[anchor=east] at (-14.6,0.3) {$\gamma^{\ast}_{T}$};
\draw [dashed] (-1.3,2.8) -- (-1.3,0);
\node[anchor=north] at (-1.3,-0.2) {$\ed_{q_{0}\bar q_{1}g_2}$};
\node[anchor=west] at (-0.3,2.5) {$0,h_0,\alpha_0$};
\node[anchor=west] at (-0.3,1.1) {$2,\sigma,a$};
\node[anchor=west] at (-0.3,0.2) {$1,h_1,\alpha_1$};
\node[anchor=west] at (-5.3,2.1) {$q, \lambda $};
\node[anchor=east] at (-6.3,2.6) {$\gamma^{\ast}_{T}$};
\node[anchor=south west] at (-14cm,1cm) {\namediag{diag:qgqbarTinst}};
\node[anchor=south west] at (-6cm,0.5cm) {\namediag{diag:qqbargTinst}};
 \end{tikzpicture}
}
\rule{0pt}{1ex}
\caption{Tree level instantaneous gluon emission diagrams \ref{diag:qgqbarTinst} and \ref{diag:qqbargTinst} contributing to the quark-antiquark-gluon component of the transverse virtual photon wave function at NLO. Imposing plus and  transverse momentum conservation at each vertex gives for diagrams \ref{diag:qgqbarTinst} and \ref{diag:qqbargTinst}: $\hat q = \hat k_0 + \hat k_1 + \hat k_2$.}
\label{fig:qgqbarTinst}
 \end{figure}

\section{Tree level gluon emission}
\label{sec:qqbargwf}

\subsection{Momentum space wavefunction}

We then move to the tree level wave functions for gluon emission from a transverse photon state, which are needed for the full cross section at NLO. As in the massless case \cite{Beuf:2017bpd,Hanninen:2017ddy}, we need to calculate four gluon emission diagrams, shown in \figs\ref{fig:qgqbarT} and \ref{fig:qgqbarTinst}.

Applying the diagrammatic LCPT rules and following the notations shown in \figs\ref{fig:qgqbarT} and \ref{fig:qgqbarTinst}, we obtain for the diagrams \ref{diag:qgqbarT} and \ref{diag:qgqbarTinst} in momentum space
\begin{equation}
\label{eq:LqgqbaT}
\begin{split}
\Psi^{\gamma^{\ast}_{T}\rightarrow q\bar{q}g}_{\ref{diag:qgqbarT}} & = \int \dkpzero  (2\pi)^{D-1}\delta^{(D-1)}(\hat k_{0'} - \hat k_2 - \hat k_0) \frac{gt^{a}_{\alpha_0\alpha_1}[\bar{u}(0)\epsl^{\ast}_{\sigma}(k_2)u(0')] ee_f[\bar{u}(0')\epsl_{\lambda}(q)v(1)]}{\ed_{q_{0'}\bar q_1}\ed_{q_0\bar q_1g_2}} \\
& = \frac{+ee_fgt^{a}_{\alpha_0\alpha_1}}{2(k^+_0 + k^+_2)} \frac{\bar{u}(0)\epsl^{\ast}_{\sigma}(k_2)u(0')\bar{u}(0')\epsl_{\lambda}(q)v(1)}{\ed_{q_{0'}\bar q_1}\ed_{q_0\bar q_1g_2}}
\end{split}
\end{equation} 
and
\begin{equation}
\label{eq:LqgqbaTinst}
\Psi^{\gamma^{\ast}_{T}\rightarrow q\bar{q}g}_{\ref{diag:qgqbarTinst}} = \frac{+ee_fgt^{a}_{\alpha_0\alpha_1}}{2(k^+_0 + k^+_2)} \frac{\bar{u}(0)\epsl^{\ast}_{\sigma}(k_2)\gamma^{+}\epsl_{\lambda}(q)v(1)}{\ed_{q_0\bar q_1g_2}},
\end{equation} 
where the energy denominators appearing in \eq\nr{eq:LqgqbaT} and \nr{eq:LqgqbaTinst} can be written as 
\begin{equation}
\begin{split}
\ed_{q_{0'}\bar q_1} & = \frac{\qt^2 - Q^2}{2q^+} - \left (\frac{\kt^2_{0'} + m^2}{2k^+_{0'}} + \frac{\kt^2_1 + m^2}{2k^+_1} \right ) \\
		 & = \frac{-q^+}{2k^+_1(k^+_0 + k^+_2)}\Biggl [\left (\kt_1 - \frac{k^+_1}{q^+}\qt\right )^2 + \overline{Q}^2_{\ref{diag:qgqbarT}} + m^2\Biggr ] \\
\end{split}
\end{equation}
\begin{equation}
\begin{split}
\label{eq:defEDqqbg1}
\ed_{q_{0}\bar q_1g_2} & = \frac{\qt^2 - Q^2}{2q^+} - \left (\frac{\kt^2_0 + m^2}{2k^+_0} + \frac{\kt^2_1 + m^2}{2k^+_1} + \frac{\kt^2_2}{2k^+_2} \right ) \\
& = \frac{-k^+_0}{2k^+_2(k^+_0 + k^+_2)}\Biggl [\left (\kt_2 - \frac{k^+_2}{k^+_0}\kt_0\right )^2 + \frac{q^+k^+_2}{k^+_0k^+_1}\Biggl \{\left (\kt_1 - \frac{k^+_1}{q^+}\qt\right )^2 + \overline{Q}^2_{\ref{diag:qgqbarT}} + m^2 + \lambda_{\ref{diag:qgqbarT}} m^2\Biggr \}\Biggr ],
\end{split}
\end{equation}
and the coefficients introduced in the denominators are defined as 
\begin{equation}
\label{eq:coeffqgbarq}
\begin{split}
\overline{Q}^2_{\ref{diag:qgqbarT}} & = \frac{k^+_1(q^+-k^+_1)}{(q^+)^2}Q^2,\quad \lambda_{\ref{diag:qgqbarT}} = \frac{k^+_1k^+_2}{q^+k^+_0}.\\
\end{split}
\end{equation}

Similarly, for the diagrams \ref{diag:qqbargT} and \ref{diag:qqbargTinst} we get 
\begin{equation}
\label{eq:qqbargT}
\begin{split}
\Psi^{\gamma^{\ast}_{T}\rightarrow q\bar{q}g}_{\ref{diag:qqbargT}} & = \int \dkpone  (2\pi)^{D-1}\delta^{(D-1)}(\hat k_{1'} - \hat k_2 - \hat k_1) \frac{ee_f[\bar{u}(0)\epsl_{\lambda}(q)v(1')] (-gt^{a}_{\alpha_0\alpha_1})[\bar{v}(1')\epsl^{\ast}_{\sigma}(k_2)v(1)]}{\ed_{q_{0}\bar q_{1'}}\ed_{q_0\bar q_1g_2}} \\
& = \frac{-ee_fgt^{a}_{\alpha_0\alpha_1}}{2(k^+_1 + k^+_2)} \frac{\bar{u}(0)\epsl_{\lambda}(q)v(1')\bar{v}(1')\epsl^{\ast}_{\sigma}(k_2)v(1)}{\ed_{q_{0}\bar q_{1'}}\ed_{q_0\bar q_1g_2}}
\end{split}
\end{equation} 
and
\begin{equation}
\label{eq:qqbargTinst}
\Psi^{\gamma^{\ast}_{T}\rightarrow q\bar{q}g}_{\ref{diag:qqbargTinst}} =     \frac{-ee_fgt^{a}_{\alpha_0\alpha_1}}{2(k^+_1 + k^+_2)} \frac{\bar{u}(0)\epsl_{\lambda}(q)\gamma^+\epsl^{\ast}_{\sigma}(k_2)v(1)}{\ed_{q_0\bar q_1g_2}},
\end{equation}
where the energy denominators appearing in \eqs\nr{eq:qqbargT} and \nr{eq:qqbargTinst} can be written as 
\begin{equation}
\begin{split}
\ed_{q_0\bar q_{1'}} & = \frac{\qt^2 - Q^2}{2q^+} - \left (\frac{\kt^2_{0} + m^2}{2k^+_{0}} + \frac{\kt^2_{1'} + m^2}{2k^+_{1'}} \right ) \\ 
	     & = \frac{-q^+}{2k^+_0(k^+_1 + k^+_2)}\Biggl [\left (\kt_0 - \frac{k^+_0}{q^+}\qt\right )^2 + \overline{Q}^2_{\ref{diag:qqbargT}} + m^2\Biggr ] \\
\end{split}
\end{equation}
\begin{equation}
\begin{split}
\label{eq:defEDqqbg2}
\ed_{q_0\bar q_1 g_2} & = \frac{\qt^2 - Q^2}{2q^+} - \left (\frac{\kt^2_0 + m^2}{2k^+_0} + \frac{\kt^2_1 + m^2}{2k^+_1} + \frac{\kt^2_2}{2k^+_2} \right ) \\
& = \frac{-k^+_1}{2k^+_2(k^+_1 + k^+_2)}\Biggl [\left (\kt_2 - \frac{k^+_2}{k^+_1}\kt_1\right )^2 + \frac{q^+k^+_2}{k^+_0k^+_1}\Biggl \{\left (\kt_0 - \frac{k^+_0}{q^+}\qt\right )^2 + \overline{Q}^2_{\ref{diag:qqbargT}} + m^2 + \lambda_{\ref{diag:qqbargT}} m^2\Biggr \}\Biggr ],
\end{split}
\end{equation}
and the coefficients introduced in the denominators are defined as 
\begin{equation}
\label{eq:coeffqbarqg}
\begin{split}
\overline{Q}^2_{\ref{diag:qqbargT}} & = \frac{k^+_0(q^+-k^+_0)}{(q^+)^2}Q^2,\quad \lambda_{\ref{diag:qqbargT}} = \frac{k^+_0k^+_2}{q^+k^+_1}.
\end{split}
\end{equation}
Note that the energy denominators \nr{eq:defEDqqbg1} and \nr{eq:defEDqqbg2} are in fact equal, but it is  convenient to write them in terms of different variables in the context of different diagrams.

The transverse momentum dependence in the massive spinors $u$ and $v$ and in the gluon or photon polarization vectors can be extracted by using the decompositions
\begin{equation}
\label{eq:decomp1}
\begin{split}
\bar{u}(0)\epsl^{\ast}_{\sigma}(k_2)u(0') = \frac{k^+_0}{k^+_2(k^+_0 + k^+_2)}\Biggl \{\Biggl [\left (1 + \frac{k^+_2}{2k^+_0}\right )\delta^{ij}_{(D_s)}\bar{u}(0)\gamma^{+}u(0') & - \left (\frac{k^+_2}{4k^+_0}\right )\bar{u}(0)\gamma^{+}[\gamma^i,\gamma^j]u(0') \Biggr ][\kt^i_2 - \frac{k^+_2}{k^+_0}\kt^i_0]\\
& - \frac{m}{2}\left (\frac{k^+_2}{k^+_0}\right )^2\bar{u}(0)\gamma^{+}\gamma^{j}u(0') \Biggr \}\epst^{\ast j}_{\sigma}
\end{split}
\end{equation}
\begin{equation}
\label{eq:decomp2}
\begin{split}
\bar{u}(0')\epsl_{\lambda}(q)v(1) = \frac{q^+}{k^+_1(q^+ - k^+_1)}\Biggl \{\Biggl [\left ( \frac{2k^+_1 - q^+}{2q^+}\right )\delta^{kl}_{(D_s)}\bar{u}(0')\gamma^{+}v(1) & - \frac{1}{4}\bar{u}(0')\gamma^{+}[\gamma^k,\gamma^l]v(1) \Biggr ][\kt^k_1 - \frac{k^+_1}{q^+}\qt^k]\\
& - \frac{m}{2}\bar{u}(0')\gamma^{+}\gamma^{l}v(1) \Biggr \}\epst^{l}_{\lambda}
\end{split}
\end{equation}
and
\begin{equation}
\label{eq:decomp3}
\begin{split}
\bar{u}(0)\epsl_{\lambda}(q)v(1') = \frac{q^+}{k^+_0(q^+ - k^+_0)}\Biggl \{\Biggl [\left ( \frac{2k^+_0 - q^+}{2q^+}\right )\delta^{kl}_{(D_s)}\bar{u}(0)\gamma^{+}v(1') & + \frac{1}{4}\bar{u}(0)\gamma^{+}[\gamma^k,\gamma^l]v(1') \Biggr ][\kt^k_0 - \frac{k^+_0}{q^+}\qt^k]\\
& - \frac{m}{2}\bar{u}(0)\gamma^{+}\gamma^{l}v(1') \Biggr \}\epst^{l}_{\lambda}
\end{split}
\end{equation}
\begin{equation}
\label{eq:decomp4}
\begin{split}
\bar{v}(1')\epsl^{\ast}_{\sigma}(k_2)v(1) = \frac{k^+_1}{k^+_2(k^+_1 + k^+_2)}\Biggl \{\Biggl [\left (1 + \frac{k^+_2}{2k^+_1}\right )\delta^{ij}_{(D_s)}\bar{v}(1')\gamma^{+}v(1) & + \left (\frac{k^+_2}{4k^+_1}\right )\bar{v}(1')\gamma^{+}[\gamma^i,\gamma^j]v(1) \Biggr ][\kt^i_2 - \frac{k^+_2}{k^+_1}\kt^i_1]\\
& - \frac{m}{2}\left (\frac{k^+_2}{k^+_1}\right )^2\bar{v}(1')\gamma^{+}\gamma^{j}v(1) \Biggr \}\epst^{\ast j}_{\sigma},
\end{split}
\end{equation}
where the remaining spinor matrix elements are independent of transverse momentum.
Inserting the decomposition in \eqs\nr{eq:decomp1} and \nr{eq:decomp2} into \eqs\nr{eq:LqgqbaT} and  \nr{eq:LqgqbaTinst}, and noting that in the light-cone gauge $\epsl_{\sigma}^{\ast}(k_2)\gamma^+\epsl_{\lambda}(q) = -\epst_{\lambda}^{l}\epst_{\sigma}^{\ast j}\gamma^+\gamma^j\gamma^l$ we find for the sum of the diagrams \ref{diag:qgqbarT} + \ref{diag:qgqbarTinst} the expression
\begin{equation}
\label{eq:finalLCWFsumjl}
\begin{split}
 \Psi^{\gamma^{\ast}_{T}\rightarrow q\bar{q}g}_{\ref{diag:qgqbarT} + \ref{diag:qgqbarTinst}} = & \frac{+ee_fgt^{a}_{\alpha_0\alpha_1}\epst_{\lambda}^{l}\epst_{\sigma}^{\ast j}}
 { \left (\kt_2 - \frac{k^+_2}{k^+_0}\kt_0\right )^2 
 + \frac{q^+k^+_2}{k^+_0k^+_1}\biggl \{\left (\kt_1 - \frac{k^+_1}{q^+}\qt\right )^2 + \overline{Q}^2_{\ref{diag:qgqbarT}} + m^2 + \lambda_{\ref{diag:qgqbarT}}m^2\biggr \} }
 \\
& \times \left \{ \frac{1}{ \left (\kt_1 - \frac{k^+_1}{q^+}\qt\right )^2 + \overline{Q}^2_{\ref{diag:qgqbarT}} + m^2 }\bar u(0)\gamma^{+}\mathcal{M}^{jl}_{\ref{diag:qgqbarT}}v(1) + \frac{k^+_2}{k^+_0}\bar u(0)\gamma^+\gamma^j\gamma^lv(1) \right\},
\end{split}
\end{equation}
where the Dirac structure $\mathcal{M}^{jl}_{\ref{diag:qgqbarT}}$ is defined as
\begin{equation}
\begin{split}
\label{eq:Mdiracqgqbar}
\mathcal{M}^{jl}_{\ref{diag:qgqbarT}}  = & \left \{\left[\left (\frac{2k^+_0 + k^+_2}{k^+_0} \right )\delta^{ij}_{(D_s)} - \left (\frac{k^+_2}{2k^+_0} \right )[\gamma^i,\gamma^j]\right] \left[\kt^i_2 - \frac{k^+_2}{k^+_0}\kt^i_0\right] - m \left (\frac{k^+_2}{k^+_0}\right )^2\gamma^j \right\}\\
& \times \left\{\left [\left (\frac{2k^+_1 - q^+}{q^+} \right )\delta^{kl}_{(D_s)} - \frac{1}{2}[\gamma^k,\gamma^l]\right]\left[\kt^k_1 - \frac{k^+_1}{q^+}\qt^k\right] - m\gamma^l \right \}.
\end{split}
\end{equation}
Similarly for the sum \ref{diag:qqbargT} + \ref{diag:qqbargTinst} inserting the decomposition in \eqs\nr{eq:decomp3} and \nr{eq:decomp4} into \eqs\nr{eq:qqbargT} and \nr{eq:qqbargTinst} yields the expression
\begin{equation}
\label{eq:finalLCWFsumkm}
\begin{split}
 \Psi^{\gamma^{\ast}_{T}\rightarrow q\bar{q}g}_{\ref{diag:qqbargT} + \ref{diag:qqbargTinst}} = & \frac{-ee_fgt^{a}_{\alpha_0\alpha_1}\epst_{\lambda}^{l}\epst_{\sigma}^{\ast j}}{\left (\kt_2 - \frac{k^+_2}{k^+_1}\kt_1\right )^2 + \frac{q^+k^+_2}{k^+_0k^+_1}\biggl \{\left (\kt_0 - \frac{k^+_0}{q^+}\qt\right )^2 + \overline{Q}^2_{\ref{diag:qqbargT}} + m^2 + \lambda_{\ref{diag:qqbargT}}m^2\biggr \} }\\
& \times \left \{ \frac{1}{ \left (\kt_0 - \frac{k^+_0}{q^+}\qt\right )^2 + \overline{Q}^2_{\ref{diag:qqbargT}} + m^2 }\bar u(0)\gamma^{+}\mathcal{M}^{jl}_{\ref{diag:qqbargT}}v(1) + \frac{k^+_2}{k^+_1}\bar u(0)\gamma^+\gamma^l\gamma^jv(1) \right \},
\end{split}
\end{equation}
where the Dirac structure $\mathcal{M}^{jl}_{\ref{diag:qqbargT}}$ is defined as
\begin{equation}
\begin{split}
\label{eq:Mdiracqqbarg}
\mathcal{M}^{jl}_{\ref{diag:qqbargT}}  = & 
\left\{\left [\left (\frac{2k^+_0 - q^+}{q^+} \right )\delta^{kl}_{(D_s)} + \frac{1}{2}[\gamma^k,\gamma^l]\right ] \left[\kt^k_0 - \frac{k^+_0}{q^+}\qt^k \right] - m\gamma^l \right \}\\
& \times \left\{ \left[\left (\frac{2k^+_1 + k^+_2}{k^+_1} \right )\delta^{ij}_{(D_s)} + \left (\frac{k^+_2}{2k^+_1} \right )[\gamma^i,\gamma^j]\right]\left[\kt^i_2 - \frac{k^+_2}{k^+_1}\kt^i_1\right] - m \left (\frac{k^+_2}{k^+_1}\right )^2\gamma^j \right\}.
\end{split}
\end{equation}

\subsection{Fourier transform to coordinate space}
\label{subsec:FTqqbargT}

The Fourier transform to mixed space of the $\gamma^{\ast}_T \to q\bar qg$ LCWF is given by the expression 
\begin{equation}
\label{eq:FTqqbarg}
\begin{split}
\widetilde{\Psi}^{\gamma^{\ast}_T\rightarrow q\bar{q}g} = (\mu^2)^{2-\frac{D}{2}}\int \frac{\ud^{D-2}\kt_0}{(2\pi)^{D-2}}\int \frac{\ud^{D-2}\kt_1}{(2\pi)^{D-2}}& \int \frac{\ud^{D-2}\kt_2}{(2\pi)^{D-2}}e^{i\kt_0\cdot \xt_0 + i\kt_1\cdot \xt_1 + i\kt_2\cdot \xt_2}\\
& \times (2\pi)^{D-2}\delta^{(D-2)}(\qt- \kt_0 - \kt_1-\kt_2) \Psi^{\gamma^{\ast}_T\rightarrow q\bar{q}g}. 
\end{split}
\end{equation}
To simplify the Fourier transformation, we first make the  change of variables $(\kt_1,\kt_2) \mapsto (\Pt,\Kt)$ in \eq\nr{eq:finalLCWFsumjl} with
\begin{equation}
\begin{split}
\Pt & = -\kt_1 + \frac{k^+_1}{q^+}\qt,\\
\Kt & = \kt_2 - \frac{k^+_2}{k^+_0 + k^+_2}(\kt_0 + \kt_2) = \left (\frac{k^+_0}{k^+_0 + k^+_2}\right )\left[\kt_2 - \frac{k^+_2}{k^+_0}\kt_0 \right],
\end{split}
\end{equation}
and correspondingly $(\kt_0,\kt_2) \mapsto (\Pt,\Kt)$ in \eq\nr{eq:finalLCWFsumkm} with
\begin{equation}
\begin{split}
\Pt & = \kt_0 - \frac{k^+_0}{q^+}\qt, \\
\Kt & = \kt_2 - \frac{k^+_2}{k^+_1 + k^+_2}(\kt_1 + \kt_2) = \left (\frac{k^+_1}{k^+_1 + k^+_2}\right )\left[\kt_2 - \frac{k^+_2}{k^+_1}\kt_1 \right].
\end{split}
\end{equation}

Here we also introduce  the following compact notation for the coordinate difference between a quark/antiquark $p$ and the center of mass of the two-particle system $m,n$ that it is recoiling against:
\begin{equation}
\label{eq:xtpnmdef}
\xt_{n+m;p} = -\xt_{p;n+m} = \left ( \frac{k^+_n\xt_n + k^+_m\xt_m}{k^+_n + k^+_m} \right ) - \xt_p,
\end{equation}
with $\xt_{nm} = \xt_n - \xt_m$.
Using these we introduce the relative coordinates corresponding to the gluon emission diagrams \ref{diag:qgqbarT} and \ref{diag:qqbargT}. For the diagram \ref{diag:qgqbarT} with emission from the quark, the natural coordinates are the separation between the gluon ($\kvec_2$) and the quark ($\kvec_0$)
\begin{equation}
\label{eq:defx2_a}
    \xt_{2;\ref{diag:qgqbarT}} \equiv  \xt_{20}
\end{equation}  
and the separation between the center of mass of the quark-gluon system and the antiquark ($\kvec_1$):
\begin{equation}
\label{eq:defx3_a}
\xt_{3;\ref{diag:qgqbarT}}\equiv   \xt_{0+2;1}  =     \left ( \frac{k^+_0\xt_0 + k^+_2\xt_2}{k^+_0 + k^+_2} \right ) - \xt_1.
\end{equation}
For the other diagram \ref{diag:qqbargT} with an emission from the antiquark, the natural coordinates that appear are the separation between the gluon and the antiquark 
\begin{equation}
\label{eq:defx2_b}
\xt_{2;\ref{diag:qqbargT}}\equiv  \xt_{21}
\end{equation}
and the separation between the quark and the center of mass of the antiquark-gluon system
\begin{equation}
\label{eq:defx3_b}
\xt_{3;\ref{diag:qqbargT}}\equiv  \xt_{0;1+2} = \xt_0 - \left ( \frac{k^+_1\xt_1 + k^+_2\xt_2}{k^+_1 + k^+_2} \right ).
\end{equation}

Next, performing the integration in \eq\nr{eq:FTqqbarg} over the delta function of transverse momenta, we obtain the expression
\begin{equation}
\label{eq:finalLCWFsumjlFT}
\widetilde{\Psi}^{\gamma^{\ast}_T\rightarrow q\bar{q}g}_{\ref{diag:qgqbarT} + \ref{diag:qgqbarTinst}} =  e^{\frac{i\qt}{q^+}\cdot \left (k^+_0\xt_0 + k^+_1\xt_1 + k^+_2\xt_2 \right )}(\mu^2)^{2-\frac{D}{2}}\int \frac{\ud^{D-2}\Pt}{(2\pi)^{D-2}}\int \frac{\ud^{D-2}\Kt}{(2\pi)^{D-2}} e^{i\Pt \cdot\xt_{3;\ref{diag:qgqbarT}}} e^{i\Kt \cdot    \xt_{2;\ref{diag:qgqbarT}} } \Psi^{\gamma^{\ast}_{T}\rightarrow q\bar{q}g}_{\ref{diag:qgqbarT} + \ref{diag:qgqbarTinst}},
\end{equation}
and 
\begin{equation}
\label{eq:finalLCWFsumkmFT}
\widetilde{\Psi}^{\gamma^{\ast}_T\rightarrow q\bar{q}g}_{\ref{diag:qqbargT} + \ref{diag:qqbargTinst}} =  e^{\frac{i\qt}{q^+}\cdot \left (k^+_0\xt_0 + k^+_1\xt_1 + k^+_2\xt_2 \right )}(\mu^2)^{2-\frac{D}{2}}\int \frac{\ud^{D-2}\Pt}{(2\pi)^{D-2}}\int \frac{\ud^{D-2}\Kt}{(2\pi)^{D-2}} e^{i\Pt \cdot \xt_{3;\ref{diag:qqbargT}}}e^{i\Kt \cdot \xt_{2;\ref{diag:qqbargT}}} \Psi^{\gamma^{\ast}_{T}\rightarrow q\bar{q}g}_{\ref{diag:qqbargT} + \ref{diag:qqbargTinst}},
\end{equation}
where the LCWF's in \eqs\nr{eq:finalLCWFsumjl} and \nr{eq:finalLCWFsumkm} are written in terms of the new variables $\Pt$ and $\Kt$ and the relative coordinate separations.

Adding the contributions in \eqs\nr{eq:finalLCWFsumjlFT} and \nr{eq:finalLCWFsumkmFT} together, we obtain the following result for the full $q\bar qg$ LCWF with massive quarks in mixed space
\begin{equation}
\label{eq:qbarqgFTfull}
\begin{split}
\widetilde{\Psi}^{\gamma^{\ast}_{T}\rightarrow q\bar{q}g} = t^{a}_{\alpha_0\alpha_1}e^{\frac{i\qt}{q^+}\cdot \left (k^+_0\xt_0 + k^+_1\xt_1 + k^+_2\xt_2 \right )}\widetilde{\psi}^{\gamma^{\ast}_{T}\rightarrow q\bar{q}g},
\end{split}
\end{equation}
where we have factorized out the dependence on the center of mass coordinate of the partonic system\footnote{See also the discussion in \cite{Beuf:2021qqa}} and the color structure from the reduced LCWF $\widetilde{\psi}^{\gamma^{\ast}_{T}\rightarrow q\bar{q}g}$. The reduced LCWF can be split into two parts: the part without light cone helicity flips (not proportional to $m$), and the part with one or two flips (proportional to $m,$ $m^2$)  as
\begin{equation}
\label{eq:reducedTCWFqqbarg}
\begin{split}
\widetilde{\psi}^{\gamma^{\ast}_{T}\rightarrow q\bar{q}g}= ee_fg\epst_{\lambda}^{l}\epst_{\sigma}^{\ast j}\biggl (\Sigma^{lj} + \Sigma^{lj}_m \biggr ).
\end{split}
\end{equation}
Here the light cone helicity nonflip contribution $\Sigma^{lj}$ is given by the following expression
\begin{equation}
\label{eq:sigmamassless}
\begin{split}
\Sigma^{lj} = & -\frac{1}{(k^+_0 + k^+_2)q^+} \bar{u}(0)\gamma^+\biggl [(2k^+_0 + k^+_2)\delta^{ij}_{(D_s)} - \frac{k^+_2}{2}[\gamma^i,\gamma^j] \biggr ]\biggl [(2k^+_1 - q^+)\delta^{kl}_{(D_s)} - \frac{q^+}{2}[\gamma^k,\gamma^l] \biggr ] v(1)\mathcal{I}^{ik}_{\ref{diag:qgqbarT}}\\
& -\frac{1}{(k^+_1 + k^+_2)q^+} \bar{u}(0)\gamma^+\biggl [(2k^+_0 - q^+)\delta^{kl}_{(D_s)} + \frac{q^+}{2}[\gamma^k,\gamma^l] \biggr ]\biggl [(2k^+_1 + k^+_2)\delta^{ij}_{(D_s)} + \frac{k^+_2}{2}[\gamma^i,\gamma^j] \biggr ] v(1)\mathcal{I}^{ik}_{\ref{diag:qqbargT}}\\
& + \frac{k^+_2k^+_0}{(k^+_0 + k^+_2)^2}\bar u(0)\gamma^+\gamma^j\gamma^l v(1)\mathcal{J}_{\ref{diag:qgqbarTinst}} - \frac{k^+_2k^+_1}{(k^+_1 + k^+_2)^2}\bar u(0)\gamma^+\gamma^l\gamma^j v(1)\mathcal{J}_{\ref{diag:qqbargTinst}}
\end{split}    
\end{equation}
and correspondingly the contribution with one or two light cone helicity flips, proportional to powers of the mass, $\Sigma^{lj}_m$ is given by 
\begin{equation}
\label{eq:sigmamassive}
\begin{split}
\Sigma^{lj}_m = & - m\frac{1}{(k^+_0 + k^+_2)}\bar u(0)\gamma^+\biggl [(2k^+_0 + k^+_2)\delta^{ij}_{(D_s)} - \frac{k^+_2}{2}[\gamma^i,\gamma^j]\biggr ]\gamma^{l}v(1)\mathcal{I}^{i}_{\ref{diag:qgqbarT}}\\
& + m\frac{(k^+_2)^2}{(k^+_0 + k^+_2)^2q^+}\bar u(0)\gamma^+\gamma^j\biggl [(2k^+_1 - q^+)\delta^{kl}_{(D_s)} - \frac{q^+}{2}[\gamma^k,\gamma^l]\biggr ]v(1) \mathcal{\hat I}^{k}_{\ref{diag:qgqbarT}}\\
& + m^2\frac{(k^+_2)^2}{(k^+_0 + k^+_2)^2}\bar u(0)\gamma^+\gamma^j\gamma^lv(1)\mathcal{I}_{\ref{diag:qgqbarT}}\\
& + m\frac{1}{(k^+_1 + k^+_2)}\bar u(0)\gamma^+\gamma^{l}\biggl [(2k^+_1 + k^+_2)\delta^{ij}_{(D_s)} + \frac{k^+_2}{2}[\gamma^i,\gamma^j]\biggr ]v(1)\mathcal{I}^{i}_{\ref{diag:qqbargT}}\\
& + m\frac{(k^+_2)^2}{(k^+_1 + k^+_2)^2q^+}\bar u(0)\gamma^+\biggl [(2k^+_0 - q^+)\delta^{kl}_{(D_s)} + \frac{q^+}{2}[\gamma^k,\gamma^l]\biggr ]\gamma^j v(1)\mathcal{\hat I}^{k}_{\ref{diag:qqbargT}}\\
& -m^2\frac{(k^+_2)^2}{(k^+_1 + k^+_2)^2}\bar u(0)\gamma^+\gamma^l\gamma^j v(1)\mathcal{I}_{\ref{diag:qqbargT}}.
\end{split}
\end{equation}
It is straightforward to check that in the massless quark limit, the expression in \eq\nr{eq:reducedTCWFqqbarg} reduces to the result obtained in \cite{Beuf:2017bpd,Hanninen:2017ddy}. The $(D-2)$-dimensional Fourier integrals of the type $\mathcal{I}, \mathcal{\hat I}$ and $\mathcal{J}$ are defined and calculated in Appendix \ref{app:FTintforqqbarg}, for which the following compact notation has been introduced:
\begin{equation}
\begin{split}
\mathcal{I}_{\ref{diag:qgqbarT}} & = \mathcal{I}(\xt_{3;\ref{diag:qgqbarT}}, \xt_{2;\ref{diag:qgqbarT}},\overline{Q}^2_{\ref{diag:qgqbarT}},\omega_{\ref{diag:qgqbarT}},\lambda_{\ref{diag:qgqbarT}}), \quad\quad 
\mathcal{I}_{\ref{diag:qqbargT}}  = \mathcal{I}(\xt_{3;\ref{diag:qqbargT}},\xt_{2;\ref{diag:qqbargT}},\overline{Q}^2_{\ref{diag:qqbargT}},\omega_{\ref{diag:qqbargT}},\lambda_{\ref{diag:qqbargT}}),
\\
\mathcal{\hat I}_{\ref{diag:qgqbarT}} & = \mathcal{\hat I}(\xt_{3;\ref{diag:qgqbarT}}, \xt_{2;\ref{diag:qgqbarT}},\overline{Q}^2_{\ref{diag:qgqbarT}},\omega_{\ref{diag:qgqbarT}},\lambda_{\ref{diag:qgqbarT}}), \quad\quad 
\mathcal{\hat I}_{\ref{diag:qqbargT}}  = \mathcal{\hat I}(\xt_{3;\ref{diag:qqbargT}},\xt_{2;\ref{diag:qqbargT}},\overline{Q}^2_{\ref{diag:qqbargT}},\omega_{\ref{diag:qqbargT}},\lambda_{\ref{diag:qqbargT}}),
\\
\mathcal{J}_{\ref{diag:qgqbarTinst}} & = \mathcal{J}(\xt_{3;\ref{diag:qgqbarT}}, \xt_{2;\ref{diag:qgqbarT}},\overline{Q}^2_{\ref{diag:qgqbarT}},\omega_{\ref{diag:qgqbarT}},\lambda_{\ref{diag:qgqbarT}}), \quad\quad 
\mathcal{J}_{\ref{diag:qqbargTinst}}  = \mathcal{J}(\xt_{3;\ref{diag:qqbargT}},\xt_{2;\ref{diag:qqbargT}},\overline{Q}^2_{\ref{diag:qqbargT}},\omega_{\ref{diag:qqbargT}},\lambda_{\ref{diag:qqbargT}}),
\\
\end{split}
\end{equation}
with the coefficients
\begin{equation}
\label{eq:defomega}
\omega_{\ref{diag:qgqbarT}}  = \frac{q^+k^+_0k^+_2}{k^+_1(k^+_0 + k^+_2)^2}, \quad\quad \omega_{\ref{diag:qqbargT}}  = \frac{q^+k^+_1k^+_2}{k^+_0(k^+_1 + k^+_2)^2},
\end{equation}
the other momentum fraction ratios $\lambda_{(x)}$ from \eqs\eqref{eq:coeffqgbarq},~\eqref{eq:coeffqbarqg},
and the coordinate differences defined in \eqs\nr{eq:defx2_a},~\nr{eq:defx3_a},~\nr{eq:defx2_b} and~\nr{eq:defx3_b}.
We now have the full expressions for the gluon emission wave functions in coordinate space.

\section{Calculating the DIS cross section}
\label{sec:xsdetails}

Let us now  compute the transverse virtual photon cross section at NLO with massive quarks in the dipole factorization framework. As in the massless case~\cite{Beuf:2016wdz,Beuf:2017bpd,Hanninen:2017ddy}, this consists of squaring separately the $q\bar{q}$ and $q\bar{q}g$-contributions, and then effectuating a cancellation of an UV-divergent contribution between them. These steps happen in a similar way here, since the additional algebraic complexity for massive quarks resides more in the UV-finite pieces that are not affected by this subtraction. Thus we will be relatively brief here and refer the reader to the massless calculation for more details. We follow here the exponential subtraction procedure of \cite{Hanninen:2017ddy}.

\subsection{Quark-antiquark contribution}

Let us first write down the $q\bar{q}$ contribution to the DIS cross section at NLO in $\as$. Applying the formula for the cross section in \eq\nr{eq:crosssectionformula}, we find the following expression 
\begin{equation}
\begin{split}
\sigma^{\gamma^{\ast}}_{T}\bigg\vert_{q\bar{q}} & = 2\nc\sum_{f}\int \frac{\ud k^+_0}{2k^+_0(2\pi)}\int \frac{\ud k^+_1}{2k^+_1(2\pi)}\frac{2\pi\delta(q^+-k^+_0 - k^+_1)}{2q^+}\int \ud^{D-2}\xt_0 \int \ud^{D-2}\xt_1\\
& \hspace{2cm} \times\frac{1}{(D_s-2)}\sum_{\lambda}\sum_{h_0,h_1}\vert \widetilde{\psi}^{\gamma^{\ast}_{T}\rightarrow q\bar{q}}\vert^2\mathrm{Re}[1-\mathcal{S}_{01}]\\
& =\frac{\nc}{(2\pi)4(q^+)^2}\sum_{f}\int_{0}^{1} \frac{\ud z}{z(1-z)}\int \ud^{D-2}\xt_0 \int \ud^{D-2}\xt_1\frac{1}{(D_s-2)}\sum_{\lambda}\sum_{h_0,h_1}\vert \widetilde{\psi}^{\gamma^{\ast}_{T}\rightarrow q\bar{q}}\vert^2\mathrm{Re}[1-\mathcal{S}_{01}],\\
\end{split}
\end{equation}
where the $\gamma^\ast_{T} \rightarrow q \bar q$ LCWF squared is summed over the quark spins $(h_0,h_1)$, transverse photon polarization $(\lambda)$ and the factor $1/(D_s-2)$ comes from averaging over the transverse photon polarization states.

We can now compute the LCWF squared by using the compact expressions for the LO and NLO wave functions in \eqs\nr{eq:psireducedLO} and \nr{eq:finalNLOmixspacereducedv2}, respectively. This leads to the following expression
\begin{equation}
    \frac{1}{(D_s-2)}\sum_{\lambda}\sum_{h_0,h_1}\vert \widetilde{\psi}^{\gamma^{\ast}_{T}\rightarrow q\bar{q}}\vert^2 = \frac{1}{(D_s-2)}  \sum_{\lambda}\sum_{h_0,h_1} \biggl [\vert \widetilde{\psi}^{\gamma^{\ast}_{T}\rightarrow q\bar{q}}_{\lo}\vert^2 + 2 (\widetilde{\psi}^{\gamma^{\ast}_{T}\rightarrow q\bar{q}}_{\lo})^{\ast}\widetilde{\psi}^{\gamma^{\ast}_{T}\rightarrow q\bar{q}}_{\nlo} \biggr ] + \mathcal{O}(\alpha_{\text{em}}^2\alpha_s^2).
\end{equation}
After performing the straightforward Dirac algebra, we obtain
\begin{equation}
\label{eq:LCWFfullsquared}
\begin{split}
&\frac{1}{(D_s-2)} \sum_{\lambda}\sum_{h_0,h_1}\vert \widetilde{\psi}^{\gamma^{\ast}_{T}\rightarrow q\bar{q}}\vert^2 =  \frac{\alpha_{\text{em}}}{\pi}e_f^2 8z(1-z)(q^+)^2 \Biggl \{\biggl [z^2 + (1-z)^2 \biggr ]\mathcal{F}^{\ast}[\Pt^i]\mathcal{F}[\Pt^i]+ m^2\mathcal{F}^{\ast}[1]\mathcal{F}[1]\\
& + \left (\frac{\alpha_s \cf}{\pi}\right )\biggl \{\bigg [z^2 + (1-z)^2  \biggr ]\mathcal{F}^{\ast}[\Pt^i]\mathcal{F}[\Pt^i \mathcal{V}^T] + \frac{(2z-1)}{2}\mathcal{F}^{\ast}[\Pt^i]\mathcal{F}[\Pt^i \mathcal{N}^T] + m^2\mathcal{F}^{\ast}[1] \mathcal{F}[\mathcal{V}^T + \mathcal{M}^T - \frac{\mathcal{S}^T}{2}] \biggr \} \Biggr \},
\end{split}
\end{equation}
where the UV finite terms has been simplified by taking the $D_s = D = 4$ limit.

Using the results obtained in \eqs\nr{eq:PiVi}, \nr{eq:PiNi} and \nr{eq:VMSscalar}, the $q\bar{q}$ contribution to the total cross section can be written as 
\begin{equation}
\label{eq:qqbarcsUVdiv}
\begin{split}
\sigma^{\gamma^{\ast}}_{T}&\bigg\vert_{q\bar{q}}  =4\nc\alpha_{em}\sum_{f}e_f^2 \int_{0}^{1} \ud z \int_{[\xt_0]}\int_{[\xt_1]} \\
& \times \Biggl \{ \Biggl [1 + \left (\frac{\alpha_s\cf}{\pi}\right )f_{\rm UV}  \Biggr ]\left ( \frac{\kaz}{2\pi\vert \xt_{01}\vert }\right )^{D-4}  \Biggl ( \biggl [z^2 + (1-z)^2 \biggr ]
\biggl [\kaz K_{\frac{D}{2} -1}\left (\vert \xt_{01}\vert \kaz \right )\biggr ]^2 +  m^2\biggl [K_{\frac{D}{2} -2}\left (\vert \xt_{01}\vert \kaz\right )\biggr ]^2 \Biggr )\\
& + \left (\frac{\alpha_s\cf}{\pi}\right )\biggl \{\biggl [z^2 + (1-z)^2 \biggr ]\kaz K_1(\vert \xt_{01}\vert \kaz)f_{\mathcal{V}} + \frac{(2z-1)}{2} \kaz K_1(\vert \xt_{01}\vert \kaz) f_{\mathcal{N}}\\
& + m^2 K_0(\vert \xt_{01}\vert \kaz) f_{\mathcal{VMS}} \biggr \}\Biggr \}\mathrm{Re}[1-\mathcal{S}_{01}]  +  \mathcal{O}(\alpha_{em}\alpha_s^2),  
\end{split}
\end{equation}
where the short-hand notation  $\int_{[\xt_i]} = \int \ud^{D-2} \xt_i/(2\pi)$ for $i=0,1$ denotes a $(D-2)$-dimensional transverse coordinate integral. The function $f_{\rm UV}$ contains the UV divergent term plus the regularization scheme dependent part, and it is given by 
\begin{equation}
\begin{split}
f_{\rm UV} = \biggl [\frac{3}{2} + \log \left ( \frac{\alpha}{z} \right ) + \log \left (\frac{\alpha}{1-z}\right ) \biggr ] \Biggl \{\frac{(4\pi)^{2-\frac{D}{2}}}{(2-\frac{D}{2})}\Gamma\left (3 - \frac{D}{2}\right ) + \log\left (\frac{\vert \xt_{01}\vert^2 \mu^2}{4}\right ) + 2\gamma_E\Biggr \} + \frac{1}{2}\frac{(D_s-4)}{(D-4)}. 
\end{split}
\end{equation}
Similarly, the UV finite functions $f_{\mathcal{V}}, f_{\mathcal{N}}$ and $f_{\mathcal{VMS}}$ are given by:
\begin{equation}
\label{eq:fVfNfVMS}
\begin{split}
f_{\mathcal{V}} & = \biggl \{\frac{5}{2} - \frac{\pi^2}{3} + \log^2\left (\frac{z}{1-z} \right ) + \Omega_{\mathcal{V}} + L\biggr \} \kaz K_{1}\left (\vert \xt_{01}\vert \kaz \right )  + \widetilde{I}_{\mathcal{V}}, \\
f_{\mathcal{N}} &  =  \Omega_{\mathcal{N}} \,  \kaz  K_{1}\left (\vert \xt_{01}\vert \kaz \right ) + \widetilde{I}_{\mathcal{N}},  \\
f_{\mathcal{VMS}}  & = \Biggl \{ 3 - \frac{\pi^2}{3}  + \log^2 \left (\frac{z}{1-z} \right )  + \Omega_{\mathcal{V}} + L  \Biggr \}K_{0}\left (\vert \xt_{01}\vert \kaz \right )   + \widetilde{I}_{\mathcal{VMS}}.
\end{split}
\end{equation}
Here we recall that $\kappa_v = \sqrt{v(1-v)Q^2 + m^2}$, 
$L$ is defined by \eq\eqref{eq:defL},  $\Omega_{\mathcal{N}}$ by \eqs\eqref{eq:degOmegaN_1}
and~\eqref{eq:degOmegaN},
$\widetilde{I}_{\mathcal{N}}$ by \eq\eqref{eq:INT},
$\widetilde{I}_{\mathcal{V}}$ by \eq\eqref{eq:IVT},
$\Omega_{\mathcal{V}}$ by \eqs\eqref{eq:OSI} and~\eqref{eq:SigmaOmega} and finally  $\widetilde{I}_{\mathcal{VMS}}$ by \eq\eqref{eq:IVMST}.

\subsection{Quark-antiquark-gluon contribution}

The $q\bar{q}g$ contribution to the DIS cross-section at NLO in $\as$ is given by the second term in \eq\nr{eq:crosssectionformula}
\begin{equation}
\label{eq:qqbargcs1}
\begin{split}
\sigma^{\gamma^{\ast}}_{T}\bigg\vert_{q\bar{q}g}  = 2\nc\cf &\sum_{f}\int_{0}^{\infty} \frac{\ud k^+_0}{2k^+_0(2\pi)}\int_{0}^{\infty} \frac{\ud k^+_1}{2k^+_1(2\pi)}\int_{0}^{\infty} \frac{\ud k^+_2}{2k^+_2(2\pi)}\frac{2\pi\delta(q^+-k^+_0 - k^+_1-k^+_2)}{2q^+}\\
&\times \int \ud^{D-2}\xt_0 \int \ud^{D-2}\xt_1\int \ud^{D-2}\xt_2 \frac{1}{(D_s-2)}\sum_{\lambda, \sigma}\sum_{h_0,h_1}\vert \widetilde{\psi}^{\gamma^{\ast}_{T}\rightarrow q\bar{q}g}\vert^2\mathrm{Re}[1-\mathcal{S}_{012}],\\
\end{split}
\end{equation}
where the $\gamma^\ast_{T} \rightarrow q\bar q g$ LCWF squared is also summed over the emitted gluon polarization $(\sigma)$. Following the straightforward but lengthy derivation shown in Appendix \ref{app:qqbargderivation}, we find the result
\begin{equation}
\label{eq:emissionTLCWFsquared}
\begin{split}
\frac{1}{(D_s-2)}\sum_{\lambda,\sigma}\sum_{h_0,h_1}\vert \widetilde{\psi}^{\gamma^{\ast}_{T}\rightarrow q\bar{q}g}\vert^2 = \alpha_{em}e_f^2\alpha_{s}(4\pi)^2 2(2k^+_0)&(2k^+_1) \biggl \{\mathcal{K}^T_{q\bar{q}g}\vert_{\text{UV}} + \mathcal{K}^T_{q\bar{q}g}\vert_{\text{UV}_m} +  \mathcal{K}^T_{q\bar{q}g}\vert_{\text{F}} + \mathcal{K}^T_{q\bar{q}g}\vert_{\text{F}_m}\biggr \}\\
& +  \mathcal{O}(\alpha_{em}\alpha_s^2).
\end{split}
\end{equation}
Here the functions $\mathcal{K}^T_{q\bar{q}g}\vert_{\text{UV}}$ and  $\mathcal{K}^T_{q\bar{q}g}\vert_{\text{UV}_m}$ are the UV divergent parts of the squares of the contributions without a light cone helicity flip and with one flip, respectively. They are given by the following expressions:
\begin{equation}
\label{eq:KTqqbargUV}
\begin{split}
\mathcal{K}^T_{q\bar{q}g}\vert_{\text{UV}} & = \frac{1}{(k_0^+ + k_2^+)^2(q^+)^2}\biggl [4k_0^+(k_0^+ + k_2^+) + 2(k_2^+)^2 + (D_s-4)(k_2^+)^2 \biggr ]\biggl [(q^+)^2 - \frac{4k_1^+}{(D_s-2)}(q^+ - k_1^+)\biggr ] \vert \mathcal{I}^{ik}_{\ref{diag:qgqbarT}}\vert^2\\
& + \frac{1}{(k_1^+ + k_2^+)^2(q^+)^2}\biggl [4k_1^+(k_1^+ + k_2^+) + 2(k_2^+)^2 + (D_s-4)(k_2^+)^2 \biggr ]\biggl [(q^+)^2 - \frac{4k_0^+}{(D_s-2)}(q^+ - k_0^+) \biggr ] \vert \mathcal{I}^{ik}_{\ref{diag:qqbargT}}\vert^2,
\end{split}
\end{equation}
and
\begin{equation}
\label{eq:KTqqbargmUV}
\begin{split}
\mathcal{K}^T_{q\bar{q}g}\vert_{\text{UV}_m} & = \frac{m^2}{(k_0^+ + k_2^+)^2}\biggl [4k_0^+(k_0^+ + k_2^+) + 2(k_2^+)^2 + (D_s-4)(k_2^+)^2 \biggr ] \vert\mathcal{I}^{i}_{\ref{diag:qgqbarT}}\vert^2\\
& + \frac{m^2}{(k_1^+ + k_2^+)^2}\biggl [4k_1^+(k_1^+ + k_2^+) + 2(k_2^+)^2 + (D_s-4)(k_2^+)^2 \biggr ] \vert\mathcal{I}^{i}_{\ref{diag:qqbargT}}\vert^2.
\end{split}
\end{equation}
Note that a flip and nonflip contribution cannot interfere and thus there are no cross terms. 

The functions $\mathcal{K}^T_{q\bar{q}g}\vert_{\text{F}}$ and $\mathcal{K}^T_{q\bar{q}g}\vert_{\text{F}_m}$ are the remaining UV finite terms, without and with light cone helicity flips  respectively, i.e. without or with explicit powers of the quark mass. The UV finite light cone helicity nonflip terms can be cast into the following form
\begin{equation}
\label{eq:KfinHNF}
\begin{split}
\mathcal{K}^T_{q\bar{q}g}\vert_{\text{F}} & =  \frac{4}{(k_0^+ +k_2^+)(k_1^+ + k_2^+)(q^+)^2}\Biggl \{q^+ k_2^+(k_0^+ - k_1^+)^2\biggl [\Re e[\mathcal{I}^{ii}_{\ref{diag:qgqbarT}}(\mathcal{I}^{kk}_{\ref{diag:qqbargT}})^{\ast}]-\Re e[\mathcal{I}^{ik}_{\ref{diag:qgqbarT}}(\mathcal{I}^{ki}_{\ref{diag:qqbargT}})^{\ast}] \biggr ]\\
& - \biggl [k_1^+(k_0^+ + k_2^+) + k_0^+(k_1^+ + k_2^+)\biggr ]\biggl [k_0^+(k_0^+ + k_2^+) + k_1^+(k_1^+ + k_2^+)\biggr ] \Re e[\mathcal{I}^{ik}_{\ref{diag:qgqbarT}}(\mathcal{I}^{ik}_{\ref{diag:qqbargT}})^{\ast}]\Biggr \}\\
& +  \frac{4(k_0^+ + k_2^+)k_1^+ k_2^+}{(k_1^+ + k_2^+)^2 q^+}\Re e[\mathcal{I}^{ii}_{\ref{diag:qgqbarT}}(\mathcal{J}_{\ref{diag:qqbargTinst}})^{\ast}] - \frac{4(k_1^+ + k_2^+)k_0^+ k_2^+}{(k_0^+ + k_2^+)^2 q^+}\Re e[\mathcal{I}^{ii}_{\ref{diag:qqbargT}}(\mathcal{J}_{\ref{diag:qgqbarTinst}})^{\ast}] \\
&   + \frac{4(k_0^+)^2 k_1^+ k_2^+}{(k_0^+ + k_2^+)^3q^+} \Re e[\mathcal{I}^{ii}_{\ref{diag:qgqbarT}}(\mathcal{J}_{\ref{diag:qgqbarTinst}})^{\ast}] - \frac{4(k_1^+)^2 k_0^+ k_2^+}{(k_1^+ + k_2^+)^3q^+} \Re e[\mathcal{I}^{ii}_{\ref{diag:qqbargT}}(\mathcal{J}_{\ref{diag:qqbargTinst}})^{\ast}] \\
& + \frac{2(k_0^+ k_2^+)^2}{(k_0^+ + k_2^+)^4} \vert \mathcal{J}_{\ref{diag:qgqbarTinst}}\vert^2 + \frac{2(k_1^+ k_2^+)^2}{(k_1^+ + k_2^+)^4}\vert \mathcal{J}_{\ref{diag:qqbargTinst}}\vert^2.
\end{split}
\end{equation}
Similarly, the UV finite contributions with one or more light cone helicity flips  are given by 
\begin{equation}
\label{eq:KfinHF}
\begin{split}
\mathcal{K}^T_{q\bar{q}g}\vert_{\text{F}_m}   = m^2 & \Biggl \{\frac{(k_2^+)^4}{(k_0^+ + k_2^+)^4 (q^+)^2}\biggl [4k_1^+(k_1^+ - q^+) + 2(q^+)^2\biggr ]\vert\mathcal{\hat I}^{i}_{\ref{diag:qgqbarT}}\vert^2  - \frac{4k_0^+k_1^+(k_2^+)^2}{(k_0^+ + k_2^+)^3 q^+} \Re e[\mathcal{I}^{i}_{\ref{diag:qgqbarT}}(\mathcal{\hat I}^{i}_{\ref{diag:qgqbarT}})^{\ast}]\\
& + \frac{(k_2^+)^4}{(k_1^+ + k_2^+)^4 (q^+)^2}\biggl [4k_0^+(k_0^+ - q^+) + 2(q^+)^2 \biggr ]\vert\mathcal{\hat I}^{i}_{\ref{diag:qqbargT}}\vert^2  + \frac{4k_0^+k_1^+(k_2^+)^2}{(k_1^+ + k_2^+)^3 q^+}  \Re e[\mathcal{I}^{i}_{\ref{diag:qqbargT}}(\mathcal{\hat I}^{i}_{\ref{diag:qqbargT}})^{\ast}] \\
 & + m^2 \frac{2(k_2^+)^4}{(k_0^+ + k_2^+)^4}\vert\mathcal{I}_{\ref{diag:qgqbarT}}\vert^2  + m^2 \frac{2(k_2^+)^4}{(k_1^+ + k_2^+)^4} \vert\mathcal{I}_{\ref{diag:qqbargT}}\vert^2\\ 
& -\frac{2}{(k_0^+ + k_2^+)(k_1^+ + k_2^+)}\biggl [(2k_0^+ + k_2^+)(2k_1^+ + k_2^+) + (k_2^+)^2\biggr ]\Re e[\mathcal{I}^{i}_{\ref{diag:qgqbarT}}(\mathcal{I}^{i}_{\ref{diag:qqbargT}})^{\ast}]\\
& + \frac{2(k_2^+)^4}{(k_0^+ + k_2^+)^2(k_1^+ + k_2^+)^2 (q^+)^2}\biggl [(2k_0^+ + k_2^+)(2k_1^+ + k_2^+) + (k_2^+)^2\biggr ]\Re e[\mathcal{\hat I}^{i}_{\ref{diag:qgqbarT}}(\mathcal{\hat I}^{i}_{\ref{diag:qqbargT}})^{\ast} ]\\
& - \frac{4(k_0^+k_2^+)^2}{(k_0^+ + k_2^+)(k_1^+ + k_2^+)^2 q^+} \Re e[\mathcal{I}^{i}_{\ref{diag:qgqbarT}}(\mathcal{\hat I}^{i}_{\ref{diag:qqbargT}})^{\ast}] + \frac{4(k_1^+k_2^+)^2}{(k_1^+ + k_2^+)(k_0^+ + k_2^+)^2 q^+}  \Re e[\mathcal{\hat I}^{i}_{\ref{diag:qgqbarT}}(\mathcal{I}^{i}_{\ref{diag:qqbargT}})^{\ast}]\\
& + \frac{4(k_0^+k_1^+)(k_2^+)^2}{(k_0^+ + k_2^+)^3 q^+}\Re e[\mathcal{I}^{ii}_{\ref{diag:qgqbarT}}(\mathcal{I}_{\ref{diag:qgqbarT}})^{\ast}] - \frac{4(k_0^+k_1^+)(k_2^+)^2}{(k_1^+ + k_2^+)^3 q^+}\Re e[\mathcal{I}^{ii}_{\ref{diag:qqbargT}}(\mathcal{I}_{\ref{diag:qqbargT}})^{\ast}]\\
& + \frac{4(k_0^+ + k_2^+)(k_2^+)^2}{(k_1^+ + k_2^+)^2 q^+} \Re e[\mathcal{I}^{ii}_{\ref{diag:qgqbarT}}(\mathcal{I}_{\ref{diag:qqbargT}})^{\ast}] - \frac{4(k_1^+ + k_2^+)(k_2^+)^2}{(k_0^+ + k_2^+)^2 q^+} \Re e[\mathcal{I}^{ii}_{\ref{diag:qqbargT}}(\mathcal{I}_{\ref{diag:qgqbarT}})^{\ast}]\\
& + \frac{4k_0^+(k_2^+)^3}{(k_0^+ + k_2^+)^4} \Re e[\mathcal{J}_{\ref{diag:qgqbarTinst}}(\mathcal{I}_{\ref{diag:qgqbarT}})^{\ast}] + \frac{4k_1^+(k_2^+)^3}{(k_1^+ + k_2^+)^4} \Re e[\mathcal{J}_{\ref{diag:qqbargTinst}}(\mathcal{I}_{\ref{diag:qqbargT}})^{\ast}]\Biggr \}. 
\end{split}
\end{equation}
Note that there are contributions proportional to $m^2$, coming from either two light cone helicity flips in the amplitude and none in the conjugate or one flip in both, as well as contributions $\sim m^4$ with three flips in the amplitude and one in the conjugate.

Finally, inserting the expression in \eq\nr{eq:emissionTLCWFsquared} into \eq\nr{eq:qqbargcs1}, we obtain the gluon emission contribution to the cross section
\begin{equation}
\label{eq:qqbargunsub}
\begin{split}
\sigma^{\gamma^{\ast}}_{T}\bigg\vert_{q\bar{q}g} = 4\nc\alpha_{em}&\left (\frac{\alpha_s\cf}{\pi}\right )\sum_{f}e_f^2 \frac{(2\pi)^4}{2}  \int_{0}^{\infty} \ud k^+_0\int_{0}^{\infty} \ud k^+_1\int_{0}^{\infty} \frac{\ud k^+_2}{k^+_2}\frac{\delta(q^+-k^+_0-k^+_1-k^+_2)}{q^+}\\
&\times   \int_{[\xt_0]}\int_{[\xt_1]} \int_{[\xt_2]}  \Biggl \{\mathcal{K}^T_{q\bar{q}g}\vert_{\text{UV}} + \mathcal{K}^T_{q\bar{q}g}\vert_{\text{UV}_m} +  \mathcal{K}^T_{q\bar{q}g}\vert_{\text{F}} + \mathcal{K}^T_{q\bar{q}g}\vert_{\text{F}_m}\Biggr \}\mathrm{Re}[1-\mathcal{S}_{012}] +  \mathcal{O}(\alpha_{em}\alpha_s^2).
\end{split}
\end{equation}

\subsection{UV subtraction}

Like in the longitudinal photon case \cite{Beuf:2021qqa}, the UV renormalization of the coupling $g$ is not relevant at the accuracy of the present calculation. Therefore, the remaining UV divergences have to cancel between the $q\bar{q}$ and  $q\bar{q}g$ Fock states contributions on the cross section level. Due to the complicate analytical structure of the gluon emission contribution in \eq\nr{eq:qqbargunsub}, the UV divergent phase-space integrals cannot be performed analytically for arbitrary dimension $D$. Hence, it is desirable to understand the cancellation of UV divergences at the integrand level.

In the expression~\nr{eq:qqbargunsub}, the first two contributions inside the curly brackets contain two terms which are UV divergent when $\xt_2 \rightarrow \xt_0$ and $\xt_2 \rightarrow \xt_1$, respectively. In addition, the remaining two contributions are UV finite and hence we have immediately taken the limit $D_s = D = 4$ at the integrand level, which simplifies the further calculation considerably. In order to subtract the UV divergences, we will follow the same steps as presented in \cite{Beuf:2021qqa} and use the following property of Wilson lines at coincident transverse coordinate points
\begin{equation}
\begin{split}
\lim_{\xt\rightarrow \yt} \biggl [t^{b}U_F(\yt) \biggr ]U_A(\xt)_{ba} &= \biggl [U_F(\yt)t^{a} \biggr ],\\
\end{split}
\end{equation}
which implies that 
\begin{equation}
\lim_{\xt_2\rightarrow \xt_0} \mathcal{S}_{012} = \lim_{\xt_2\rightarrow \xt_1} \mathcal{S}_{012} =  \mathcal{S}_{01}.  
\end{equation}
Thus, the UV divergences in \eq\nr{eq:qqbargunsub} are subtracted by replacing: 
\begin{equation}
\label{eq:subprocedure1}
\begin{split}
\vert \mathcal{I}^{ik}_{\ref{diag:qgqbarT}}\vert^2\mathrm{Re}[1-\mathcal{S}_{012}] & \mapsto \Biggl \{\vert \mathcal{I}^{ik}_{\ref{diag:qgqbarT}}\vert^2\mathrm{Re}[1-\mathcal{S}_{012}] - \vert \mathcal{I}^{ik}_{\ref{diag:qgqbarT} \rm UV}\vert^2\mathrm{Re}[1-\mathcal{S}_{01}] \Biggr \} + \vert \mathcal{I}^{ik}_{\ref{diag:qgqbarT}\rm UV}\vert^2\mathrm{Re}[1-\mathcal{S}_{01}]\\
\vert \mathcal{I}^{i}_{\ref{diag:qgqbarT}}\vert^2\mathrm{Re}[1-\mathcal{S}_{012}] & \mapsto \Biggl \{\vert \mathcal{I}^{i}_{\ref{diag:qgqbarT}}\vert^2\mathrm{Re}[1-\mathcal{S}_{012}] - \vert \mathcal{I}^{i}_{\ref{diag:qgqbarT} \rm UV}\vert^2\mathrm{Re}[1-\mathcal{S}_{01}] \Biggr \} + \vert \mathcal{I}^{i}_{\ref{diag:qgqbarT}\rm UV}\vert^2\mathrm{Re}[1-\mathcal{S}_{01}]
\end{split}
\end{equation}
and
\begin{equation}
\label{eq:subprocedure2}
\begin{split}
\vert \mathcal{I}^{ik}_{\ref{diag:qqbargT}}\vert^2\mathrm{Re}[1-\mathcal{S}_{012}] &\mapsto \Biggl \{\vert \mathcal{I}^{ik}_{\ref{diag:qqbargT}}\vert^2\mathrm{Re}[1-\mathcal{S}_{012}] - \vert \mathcal{I}^{ik}_{\ref{diag:qqbargT}\rm UV}\vert^2\mathrm{Re}[1-\mathcal{S}_{01}] \Biggr \} + \vert \mathcal{I}^{ik}_{\ref{diag:qqbargT}\rm UV}\vert^2\mathrm{Re}[1-\mathcal{S}_{01}]\\
\vert \mathcal{I}^{i}_{\ref{diag:qqbargT}}\vert^2\mathrm{Re}[1-\mathcal{S}_{012}] &\mapsto \Biggl \{\vert \mathcal{I}^{i}_{\ref{diag:qqbargT}}\vert^2\mathrm{Re}[1-\mathcal{S}_{012}] - \vert \mathcal{I}^{i}_{\ref{diag:qqbargT}\rm UV}\vert^2\mathrm{Re}[1-\mathcal{S}_{01}] \Biggr \} + \vert \mathcal{I}^{i}_{\ref{diag:qqbargT}\rm UV}\vert^2\mathrm{Re}[1-\mathcal{S}_{01}],
\end{split}
\end{equation}
where the subtraction terms are given in terms of a two functions $\mathcal{I}^{ik}_{\rm UV}$ and $\mathcal{I}^{i}_{\rm UV}$:
\begin{equation}
\begin{split}
\mathcal{I}^{ik}_{\ref{diag:qgqbarT}\rm UV} &=
\mathcal{I}^{ik}_{\rm UV}  (\xt_{01},\xt_{20},\overline{Q}^2_{\ref{diag:qgqbarT}},\omega_{\ref{diag:qgqbarT}},\lambda_{\ref{diag:qgqbarT}}),\\
\mathcal{I}^{i}_{\ref{diag:qgqbarT}\rm UV} &=
\mathcal{I}^{i}_{\rm UV}  (\xt_{01},\xt_{20},\overline{Q}^2_{\ref{diag:qgqbarT}},\omega_{\ref{diag:qgqbarT}},\lambda_{\ref{diag:qgqbarT}}),
\\
\mathcal{I}^{ik}_{\ref{diag:qqbargT}\rm UV} &=
\mathcal{I}^{ik}_{\rm UV}  (\xt_{01},\xt_{21},\overline{Q}^2_{\ref{diag:qqbargT}},\omega_{\ref{diag:qqbargT}},\lambda_{\ref{diag:qqbargT}}),\\
\mathcal{I}^{i}_{\ref{diag:qqbargT}\rm UV} &=
\mathcal{I}^{i}_{\rm UV}  (\xt_{01},\xt_{21},\overline{Q}^2_{\ref{diag:qqbargT}},\omega_{\ref{diag:qqbargT}},\lambda_{\ref{diag:qqbargT}}).
\end{split}
\end{equation}
Now, for the function $\mathcal{I}^{ik}_{\rm UV}$ be a good UV approximation of the full integral, it must satisfy
\begin{equation}
\label{eq:subtractionintdef}
\lim_{\xt_2\rightarrow \xt_0}\mathcal{I}^{ik} = 
 \lim_{\xt_2\rightarrow \xt_0}\mathcal{I}^{ik}_{\rm UV} 
\end{equation}
from which it follows that
\begin{equation}
\begin{split}
\lim_{\xt_2\rightarrow \xt_0}\mathcal{I}^{ik}_{\ref{diag:qgqbarT}} &= 
 \lim_{\xt_2\rightarrow \xt_0}\mathcal{I}^{ik}_{\ref{diag:qgqbarT}\rm UV} 
\\
\lim_{\xt_2\rightarrow \xt_1}\mathcal{I}^{ik}_{\ref{diag:qqbargT}} &= 
 \lim_{\xt_2\rightarrow \xt_1}\mathcal{I}^{ik}_{\ref{diag:qqbargT}\rm UV} .
\end{split}
\end{equation}
An identical analysis holds for the  function $\mathcal{I}^{i}_{\rm UV}$.
It is important to note that there is no unique choice for the UV divergent subtraction in \eq\nr{eq:subtractionintdef}. The only requirement for the subtraction is that the UV divergence needs to cancel between the $q\bar{q}$ and $q\bar{q}g$ contributions. Thus, it is sufficient for the subtraction to approximate the original integrals  by any function that has the same value in the UV limits (for any $D$). Because of this cancellation, the integrals of the expressions inside the curly brackets in \eqs\nr{eq:subprocedure1} and \nr{eq:subprocedure2} are finite, and one can safely take the limit $D_s = D = 4$ under the $\xt_2$ integral.

In an arbitrary dimension $D$, the integrals $\mathcal{I}^{ik}$ and $\mathcal{I}^{i}$ (see \eqs\nr{eq:Iranktwofinal} and \nr{eq:Irankonefinal}) are given by 
\begin{equation}
\label{eq:generalIik}
\mathcal{I}^{ik}(\bt,\rt,\overline{Q}^2,\omega,\lambda) = -\frac{\mu^{2-D/2}}{4(4\pi)^{D-2}}\bt^i\rt^k\int_{0}^{\infty} \ud u\, u^{-D/2}e^{-u[\overline{Q}^2 + m^2]}e^{-\frac{\vert\bt\vert^2}{4u}}\int_{0}^{u/\omega}\ud t t^{-D/2}e^{-t\omega \lambda m^2}e^{-\frac{\vert\rt\vert^2}{4t}},
\end{equation}
\begin{equation}
\label{eq:generalIi}
\mathcal{I}^i(\bt,\rt,\overline{Q}^2,\omega,\lambda) = \frac{i\mu^{2-D/2}}{2(4\pi)^{D-2}}\rt^i\int_{0}^{\infty} \ud u u^{1-D/2}e^{-u[\overline{Q}^2 + m^2]}e^{-\frac{\vert\bt\vert^2}{4u}}\int_{0}^{u/\omega}\ud t t^{-D/2}e^{-t\omega \lambda m^2}e^{-\frac{\vert\rt\vert^2}{4t}}.
\end{equation}
It is straightforward to see that to get the leading behavior in the limit $\vert\rt\vert^2\rightarrow 0$ we can set $\lambda=0$. This leads us to 
\begin{equation}
\label{eq:uvfunctiongeneral1}
\mathcal{I}^{ik}(\bt,\rt,\overline{Q}^2,\omega)  \stackrel{\rt^2 \to 0}{=}
 -\frac{\mu^{2-D/2}}{4(4\pi)^{D-2}}\bt^i\rt^k\left (\frac{\vert\rt\vert^2}{4}\right )^{1-D/2}\int_{0}^{\infty}\ud u u^{-D/2}e^{-u[\overline{Q}^2 + m^2]}e^{-\frac{\vert\bt\vert^2}{4u}}\Gamma\left (\frac{D}{2}-1, \frac{\vert\rt\vert^2\omega}{4u}\right ),
\end{equation}
and
\begin{equation}
\label{eq:uvfunctiongeneral2}
\mathcal{I}^i(\bt,\rt,\overline{Q}^2,\omega)  \stackrel{\rt^2 \to 0}{=}
 \frac{i\mu^{2-D/2}}{2(4\pi)^{D-2}}\rt^i\left (\frac{\vert\rt\vert^2}{4}\right )^{1-D/2}\int_{0}^{\infty}\ud u u^{1-D/2}e^{-u[\overline{Q}^2 + m^2]}e^{-\frac{\vert\bt\vert^2}{4u}}\Gamma\left (\frac{D}{2}-1, \frac{\vert\rt\vert^2\omega}{4u}\right ),
\end{equation}
where we have suppressed the dependence on the variable $\lambda$ in the notation.

Now there are several possible ways of performing the UV subtraction. Using the exponential subtraction procedure
introduced in \cite{Hanninen:2017ddy}, we approximate the incomplete gamma function with
\begin{equation}
\label{eq:uvappr1}
\Gamma\left (\frac{D}{2}-1, \frac{\vert\rt\vert^2\omega}{4u}\right ) \mapsto \Gamma\left (\frac{D}{2}-1\right )e^{-\frac{\vert\rt\vert^2}{2\vert\bt\vert^2e^{\gamma_E}}},
\end{equation}
where the exponential is independent of $u$ allowing for an analytical calculation of the $u$-integral. This replacement has the correct behavior in the UV limit $\vert\rt\vert^2\to 0$, but also is regular in the  IR limit of large $\vert\rt\vert^2 \rightarrow \infty$. Here we note that another option would be to follow the polynomial subtraction scheme used in \cite{Beuf:2017bpd}. Here the subtraction function is polynomial in $r$. This however, introduces a new IR divergence, which must be compensated with another subtraction\footnote{See the discussion in Appendix E of \cite{Hanninen:2017ddy}}. Finally, substituting \eq\nr{eq:uvappr1} into \eqs\nr{eq:uvfunctiongeneral1} and \nr{eq:uvfunctiongeneral2} gives 
\begin{equation}
\label{eq:uvsubtexp1}
\hspace{-0.2cm}\mathcal{I}^{ik}_{\rm UV}(\bt,\rt,\overline{Q}^2,\omega)= -\frac{\mu^{2-D/2}\bt^i\rt^k}{2\pi^{D/2-1}} (\vert\rt\vert^2 )^{1-D/2} \Gamma\left (\frac{D}{2}-1\right )e^{-\frac{\vert\rt\vert^2}{2\vert\bt\vert^2e^{\gamma_E}}} \left ( \frac{\sqrt{\overline{Q}^2 + m^2}}{2\pi\vert \bt \vert }\right )^{\frac{D}{2}-1}\!\!K_{\frac{D}{2} -1}\left (\vert \bt\vert\sqrt{\overline{Q}^2 + m^2}\right ),
\end{equation}
and
\begin{equation}
\label{eq:uvsubtexp2}
\mathcal{I}^i_{\rm UV}(\bt,\rt,\overline{Q}^2,\omega)= \frac{i\mu^{2-D/2}\rt^i}{4\pi^{D/2}} (\vert\rt\vert^2 )^{1-D/2} \Gamma\left (\frac{D}{2}-1\right )e^{-\frac{\vert\rt\vert^2}{2\vert\bt\vert^2e^{\gamma_E}}} \left ( \frac{\sqrt{\overline{Q}^2 + m^2}}{2\pi\vert \bt \vert }\right )^{\frac{D}{2}-2}\!\!K_{\frac{D}{2} -2}\left (\vert \bt\vert\sqrt{\overline{Q}^2 + m^2}\right ).
\end{equation}

The following master integral will be used to compute the UV divergent $\rt$-integral:
\begin{equation}
\begin{split}
\frac{(\mu^2)^{2-\frac{D}{2}}\Gamma\left (\frac{D}{2}-1 \right )^2}{\pi^D}\int \ud^{D-2}\rt \,(\vert\rt\vert^2 )^{3-D}e^{-\frac{\vert\rt\vert^2}{\vert\bt\vert^2e^{\gamma_E}}}
 = \frac{1}{\pi^3}\Biggl \{&\frac{(4\pi)^{2-\frac{D}{2}}}{(2-\frac{D}{2})}\Gamma\left (3 - \frac{D}{2} \right ) + \log\left (\frac{\vert\xt_{01}\vert^2\mu^2}{4} \right ) \\
 & + 2\gamma_E + \mathcal{O}(D-4)\Biggr \}.
\end{split}
\end{equation}

\subsection{UV subtracted results}

For the UV subtraction terms coming from the diagram \ref{diag:qgqbarT}, we find 
\begin{equation}
\label{eq:qqbaruvsubtexpj2v2}
\begin{split}
\sigma^{\gamma^{\ast}}_{T}&\bigg\vert^{\vert \ref{diag:qgqbarT}\vert^2_{\rm UV}}_{q\bar qg} = 4\nc\alpha_{em}\left (\frac{\alpha_s\cf}{\pi}\right )\sum_{f} \frac{e_f^2}{q^+} \int_{[\xt_0]}\int_{[\xt_1]}  \left ( \frac{\kaz}{2\pi\vert \xt_{01} \vert }\right )^{D-4}\biggl [\kaz K_{\frac{D}{2} -1}\left (\vert \xt_{01}\vert \kaz\right )\biggr ]^2\\
&\times \int_{0}^{q^+} \ud k^+_1 \Biggl \{\biggl [1 - \frac{2k_1^+(q^+-k_1^+)}{(q^+)^2} \biggr ]\biggl [-\frac{3}{4}-\log\left (\frac{k^+_{2,min}}{q^+-k^+_1}\right ) \biggr ]\biggl \{ \frac{(4\pi)^{2-\frac{D}{2}} }{ (2-\frac{D}{2})} \Gamma \left (3-\frac{D}{2} \right ) + \log\left (\frac{\xt^2_{01}\mu^2}{4}\right )\\
& + 2\gamma_E\biggr  \} - \frac{1}{4}\frac{(D_s-4)}{(D-4)} \Biggr \} \mathrm{Re}[1-\mathcal{S}_{01}]  + \mathcal{O}(D-4)\\
\end{split}
\end{equation}
and
\begin{equation}
\label{eq:qqbaruvsubtexpj2v2m}
\begin{split}
\sigma^{\gamma^{\ast}}_{T}&\bigg\vert^{\vert \ref{diag:qgqbarT}\vert^2_{\rm{UV}_m}}_{q\bar qg} = 4\nc\alpha_{em}\left (\frac{\alpha_s\cf}{\pi}\right )\sum_{f}e_f^2 \frac{m^2}{q^+} \int_{[\xt_0]}\int_{[\xt_1]}  \left ( \frac{\kaz}{2\pi\vert \xt_{01} \vert }\right )^{D-4}\biggl [K_{\frac{D}{2} -2}\left (\vert \xt_{01}\vert \kaz\right )\biggr ]^2\\
&\times \int_{0}^{q^+} \ud k^+_1  \Biggl \{\biggl [-\frac{3}{4}-\log\left (\frac{k^+_{2,min}}{q^+-k^+_1}\right ) \biggr ]\biggl \{ \frac{(4\pi)^{2-\frac{D}{2}} }{ (2-\frac{D}{2})} \Gamma \left (3-\frac{D}{2} \right ) + \log\left (\frac{\xt^2_{01}\mu^2}{4}\right )\\
& + 2\gamma_E\biggr  \} - \frac{1}{4}\frac{(D_s-4)}{(D-4)} \Biggr \} \mathrm{Re}[1-\mathcal{S}_{01}]  + \mathcal{O}(D-4)\\
\end{split}
\end{equation}
Similarly, for the diagram \ref{diag:qqbargT}, we find 
\begin{equation}
\label{eq:qqbaruvsubtexpk2v2}
\begin{split}
\sigma^{\gamma^{\ast}}_{T}&\bigg\vert^{\vert \ref{diag:qqbargT}\vert^2_{\rm UV}}_{q\bar qg} = 4\nc\alpha_{em}\left (\frac{\alpha_s\cf}{\pi}\right )\sum_{f} \frac{e_f^2}{q^+} \int_{[\xt_0]}\int_{[\xt_1]}  \left ( \frac{\kaz}{2\pi\vert \xt_{01} \vert }\right )^{D-4}\biggl [\kaz K_{\frac{D}{2} -1}\left (\vert \xt_{01}\vert \kaz \right )\biggr ]^2\\
&\times \int_{0}^{q^+} \ud k^+_0 \Biggl \{\biggl [1 - \frac{2k_0^+(q^+-k_0^+)}{(q^+)^2} \biggr ]\biggl [-\frac{3}{4}-\log\left (\frac{k^+_{2,min}}{q^+-k^+_0}\right ) \biggr ]\biggl \{ \frac{(4\pi)^{2-\frac{D}{2}} }{ (2-\frac{D}{2})} \Gamma \left (3-\frac{D}{2} \right ) + \log\left (\frac{\xt^2_{01}\mu^2}{4}\right )\\
& + 2\gamma_E\biggr  \} - \frac{1}{4}\frac{(D_s-4)}{(D-4)} \Biggr \} \mathrm{Re}[1-\mathcal{S}_{01}]  + \mathcal{O}(D-4)\\
\end{split}
\end{equation}
and
\begin{equation}
\label{eq:qqbaruvsubtexpk2v2m}
\begin{split}
\sigma^{\gamma^{\ast}}_{T}&\bigg\vert^{\vert \ref{diag:qqbargT}\vert^2_{\rm{UV}_m}}_{q\bar qg} = 4\nc\alpha_{em}\left (\frac{\alpha_s\cf}{\pi}\right )\sum_{f}e_f^2 \frac{m^2}{q^+} \int_{[\xt_0]}\int_{[\xt_1]}  \left ( \frac{\kaz}{2\pi\vert \xt_{01} \vert }\right )^{D-4}\biggl [K_{\frac{D}{2} -2}\left (\vert \xt_{01}\vert \kaz\right )\biggr ]^2\\
&\times \int_{0}^{q^+} \ud k^+_0  \Biggl \{\biggl [-\frac{3}{4}-\log\left (\frac{k^+_{2,min}}{q^+-k^+_0}\right ) \biggr ]\biggl \{ \frac{(4\pi)^{2-\frac{D}{2}} }{ (2-\frac{D}{2})} \Gamma \left (3-\frac{D}{2} \right ) + \log\left (\frac{\xt^2_{01}\mu^2}{4}\right )\\
& + 2\gamma_E\biggr  \} - \frac{1}{4}\frac{(D_s-4)}{(D-4)} \Biggr \} \mathrm{Re}[1-\mathcal{S}_{01}]  + \mathcal{O}(D-4).\\
\end{split}
\end{equation}
In order to simplify our subtraction terms further, we introduce the same parametrization as in the $q\bar q$ contribution: $k^+_0 = zq^+$, $k^+_1 = (1-z)q^+$ and $k^+_{2,min} = \alpha q^+$ and change of variables from $(k^+_1,k^+_0) \mapsto z$. Hence, the sum of \eqs\nr{eq:qqbaruvsubtexpj2v2} and \nr{eq:qqbaruvsubtexpk2v2} yields the following expression for the UV divergent light cone helicity nonflip subtraction contribution
\begin{equation}
\label{eq:uvsubtractiontermhnonflip}
\begin{split}
\sigma^{\gamma^{\ast}}_{T}&\bigg\vert^{\vert \ref{diag:qgqbarT}\vert^2_{\rm UV} + \vert \ref{diag:qqbargT}\vert^2_{\rm UV}}_{q\bar qg}   = -4\nc\alpha_{em}\left (\frac{\alpha_s\cf}{\pi}\right )\sum_{f}e_f^2 \int_{0}^{1} \ud z \int_{[\xt_0]}\int_{[\xt_1]}  \left ( \frac{\kaz}{2\pi\vert \xt_{01} \vert }\right )^{D-4}\biggl [\kaz K_{\frac{D}{2} -1}\left (\vert \xt_{01}\vert \kaz\right )\biggr ]^2\\
&\times   \biggl [z^2 + (1-z)^2\biggr ]\Biggl \{\biggl [\frac{3}{2} + \log\left (\frac{\alpha}{z}\right ) + \log\left (\frac{\alpha}{1-z}\right ) \biggr ]\biggl \{  \frac{(4\pi)^{2-\frac{D}{2}} }{ (2-\frac{D}{2})} \Gamma \left (3-\frac{D}{2} \right ) + \log \left (\frac{\xt^2_{01} \mu^2}{4}\right )\\
& +  2\gamma_E  \biggr \}  + \frac{1}{2}\frac{(D_s - 4)}{(D - 4)} \Biggr \}\mathrm{Re}[1-\mathcal{S}_{01}] + \mathcal{O}(D-4)\\
\end{split}
\end{equation}
and for the light cone helicity flip contribution 
\begin{equation}
\label{eq:uvsubtractiontermhflip}
\begin{split}
\sigma^{\gamma^{\ast}}_{T}&\bigg\vert^{\vert \ref{diag:qgqbarT}\vert^2_{\rm{UV}_m} + \vert \ref{diag:qqbargT}\vert^2_{\rm{UV}_m}}_{q\bar qg}   = -4\nc\alpha_{em}\left (\frac{\alpha_s\cf}{\pi}\right )\sum_{f}e_f^2 \int_{0}^{1} \ud z  \int_{[\xt_0]}\int_{[\xt_1]} m^2 \left ( \frac{\kaz}{2\pi\vert \xt_{01} \vert }\right )^{D-4}\biggl [K_{\frac{D}{2} -2}\left (\vert \xt_{01}\vert \kaz\right )\biggr ]^2\\
&\times   \Biggl \{\biggl [\frac{3}{2} + \log\left (\frac{\alpha}{z}\right ) + \log\left (\frac{\alpha}{1-z}\right ) \biggr ]\biggl \{  \frac{(4\pi)^{2-\frac{D}{2}} }{ (2-\frac{D}{2})} \Gamma \left (3-\frac{D}{2} \right ) + \log \left (\frac{\xt^2_{01} \mu^2}{4}\right )\\
& +  2\gamma_E  \biggr \}  + \frac{1}{2}\frac{(D_s - 4)}{(D - 4)} \Biggr \}\mathrm{Re}[1-\mathcal{S}_{01}] + \mathcal{O}(D-4).\\
\end{split}
\end{equation}
These expressions precisely cancels the  UV-divergent  part in the $q\bar{q}$ contribution \eq\nr{eq:qqbarcsUVdiv}, including the scheme-dependent part. The remaining terms in \eq\nr{eq:qqbarcsUVdiv} are UV finite and regularization scheme independent.  

The remaining finite terms after the UV subtraction of the cross section are obtained by using the parts inside the curly brackets of \eqs\nr{eq:subprocedure1} and \nr{eq:subprocedure2} in \eqs\eqref{eq:KTqqbargUV} and~\eqref{eq:KTqqbargmUV}. They are a part of our final result for the cross section.

\section{Result: transverse photon cross section}
\label{sec:xsfinal}

We can now gather here the main result of our paper, which is the transverse virtual photon total cross section
at NLO with massive quarks. The cross section can be written as a sum of two UV finite terms
\begin{equation}
\label{eq:fullcslongitudinal}
\sigma^{\gamma^{\ast}}_{T} = \sigma^{\gamma^{\ast}}_{T}\bigg\vert^{\rm subt}_{q\bar{q}} +  \sigma^{\gamma^{\ast}}_{T}\bigg\vert^{\rm subt}_{q\bar{q}g} + \mathcal{O}(\alpha_{em}\alpha_s^2),
\end{equation}
where the first term in \eq\nr{eq:fullcslongitudinal} is the mass renormalized and the UV subtracted $q\bar{q}$ contribution, which is obtained by adding the UV subtraction terms in \eqs\nr{eq:uvsubtractiontermhnonflip} and \nr{eq:uvsubtractiontermhflip} into \eq\nr{eq:qqbarcsUVdiv}. This gives 
\begin{equation}
\label{eq:qqbarcsUVsubt}
\begin{split}
\sigma^{\gamma^{\ast}}_{T}\bigg\vert^{\rm subt}_{q\bar{q}}  & = 4\nc\alpha_{em}\sum_{f}e_f^2 \int_{0}^{1} \ud z \int_{\xt_0}\int_{\xt_1} \Biggl \{ \biggl [z^2 + (1-z)^2 \biggr ]
\biggl [\kaz K_{1}\left (\vert \xt_{01}\vert \kaz \right )\biggr ]^2 +  m^2\biggl [K_{0}\left (\vert \xt_{01}\vert \kaz\right )\biggr ]^2 \\
& + \left (\frac{\alpha_s\cf}{\pi}\right )\biggl \{\biggl [z^2 + (1-z)^2 \biggr ]\kaz K_1(\vert \xt_{01}\vert \kaz)f_{\mathcal{V}} + \frac{(2z-1)}{2} \kaz K_1(\vert \xt_{01}\vert \kaz) f_{\mathcal{N}}\\
& + m^2 K_0(\vert \xt_{01}\vert \kaz) f_{\mathcal{VMS}} \biggr \}\Biggr \}\mathrm{Re}[1-\mathcal{S}_{01}]  +  \mathcal{O}(\alpha_{em}\alpha_s^2),  
\end{split}
\end{equation}
where the short-hand notation  $\int_{\xt_i} = \int \ud^{2} \xt_i/(2\pi)$ for $i=0,1$ denotes a 2-dimensional transverse coordinate integral and  we recall that $\kappa_v = \sqrt{v(1-v)Q^2 + m^2}$. The functions $f_{\mathcal{V}}, f_{\mathcal{N}}$ and $f_{\mathcal{VMS}}$ are explicitly written down in \eq\nr{eq:fVfNfVMS}. Evaluating the expressions $f_{\mathcal{V}}, f_{\mathcal{N}}$ and $f_{\mathcal{VMS}}$ also requires  the definitions of 
$L$ from \eq\eqref{eq:defL},  $\Omega_{\mathcal{N}}$ from \eqs\eqref{eq:degOmegaN_1}
and~\eqref{eq:degOmegaN},
$\widetilde{I}_{\mathcal{N}}$ from \eq\eqref{eq:INT},
$\widetilde{I}_{\mathcal{V}}$ from \eq\eqref{eq:IVT},
$\Omega_{\mathcal{V}}$ from \eqs\eqref{eq:OSI} and~\eqref{eq:SigmaOmega} and   $\widetilde{I}_{\mathcal{VMS}}$ from \eq\eqref{eq:IVMST}.
Here the first line in \eq\nr{eq:qqbarcsUVsubt}, not proportional to $\alpha_s$, is explicitly the known leading order cross section for massive quarks and the functions $f_{i}$ encode the order $\alpha_s$ NLO corrections. We have checked that in the limit of zero quark mass these expressions reduce to the known results in Refs.~\cite{Hanninen:2017ddy,Beuf:2016wdz}.

The second term in \eq\nr{eq:fullcslongitudinal} is the UV subtracted $q\bar{q}g$ contribution. It contains both the UV-subtracted terms corresponding to the parts inside the curly brackets of \eqs\nr{eq:subprocedure1} and \nr{eq:subprocedure2}, and the parts of the $q\bar{q}g$-contribution that were finite to begin with:
\begin{equation}
\label{eq:csqqbarg4terms}
\sigma^{\gamma^{\ast}}_{T}\bigg\vert^{\rm subt}_{q\bar{q}g} =  \sigma^{\gamma^{\ast}}_{T}\bigg\vert^{\vert \ref{diag:qgqbarT}\vert^2 + \vert \ref{diag:qqbargT}\vert^2}_{q\bar qg} + \sigma^{\gamma^{\ast}}_{T}\bigg\vert^{\vert \ref{diag:qgqbarT}\vert^2_m + \vert \ref{diag:qqbargT}\vert^2_m}_{q\bar qg} + \sigma^{\gamma^{\ast}}_{T}\bigg\vert^{\text{F}}_{q\bar{q}g} + \sigma^{\gamma^{\ast}}_{T}\bigg\vert^{ \text{F}_m}_{q\bar{q}g}, 
\end{equation}
where the last two terms correspond a contribution coming from the finite terms without a light cone helicity flip \eqs\nr{eq:KfinHNF} and the ones with one or more helicity flip terms, proportional to powers of the mass, \nr{eq:KfinHF}.

The  final result for the $q\bar{q}g$-part contains a significant number of terms. In order to clarify the structure we use  the relative coordinates corresponding to the gluon emission diagrams \ref{diag:qgqbarT} and \ref{diag:qqbargT}: 
 for the diagram \ref{diag:qgqbarT} with emission from the quark the $q(\kvec_0)-g(\kvec_2)$ relative separation $\xt_{2;\ref{diag:qgqbarT}} \equiv  \xt_{20}$ defined in \eq\nr{eq:defx2_a} and  the separation of the $qg$ system to the recoiling  
 antiquark ($\bar{q}(\kvec_1)$) $\xt_{3;\ref{diag:qgqbarT}}\equiv   \xt_{0+2;1}$ from \eq\nr{eq:defx3_a}, and conversely 
 for the diagram \ref{diag:qqbargT} with emission from the antiquark the $\bar{q}(\kvec_1)-g(\kvec_2)$ relative separation $\xt_{2;\ref{diag:qqbargT}} \equiv  \xt_{21}$ defined in \eq\nr{eq:defx2_b} and  the separation of the recoiling quark to the $\bar{q}g$ system  ($q(\kvec_0)$) $\xt_{3;\ref{diag:qqbargT}}\equiv   \xt_{0;1+2}$ from \eq\nr{eq:defx3_b}.

In the case of the exponential subtraction scheme developed \cite{Hanninen:2017ddy}, i.e. using \eq\eqref{eq:uvappr1}, the sum of two UV subtracted terms in  can be simplified to 
\begin{equation}
\label{eq:exptsubtqbarqghnf}
\begin{split}
&\sigma^{\gamma^{\ast}}_{T}\bigg\vert^{\vert \ref{diag:qgqbarT}\vert^2 + \vert \ref{diag:qqbargT}\vert^2}_{q\bar qg}   =  4\nc\alpha_{em} \left (\frac{\alpha_s\cf}{\pi}\right )\sum_{f}e_f^2\int_{\xt_0}\int_{\xt_1}\int_{\xt_2} \int_{0}^{\infty} \ud k^+_0\int_{0}^{\infty} \ud k^+_1\int_{0}^{\infty} \frac{\ud k^+_2}{k^+_2}\frac{\delta(q^+-\sum_{i=0}^{2}k^+_i)}{q^+}\\
&\times \Biggl \{\frac{1}{(k_0^+ + k_2^+)^2}\biggl [2k^+_0(k^+_0 + k^+_2) + (k^+_2)^2 \biggr ]\biggl [1 - \frac{2k_1^+(q^+-k_1^+)}{(q^+)^2} \biggr ]\\
&  \times \biggl \{   \frac{\vert \xt_{3;\ref{diag:qgqbarT}}\vert^2\vert\xt_{2;\ref{diag:qgqbarT}}\vert^2}{256}[\mathcal{G}_{\ref{diag:qgqbarT}}^{(2;2)}]^2 \,\mathrm{Re}[1-\mathcal{S}_{012}]  - \frac{e^{-\vert\xt_{2;\ref{diag:qgqbarT}}\vert^2/(\vert\xt_{01}\vert^2 e^{\gamma_E})    }}{\vert\xt_{2;\ref{diag:qgqbarT}}\vert^2} \biggl [\sqrt{\overline{Q}^2_{\ref{diag:qgqbarT}} + m^2}K_1 \left (\vert \xt_{01}\vert \sqrt{\overline{Q}^2_{\ref{diag:qgqbarT}} + m^2} \right )\biggr ]^2   \mathrm{Re}[1-\mathcal{S}_{01}]           \biggr \} \\
& + \frac{1}{(k_1^+ + k_2^+)^2}\biggl [2k^+_1(k^+_1 + k^+_2) + (k^+_2)^2 \biggr ]\biggl [1 - \frac{2k_0^+(q^+-k_0^+)}{(q^+)^2} \biggr ]\\
&  \times \biggl \{    \frac{\vert \xt_{3;\ref{diag:qqbargT}}\vert^2\vert\xt_{2;\ref{diag:qqbargT}}\vert^2}{256}[\mathcal{G}_{\ref{diag:qqbargT}}^{(2;2)}]^2 \,\mathrm{Re}[1-\mathcal{S}_{012}]  - \frac{e^{-\vert\xt_{2;\ref{diag:qqbargT}}\vert^2/(\vert\xt_{01}\vert^2 e^{\gamma_E})  }}{\vert\xt_{2;\ref{diag:qqbargT}}\vert^2} \biggl [\sqrt{\overline{Q}^2_{\ref{diag:qqbargT}} + m^2}  K_1 \left (\vert \xt_{01}\vert \sqrt{\overline{Q}^2_{\ref{diag:qqbargT}} + m^2} \right )\biggr ]^2   \mathrm{Re}[1-\mathcal{S}_{01}]   \biggr \}\Biggr \},
\end{split}
\end{equation}
and
\begin{equation}
\label{eq:exptsubtqbarqghf}
\begin{split}
&\sigma^{\gamma^{\ast}}_{T}\bigg\vert^{\vert \ref{diag:qgqbarT}\vert^2_m + \vert \ref{diag:qqbargT}\vert^2_m}_{q\bar qg}   =  4\nc\alpha_{em} m^2\left (\frac{\alpha_s\cf}{\pi}\right )\sum_{f}e_f^2\int_{\xt_0}\int_{\xt_1}\int_{\xt_2} \int_{0}^{\infty} \ud k^+_0\int_{0}^{\infty} \ud k^+_1\int_{0}^{\infty} \frac{\ud k^+_2}{k^+_2}\frac{\delta(q^+-\sum_{i=0}^{2}k^+_i)}{q^+}\\
&\times \Biggl \{\frac{1}{(k_0^+ + k_2^+)^2}\biggl [2k^+_0(k^+_0 + k^+_2) + (k^+_2)^2 \biggr ]\\
&  \times \biggl \{   \frac{\vert\xt_{2;\ref{diag:qgqbarT}}\vert^2}{64}[\mathcal{G}_{\ref{diag:qgqbarT}}^{(1;2)}]^2 \,\mathrm{Re}[1-\mathcal{S}_{012}]  - \frac{e^{-\vert\xt_{2;\ref{diag:qgqbarT}}\vert^2/(\vert\xt_{01}\vert^2 e^{\gamma_E})    }}{\vert\xt_{2;\ref{diag:qgqbarT}}\vert^2} \biggl [K_0\left (\vert \xt_{01}\vert \sqrt{\overline{Q}^2_{\ref{diag:qgqbarT}} + m^2} \right )\biggr ]^2   \mathrm{Re}[1-\mathcal{S}_{01}]           \biggr \} \\
& + \frac{1}{(k_1^+ + k_2^+)^2}\biggl [2k^+_1(k^+_1 + k^+_2) + (k^+_2)^2 \biggr ]\\
&  \times \biggl \{    \frac{\vert\xt_{2;\ref{diag:qqbargT}}\vert^2}{64}[\mathcal{G}_{\ref{diag:qqbargT}}^{(1;2)}]^2 \,\mathrm{Re}[1-\mathcal{S}_{012}]  - \frac{e^{-\vert\xt_{2;\ref{diag:qqbargT}}\vert^2/(\vert\xt_{01}\vert^2 e^{\gamma_E})  }}{\vert\xt_{2;\ref{diag:qqbargT}}\vert^2} \biggl [K_0 \left (\vert \xt_{01}\vert \sqrt{\overline{Q}^2_{\ref{diag:qqbargT}} + m^2} \right )\biggr ]^2   \mathrm{Re}[1-\mathcal{S}_{01}] \biggr \}\Biggr \}.
\end{split}
\end{equation}
Here we have introduced the compact notation  
\begin{equation}
\label{eq:defG_a}
\mathcal{G}_{\ref{diag:qgqbarT}}^{(a;b)} = \int_{0}^{\infty} \frac{\ud u}{u^a} 
\exp\left\{-u\left(\overline{Q}^2_{\ref{diag:qgqbarT}} + m^2\right) \right\} \exp\left\{-\frac{\left|\xt_{3;\ref{diag:qgqbarT}}\right|^2}{4u}\right\}
\int_{0}^{u/\omega_{\ref{diag:qgqbarT}}} \frac{\ud t}{t^b} \exp\left\{-t\left(\omega_{\ref{diag:qgqbarT}}\lambda_{\ref{diag:qgqbarT}}m^2\right)\right\}
\exp\left\{-\frac{\left|\xt_{2;\ref{diag:qgqbarT}}\right|^2}{4t}\right\}
\end{equation}
and
\begin{equation}
\label{eq:defG_b}
\mathcal{G}_{\ref{diag:qqbargT}}^{(a;b)} = \int_{0}^{\infty} \frac{\ud u}{u^a} 
\exp\left\{-u\left(\overline{Q}^2_{\ref{diag:qqbargT}} + m^2\right)\right\} \exp\left\{-\frac{\left|\xt_{3;\ref{diag:qqbargT}}\right|^2}{4u}\right\}
\int_{0}^{u/\omega_{\ref{diag:qqbargT}}} \frac{\ud t}{t^b} \exp\left\{-t\left(\omega_{\ref{diag:qqbargT}}\lambda_{\ref{diag:qqbargT}}m^2\right)\right\}
\exp\left\{-\frac{\left|\xt_{2;\ref{diag:qqbargT}}\right|^2}{4t}\right\}. 
\end{equation}
The notations $\overline{Q}^2_{\ref{diag:qgqbarT}},
\overline{Q}^2_{\ref{diag:qqbargT}},
  \lambda_{\ref{diag:qgqbarT}} ,
 \lambda_{\ref{diag:qqbargT}}$ are defined in 
\eqs\eqref{eq:coeffqgbarq},~\eqref{eq:coeffqbarqg}
and
$\omega_{\ref{diag:qgqbarT}}, \omega_{\ref{diag:qqbargT}}$ in \eq\eqref{eq:defomega}.
As discussed also in the longitudinal case \cite{Beuf:2021qqa}, these integrals could be seen as generalizations of the integral representation of Bessel
functions that appear in the massless case. In addition, these
integrals are very rapidly converging at both small and large values of the integration variable, and hence they should be well suited for numerical evaluation as is.


Let us finally turn to the last two terms in \eq\nr{eq:csqqbarg4terms}, which are finite from the beginning. They can be simplified starting from the intermediate expressions in \eqs\nr{eq:KfinHNF} and~\nr{eq:KfinHF}, and using the Fourier transform integrals in Appendix~\ref{app:FTintforqqbarg}. 
After some intermediate stages of algebra, the third term in  \eq\nr{eq:csqqbarg4terms} can be simplified into the following form 
 \begin{equation}
\begin{split}
\sigma&^{\gamma^{\ast}}_{T}\bigg\vert^{\text{F}}_{q\bar{q}g}    =  4\nc\alpha_{em} \frac{1}{2}\left (\frac{\alpha_s\cf}{\pi}\right )\sum_{f}e_f^2\int_{\xt_0}\int_{\xt_1}\int_{\xt_2} \int_{0}^{\infty} \ud k^+_0\int_{0}^{\infty} \ud k^+_1\int_{0}^{\infty} \frac{\ud k^+_2}{k^+_2}\frac{\delta(q^+-\sum_{i=0}^{2}k^+_i)}{q^+}\\
&\times \Biggl \{\frac{1}{64 (k_0^+ +k_2^+)(k_1^+ + k_2^+)(q^+)^2}\biggl \{q^+ k_2^+(k_0^+ - k_1^+)^2\biggl [(\xt_{3;\ref{diag:qgqbarT}} \cdot \xt_{2;\ref{diag:qgqbarT}})(\xt_{3;\ref{diag:qqbargT}}\cdot\xt_{2;\ref{diag:qqbargT}}) - (\xt_{3;\ref{diag:qgqbarT}}\cdot\xt_{2;\ref{diag:qqbargT}})(\xt_{3;\ref{diag:qqbargT}}\cdot \xt_{2;\ref{diag:qgqbarT}}) \biggr ]\\
& - \biggl [k_1^+(k_0^+ + k_2^+) + k_0^+(k_1^+ + k_2^+)\biggr ]\biggl [k_0^+(k_0^+ + k_2^+) + k_1^+(k_1^+ + k_2^+)\biggr ] (\xt_{3;\ref{diag:qgqbarT}}\cdot \xt_{3;\ref{diag:qqbargT}})(\xt_{2;\ref{diag:qgqbarT}}\cdot\xt_{2;\ref{diag:qqbargT}})\biggr \} \mathcal{G}_{\ref{diag:qgqbarT}}^{(2;2)} \mathcal{G}_{\ref{diag:qqbargT}}^{(2;2)} \\
& -  \frac{(k_0^+ + k_2^+)k_1^+ k_2^+}{16(k_1^+ + k_2^+)^2 q^+}(\xt_{3;\ref{diag:qgqbarT}}\cdot \xt_{2;\ref{diag:qgqbarT}}) \mathcal{G}_{\ref{diag:qgqbarT}}^{(2;2)} \mathcal{H}_{\ref{diag:qqbargT}} 
+ \frac{(k_1^+ + k_2^+)k_0^+ k_2^+}{16(k_0^+ + k_2^+)^2 q^+}(\xt_{3;\ref{diag:qqbargT}}\cdot\xt_{2;\ref{diag:qqbargT}}) \mathcal{G}_{\ref{diag:qqbargT}}^{(2;2)}\mathcal{H}_{\ref{diag:qgqbarT}} \\
&   - \frac{ (k_0^+)^2 k_1^+ k_2^+}{16(k_0 + k_2^+)^3q^+} (\xt_{3;\ref{diag:qgqbarT}}\cdot \xt_{2;\ref{diag:qgqbarT}}) \mathcal{G}_{\ref{diag:qgqbarT}}^{(2;2)}\mathcal{H}_{\ref{diag:qgqbarT}}
+ \frac{(k_1^+)^2 k_0^+ k_2^+}{16(k_1 + k_2^+)^3q^+} (\xt_{3;\ref{diag:qqbargT}}\cdot\xt_{2;\ref{diag:qqbargT}}) \mathcal{G}_{\ref{diag:qqbargT}}^{(2;2)}\mathcal{H}_{\ref{diag:qqbargT}}  \\
& + \frac{(k_0^+ k_2^+)^2}{8(k_0^+ + k_2^+)^4}  [\mathcal{H}_{\ref{diag:qgqbarT}}]^2
+ \frac{(k_1^+ k_2^+)^2}{8(k_1^+ + k_2^+)^4} [\mathcal{H}_{\ref{diag:qqbargT}}]^2 \Biggr \}
\mathrm{Re}[1-\mathcal{S}_{012}].
\end{split}
\end{equation}
Similarly, the last term in \eq\nr{eq:csqqbarg4terms} reads
\begin{equation}
\begin{split}
&\sigma^{\gamma^{\ast}}_{T}\bigg\vert^{\text{F}_m}_{q\bar{q}g}    =  4\nc\alpha_{em} \frac{m^2}{2} \left (\frac{\alpha_s\cf}{\pi}\right )\sum_{f}e_f^2\int_{\xt_0}\int_{\xt_1}\int_{\xt_2} \int_{0}^{\infty} \ud k^+_0\int_{0}^{\infty} \ud k^+_1\int_{0}^{\infty} \frac{\ud k^+_2}{k^+_2}\frac{\delta(q^+-\sum_{i=0}^{2}k^+_i)}{q^+}  \\
&\times \Biggl \{\frac{(k_2^+)^4}{64(k_0^+ + k_2^+)^4 (q^+)^2}\biggl [4k_1^+(k_1^+ - q^+) + 2(q^+)^2\biggr ] \vert\xt_{3;\ref{diag:qgqbarT}}\vert^2 [\mathcal{G}_{\ref{diag:qgqbarT}}^{(2;1)}]^2 
- \frac{k_0^+k_1^+(k_2^+)^2}{16(k_0^+ + k_2^+)^3 q^+}  (\xt_{3;\ref{diag:qgqbarT}}\cdot \xt_{2;\ref{diag:qgqbarT}})  \mathcal{G}_{\ref{diag:qgqbarT}}^{(1;2)} \mathcal{G}_{\ref{diag:qgqbarT}}^{(2;1)} \\
& + \frac{(k_2^+)^4}{64(k_1^+ + k_2^+)^4 (q^+)^2}\biggl [4k_0^+(k_0^+ - q^+) + 2(q^+)^2 \biggr ] \vert\xt_{3;\ref{diag:qqbargT}}\vert^2 [\mathcal{G}_{\ref{diag:qqbargT}}^{(2;1)}]^2 
+ \frac{k_0^+k_1^+(k_2^+)^2}{16(k_1^+ + k_2^+)^3 q^+}  (\xt_{3;\ref{diag:qqbargT}}\cdot\xt_{2;\ref{diag:qqbargT}})\mathcal{G}_{\ref{diag:qqbargT}}^{(1;2)}\mathcal{G}_{\ref{diag:qqbargT}}^{(2;1)} \\
& -\frac{1}{32(k_0^+ + k_2^+)(k_1^+ + k_2^+)}\biggl [(2k_0^+ + k_2^+)(2k_1^+ + k_2^+) + (k_2^+)^2\biggr ] (\xt_{2;\ref{diag:qgqbarT}} \cdot\xt_{2;\ref{diag:qqbargT}}) \mathcal{G}_{\ref{diag:qgqbarT}}^{(1;2)} \mathcal{G}_{\ref{diag:qqbargT}}^{(1;2)} \\
& + \frac{(k_2^+)^4}{32(k_0^+ + k_2^+)^2(k_1^+ + k_2^+)^2 (q^+)^2}\biggl [(2k_0^+ + k_2^+)(2k_1^+ + k_2^+) + (k_2^+)^2\biggr ] (\xt_{3;\ref{diag:qgqbarT}} \cdot \xt_{3;\ref{diag:qqbargT}}) \mathcal{G}_{\ref{diag:qgqbarT}}^{(2;1)}\mathcal{G}_{\ref{diag:qqbargT}}^{(2;1)}\\
& + \frac{m^2}{16} \frac{2(k_2^+)^4}{(k_0^+ + k_2^+)^4}[\mathcal{G}_{\ref{diag:qgqbarT}}^{(1;1)}]^2  + \frac{m^2}{16} \frac{2(k_2^+)^4}{(k_1^+ + k_2^+)^4} [\mathcal{G}_{\ref{diag:qqbargT}}^{(1;1)}]^2\\
& - \frac{(k_0^+k_2^+)^2}{16(k_0^+ + k_2^+)(k_1^+ + k_2^+)^2 q^+}  (\xt_{2;\ref{diag:qgqbarT}} \cdot \xt_{3;\ref{diag:qqbargT}}) \mathcal{G}_{\ref{diag:qgqbarT}}^{(1;2)}\mathcal{G}_{\ref{diag:qqbargT}}^{(2;1)} 
+ \frac{(k_1^+k_2^+)^2}{16(k_1^+ + k_2^+)(k_0^+ + k_2^+)^2 q^+} (\xt_{3;\ref{diag:qgqbarT}} \cdot\xt_{2;\ref{diag:qqbargT}})\mathcal{G}_{\ref{diag:qgqbarT}}^{(2;1)}\mathcal{G}_{\ref{diag:qqbargT}}^{(1;2)}\\
& - \frac{(k_0^+k_1^+)(k_2^+)^2}{16(k_0^+ + k_2^+)^3 q^+} (\xt_{2;\ref{diag:qgqbarT}} \cdot \xt_{3;\ref{diag:qgqbarT}}) \mathcal{G}_{\ref{diag:qgqbarT}}^{(2;2)} \mathcal{G}_{\ref{diag:qgqbarT}}^{(1;1)}
+ \frac{(k_0^+k_1^+)(k_2^+)^2}{16(k_1^+ + k_2^+)^3 q^+} (\xt_{2;\ref{diag:qqbargT}} \cdot \xt_{3;\ref{diag:qqbargT}}) \mathcal{G}_{\ref{diag:qqbargT}}^{(2;2)}][\mathcal{G}_{\ref{diag:qqbargT}}^{(1;1)}]\\
& - \frac{(k_0^+ + k_2^+)(k_2^+)^2}{16(k_1^+ + k_2^+)^2 q^+} (\xt_{2;\ref{diag:qgqbarT}} \cdot \xt_{3;\ref{diag:qgqbarT}}) \mathcal{G}_{\ref{diag:qgqbarT}}^{(2;2)} \mathcal{G}_{\ref{diag:qqbargT}}^{(1;1)} 
+ \frac{(k_1^+ + k_2^+)(k_2^+)^2}{16(k_0^+ + k_2^+)^2 q^+}  (\xt_{2;\ref{diag:qqbargT}} \cdot \xt_{3;\ref{diag:qqbargT}}) \mathcal{G}_{\ref{diag:qqbargT}}^{(2;2)} \mathcal{G}_{\ref{diag:qgqbarT}}^{(1;1)} \\
& + \frac{k_0^+(k_2^+)^3}{4(k_0^+ + k_2^+)^4} \mathcal{H}_{\ref{diag:qgqbarT}} \mathcal{G}_{\ref{diag:qgqbarT}}^{(1;1)}  + \frac{k_1^+(k_2^+)^3}{4(k_1^+ + k_2^+)^4} \mathcal{H}_{\ref{diag:qqbargT}} \mathcal{G}_{\ref{diag:qqbargT}}^{(1;1)} \Biggr \}\mathrm{Re}[1-\mathcal{S}_{012}].
\end{split}
\end{equation}
Here we have also introduced the functions $\mathcal{H}_{\ref{diag:qgqbarT}}$ and $\mathcal{H}_{\ref{diag:qqbargT}}$, which are defined as
\begin{eqnarray}
\label{eq:defH_a}
\mathcal{H}_{\ref{diag:qgqbarT}} &=& \int_{0}^{\infty} \frac{\ud u}{u^2} 
\exp\left\{-u \left[\overline{Q}^2_{\ref{diag:qgqbarT}} + m^2 + \lambda_{\ref{diag:qgqbarT}}m^2\right]\right\}
\exp\left\{ -\frac{\vert \xt_{3;\ref{diag:qgqbarT}}\vert^2 + \omega_{\ref{diag:qgqbarT}}\vert\xt_{2;\ref{diag:qgqbarT}}\vert^2}{4u}\right\} 
\\
\label{eq:defH_b}
\mathcal{H}_{\ref{diag:qqbargT}} &=& \int_{0}^{\infty} \frac{\ud u}{u^2} 
\exp\left\{-u \left[\overline{Q}^2_{\ref{diag:qqbargT}} + m^2 + \lambda_{\ref{diag:qqbargT}}m^2\right]\right\}
\exp\left\{-\frac{\vert \xt_{3;\ref{diag:qqbargT}}\vert^2 + \omega_{\ref{diag:qqbargT}}\vert\xt_{2;\ref{diag:qqbargT}}\vert^2}{4u}\right\}. 
\end{eqnarray}
To summarize the abbreviated notations, we recall that the definitions needed for an evaluation of the $q\bar{q}g$ results can be found in 
\begin{itemize}
    \item For the gluon emission relative coordinates $\xt_{2;\ref{diag:qgqbarT}}$, $\xt_{2;\ref{diag:qqbargT}}$ in \eqs\eqref{eq:defx2_a}, \eqref{eq:defx2_b} and 3-particle center of mass coordinates 
    $\xt_{3;\ref{diag:qgqbarT}}$ and $\xt_{3;\ref{diag:qqbargT}}$ in \eqs\eqref{eq:defx3_a}, \eqref{eq:defx3_b}.
\item Generalized Bessel functions $\mathcal{G}_{(x)}^{(a;b)}$ in  \eqs\eqref{eq:defG_a}, \eqref{eq:defG_b}
     and 
$\mathcal{H}_{(x)}$ in \eqs\eqref{eq:defH_a}, \eqref{eq:defH_b}.
    \item The momentum scales $\overline{Q}^2_{\ref{diag:qgqbarT}},
\overline{Q}^2_{\ref{diag:qqbargT}}$ and longitudinal momentum fraction ratios 
$  \lambda_{\ref{diag:qgqbarT}} ,
 \lambda_{\ref{diag:qqbargT}}$  in 
\eqs\eqref{eq:coeffqgbarq},~\eqref{eq:coeffqbarqg}
and the momentum fraction ratios
$\omega_{\ref{diag:qgqbarT}}, \omega_{\ref{diag:qqbargT}}$ in \eq\eqref{eq:defomega}.
\end{itemize}

\section{Conclusion}
\label{sec:conc}

In this paper, we have calculated, for the first time in the literature, the one-loop light cone wave function for a transverse photon splitting into a quark-antiquark pair including quark masses. After obtaining the one loop LCWF we also Fourier-transformed our result to mixed transverse coordinate–longitudinal momentum space. Combined with the tree level wave function for a quark-antiquark-gluon state this enabled us to obtain an explicit expression for the total transverse photon cross section in the dipole picture, suited for cross section calculations in the nonlinear gluon saturation regime.

We believe that our results are a significant in several ways. Firstly, we have provided the first calculation of the total DIS cross section for massive quarks in the dipole picture at NLO accuracy. This paves the way for a description of existing and future total DIS cross sections in the small-$x$ saturation regime, following the recent first successful NLO fit for massless quarks in \cite{Beuf:2020dxl}. We provide in this paper the explicit expressions for the cross section which, together with a BK-factorization procedure,  e.g. one of the three used in \cite{Beuf:2020dxl}, can directly be used in similar fits. More generally, the light cone wave function is a central ingredient in any NLO calculation in the small-$x$ dipole factorization formulation for DIS or photoproduction processes involving heavy quarks. Thus it is an essential ingredient in bringing the phenomenology of gluon saturation in high energy deep inelastic scattering to NLO accuracy. The light cone wave function is needed for calculations of, e.g. exclusive vector meson production, dijet production or  diffractive structure functions in DIS. All of these processes can now be accessed at NLO accuracy in the saturation regime, also for massive quarks. On yet a more fundamental level, the light cone wavefunction, independently of the needs of a specific cross section calculation, is a fundamental universal quantity in perturbative QCD. It encodes the structure of a photon in light cone gauge, which is required for a proper partonic interpretation. In calculating the wave function to one loop accuracy with massive quarks, we have also had to address explicitly the issue of quark mass renormalization in light cone perturbation theory. This is an issue that had previously been discussed in the literature at the level of the divergent terms. To our knowledge our calculation is, however, the first LCPT loop calculation to perform a full quark mass renormalization in a specific scheme, including the finite terms. We will discuss the details of different regularization procedures and their relation to mass renormalization in more detail in a separate publication. However, already here we have demonstrated in practice the relation between  mass renormalization in LCPT and Lorentz-invariance, encoded in the structure of the quark form factor. The final result for the wavefunction and cross section obtained in this paper are also summarized in a shorter companion paper~\cite{Beuf:2021srj}.

\subsection*{Acknowledgments} 

We thank H. M\"{a}ntysaari and J. Penttala for useful discussions. This work has been supported by the Academy of Finland  Centre of Excellence in Quark Matter (project 346324) and projects  321840 and 322507, under the European Union’s Horizon 2020 research and innovation programme by the STRONG-2020 project (grant agreement No 824093), 
by the European Research Council,  grant agreements ERC-2015-CoG-681707, ERC-2016-CoG-725369 and ERC-2018-AdG-835105
and by the National Science Centre (Poland) under the research Grant No. 2020/38/E/ST2/00122 (SONATA BIS 10).  The content of this article does not reflect the official opinion of the European Union and responsibility for the information and views expressed therein lies entirely with the authors.
\appendix

\section{Calculation of the numerators}
\label{app:nums}

In this Appendix we evaluate the numerators for the self-energy diagram  \ref{diag:oneloopSEUPT}, the instantaneous diagrams \ref{diag:oneloopSEUPTinst}, \ref{diag:oneloopvertexinst1}, $\ref{diag:gluoninst}_1$, and the vertex correction diagram \ref{diag:vertexqbaremT}. Summation notation over the internal helicities, gluon polarization and color is always implicit.

\subsection{Numerator for the quark self-energy diagram \ref{diag:oneloopSEUPT}}
\label{app:numa}

The numerator for the quark self-energy diagram \ref{diag:oneloopSEUPT} can be written as   
\begin{equation}
\label{eq:numaapp}
\begin{split}
N^T_{\ref{diag:oneloopSEUPT}} & = 
ee_fg^2t^{a}_{\alpha_0\alpha_{0'}}t^{a}_{\alpha_{0'}\alpha_1}\biggl [\bar{u}(0)\epsl_{\sigma}(k)u(0')\biggr ]\biggl [\bar{u}(0')\epsl^{\ast}_{\sigma}(k)u(0'')\biggr ]\biggl [\bar{u}(0'')\epsl_{\lambda}(q)v(1)\biggr ]\\
& = ee_fg^2\delta_{\alpha_0\alpha_1}\cf \biggl [\bar{u}(0)\epsl_{\sigma}(k)u(0')\biggr ]\biggl [\bar{u}(0')\epsl^{\ast}_{\sigma}(k)u(0)\biggr ]\biggl [\bar{u}(0)\epsl_{\lambda}(q)v(1)\biggr ],
\end{split}
\end{equation}
where the four-vectors $k_0$ and $k_{0''}$ are the same since the plus and the transverse momentum are conserved and both $k_0$ and $ k_{0''}$ are on-shell. The spinor structures inside the first two square brackets can be written in the helicity basis as:
\begin{equation}
\label{eq:vertex1a}
\begin{split}
\bar{u}(0)\epsl_{\sigma}(k)u(0') = \frac{1}{k^+_0 \left (\frac{k^+}{k^+_0}\right )\left (1 -\frac{k^+}{k^+_0}\right ) }\Biggl \{\Biggl [\left (1-\frac{k^+}{2k^+_0}\right )\delta^{ij}_{(D_s)}\bar{u}(0)\gamma^{+}u(0') & + \left (\frac{k^+}{4k^+_0}\right )\bar{u}(0)\gamma^{+}[\gamma^i,\gamma^j]u(0')  \Biggr ]\Kt^i\\
& + \frac{m}{2}\left (\frac{k^+}{k^+_0}\right )^2\bar{u}(0)\gamma^+\gamma^ju(0')\Biggr \}\epst_{\sigma}^j,
\end{split}
\end{equation}
\begin{equation}
\label{eq:vertex2a}
\begin{split}
\bar{u}(0')\epsl^{\ast}_{\sigma}(k)u(0) = \frac{1}{k^+_0 \left (\frac{k^+}{k^+_0}\right )\left (1 -\frac{k^+}{k^+_0}\right ) }\Biggl \{\Biggl [\left (1-\frac{k^+}{2k^+_0}\right )\delta^{kl}_{(D_s)}\bar{u}(0')\gamma^{+}u(0) & - \left (\frac{k^+}{4k^+_0}\right )\bar{u}(0')\gamma^{+}[\gamma^k,\gamma^l]u(0)  \Biggr ]\Kt^k\\
& - \frac{m}{2}\left (\frac{k^+}{k^+_0}\right )^2\bar{u}(0')\gamma^+\gamma^lu(0)\Biggr \}\epst^{\ast l}_{\sigma},
\end{split}
\end{equation}
where the variable $\Kt$ is defined in \eq\nr{eq:KTdef}. Using the completeness relation for the spinors
\begin{equation}
\label{eq:sumover0}
\sum_{h_{0'}}u(0')\bar{u}(0') = \ksl_{0'} + m
\end{equation}
and noting that $\gamma^+(\ksl_{0'} + m)\gamma^+ = 2(k^+_0-k^+)\gamma^+$, we obtain
\begin{equation}
\label{eq:numaappv2}
\begin{split}
N^T_{\ref{diag:oneloopSEUPT}} = &\frac{2ee_fg^2\delta_{\alpha_0\alpha_1}\cf}{k_0^+\left (\frac{k^+}{k^+_0}\right )^2\left (1 -\frac{k^+}{k^+_0}\right )}\biggl [\bar{u}(0)\epsl_{\lambda}(q)v(1)\biggr ]\Biggl \{\Biggl [\left (1-\frac{k^+}{2k^+_0}\right )^2\delta^{ij}_{(D_s)}\delta^{kl}_{(D_s)}\bar{u}(0)\gamma^{+}u(0)\\
& - \left (\frac{k^+}{4k^+_0}\right )^2\bar{u}(0)\gamma^{+}[\gamma^i,\gamma^j][\gamma^k,\gamma^l]u(0) \Biggr ]\Kt^i\Kt^k - \frac{m^2}{4}\left (\frac{k^+}{k^+_0} \right )^4 \bar{u}(0)\gamma^+\gamma^j\gamma^lu(0)\Biggr \}\epst_{\sigma}^j\epst^{\ast l}_{\sigma}.
\end{split}
\end{equation}
Here the terms linear in the transverse integration variable $\Kt$ vanish due to rotational symmetry. Summing over the gluon helicity states yields
\begin{equation}
\label{eq:numaappv3}
\begin{split}
N^T_{\ref{diag:oneloopSEUPT}} = \frac{4g^2\cf}{\left (\frac{k^+}{k^+_0}\right )^2\left (1 -\frac{k^+}{k^+_0}\right )}\delta_{\alpha_0\alpha_1}V^{\gamma^{\ast}_T \rightarrow q\bar{q}}_{h_0;h_1}\Biggl \{\Biggl [\left (1-\frac{k^+}{2k^+_0}\right )^2  + 4(D_s-3)\left (\frac{k^+}{4k^+_0}\right )^2\Biggr ]\Kt^2  + \frac{m^2}{4}(D_s-2)\left (\frac{k^+}{k^+_0} \right )^4 \Biggr \},
\end{split}
\end{equation}
where the definition of the leading order QED photon splitting vertex $V^{\gamma^{\ast}_T \rightarrow q\bar{q}}_{h_0;h_1}$ is given in  \eq\nr{eq:LOvertex}. Finally, using the parametrization $k^+/k^+_0 = \xi$, we find the following expression
\begin{equation}
\label{eq:numaappvfinal}
N^T_{\ref{diag:oneloopSEUPT}} =  \frac{2g^2\cf}{\xi^2(1-\xi)}\delta_{\alpha_0\alpha_1}V^{\gamma^{\ast}_T \rightarrow q\bar{q}}_{h_0;h_1}\Biggl \{\biggl [1 + (1-\xi)^2\biggr ]\Kt^2  + m^2\xi^4 + \frac{(D_s-4)}{2}\xi^2\biggl [\Kt^2 + m^2\xi^2\biggr ] \Biggr \}.
\end{equation}

\subsection{Numerator for the instantaneous diagrams \ref{diag:oneloopSEUPTinst}, \ref{diag:oneloopvertexinst1} and $\ref{diag:gluoninst}_1$}

The numerator for the instantaneous diagram \ref{diag:oneloopSEUPTinst} can be written as  
\begin{equation}
\label{eq:appnumc}
\begin{split}
N^T_{\ref{diag:oneloopSEUPTinst}} & = ee_fg^2t^{a}_{\alpha_0\alpha_{0'}}t^{a}_{\alpha_{0'}\alpha_1} \biggl [\bar{u}(0)\epsl_{\sigma}(k)u(0')\biggr ]\biggl [\bar{u}(0')\epsl^{\ast}_{\sigma}(k) \gamma^+ \epsl_\lambda(q)v(1)\biggr ]\\
& = -ee_fg^2 \delta_{\alpha_0\alpha_1}\cf\biggl [\bar{u}(0)\epsl_{\sigma}(k)u(0')\biggr ]\biggl [\bar{u}(0')\gamma^+\epsl^{\ast}_{\sigma}(k) \epsl_\lambda(q)v(1)\biggr ].
\end{split}
\end{equation}
To simplify this expression, we first use the completeness relation (see e.g. Appendix A in Ref.~\cite{Beuf:2016wdz})
\begin{equation}
\sum_\sigma \varepsilon_{\sigma,\mu}(k)\varepsilon_{\sigma,\nu}^{\ast}(k) = -g_{\mu\nu} + \frac{k_\mu n_\nu + k_\nu n_\mu}{k^+}, 
\end{equation}
where $n^\mu a_\mu = a^+$ and \eq\nr{eq:sumover0}. This gives
\begin{equation}
\label{eq:appnumc2}
\begin{split}
N^T_{\ref{diag:oneloopSEUPTinst}} & = -ee_fg^2 \delta_{\alpha_0\alpha_1}\cf \biggl [-g_{\mu\nu} + \frac{k_\mu n_\nu + k_\nu n_\mu}{k^+} \biggr ]\biggl [\bar{u}(0)\gamma^\mu (\ksl_{0'} + m)\gamma^+\gamma^\nu \epsl_\lambda(q)v(1)\biggr ] \\
& =  -ee_fg^2 \delta_{\alpha_0\alpha_1}\cf \Biggl \{-4(k_0^+ - k^+)\bar u(0)\epsl_\lambda(q)v(1) - (D_s-4)\bar u(0)(\ksl_0 - \ksl)\gamma^+\epsl_\lambda(q)v(1) \\
&  + 2\frac{(k_0^+ - k^+)}{k^+}\bar{u}(0)\gamma^+\ksl\epsl_\lambda(q)v(1) + (D_s-2) m \bar{u}(0) \gamma^+ \epsl_\lambda(q) v(1) \Biggr \},
\end{split}
\end{equation}
where the standard $\gamma$-matrix relations have been used, as well as the plus and transverse momentum conservation relation $\hat k_{0'} = \hat k_0 - \hat k$. Note that this implies $\ksl_{0'} \gamma^ + = (\ksl_0 - \ksl)\gamma^+$. 

Finally, using the Dirac equation $\bar{u}(0)\ksl_0 = m \bar u(0) $, we get
\begin{equation}
\label{eq:appnumc3}
\begin{split}
N^T_{\ref{diag:oneloopSEUPTinst}}
& =  -ee_fg^2 \delta_{\alpha_0\alpha_1}\cf \Biggl \{\biggl [- 2\frac{k_0^+}{k^+} + D_s -2\biggr ]\bar{u}(0)\ksl\gamma^+\epsl_\lambda(q)v(1)  +2m \bar{u}(0) \gamma^+ \epsl_\lambda(q) v(1) \Biggr \} \\
& =   -ee_fg^2 \delta_{\alpha_0\alpha_1}\cf \Biggl \{\biggl [- 2\frac{k_0^+}{k^+} + D_s -2\biggr ]\bar{u}(0)\biggl [\ksl - \frac{k^+}{k_0^+}\ksl_0\biggr ]\gamma^+\epsl_\lambda(q)v(1) + \frac{k^+}{k_0^+}(D_s-2)m \bar{u}(0) \gamma^+ \epsl_\lambda(q) v(1) \Biggr \}.
\end{split}
\end{equation}
Note that in the shift $\ksl - (k^+/k_0^+)\ksl_0$ the $\gamma^-$ term vanishes,  leaving only the terms involving $\gamma^i$. Thus, we can rewrite $[\ksl - \frac{k^+}{k_0^+}\ksl_0]\gamma^+ = [\kt^i - \frac{k^+}{k_0^+}\kt_0^i]\gamma^+\gamma^i$.

Using the variables $\Kt$ and $\xi$ introduced in \eq\nr{eq:KTdef} the expression in \eq\nr{eq:appnumc3} can be rewritten as 
\begin{equation}
\label{eq:appnumc4}
\begin{split}
N^T_{\ref{diag:oneloopSEUPTinst}}
& =   -ee_fg^2 \delta_{\alpha_0\alpha_1}\cf \Biggl \{\biggl [-\frac{2}{\xi} + D_s -2\biggr ]\Kt^i\, \bar{u}(0) \gamma^+\gamma^i\epsl_\lambda(q)v(1) + \xi(D_s-2)m \bar{u}(0) \gamma^+ \epsl_\lambda(q) v(1) \Biggr \}.
\end{split}
\end{equation}

The numerator for the instantaneous diagram \ref{diag:oneloopvertexinst1} can be written as
\begin{equation}
\label{eq:appnumg}
N^T_{\ref{diag:oneloopvertexinst1} } = -ee_fg^2t^{a}_{\alpha_0\alpha_{0'}}t^{a}_{\alpha_{0'}\alpha_1} \biggl [\bar{u}(0)\epsl_{\sigma}(k)u(0')\biggr ]\biggl [\bar{u}(0')  \epsl_\lambda(q)\gamma^+\epsl^{\ast}_{\sigma}(k)v(1)\biggr ].
\end{equation}
Following the same steps as above we find
\begin{equation}
\begin{split}
N^T_{\ref{diag:oneloopvertexinst1}} = -ee_fg^2\cf \delta_{\alpha_0\alpha_1}\Biggl \{\biggl (\frac{2}{\xi} \bar u(0)\gamma^+\epsl_\lambda(q)  \gamma^i v(1)  + (D_s -4) \bar u(0)\gamma^+\gamma^i\epsl_\lambda v(1) \biggr )\Kt^i  + \xi(D_s-4)m\bar u(0)\gamma^+\epsl_\lambda(q)v(1)\Biggr \}. 
\end{split}
\end{equation}

Finally, the numerator for the instantaneous diagram $\ref{diag:gluoninst}_1$ can be written as  
\begin{equation}
\begin{split}
N^T_{\ref{diag:gluoninst}_1} & = -ee_fg^2t^{a}_{\alpha_0\alpha_{0'}}t^{a}_{\alpha_{0'}\alpha_1} \biggl [\bar{u}(0)\gamma^+u(0')\biggr ]\biggl [\bar{u}(0')\epsl_\lambda(q) v(1') \biggr ]\biggl [\bar{v}(1')\gamma^+v(1)\biggr ]\\
& = -ee_fg^2 \cf \delta_{\alpha_0\alpha_1} \biggl [\bar{u}(0)\gamma^+u(0')\biggr ]\biggl [\bar{u}(0')\epsl_\lambda(q) v(1') \biggr ]\biggl [\bar{v}(1')\gamma^+v(1)\biggr ].
\end{split}
\end{equation}
The simplifications here again follow the same steps as above leading to 
\begin{equation}
\begin{split}
N^T_{\ref{diag:gluoninst}_1} 
 = -ee_fg^2 \cf \delta_{\alpha_0\alpha_1} \biggl \{-4k_{1'}^+\, \epst_\lambda \cdot \biggl [\ktpzero - \frac{k_{0'}^+}{q^+}\qt \biggr ]\bar{u}(0)\gamma^+ v(1)  & - 2q^+ \biggl [\ktpzero^i - \frac{k_{0'}^+}{q^+} \qt^i\biggr ] \bar{u}(0)\epsl_\lambda(q)\gamma^+\gamma^i v(1)\\
& + 2q^+ m\bar{u}(0)\gamma^+ \epsl_\lambda(q) v(1) \biggr \}.
\end{split}
\end{equation}
Noting that $\ktpzero - \frac{k_{0'}^+}{q^+}\qt = -(\Kt + \Lt)$, we get 
\begin{equation}
\label{eq:appnumi2}
\begin{split}
N^T_{\ref{diag:gluoninst}_1} 
 = -ee_fg^2 \cf \delta_{\alpha_0\alpha_1} \biggl \{\biggl (4q^+[1-z(1-\xi)] \epst_\lambda^i \bar{u}(0)\gamma^+ v(1)  & + 2q^+ \bar{u}(0)\epsl_\lambda(q)\gamma^+\gamma^i v(1)\biggr ) \biggl [\Kt^i + \Lt^i\biggr ]\\
& + 2q^+ m\bar{u}(0)\gamma^+ \epsl_\lambda(q) v(1) \biggr \}.
\end{split}
\end{equation}

\subsection{Numerator for the quark vertex correction diagram \ref{diag:vertexqbaremT}}
\label{app:numc}

The numerator for the diagram \ref{diag:vertexqbaremT} can be written as  
\begin{equation}
\label{eq:numcapp}
N^T_{\ref{diag:vertexqbaremT}} = -ee_fg^2\delta_{\alpha\beta}\cf  \biggl [\bar{u}(0)\epsl_{\sigma}(k)u(0')\biggr ]\biggl [\bar{u}(0')\epsl_{\lambda}(q)v(1')\biggr ]\biggl [\bar{v}(1')\epsl^{\ast}_{\sigma}(k)v(1)\biggr ].
\end{equation}
With the kinematical variables as in \fig\ref{fig:vertexT}\ref{diag:vertexqbaremT}, the independent spinor structures can be decomposed as 
\begin{equation}
\label{eq:vertex1c}
\begin{split}
\bar{u}(0)\epsl_{\sigma}(k)u(0') = \frac{1}{k^+_0 \left (\frac{k^+}{k^+_0}\right )\left (1 -\frac{k^+}{k^+_0}\right ) }\Biggl \{\Biggl [\left (1-\frac{k^+}{2k^+_0}\right )\delta^{ij}_{(D_s)}\bar{u}(0)\gamma^{+}u(0') & + \left (\frac{k^+}{4k^+_0}\right )\bar{u}(0)\gamma^{+}[\gamma^i,\gamma^j]u(0')  \Biggr ]\Kt^i\\
& + \frac{m}{2}\left (\frac{k^+}{k^+_0}\right )^2\bar{u}(0)\gamma^+\gamma^ju(0')\Biggr \}\epst_{\sigma}^j
\end{split}
\end{equation}
\begin{equation}
\begin{split}
\label{eq:vertex3c}
\bar{u}(0')\epsl_{\lambda}(q)v(1') = -\left (\frac{q^+}{2k^+_{0'}k^+_{1'}} \right )\Biggl \{\Biggl [\left (\frac{k^+_{0'}-k^+_{1'}}{q^+}\right )\delta^{kl}_{(D_s)}\bar{u}(0')\gamma^{+}v(1') & + \frac{1}{2}\bar{u}(0')\gamma^{+}[\gamma^k,\gamma^l]v(1')  \Biggr ]\Ht^k\\
& + m\bar{u}(0')\gamma^+\gamma^lv(1')\Biggr \}\epst_{\lambda}^l
\end{split}
\end{equation}
and 
\begin{equation}
\label{eq:vertex2c}
\begin{split}
\bar{v}(1')\epsl^{\ast}_{\sigma}(k)v(1) = \frac{1}{k^+_1 \left (\frac{k^+}{k^+_1}\right )\left (1 + \frac{k^+}{k^+_1}\right ) }\Biggl \{\Biggl [\left (1+\frac{k^+}{2k^+_1}\right )\delta^{nm}_{(D_s)}\bar{v}(1')\gamma^{+}v(1) & + \left (\frac{k^+}{4k^+_1}\right )\bar{v}(1')\gamma^{+}[\gamma^n,\gamma^m]v(1)  \Biggr ]\Rt^n\\
& - \frac{m}{2}\left (\frac{k^+}{k^+_1}\right )^2\bar{v}(1')\gamma^+\gamma^mv(1)\Biggr \}\epst^{\ast m}_{\sigma},
\end{split}
\end{equation}
where we have introduced the variables
\begin{equation}
\label{eq:RtdefandHtdef}
\Ht = \Kt - \frac{(k_0^+ - k^+)}{k_0^+}\Pt = \Kt + \Lt, \quad\quad \Rt = \Kt + \frac{k^+q^+}{k^+_0k^+_1}\Pt = \Kt + \frac{\omega_1}{\omega_2}\Lt
\end{equation}
with $\Lt = \omega_2 \Pt$ and 
\begin{equation}
\omega_1 =\frac{k^+q^+}{k^+_0k^+_1}, \quad \omega_2 = -\left (1 - \frac{k^+}{k^+_0} \right ).
\end{equation}
Substituting \eqs\nr{eq:vertex1c}, \nr{eq:vertex3c}  and \nr{eq:vertex2c} into \nr{eq:numcapp}, and following the same steps as before, we find the expression
\begin{equation}
\label{eq:numcappv2}
\begin{split}
N^T_{\ref{diag:vertexqbaremT}}  = \frac{2ee_fg^2\delta_{\alpha\beta}\cf}{(k^+)^2\left (1 -\frac{k^+}{k^+_0}\right )\left (1 + \frac{k^+}{k^+_1}\right )}q^+\bar{u}(0) \gamma^{+}\biggl [a^{ij}\Kt^i + m a_m \gamma^j \biggr ]\biggl [b^{kl}\Ht^k + m \gamma^l \biggr ]\biggl [c^{nj}\Rt^n - m c_m \gamma^j \biggr ] v(1)\epst^{l}_{\lambda},
\end{split}
\end{equation}
where we have introduced the following compact notation
\begin{equation}
\begin{split}
a^{ij} & =  a \delta^{ij}_{(D_s)}  + b[\gamma^i,\gamma^j]    \\
b^{kl}& =  e \delta^{kl}_{(D_s)}  + f[\gamma^k,\gamma^l]  \\
c^{nj}& =  g \delta^{nj}_{(D_s)}  + h[\gamma^n,\gamma^j]  \\
\end{split}
\end{equation}
and defined the coefficients $a,b,e,f,g,h$ as
\begin{equation}
\begin{split}
a & = 1-\frac{k^+}{2k^+_0}, \quad b  = \frac{k^+}{4k^+_0}, \quad e  =  -\frac{2k^+}{q^+} + \frac{k^+_0 - k^+_1	}{q^+} \\
f & = \frac{1}{2} , \quad g  = 1+\frac{k^+}{2k^+_1}, \quad h  = \frac{k^+}{4k^+_1} \\
\end{split}
\end{equation}
and the coefficient $a_m, c_m$ as  
\begin{equation}
a_m = \frac{1}{2}\left (\frac{k^+}{k^+_0}\right )^2, \quad c_m = \frac{1}{2}\left (\frac{k^+}{k^+_1}\right )^2.
\end{equation}

To perform the contractions in \eq\nr{eq:numcappv2} requires tedious but straightforward algebraic manipulations. To this end, we have automatized the computation of the Lorentz contractions with FORM~\cite{Vermaseren:2000nd,Kuipers:2012rf}, and we have also cross-checked our results manually by pen and paper. The one-loop vertex correction diagram \ref{diag:vertexqbaremT} can be written in the spinor basis of \eq\nr{NLOformfactors}, and the result reads
\begin{equation}
\label{eq:vertexcvint}
\begin{split}
\Psi^{\gamma^{\ast}_T\rightarrow q\bar{q}}_{\ref{diag:vertexqbaremT}} & = \frac{ee_f\delta_{\alpha\beta}}{\ed_{\lo}}\left (\frac{\alpha_s\cf}{2\pi}\right )\biggl \{\bar u(0) \epsl_{\lambda}(q)v(1) \mathcal{V}^T_{\ref{diag:vertexqbaremT}}   +   \left ( \frac{2k^{+}_0k^{+}_1}{q^+} \right ) \Pt \cdot \epst_\lambda \bar u(0)\gamma^{+} v(1)\mathcal{N}^T_{\ref{diag:vertexqbaremT}}  \\
& +  \left ( \frac{2k^{+}_0k^{+}_1}{q^+} \right )\frac{\Pt\cdot \epst_{\lambda}}{\Pt^2}\Pt^j m\bar u(0)\gamma^{+}\gamma^{j}v(1) \mathcal{S}^T_{\ref{diag:vertexqbaremT}}    +  \left ( \frac{2k^{+}_0k^{+}_1}{q^+} \right ) m\bar u(0)\gamma^{+}\epsl_{\lambda}(q)v(1) \mathcal{M}^T_{\ref{diag:vertexqbaremT}}   \biggr \}, 
\end{split}
\end{equation}
where the form factors $\mathcal{V}^T_{\ref{diag:vertexqbaremT}},  \mathcal{N}^T_{\ref{diag:vertexqbaremT}},  \mathcal{S}^T_{\ref{diag:vertexqbaremT}}$ and $\mathcal{M}^T_{\ref{diag:vertexqbaremT}}$ are given in \eqs\nr{eq:VT}, \nr{eq:NT}, \nr{eq:ST} and \nr{eq:MT}, respectively.

\section{Useful integrals}
\label{app:logintegrals}

In this Appendix we calculate the functions $\mathcal{I}_{\xi;i}$ and $\mathcal{J}_{\xi;i}$ that are introduced in Section~\ref{vertexcontribution}. Let us first consider the functions $\mathcal{I}_{\xi;i}$ which are independent of the transverse momenta $\Pt$. These functions are defined as follows:
\begin{equation}
\label{eq:xiint1}
\mathcal{I}_{\xi;1} \equiv \int^{1}_{\alpha/z}  \frac{\ud \xi}{\xi}\biggl [-\log\left (\frac{\Delta_2}{\mu^2}\right )\biggr ] = \log\left (\frac{\alpha}{z}\right )\log\left (\frac{\overline{Q}^2 + m^2}{\mu^2}\right ) + \mathrm{Li}_2\left (\frac{1}{1-\frac{1}{2z}(1-\gamma)} \right ) + \mathrm{Li}_2\left (\frac{1}{1-\frac{1}{2z}(1+\gamma)} \right ),
\end{equation}
\begin{equation}
\label{eq:xiint2}
\begin{split}
\mathcal{I}_{\xi;2} \equiv \int^{1}_{\alpha/z} \ud \xi\biggl [-\log\left (\frac{\Delta_2}{\mu^2}\right )\biggr ] = -\log\left (\frac{\overline{Q}^2 + m^2}{\mu^2}\right ) + 2 &- \frac{1}{z}\gamma\log\left (\frac{1+\gamma}{1+\gamma - 2z}\right ) - \frac{1}{2z}(\gamma-1 )\log\left (\frac{\overline{Q}^2+m^2}{m^2}\right ),
\end{split}
\end{equation}
and
\begin{equation}
\label{eq:xiint3}
\begin{split}
\mathcal{I}_{\xi;3} \equiv \int^{1}_{\alpha/z} \ud \xi \xi\biggl [-\log\left (\frac{\Delta_2}{\mu^2}\right )\biggr ] = & -\frac{1}{2}\log\left (\frac{\overline{Q}^2 + m^2}{\mu^2}\right ) + \frac{3}{2}-\frac{1}{2z} - \frac{1}{z}\left (1-\frac{1}{2z}\right )\gamma\log\left (\frac{1+\gamma}{1+\gamma - 2z}\right )\\
& - \frac{1}{2z}\left (1-\frac{1}{2z}\right ) (\gamma-1 )\log\left (\frac{\overline{Q}^2+m^2}{m^2}\right ) - \frac{m^2}{2z^2Q^2}\log\left (\frac{\overline{Q}^2+m^2}{m^2}\right ).
\end{split}
\end{equation}
Here we have used the following compact notation $\gamma = \sqrt{1 + 4m^2/Q^2}$ introduced in \eq\nr{eq:xizeros}. 

Similarly, the functions $\mathcal{J}_{\xi;i}$  which depend on the transverse momenta $\Pt$ are defined as follows:
    \begin{equation}
\begin{split}
\mathcal{J}_{\xi;1} \equiv \int^{1}_{\alpha/z} \frac{\ud \xi}{\xi}\biggl [\log\left (\frac{\Delta_1}{\mu^2}\right )\biggr ] =   & -\log\left (\frac{\alpha}{z}\right ) \log\left (\frac{\Pt^2 + \overline{Q}^2 + m^2}{(1-z)\mu^2}\right )  - \frac{1}{2}\log^2\left (\frac{\alpha}{z} \right ) - \frac{\pi^2}{6} \\
& - m^2 \int_{0}^{1} \ud \xi \frac{ \log(\xi)}{(1-\xi)^2} \frac{(1-z)}{\Pt^2 + \overline{Q}^2 + m^2 + \frac{\xi(1-z)}{(1-\xi)}m^2},
\end{split}
\end{equation}
\begin{equation}
\begin{split}
\mathcal{J}_{\xi;2} \equiv  \int^{1}_{\alpha/z} \ud \xi \biggl [\log\left (\frac{\Delta_1}{\mu^2}\right )\biggr ] =  -2 + \log \left (   \frac{m^2}{\mu^2}\right )  + \frac{\Pt^2 + \overline{Q}^2 + m^2}{\Pt^2 + \overline{Q}^2 + z m^2}\log \left ( \frac{\Pt^2 + \overline{Q}^2 + m^2}{(1-z)m^2}\right ), 
\end{split}
\end{equation}
and
\begin{equation}
\begin{split}
\mathcal{J}_{\xi;3} \equiv \int^{1}_{\alpha/z} \ud \xi \xi\biggl [\log\left (\frac{\Delta_1}{\mu^2}\right )\biggr ] =  & -1 - \frac{(1-z)m^2}{2(\Pt^2 + \overline{Q}^2 + z m^2)}  + \frac{1}{2}\log \left ( \frac{m^2}{\mu^2}\right )\\
& + \frac{(\Pt^2 + \overline{Q}^2 + m^2)^2}{2(\Pt^2 + \overline{Q}^2 + z m^2)^2}\log \left ( \frac{\Pt^2 + \overline{Q}^2 + m^2}{(1-z)m^2}\right ).
\end{split}
\end{equation}

Next we provide two important intermediate results used for the computation shown in Section~\ref{vertexcontribution}. After integration over the $\xi$, we obtain the following result
\begin{equation}
\label{eq:Ixisum}
\begin{split}
-\mathcal{I}_{\xi;1}  + z\biggl (2 \mathcal{I}_{\xi;2} & - \mathcal{I}_{\xi;3} \biggr )    =  -\biggl [\frac{3z}{2}  + \log\left (\frac{\alpha}{z}\right ) \biggr ]\log \left (\frac{\overline{Q}^2 + m^2}{\mu^2} \right ) + \frac{1 + 5z}{2}\\
& + \frac{1}{2}\biggl [\frac{m^2}{zQ^2} + \left (1 + \frac{1}{2z}\right )(1-\gamma ) \biggr ]\log \left (\frac{\overline{Q}^2 + m^2}{m^2} \right ) - \left (1+ \frac{1}{2z}\right )\gamma\log \left (\frac{1+\gamma}{1+\gamma - 2z} \right ) \\
& - \mathrm{Li}_2\left (\frac{1}{1-\frac{1}{2z}(1-\gamma)} \right ) - \mathrm{Li}_2\left (\frac{1}{1-\frac{1}{2z}(1+\gamma)} \right ).\\
\end{split}
\end{equation}
Similarly, we find 
\begin{equation}
\label{eq:Jxisum}
\begin{split}
-\mathcal{J}_{\xi;1}  + z\biggl (2\mathcal{J}_{\xi;2}  & - \mathcal{J}_{\xi;3} \biggr )  = \biggl [\frac{3z}{2} + \log\left (\frac{\alpha}{z}\right )\biggr ]\Biggl \{\log \left (\frac{\overline{Q}^2 + m^2}{\mu^2} \right )  + \log\left (\frac{\Pt^2 + \overline{Q}^2 + m^2}{\overline{Q}^2 + m^2} \right ) - \log(1-z)\Biggr \}\\
& + \frac{1}{2}\log^2\left (\frac{\alpha}{z}\right ) - 3z + \frac{\pi^2}{6} + m^2 \int_{0}^{1} \ud \xi \biggl [ \frac{ \log(\xi)}{(1-\xi)^2} + \frac{z}{1-\xi} + \frac{z}{2} \biggr ]\frac{(1-z)}{\Pt^2 + \overline{Q}^2 +  m^2 + \frac{\xi(1-z)}{(1-\xi)}m^2}.
\end{split}
\end{equation}

\section{Fourier transforms for $q\bar{q}$ states}
\label{app:FTSqqbarcase}

In this Appendix we present the relevant integrals that are needed to calculate the Fourier transformed $\gamma^{\ast}_T \rightarrow q\bar q$ LCWF’s in mixed space up to NLO. 

We start by considering the following integral
\begin{equation}
\label{eq:generalFTint}
 \int\frac{\ud^{D-2}\Pt}{(2\pi)^{D-2}}\frac{e^{i\Pt\cdot \xt_{01}}}{[\Pt^2 + \Delta^2]},
\end{equation}
where $\xt_{01} = \xt_0 - \xt_1$ and $\Pt^2 + \Delta^2 >0$. The standard technique to evaluate these type of integrals is to first introduce a Schwinger parametrisation for each denominator
\begin{equation}
\label{eq:SParam}
\frac{1}{A^{\beta}} = \frac{1}{\Gamma(\beta)}\int_{0}^{\infty}\ud t t^{\beta - 1} e^{-t A}, \quad A,\beta >0,  
\end{equation}
which turns the integral in \eq\nr{eq:generalFTint} into a simple Gaussian form in $\Pt$. The resulting expression can be evaluated by using the standard Gaussian integral:
\begin{equation}
\int \frac{\ud^n \yt}{(2\pi)^n} \exp \biggl [-\sum_{i,j=1}^{n} a_{ij}y_i y_j \biggr ]\biggl [\sum_{i=1}^{n}b_i y_i\biggr ] = \sqrt{\frac{1}{(4\pi)^{n}\det a}}\exp \biggl [ {\frac{1}{4}(a^{-1})_{ij}b_ib_j} \biggr ],
\end{equation}
where $\yt = (y_1,y_2,\ldots,y_n)$, $\mathbf{b} = (b_1,b_2,\ldots, b_n)$ and $a_{ij}$ is a symmetric, non-singular and positive defined $n\times n$ matrix.

These steps lead to the result
\begin{equation}
\label{eq:generalFTintSPG}
 \int\frac{\ud^{D-2}\Pt}{(2\pi)^{D-2}}\frac{e^{i\Pt\cdot \xt_{01}}}{[\Pt^2 + \Delta^2]} = (4\pi)^{1-\frac{D}{2}} \int_{0}^{\infty} \ud t\, t^{1 - \frac{D}{2}}e^{-t \Delta^2}e^{-\frac{\vert \xt_{01}\vert^2}{4t}}.
\end{equation}
The remaining one-dimensional integral can be performed by using the general formula
\begin{equation}
\int_{0}^{\infty} \ud t\, t^{\beta -1}e^{-tA} e^{-\frac{B}{t}} = 2\left (\frac{B}{A} \right )^{\beta/2} K_{-\beta}(2\sqrt{AB}), \quad A,B > 0.  
\end{equation}
Here $K_{-\beta}(x)$ is the modified Bessel function. Notice that for positive values of $x$  $K_{\beta}(x)$ is an analytic function of $\beta$ and that $K_{\beta}(x)$ is even in $\beta$, i.e. $K_{\beta}(x) = K_{-\beta}(x)$. 

Hence, the integral in \eq\nr{eq:generalFTintSPG} yields the result
\begin{equation}
\label{eq:F1}
 \int\frac{\ud^{D-2}\Pt}{(2\pi)^{D-2}}\frac{e^{i\Pt\cdot \xt_{01}}}{[\Pt^2 + \Delta^2]} = \frac{1}{(2\pi)}\left (\frac{\Delta}{2\pi\vert \xt_{01}\vert} \right )^{\frac{D}{2}-2}K_{\frac{D}{2}-2}\left (\vert \xt_{01}\vert \Delta\right ).
\end{equation}
We also note that if the expression in \eq\nr{eq:generalFTint} contains a logarithmic function, one can first use the relation $\log(A) = \lim_{\alpha \rightarrow 0} \partial_{\alpha} A^{\alpha}$ and then apply the Schwinger parametrization formula in \eq\nr{eq:SParam}. 

It is then easy to show that for the rank-one tensor integral
\begin{equation}
\label{eq:F2}
 \int\frac{\ud^{D-2}\Pt}{(2\pi)^{D-2}}\frac{\Pt^j e^{i\Pt\cdot \xt_{01}}}{[\Pt^2 + \Delta^2]} =  \frac{i\xt^{j}_{01}}{(2\pi)\vert \xt_{01}\vert}\left (\frac{\Delta}{2\pi\vert \xt_{01}\vert} \right )^{\frac{D}{2}-2} \Delta \, K_{\frac{D}{2}-1}\left (\vert \xt_{01}\vert \Delta\right ).
\end{equation}
In addition, we need to compute the following symmetric and traceless rank-two tensor integral in $D=4$ 
\begin{equation}
 \int\frac{\ud^{2}\Pt}{(2\pi)^{2}}\frac{e^{i\Pt\cdot \xt_{01}}
 (\Pt^i\Pt^j-\frac{1}{2}\delta^{ij}\Pt^2) }{[\Pt^2 + \Delta^2]} = 
-\frac{\Delta^2}{(2\pi)}\left(\frac{\xt_{01}^i\xt_{01}^j}{\vert\xt_{01}\vert^2}-\frac{\delta^{ij}}{2}\right)K_2(\vert\xt_{01}\vert\Delta ) 
\end{equation}
plus delta functions in $\xt_{01}$, which we ignore.

Furthermore, Fourier transforms of a product can be treated by a partial fraction decomposition before integrating: 
\begin{equation}
\frac{1}{[\Pt^2 + \Delta_1^2][\Pt^2 + \Delta_2^2]} 
= 
\frac{1}{\Delta_2^2-\Delta_1^2}\left[
\frac{1}{\Pt^2 + \Delta_1^2} 
-
\frac{1}{\Pt^2 + \Delta_2^2} 
\right],
\end{equation}
which leads to the Fourier-transform being a difference of two Bessel K functions with different arguments. 
Doing a double Schwinger parametrization of the denominators, which one would also be tempted to do, ends up expressing the result in the same form. Here the two terms come from the integration limits of the innermost Schwinger parameter integral after the usual shift, with then the outermost Schwinger parameter integration transforming the terms to Bessel K functions. One can also first combine the denominators with a Feynman parametrization and then perform the integral with a single Schwinger parameter. This gives a result as a Feynman parameter integral of a single Bessel K function. However, this Feynman parameter integral can be performed using the Bessel equation satisfied by the K functions. This again results in the same difference of two terms as the partial fraction method, coming from the upper and lower integration limits of the Feynman parameter integral. 
In some cases we have been able to express the differences of Bessel K functions as a single Bessel function, using a partial integration in some other Feynman parameter. However, this is only possible on a case-by-case basis, not in general.

\section{Squaring the gluon emission contribution}
\label{app:qqbargderivation}

In this appendix we present the detailed computation of  the $\gamma^\ast_T \to q\bar{q}g$ LCWF squared in \eq\nr{eq:emissionTLCWFsquared}. As a first step, we need to square \eq\nr{eq:reducedTCWFqqbarg} which is then averaged and summed over the transverse photon polarization $(\lambda)$ and summed over the quark spin $(h_0,h_1)$ and gluon polarization $(\sigma)$ states. This gives the following starting point:
\begin{equation}
\label{eq:qqgkernelstep1}
\begin{split}
\frac{1}{(D_s-2)}\sum_{\lambda,\sigma}\sum_{h_0,h_1}\bigg\vert \epst_{\lambda}^{l}\epst_{\sigma}^{\ast j}\left (\Sigma^{lj} + \Sigma^{lj}_m \right )\bigg\vert^2 & = \frac{1}{(D_s-2)}\sum_{\lambda,\sigma}\sum_{h_0,h_1}\left ( \Sigma^{lj} + \Sigma^{lj}_m \right ) \left ( \Sigma^{l'j'} + \Sigma^{l'j'}_m\right )^{\ast} \epst^{l}_{\lambda}\epst^{\ast l'}_{\lambda} \epst^{\ast j}_{\sigma}\epst^{j'}_{\sigma}\\
& = \frac{1}{(D_s-2)}\sum_{h_0,h_1}\left ( \Sigma^{lj} + \Sigma^{lj}_m \right ) \left ( \Sigma^{l'j'} + \Sigma^{l'j'}_m\right )^{\ast} \delta^{ll'}_{(D_s)}\delta^{jj'}_{(D_s)}\\
& = \frac{1}{(D_s-2)}\sum_{h_0,h_1} \biggl (\vert \Sigma^{lj}\vert^2 + \vert \Sigma^{lj}_m\vert^2 + 2\Re e[\Sigma^{lj}(\Sigma^{lj}_m)^{\ast}]  \biggr ).
\end{split}
\end{equation}

To begin with, we concentrate on the light cone helicity nonflip contribution $\vert \Sigma^{lj}\vert^2$, which is given by the first term (in the third line) on the right-hand side of \eq\nr{eq:qqgkernelstep1}. To simplify the algebra, we divide the function $\Sigma^{lj}$ in \eq\nr{eq:sigmamassless} into the four terms
\begin{equation}
\label{eq:sigmassplit}
\Sigma^{lj}  = A^{lj}_{\ref{diag:qgqbarT}} + B^{lj}_{\ref{diag:qqbargT}} + C^{lj}_{\ref{diag:qgqbarTinst}} + C^{lj}_{\ref{diag:qqbargTinst}}, 
\end{equation}
where we have defined the set of four functions as:
\begin{equation}
A^{lj}_{\ref{diag:qgqbarT}} = -a_{\ref{diag:qgqbarT}}\bar{u}(0)\gamma^+\biggl [(2k^+_0 + k^+_2)\delta^{ij}_{(D_s)} - \frac{k^+_2}{2}[\gamma^i,\gamma^j] \biggr ]\biggl [(2k^+_1 - q^+)\delta^{kl}_{(D_s)} - \frac{q^+}{2}[\gamma^k,\gamma^l] \biggr ] v(1)\mathcal{I}^{ik}_{\ref{diag:qgqbarT}},     
\end{equation}
\begin{equation}
B^{lj}_{\ref{diag:qqbargT}}  = -b_{\ref{diag:qqbargT}}\bar{u}(0)\gamma^+\biggl [(2k^+_0 - q^+)\delta^{kl}_{(D_s)} + \frac{q^+}{2}[\gamma^k,\gamma^l] \biggr ]\biggl [(2k^+_1 + k^+_2)\delta^{ij}_{(D_s)} + \frac{k^+_2}{2}[\gamma^i,\gamma^j] \biggr ] v(1)\mathcal{I}^{ik}_{\ref{diag:qqbargT}},    
\end{equation}
and
\begin{equation}
\begin{split}
C^{lj}_{\ref{diag:qgqbarTinst}} & = +c_{\ref{diag:qgqbarTinst}}\bar u(0)\gamma^+\gamma^j\gamma^l v(1)\mathcal{J}_{\ref{diag:qgqbarTinst}},\\ 
C^{lj}_{\ref{diag:qqbargTinst}} & = -c_{\ref{diag:qqbargTinst}}\bar u(0)\gamma^+\gamma^l\gamma^j v(1)\mathcal{J}_{\ref{diag:qqbargTinst}},
\end{split}
\end{equation}
with the coefficients 
\begin{equation}
a_{\ref{diag:qgqbarT}} = \frac{1}{(k^+_0 + k^+_2)q^+} , \quad b_{\ref{diag:qqbargT}} = \frac{1}{(k^+_1 + k^+_2)q^+}, \quad c_{\ref{diag:qgqbarTinst}} = \frac{k^+_2k^+_0}{(k^+_0 + k^+_2)^2}, \quad c_{\ref{diag:qqbargTinst}} = \frac{k^+_2k^+_1}{(k^+_1 + k^+_2)^2}.   
\end{equation}
Squaring the expression in \eq\nr{eq:sigmassplit} yields
\begin{equation}
\label{eq:qqgkernelmassless}
\begin{split}
\sum_{h_0,h_1} \vert \Sigma^{lj}\vert^2 = &\sum_{h_0,h_1} \biggl ( \vert A^{lj}_{\ref{diag:qgqbarT}} \vert^2 + \vert B^{lj}_{\ref{diag:qqbargT}}\vert^2  + 2\Re e[ A^{lj}_{\ref{diag:qgqbarT}}( B^{lj}_{\ref{diag:qqbargT}})^{\ast}] + 2\Re e[C^{lj}_{\ref{diag:qgqbarTinst}}(C^{lj}_{\ref{diag:qqbargTinst}})^{\ast}]\\
& + \Re e[(C^{lj}_{\ref{diag:qgqbarTinst}})^{\ast}(C^{lj}_{\ref{diag:qgqbarTinst}} + 2A^{lj}_{\ref{diag:qgqbarT}} + 2B^{lj}_{\ref{diag:qqbargT}})] + \Re e[(C^{lj}_{\ref{diag:qqbargTinst}})^{\ast}(C^{lj}_{\ref{diag:qqbargTinst}} + 2A^{lj}_{\ref{diag:qgqbarT}} + 2B^{lj}_{\ref{diag:qqbargT}})] \biggr ).
\end{split}
\end{equation}
In \eq\nr{eq:qqgkernelmassless} only the first two terms, which are proportional to the Fourier integrals\footnote{See the discussion in Appendix \ref{app:FTintforqqbarg}} $\vert \mathcal{I}^{ik}_{\ref{diag:qgqbarT}}\vert^2$ and $\vert \mathcal{I}^{ik}_{\ref{diag:qqbargT}}\vert^2$, contain the UV divergences at $D=4$. Hence, the remaining terms can be directly computed in $D_s = D =4$, which is an enormous technical simplification. The computation of individual terms in \eq\nr{eq:qqgkernelmassless} closely follows the detailed derivation presented in the case of massless quarks. Therefore, for a detailed discussion, we refer the reader to our previous work \cite{Beuf:2017bpd, Hanninen:2017ddy}.  After some amount of algebra, we obtain: 
\begin{equation}
\label{eq:AandBsquared}
\begin{split}
\sum_{h_0,h_1} \vert A^{lj}_{\ref{diag:qgqbarT}}\vert^2 & = 8k_0^+k_1^+ a^2_{\ref{diag:qgqbarT}} \biggl [4k_0^+(k_0^+ + k_2^+) + (D_s-2)(k_2^+)^2 \biggr ]\biggl [(D_s-2)(q^+)^2 - 4k_1^+(q^+ - k_1^+)\biggr ] \vert \mathcal{I}^{ik}_{\ref{diag:qgqbarT}}\vert^2,\\
\sum_{h_0,h_1} \vert B^{lj}_{\ref{diag:qqbargT}}\vert^2 & = 8k_0^+k_1^+ b^2_{\ref{diag:qqbargT}} \biggl [4k_1^+(k_1^+ + k_2^+) + (D_s-2)(k_2^+)^2 \biggr ]\biggl [(D_s-2)(q^+)^2 - 4k_0^+(q^+ - k_0^+) \biggr ] \vert \mathcal{I}^{ik}_{\ref{diag:qqbargT}}\vert^2,\\
\end{split}
\end{equation}
and 
\begin{equation}
\label{eq:ABcross}
\begin{split}
\sum_{h_0,h_1} 2\Re e[ A^{lj}_{\ref{diag:qgqbarT}}(B^{lj}_{\ref{diag:qqbargT}})^{\ast}] & = 8k_0^+k_1^+ 8a_{\ref{diag:qgqbarT}}b_{\ref{diag:qqbargT}}\biggl \{q^+ k_2^+(k_0^+ - k_1^+)^2\biggl [\Re e[\mathcal{I}^{ii}_{\ref{diag:qgqbarT}}(\mathcal{I}^{kk}_{\ref{diag:qqbargT}})^{\ast}]-\Re e[\mathcal{I}^{ik}_{\ref{diag:qgqbarT}}(\mathcal{I}^{ki}_{\ref{diag:qqbargT}})^{\ast}] \biggr ]\\
& - \biggl [k_1^+(k_0^+ + k_2^+) + k_0^+(k_1^+ + k_2^+)\biggr ]\biggl [k_0^+(k_0^+ + k_2^+) + k_1^+(k_1^+ + k_2^+)\biggr ] \Re e[\mathcal{I}^{ik}_{\ref{diag:qgqbarT}}(\mathcal{I}^{ik}_{\ref{diag:qqbargT}})^{\ast}]\biggr \} .
\end{split}
\end{equation}
Furthermore, for the terms which contain instantaneous diagrams, we obtain the following expressions:
\begin{equation}
\begin{split}
\sum_{h_0,h_1} \Re e[(C^{lj}_{\ref{diag:qgqbarTinst}})^{\ast}(C^{lj}_{\ref{diag:qgqbarTinst}} + 2A^{lj}_{\ref{diag:qgqbarT}} + 2B^{lj}_{\ref{diag:qqbargT}})]   = 8k_0^+k_1^+ \biggl \{4c_{\ref{diag:qgqbarTinst}}^2 \vert \mathcal{J}_{\ref{diag:qgqbarTinst}}\vert^2 & + 8k_0^+k_1^+ a_{\ref{diag:qgqbarT}}c_{\ref{diag:qgqbarTinst}} \Re e[\mathcal{I}^{ii}_{\ref{diag:qgqbarT}}(\mathcal{J}_{\ref{diag:qgqbarTinst}})^{\ast}]\\
&- 8(k_1^+ + k_2^+)^2b_{\ref{diag:qqbargT}}c_{\ref{diag:qgqbarTinst}}\Re e[\mathcal{I}^{ii}_{\ref{diag:qqbargT}}(\mathcal{J}_{\ref{diag:qgqbarTinst}})^{\ast}] \biggr \}, \\
\end{split}
\end{equation}
\begin{equation}
\begin{split}
\sum_{h_0,h_1} \Re e[(C^{lj}_{\ref{diag:qqbargTinst}})^{\ast}(C^{lj}_{\ref{diag:qqbargTinst}} + 2A^{lj}_{\ref{diag:qgqbarT}} + 2B^{lj}_{\ref{diag:qqbargT}})]   = 8k_0^+k_1^+ \biggl \{4c_{\ref{diag:qqbargTinst}}^2 \vert \mathcal{J}_{\ref{diag:qqbargTinst}}\vert^2 & + 8(k_0^+ + k_2^+)^2 a_{\ref{diag:qgqbarT}}c_{\ref{diag:qqbargTinst}} \Re e[\mathcal{I}^{ii}_{\ref{diag:qgqbarT}}(\mathcal{J}_{\ref{diag:qqbargTinst}})^{\ast}]\\
& - 8k_0^+k_1^+ b_{\ref{diag:qqbargT}}c_{\ref{diag:qqbargTinst}}\Re e[\mathcal{I}^{ii}_{\ref{diag:qqbargT}}(\mathcal{J}_{\ref{diag:qqbargTinst}})^{\ast}] \biggr \}, \\
\end{split}
\end{equation}
and
\begin{equation}
\sum_{h_0,h_1} 2\Re e[C^{lj}_{\ref{diag:qgqbarTinst}}(C^{lj}_{\ref{diag:qqbargTinst}})^{\ast}] = 0.
\end{equation}

Next, we concentrate on the contributions with one or two light cone helicity flips in the amplitude. We start by computing the second term $\vert \Sigma^{lj}_m\vert^2$ in \eq\nr{eq:qqgkernelstep1}. Again, to simplify the algebra, we divide the function $\Sigma^{lj}_m$ in \eq\nr{eq:sigmamassive} into the two parts
\begin{equation}
\Sigma^{lj}_m  = A^{lj}_m + B^{lj}_m, 
\end{equation}
where we have defined the functions $A^{lj}_m$ and $B^{lj}_m$ as:
\begin{equation}
\begin{split}
A^{lj}_m  =  & - m a_1\bar u(0)\gamma^+\biggl [(2k^+_0 + k^+_2)\delta^{ij}_{(D_s)} - \frac{k^+_2}{2}[\gamma^i,\gamma^j]\biggr ]\gamma^{l}v(1)\mathcal{I}^{i}_{\ref{diag:qgqbarT}}\\
& + m a_2\bar u(0)\gamma^+\gamma^j\biggl [(2k^+_1 - q^+)\delta^{kl}_{(D_s)} - \frac{q^+}{2}[\gamma^k,\gamma^l]\biggr ]v(1)\mathcal{\hat I}^{k}_{\ref{diag:qgqbarT}}\\
& + m^2a_3\bar u(0)\gamma^+\gamma^j\gamma^lv(1)\mathcal{I}_{\ref{diag:qgqbarT}},
\end{split}
\end{equation}
and 
\begin{equation}
\begin{split}
B^{lj}_m = & + m b_1\bar u(0)\gamma^+\gamma^{l}\biggl [(2k^+_1 + k^+_2)\delta^{ij}_{(D_s)} + \frac{k^+_2}{2}[\gamma^i,\gamma^j]\biggr ]v(1)\mathcal{I}^{i}_{\ref{diag:qqbargT}}\\
& + m b_2\bar u(0)\gamma^+\biggl [(2k^+_0 - q^+)\delta^{kl}_{(D_s)} + \frac{q^+}{2}[\gamma^k,\gamma^l]\biggr ]\gamma^j v(1)\mathcal{\hat I}^{k}_{\ref{diag:qqbargT}}\\
& -m^2 b_3\bar u(0)\gamma^+\gamma^l\gamma^j v(1)\mathcal{I}_{\ref{diag:qqbargT}}.
\end{split}
\end{equation}
Here the coefficients $a_i$ and $b_i$ for $i=1,2,3$ are defined as 
\begin{equation}
\begin{split}
a_1 & = \frac{1}{(k_0^+ + k_2^+)}, \quad a_2 = \frac{(k_2^+)^2}{(k_0^+ + k_2^+)^2q^+}, \quad a_3 = \frac{(k_2^+)^2}{(k_0^+ + k_2^+)^2} \\
b_1 & = \frac{1}{(k_1^+ + k_2^+)}, \quad b_2 = \frac{(k_2^+)^2}{(k_1^+ + k_2^+)^2q^+}, \quad b_3 = \frac{(k_2^+)^2}{(k_1^+ + k_2^+)^2}.  
\end{split}
\end{equation}
This yields the following expression 
\begin{equation}
\label{eq:qqgkernelterm2}
\begin{split}
\sum_{h_0,h_1} \vert \Sigma^{lj}_m\vert^2 = \sum_{h_0,h_1}\biggl (\vert A^{lj}_m\vert^2 + \vert B^{lj}_m\vert^2 + 2\Re e[A^{lj}_m(B^{lj}_m)^{\ast}]  \biggr ).
\end{split}
\end{equation}
By taking the complex conjugate of the function $A^{lj}_m$, we find
for the first term on the right-hand side of \eq\nr{eq:qqgkernelterm2} the following expression
\begin{equation}
\label{eq:Amsquared}
\begin{split}
\sum_{h_0,h_1}\vert A^{lj}_m\vert^2 = m^2& \sum_{h_0,h_1}\biggl \{ - a_1\bar u(0)\gamma^+\biggl [(2k^+_0 + k^+_2)\delta^{ij}_{(D_s)} - \frac{k^+_2}{2}[\gamma^i,\gamma^j]\biggr ]\gamma^{l}v(1)\mathcal{I}^{i}_{\ref{diag:qgqbarT}}\\
& + a_2\bar u(0)\gamma^+\gamma^j\biggl [(2k^+_1 - q^+)\delta^{kl}_{(D_s)} - \frac{q^+}{2}[\gamma^k,\gamma^l]\biggr ]v(1)\mathcal{\hat I}^{k}_{\ref{diag:qgqbarT}}
 + m a_3\bar u(0)\gamma^+\gamma^j\gamma^lv(1)\mathcal{I}_{\ref{diag:qgqbarT}} \biggr \}\\
 \times \biggl \{& - a_1\bar v(1)\gamma^{l}\biggl [(2k^+_0 + k^+_2)\delta^{i'j}_{(D_s)} + \frac{k^+_2}{2}[\gamma^{i'},\gamma^j]\biggr ]\gamma^{+}u(0) (\mathcal{I}^{i'}_{\ref{diag:qgqbarT}})^{\ast}\\
& + a_2\bar v(1)\biggl [(2k^+_1 - q^+)\delta^{k'l}_{(D_s)} + \frac{q^+}{2}[\gamma^{k'},\gamma^l]\biggr ]\gamma^{j}\gamma^{+}u(0)(\mathcal{\hat I}^{k'}_{\ref{diag:qgqbarT}})^{\ast}
 + m a_3\bar v(1)\gamma^{l}\gamma^{j}\gamma^{+}u(0)(\mathcal{I}_{\ref{diag:qgqbarT}})^{\ast} \biggr \}.
\end{split}
\end{equation}
Let us then perform further simplifications for the terms $\propto a_1^2$ from the above expression. This part is given by
\begin{equation}
\label{eq:examplterm}
\begin{split}
\sum_{h_0,h_1} & \biggl \{ - a_1\bar u(0)\gamma^+\biggl [(2k^+_0 + k^+_2)\delta^{ij}_{(D_s)} - \frac{k^+_2}{2}[\gamma^i,\gamma^j]\biggr ]\gamma^{l}v(1)\mathcal{I}^{i}_{\ref{diag:qgqbarT}} \biggr \}\\
& \times \biggl \{ - a_1\bar v(1)\gamma^{l}\biggl [(2k^+_0 + k^+_2)\delta^{i'j}_{(D_s)} + \frac{k^+_2}{2}[\gamma^{i'},\gamma^j]\biggr ]\gamma^{+}u(0) (\mathcal{I}^{i'}_{\ref{diag:qgqbarT}} )^{\ast}\biggr \}.
\end{split}
\end{equation}
Performing the sum over $h_1$ with $\gamma^{+}(\ksl_1 - m)\gamma^{+} = 2k_1^{+}\gamma^{+}$, and then using the relations
\begin{equation}
\label{eq:relations}
\begin{split}
\gamma^{l}\gamma^{l} & = 2-D_s \\
\{\gamma^{+},\gamma^{l} \} & =0,\\
[\gamma^i,\gamma^j] & = 2\left (\delta^{ij}_{(D_s)} + \gamma^{i}\gamma^{j} \right ),
\end{split}
\end{equation}
we find that \eqs\nr{eq:examplterm} can be simplified to
\begin{equation}
\label{eq:examplterm1}
\begin{split}
\text{\nr{eq:examplterm}} = (D_s-2)a_1^2 2k_1^+\sum_{h_0} \bar u(0)\gamma^{+}\biggl \{& \biggl [(2k_0^{+} + k_2^+)^2 + (k_2^+)^2(D_s-3) \biggr ]\delta^{ii'}_{(D_s)}\\
& + \biggl [k_2^+(2k_0^+ + k_2^+) - \frac{1}{2}(D_s-4)(k_2^+)^2 \biggr ][\gamma^i,\gamma^{i'}]\biggr \}u(0)\mathcal{I}^{i}_{\ref{diag:qgqbarT}} (\mathcal{I}^{i'}_{\ref{diag:qgqbarT}})^{\ast}.
\end{split}
\end{equation}
Next, utilizing the fact that $\mathcal{I}^{i}_{\ref{diag:qgqbarT}} \propto \xt^i_{20}$ (see Appendix \ref{app:FTintforqqbarg}) so that only the term symmetric  in $i,i'$ in the above expression survives, and finally performing a simple trace
\begin{equation}
\label{eq:trace1}
\sum_{h_0} \bar u(0)\gamma^{+} u(0) = 4k_{0,\mu}g^{\mu+} = 4k_0^+
\end{equation}
then gives for \eq\nr{eq:examplterm} the following result
\begin{equation}
\label{eq:examplterm1v2}
\begin{split}
\text{\nr{eq:examplterm}} = (D_s-2)a_1^2 8k_0^+k_1^+\biggl [4k_0^+(k_0^+ + k_2^+) + (k_2^+)^2(D_s - 2) \biggr ]\vert\mathcal{I}^{i}_{\ref{diag:qgqbarT}}\vert^2.
\end{split}
\end{equation}

For the cross term $\propto a_1a_2$ in \eq\nr{eq:Amsquared} we have the following starting point 
\begin{equation}
\label{eq:exampltermcross}
\begin{split}
2\Re e\biggl [\sum_{h_0,h_1} & \biggl \{ - a_1\bar u(0)\gamma^+\biggl [(2k^+_0 + k^+_2)\delta^{ij}_{(D_s)} - \frac{k^+_2}{2}[\gamma^i,\gamma^j]\biggr ]\gamma^{l}v(1)\mathcal{I}^{i}_{\ref{diag:qgqbarT}} \biggr \}\\
& \times \biggl \{ + a_2\bar v(1)\biggl [(2k^+_1 - q^+)\delta^{k'l}_{(D_s)} + \frac{q^+}{2}[\gamma^{k'},\gamma^l]\biggr ]\gamma^j\gamma^{+}u(0)(\mathcal{\hat I}^{k'}_{\ref{diag:qgqbarT}})^{\ast}\biggr \} \biggr ].
\end{split}
\end{equation}
First, carrying out the sum over $h_1$ and using the relations in \eq\nr{eq:relations}, we obtain after some algebra
\begin{equation}
\label{eq:exampltermcrossv2}
\begin{split}
\text{\nr{eq:exampltermcross}} = - a_1a_2 2k_1^+\sum_{h_0} \bar u(0)\gamma^+\biggl [-2k_2^+(D_s-4)\delta_{(D_s)}^{k' i} - \left (2k_0^+ + k_2^+(D_s-4)\right )\gamma^{k'}\gamma^i\biggr ] &\biggl [2k_1^+ + q^+(D_s-4)\biggr ]u(0)\\
& \times 2\Re e[\mathcal{I}^{i}_{\ref{diag:qgqbarT}} (\mathcal{\hat I}^{k'}_{\ref{diag:qgqbarT}})^{\ast}].
\end{split}
\end{equation}
Note that since the integrals $\mathcal{I}^{i}_{\ref{diag:qgqbarT}} \propto \xt_{20}^i$ and $\mathcal{\hat I}^{k'}_{\ref{diag:qgqbarT}}\propto \xt_{0+2;1}^{k'}$ (see Appendix \ref{app:FTintforqqbarg}), also the antisymmetric term in $i,k'$ contributes. Next, simplifying the above expression further by performing a trace with \eq\nr{eq:trace1} and \begin{equation}
\label{eq:trace2}
\sum_{h_0} \bar u(0)\gamma^{+}\gamma^{k'}\gamma^i u(0) =  -4k_0^+\delta_{(D_s)}^{k' i}
\end{equation}
we find the result
\begin{equation}
\text{\nr{eq:exampltermcross}} = - a_1a_2 8k_0^+k_1^+ \biggl [2k_0^+ - k_2^+(D_s-4)\biggr ] \biggl [2k_1^+ + q^+(D_s-4)\biggr ]2\Re e[\mathcal{I}^{i}_{\ref{diag:qgqbarT}} (\mathcal{\hat I}^{i}_{\ref{diag:qgqbarT}})^{\ast}].
\end{equation}
Following the same steps for the other terms, \eq\nr{eq:examplterm} yields the result
\begin{equation}
\label{eq:Amsquaredfinal}
\begin{split}
\sum_{h_0,h_1}\vert A^{lj}_m\vert^2 = m^2 (D_s-2)8k_0^{+}&k_1^{+} \biggl \{ a_1^2\biggl [4k_0^+(k_0^+ + k_2^+) + (k_2^+)^2(D_s-2) \biggr ]\vert\mathcal{I}^{i}_{\ref{diag:qgqbarT}}\vert^2 \\
& + a_2^2\biggl [4k_1^+(k_1^+ - q^+) + (q^+)^2(D_s-2) \biggr ]\vert\mathcal{\hat I}^{k}_{\ref{diag:qgqbarT}}\vert^2 + m^2 a_3^2(D_s-2)\vert\mathcal{I}_{\ref{diag:qgqbarT}}\vert^2\\
& - \frac{2a_1a_2}{(D_s-2)}\biggl [2k_0^+  - k_2^+(D_s-4) \biggr ]\biggl [2k_1^+ + q^+(D_s-4) \biggr ]\Re e[\mathcal{I}^{i}_{\ref{diag:qgqbarT}}(\mathcal{\hat I}^{i}_{\ref{diag:qgqbarT}})^{\ast}]\biggr \}.
\end{split}
\end{equation}
Note that the terms $\propto m^3$ in \eq\nr{eq:Amsquared} produce a trace over an odd number of gamma matrices, which consequently results in zero. Furthermore, by noting that only the term $\propto \vert\mathcal{I}^{i}_{\ref{diag:qgqbarT}}\vert^2$ is UV divergent, we get
\begin{equation}
\label{eq:Amsquaredfinal2}
\begin{split}
\sum_{h_0,h_1}\vert A^{lj}_m\vert^2 = m^2 (D_s-2)8k_0^{+}&k_1^{+} \biggl \{ a_1^2\biggl [4k_0^+(k_0^+ + k_2^+) + 2(k_2^+)^2 + (k_2^+)^2(D_s-4) \biggr ]\vert\mathcal{I}^{i}_{\ref{diag:qgqbarT}}\vert^2 \\
& + a_2^2\biggl [4k_1^+(k_1^+ - q^+) + 2(q^+)^2\biggr ]\vert\mathcal{\hat I}^{k}_{\ref{diag:qgqbarT}}\vert^2 + 2m^2 a_3^2\vert\mathcal{I}_{\ref{diag:qgqbarT}}\vert^2\\
& - a_1a_24k_0^+k_1^+ \Re e[\mathcal{I}^{i}_{\ref{diag:qgqbarT}}(\mathcal{\hat I}^{i}_{\ref{diag:qgqbarT}})^{\ast}]\biggr \}.
\end{split}
\end{equation}
The computation of the second term on the right-hand side of \eq\nr{eq:qqgkernelterm2} follows the same steps as above and it yields 
\begin{equation}
\label{eq:Bmsquaredfinal}
\begin{split}
\sum_{h_0,h_1}\vert B^{lj}_m\vert^2 = m^2 (D_s-2)8k_0^{+}&k_1^{+} \biggl \{ b_1^2\biggl [4k_1^+(k_1^+ + k_2^+) + 2(k_2^+)^2 + (k_2^+)^2(D_s-4) \biggr ]\vert\mathcal{I}^{i}_{\ref{diag:qqbargT}}\vert^2 \\
& + b_2^2\biggl [4k_0^+(k_0^+ - q^+) + 2(q^+)^2 \biggr ]\vert\mathcal{\hat I}^{k}_{\ref{diag:qqbargT}}\vert^2 + 2m^2 b_3^2 \vert\mathcal{I}_{\ref{diag:qqbargT}}\vert^2\\
& + b_1b_2 4k_0^+k_1^+ \Re e[\mathcal{I}^{i}_{\ref{diag:qqbargT}}(\mathcal{\hat I}^{i}_{\ref{diag:qqbargT}})^{\ast}]\biggr \}.
\end{split}
\end{equation}

The third term on the right-hand side of \eq\nr{eq:qqgkernelterm2} is given by the expression 
\begin{equation}
\label{eq:AmBmcrossterm}
\begin{split}
\sum_{h_0,h_1} 2\Re e[A^{lj}_m &(B^{lj}_m)^{\ast}]  = 2m^2\sum_{h_0,h_1}\Re e \Biggl [  \biggl \{ - a_1\bar u(0)\gamma^+\biggl [(2k^+_0 + k^+_2)\delta^{ij}_{(D_s)} - \frac{k^+_2}{2}[\gamma^i,\gamma^j]\biggr ]\gamma^{l}v(1)\mathcal{I}^{i}_{\ref{diag:qgqbarT}}\\
& + a_2\bar u(0)\gamma^+\gamma^j\biggl [(2k^+_1 - q^+)\delta^{kl}_{(D_s)} - \frac{q^+}{2}[\gamma^k,\gamma^l]\biggr ]v(1)\mathcal{\hat I}^{k}_{\ref{diag:qgqbarT}}
 + m a_3\bar u(0)\gamma^+\gamma^j\gamma^lv(1)\mathcal{I}_{\ref{diag:qgqbarT}} \biggr \}\\
& \times \biggl \{b_1 \bar v(1)\biggl [(2k^+_1 + k^+_2)\delta^{i'j}_{(D_s)} - \frac{k^+_2}{2}[\gamma^{i'},\gamma^j]\biggr ]\gamma^{l}\gamma^+u(0)(\mathcal{I}^{i'}_{\ref{diag:qqbargT}})^{\ast}\\
& + b_2\bar v(1)\gamma^{j}\biggl [(2k^+_0 - q^+)\delta^{k'l}_{(D_s)} - \frac{q^+}{2}[\gamma^{k'},\gamma^l]\biggr ]\gamma^+ u(0)(\mathcal{\hat I}^{k'}_{\ref{diag:qqbargT}})^{\ast} -m b_3\bar v(1)\gamma^j\gamma^l\gamma^+ u(0)(\mathcal{I}_{\ref{diag:qqbargT}})^{\ast}\biggr \} \Biggr ]. 
\end{split}
\end{equation}
This term is fully UV finite, and therefore we can immediately set $D_s = D =4$ everywhere. After performing the spinor and gamma-matrix algebra, we obtain the following expression
\begin{equation}
\begin{split}
\sum_{h_0,h_1} 2\Re e[A^{lj}_m (B^{lj}_m)^{\ast}]  =  4m^2 8k_0^{+}k_1^{+} \Biggl \{& -a_1b_1\biggl [(2k_0^+ + k_2^+)(2k_1^+ + k_2^+) + (k_2^+)^2\biggr ]\Re e[\mathcal{I}^{i}_{\ref{diag:qgqbarT}}(\mathcal{I}^{i}_{\ref{diag:qqbargT}})^{\ast}]\\
& + a_2b_2\biggl [(2k_1^+ -q^+)(2k_0^+ - q^+) + (q^+)^2 \biggr ]\Re e[\mathcal{\hat I}^{i}_{\ref{diag:qgqbarT}}(\mathcal{\hat I}^{i}_{\ref{diag:qqbargT}})^{\ast} ]\\
& -2a_1b_2 (k_0^+)^2 \Re e[\mathcal{I}^{i}_{\ref{diag:qgqbarT}}(\mathcal{\hat I}^{i}_{\ref{diag:qqbargT}})^{\ast}] + 2a_2b_1 (k_1^+)^2 \Re e[\mathcal{\hat I}^{i}_{\ref{diag:qgqbarT}}(\mathcal{I}^{i}_{\ref{diag:qqbargT}})^{\ast}]\Biggr \}.
\end{split}
\end{equation}

 Finally, we concentrate on the third contribution in \eq\nr{eq:qqgkernelstep1}, which is the cross term between the helicity nonflip and the helicity one and two flip contributions. It is straightforward to show that terms coming from the helicity nonflip and single helicity flip contributions yield a vanishing result, since they involve a trace over an odd number of gamma matrices. Therefore, we are left with the terms which involve only a product of a helicity nonflip contribution and a two helicity flip one. These terms give the following result 
\begin{equation}
\begin{split}
\sum_{h_0,h_1} 2\Re e[\Sigma^{lj}(\Sigma^{lj}_m)^{\ast}]  = 8m^2 8k_0^+k_1^+ &\Biggl \{a_{\ref{diag:qgqbarT}} a_3 (k_0^+k_1^+)\Re e[\mathcal{I}^{ii}_{\ref{diag:qgqbarT}}(\mathcal{I}_{\ref{diag:qgqbarT}})^{\ast}] - b_{\ref{diag:qqbargT}}  b_3(k_0^+k_1^+)\Re e[\mathcal{I}^{ii}_{\ref{diag:qqbargT}}(\mathcal{I}_{\ref{diag:qqbargT}})^{\ast}]\\
& + a_{\ref{diag:qgqbarT}}b_3 (k_0^+ + k_2^+)^2 \Re e[\mathcal{I}^{ii}_{\ref{diag:qgqbarT}}(\mathcal{I}_{\ref{diag:qqbargT}})^{\ast}] - b_{\ref{diag:qqbargT}} a_3 (k_1^+ + k_2^+)^2 \Re e[\mathcal{I}^{ii}_{\ref{diag:qqbargT}}(\mathcal{I}_{\ref{diag:qgqbarT}})^{\ast}]\\
& + c_{\ref{diag:qgqbarTinst}} a_3 \Re e[\mathcal{J}_{\ref{diag:qgqbarTinst}}(\mathcal{I}_{\ref{diag:qgqbarT}})^{\ast}] + c_{\ref{diag:qqbargTinst}} b_3 \Re e[\mathcal{J}_{\ref{diag:qqbargTinst}}(\mathcal{I}_{\ref{diag:qqbargT}})^{\ast}]\Biggr \}.
\end{split}
\end{equation}

\section{Fourier transforms for gluon emission}
\label{app:FTintforqqbarg}

For the gluon emission diagrams from a transverse photon state, we need to calculate the following set of Fourier integrals: 
\begin{equation}
\label{eq:Iijint}
\mathcal{I}^{ij}(\bt,\rt,\overline{Q}^2,\omega,\lambda)  = \mu^{2-\frac{D}{2}}\int \frac{\ud^{D-2}\Pt}{(2\pi)^{D-2}}\int \frac{\ud^{D-2}\Kt}{(2\pi)^{D-2}} \frac{\Pt^i \Kt^j e^{i\Pt \cdot \bt}e^{i\Kt \cdot \rt}}{\left [\Pt^2 + \overline{Q}^2 + m^2\right ]\left [\Kt^2 +\omega\left (\Pt^2 + \overline{Q}^2+ m^2 + \lambda m^2 \right )\right ]}
\end{equation}
\begin{equation}
\label{eq:Iiint}
\mathcal{I}^{i}(\bt,\rt,\overline{Q}^2,\omega,\lambda)  = \mu^{2-\frac{D}{2}}\int \frac{\ud^{D-2}\Pt}{(2\pi)^{D-2}}\int \frac{\ud^{D-2}\Kt}{(2\pi)^{D-2}} \frac{\Kt^i e^{i\Pt \cdot \bt}e^{i\Kt \cdot \rt}}{\left [\Pt^2 + \overline{Q}^2 + m^2\right ]\left [\Kt^2 +\omega\left (\Pt^2 + \overline{Q}^2+ m^2 + \lambda m^2 \right )\right ]}
\end{equation}
\begin{equation}
\label{eq:Iihatint}
\mathcal{\hat I}^{i}(\bt,\rt,\overline{Q}^2,\omega,\lambda)  = \mu^{2-\frac{D}{2}}\int \frac{\ud^{D-2}\Pt}{(2\pi)^{D-2}}\int \frac{\ud^{D-2}\Kt}{(2\pi)^{D-2}} \frac{\Pt^i e^{i\Pt \cdot \bt}e^{i\Kt \cdot \rt}}{\left [\Pt^2 + \overline{Q}^2 + m^2\right ]\left [\Kt^2 +\omega\left (\Pt^2 + \overline{Q}^2+ m^2 + \lambda m^2 \right )\right ]}
\end{equation}
\begin{equation}
\label{eq:Iscalarint}
\mathcal{I}(\bt,\rt,\overline{Q}^2,\omega,\lambda)  = \mu^{2-\frac{D}{2}}\int \frac{\ud^{D-2}\Pt}{(2\pi)^{D-2}}\int \frac{\ud^{D-2}\Kt}{(2\pi)^{D-2}} \frac{e^{i\Pt \cdot \bt}e^{i\Kt \cdot \rt}}{\left [\Pt^2 + \overline{Q}^2 + m^2\right ]\left [\Kt^2 +\omega\left (\Pt^2 + \overline{Q}^2+ m^2 + \lambda m^2 \right )\right ]}
\end{equation}
and
\begin{equation}
\label{eq:Jscalarint}
\mathcal{J}(\bt,\rt,\overline{Q}^2,\omega,\lambda)  = \mu^{2-\frac{D}{2}}\int \frac{\ud^{D-2}\Pt}{(2\pi)^{D-2}}\int \frac{\ud^{D-2}\Kt}{(2\pi)^{D-2}} \frac{e^{i\Pt \cdot \bt}e^{i\Kt \cdot \rt}}{\left [\Kt^2 +\omega\left (\Pt^2 + \overline{Q}^2+ m^2 + \lambda m^2 \right )\right ]}.
\end{equation}
To perform the above integrals, we use the Schwinger parametrization introduced in \eq\nr{eq:SParam} for each denominator, and then perform the $(D-2)$-dimensional transverse momentum integrals, which are then Gaussian.

For the integrals in  \eqs\nr{eq:Iijint} - \nr{eq:Iscalarint}, we obtain 
\begin{equation}
\label{eq:Iranktwofinal}
\mathcal{I}^{ij}(\bt,\rt,\overline{Q}^2,\omega,\lambda) = -\frac{\mu^{2-D/2}}{4(4\pi)^{D-2}}\bt^i\rt^j\int_{0}^{\infty} \ud u\, u^{-D/2}e^{-u[\overline{Q}^2 + m^2]}e^{-\frac{\vert\bt\vert^2}{4u}}\int_{0}^{u/\omega}\ud t \,t^{-D/2}e^{-t\omega \lambda m^2}e^{-\frac{\vert\rt\vert^2}{4t}}
\end{equation}
\begin{equation}
\label{eq:Irankonefinal}
\mathcal{I}^i(\bt,\rt,\overline{Q}^2,\omega,\lambda) = \frac{i\mu^{2-D/2}}{2(4\pi)^{D-2}}\rt^i\int_{0}^{\infty} \ud u \,u^{1-D/2}e^{-u[\overline{Q}^2 + m^2]}e^{-\frac{\vert\bt\vert^2}{4u}}\int_{0}^{u/\omega}\ud t \,t^{-D/2}e^{-t\omega \lambda m^2}e^{-\frac{\vert\rt\vert^2}{4t}}
\end{equation}
\begin{equation}
\label{eq:Irankonehatfinal}
\hat{\mathcal{I}}^i(\bt,\rt,\overline{Q}^2,\omega,\lambda) = \frac{i\mu^{2-D/2}}{2(4\pi)^{D-2}}\bt^i\int_{0}^{\infty} \ud u \,u^{-D/2}e^{-u[\overline{Q}^2 + m^2]}e^{-\frac{\vert\bt\vert^2}{4u}}\int_{0}^{u/\omega}\ud t \,t^{1-D/2}e^{-t\omega \lambda m^2}e^{-\frac{\vert\rt\vert^2}{4t}}
\end{equation}
and
\begin{equation}
\label{eq:Iscalarfinal}
\mathcal{I}(\bt,\rt,\overline{Q}^2,\omega,\lambda) = \frac{\mu^{2-D/2}}{(4\pi)^{D-2}}\int_{0}^{\infty} \ud u \,u^{1-D/2}e^{-u[\overline{Q}^2 + m^2]}e^{-\frac{\vert\bt\vert^2}{4u}} \int_{0}^{u/\omega}\ud t \,t^{1-D/2}e^{-t\omega \lambda m^2}e^{-\frac{\vert\rt\vert^2}{4t}}.
\end{equation}
Note that in an arbitrary dimension $D$, the $t$-integrals above would give a $u$ dependent incomplete Gamma function. Hence, we are not able to express the final results in terms of a familiar special functions (e.g. the modified Bessel $K$- functions).

In the case of \eq\nr{eq:Jscalarint}, we are only left with  an  integral  over  a  single Schwinger  parameter  
\begin{equation}
\mathcal{J}(\bt,\rt,\overline{Q}^2,\omega,\lambda) = \mu^{2-D/2} (4\pi)^{2-D} \omega^{1-\frac{D}{2}} \int_{0}^{\infty} \ud u\, u^{2-D}e^{-u\omega[\overline{Q}^2 + m^2 + \lambda m^2]}e^{-\frac{1}{4u\omega} [\vert\bt\vert^2 + \omega \vert \rt\vert^2]}.
\end{equation}
The remaining $u$-integral can  be performed for generic dimension $D$, yielding the following results
\begin{equation}
\label{eq:Jscalarintfinal}
\mathcal{J}(\bt,\rt,\overline{Q}^2,\omega,\lambda) = (2\pi)^{2-D}\left (\frac{\mu}{\omega}\right)^{2-\frac{D}{2}}\left ( \frac{\sqrt{\overline{Q}^2 + m^2 + \lambda m^2}}{\sqrt{\vert\bt\vert^2 + \omega\vert\rt\vert^2}}\right )^{D-3}K_{D-3}\left (\sqrt{\overline{Q}^2 + m^2 + \lambda m^2}\sqrt{\vert\bt\vert^2 + \omega\vert\rt\vert^2} \right ).
\end{equation}

\section{Form factors and Lorentz-invariance}
\label{sec:Pauliform}

In this appendix we discuss some of the constraints from the Lorentz-invariance on the  momentum space scalar form factors ${\cal S}^{T}$, ${\cal N}^{T}$ and ${\cal V}^{T}$, and the longitudinal photon ${\cal V}^{L}$ from our earlier paper \cite{Beuf:2021qqa}, at the on-shell point, as discussed in Sec.~\ref{sec:spinstuctureandmrenorm}.

Our starting point is the usual parametrization of the $\gamma q \bar{q}$ vertex function (based on Lorentz and gauge invariance),
\begin{align}
\Gamma^{\mu}(q)
= & F_D(q^2/m^2)\; \gamma^{\mu} + F_P(q^2/m^2)\; \frac{q_{\nu}}{2m}\; i \sigma^{\mu\nu}
\, ,
\label{eq:1PI_vertex_generic}
\end{align}
which introduces the Dirac and Pauli form factors $F_D(q^2/m^2)$ and $F_P(q^2/m^2)$. This form should now be compared with the general parametrization of the spinor structure of the wavefunction in \eq\eqref{NLOformfactors}. 
 The difference between vertex corrections as encoded in the form factors $F_D$ and $F_P$ and the LCWF   is that in the former the ``outgoing'' $q\bar{q}$ is an asymptotic outgoing state in the scattering process, but in the latter it is an intermediate state that will scatter off the target shockwave. This means that in the form factors the quarks are on mass shell, which in our case means  
$\Pt^2= -\overline{Q}^2 -m^2$. Note that for a quark-antiquark pair in the final state, which is the situation here, this means that the virtual photon must be timelike, $q^2 = -Q^2 > 4m^2$. Relatedly, the $q\bar{q}$ state in the LCWF has its own energy denominator $\ed_{\lo}$. In the comparing the LCWF with the form factors this energy denominator (which becomes zero at the on-shell point) must be left out.  Leaving out the additional color and electric charge factors and contracting with the physical polarization vector for the photon, the condition that should be satisfied is
\begin{multline}
\label{eq:formfactorconstraint}
  \bar u(0) \epsl_{\lambda}(\qvec)v(1) F_D(q^2/m^2)+  
\frac{q_{\nu} \varepsilon_{\lambda,\mu} (\qvec) }{2m}\;
i \bar u(0)\sigma^{\mu\nu} v(1)F_P(q^2/m^2)
= 
\Bigg\{
\bar u(0) \epsl_{\lambda}(q)v(1)\left [1 + \left (\frac{\alpha_s\cf}{2\pi}\right )\mathcal{V}^T\right ] 
\\
+ \frac{q^+}{2k^+_0k^+_1}(\Pt \cdot \epst_{\lambda}) \bar u(0)\gamma^+ v(1) \left (\frac{\alpha_s\cf}{2\pi}\right )\mathcal{N}^T
 +  \frac{q^+}{2k^+_0k^+_1} \frac{(\Pt \cdot \epst_{\lambda})}{\Pt^2}\Pt^j m \bar u(0)\gamma^+\gamma^j v(1) \left (\frac{\alpha_s\cf}{2\pi}\right )\mathcal{S}^T 
 \\
 + \frac{q^+}{2k^+_0k^+_1}m \bar u(0)\gamma^+ \epsl_{\lambda}(q) v(1)\left (\frac{\alpha_s\cf}{2\pi}\right )\mathcal{M}^T
\left. \Bigg\} \right|_{\Pt^2= -\overline{Q}^2 -m^2} .
\end{multline}
This same constraint  with a longitudinal photon polarization (the $\Gamma^+$ component of the constraint) was already considered in our earlier paper \cite{Beuf:2021qqa}.
For a transverse photon polarization one can use  the relation 
\begin{align}
&\varepsilon_{\lambda,\mu} (\qvec) q_{\nu}\, \overline{u}(0)\, i \sigma^{\mu\nu}\, v(1) 
 = 2 m\; \overline{u}(0)\, \epsl_{\lambda} v(1) 
-2 m\frac{(k_0^+\!-\!k_1^+)}{q^+}\;    \frac{q^+}{2 k_0^+ k_1^+}\, \left(\Pt\cdot \epst_{\lambda}\right)\; 
     \overline{u}(0)\, \gamma^+\, v(1)
\\
& 
-2\Pt^2\; \frac{q^+}{2 k_0^+ k_1^+}\, \frac{\left(\Pt\cdot \epst_{\lambda}\right)}{\Pt^2}\, \Pt^j\, 
     \, \overline{u}(0)\, \gamma^+ \gamma^j\, v(1)  
-\Big[\Pt^2\!+\! \overline{Q}^2\!+\! m^2\Big]\;  \frac{q^+}{2 k_0^+ k_1^+}\,  
     \, \overline{u}(0)\, \gamma^+\epsl_{\lambda}(\qvec)\; v(1) 	
\end{align}
to  arrive at the constraints \eqref{FP_ST_correspondence}, \eqref{FP_NT_correspondence},
\eqref{FD_VT_correspondence} and~\eqref{MT_condition}. In the remainder of this appendix, we will verify that our results for the form factors satisfy the constraints  \eqref{FP_ST_correspondence}, \eqref{FP_NT_correspondence} and show how to use \eqref{MT_condition} for quark mass renormalization. We will also check an additional relation that relates the transverse and longitudinal polarization states.
 It would also be interesting to check what happens with the $\Gamma^-$ component, which does not  couple to photons in the light cone gauge, but we leave such a discussion for another day. 
 
\subsection{From ${\cal S}^{T}$ to $F_P$}

The form factor ${\cal S}^{T}$ is related to the Pauli form factor by \eq\eqref{FP_ST_correspondence}. It gets contributions only from the two non-instantaneous vertex correction diagrams \ref{diag:vertexqbaremT} and \ref{diag:vertexqemT}, calculated in Sec.~\ref{vertexcontribution}. 
At the on-shell point $\Pt^2= -\overline{Q}^2 -m^2$ the contribution from diagram~\ref{diag:vertexqbaremT} in \eq\eqref{ST_result_y_chi} is such that
\begin{align}
\left. \frac{1}{\Pt^2}\, {\cal S}^{T}_{\ref{diag:vertexqbaremT}}\right|_{\Pt^2= -\overline{Q}^2 -m^2} = &  -2 \, \int_{0}^{1} \ud y \int_{0}^{z} \ud \chi \frac{1-y}{\left[m^2+\frac{\chi(1-\chi)}{z(1-z)} \overline{Q}^2\right]}.
\end{align}
At this point it is trivial to integrate over $y$ and, writing the photon virtuality in terms of the 4-vector $q^2=-Q^2$ one has
\begin{align}
\left. \frac{1}{\Pt^2}\,{\cal S}^{T}_{\ref{diag:vertexqbaremT}}\right|_{\Pt^2= -\overline{Q}^2 -m^2} = &  
-   \int_{0}^{z} \ud \chi \frac{1}{\left[m^2-\chi(1-\chi)q^2 \right]}.
\end{align}
The result from the other diagram \ref{diag:vertexqemT} is the same up to an exchange $z\leftrightarrow 1-z$, and thus the total contribution is 
\begin{align}
\left. \frac{1}{\Pt^2}\, {\cal S}^{T}\right|_{\Pt^2= -\overline{Q}^2 -m^2} = &  
-   \int_{0}^{z} \ud \chi \frac{1}{\left[m^2-\chi(1-\chi)q^2 \right]} 
-   \int_{0}^{1-z} \ud \chi \frac{1}{\left[m^2-\chi(1-\chi)q^2 \right]} 
= -   \int_{0}^{1} \ud \chi \frac{1}{\left[m^2-\chi(1-\chi)q^2 \right]},
\end{align}
where one uses the symmetry of the $\chi$ integrand under $\chi \leftrightarrow 1-\chi$.
Thus the l.h.s of the condition \eq\eqref{FP_ST_correspondence} is
\begin{align}
\left.  -\frac{m^2}{\Pt^2} \left (\frac{\alpha_s \cf}{2\pi}\right ) {\cal S}^{T}\right|_{\Pt^2= -\overline{Q}^2 -m^2} 
& = \left (\frac{\alpha_s \cf}{2\pi}\right ) \int_{0}^{1} d\chi\, \frac{m^2}{\left\{m^2 -\chi (1\!-\!\chi) q^2\right\}}
+ O(D\!-\!4)
\label{SL1_FF_2}
\, ,
\end{align}
which is indeed the one-loop Pauli form factor.


\subsection{From ${\cal N}^{T}$ to $F_P$}

For the form factor ${\cal N}^{T}$ it is best to start from 
\begin{equation}
\mathcal{N}^T = I_{{\mathcal{N}}_{\ref{diag:vertexqbaremT}}} +  \Omega_{\mathcal{N}_{\ref{diag:vertexqbaremT}}}  -  \biggl [z \leftrightarrow 1-z \biggr ] ,
\end{equation}
with the expression for $\Omega_{\mathcal{N}_{\ref{diag:vertexqbaremT}}}$ given in 
\eq\eqref{eq:degOmegaN} and for $I_{{\mathcal{N}}_{\ref{diag:vertexqbaremT}}}$ in 
\eq\eqref{eq:INev2}. At the on-shell point the first contribution reduces to 
\begin{equation}
\label{eq:INev2_onshell}
\left. I_{\mathcal{N}_{\ref{diag:vertexqbaremT}}} 
\right|_{\Pt^2= -\overline{Q}^2 -m^2} 
=  \int_{0}^{1} \ud y \int_{0}^{z} \ud \chi \frac{-2 m^2 \left [2z^2 \chi (1-y) + \chi (1-2\chi)y + z(1-y-3\chi + 2\chi (1+\chi)y)\right ]
}{ z(1-\chi)\left [m^2 + \chi(1-\chi)Q^2 \right ]  }.
\end{equation}
A straightforward integration over $y$ gives
\begin{equation}
\label{eq:INev2_onshell2}
 \left. I_{\mathcal{N}_{\ref{diag:vertexqbaremT}}} \right|_{\Pt^2= -\overline{Q}^2 -m^2} =   \int_{0}^{z} \ud \chi \frac{- m^2 \left [2z^2 \chi  + \chi  (1-2\chi) + z(1-4\chi + 2\chi^2)\right ]
}{ z(1-\chi)\left [m^2 + \chi(1-\chi)Q^2 \right ]  }.
\end{equation}
In order to arrive at the same manipulations as for ${\cal S}^{T}$ above, we split this into 
\begin{equation}
\label{eq:INev2_onshell3}
\left. I_{\mathcal{N}_{\ref{diag:vertexqbaremT}}} \right|_{\Pt^2= -\overline{Q}^2 -m^2} =  
 I_{\mathcal{N}_{\ref{diag:vertexqbaremT}}}^{\text{on-shell } a} + I_{\mathcal{N}_{\ref{diag:vertexqbaremT}}}^{\text{on-shell } b}
 \end{equation}
 with 

\begin{equation}
 I_{\mathcal{N}_{\ref{diag:vertexqbaremT}}}^{\text{on-shell }a} 
= -m^2(2z-1)\int_{0}^{z} \ud \chi 
\frac{1 
}{ m^2 + \chi(1-\chi)Q^2  }
\end{equation}
and 
\begin{equation}
 I_{\mathcal{N}_{\ref{diag:vertexqbaremT}}}^{\text{on-shell }b} 
= -m^2\int_{0}^{z} \ud \chi 
\frac{(2\chi-1)(1-z)(\chi+2z)
}{ z(1-\chi)\left [m^2 + \chi(1-\chi)Q^2 \right ]  }.
\end{equation}
The first terms yield the Pauli form factor, following the same procedure as above:
\begin{equation}
\frac{-1}{2z-1}\left[I_{\mathcal{N}_{\ref{diag:vertexqbaremT}}}^{\text{on-shell }a} - 
 \biggl [z \leftrightarrow 1-z \biggr ]
 \right]
 = \int_{0}^{1} d\chi\, \frac{m^2}{\left\{m^2 -\chi (1\!-\!\chi) q^2\right\}},
\end{equation}
agreeing with the condition \eqref{FP_NT_correspondence}.
The second term can be shown, after a somewhat lengthy calculation, to cancel against the $\Omega_{\mathcal{N}_{\ref{diag:vertexqbaremT}}}$ term even without antisymmetrization with respect to $z\leftrightarrow 1-z$:
\begin{equation}
I_{\mathcal{N}_{\ref{diag:vertexqbaremT}}}^{\text{on-shell }b} 
 + \Omega_{{\mathcal{N}}_{\ref{diag:vertexqbaremT}}}  =0.
\end{equation}
Thus our result for $\mathcal{N}^T$  indeed satisfies the condition \eqref{FP_NT_correspondence}.


\subsection{${\cal M}^{T}$ and vertex mass renormalization}
\label{app:Mct}

Using the form \eqref{eq:MTct} for the ${\cal M}^{T}$  counterterm and 
\begin{equation}
\mathcal{A}_0(\Delta) =
 \Gamma\left(2-\frac{D}{2}\right)\,  \left[\frac{\Delta}{4\pi\, \mu^2}\right]^{\frac{D}{2}\!-\!2}
+ O(D-4)
\end{equation}
one can easily perform the $\xi$-integral to obtain
\begin{equation}
\mathcal{M}^T_{\text{c.t.}} = \frac{1}{2} \frac{(D_s-4)}{(D-4)} -\frac{5}{2}
-\frac{3}{2}\, \Gamma\left(2\!-\!\frac{D}{2}\right)\,  \left[\frac{m^2}{4\pi\, \mu^2}\right]^{\frac{D}{2}\!-\!2}+ O(D\!-\!4)
\end{equation}
In terms of the mass renormalization coefficient $Z_m $ in the on-shell scheme this corresponds to 
\begin{align}
Z_m 
= & 1   -\left[\frac{\alpha_s\, C_F}{2\pi}\right]\, 
\left\{\frac{3}{2}\, \Gamma\left(2\!-\!\frac{D}{2}\right)\,  \left[\frac{m^2}{4\pi\, \mu^2}\right]^{\frac{D}{2}\!-\!2}
+\frac{5}{2} -\frac{1}{2}\, \frac{(D_s\!-\!4)}{(D\!-\!4)}  + O(D\!-\!4)\right\}
\, .
\label{eq:Zm_result}
\end{align}
This is indeed the known result (including the finite piece) for the mass renormalization in the on-shell scheme in CDR for $D_s=D$ and in FDH or DRED for $D_s=4$ (see e.g.~\cite{Marquard:2007uj}), and also in QED up to the replacement $\alpha_s\, C_F \rightarrow \alpha_{em}\, e_f^2$. Indeed, the expression \eqref{eq:1PI_vertex_generic} for the $\gamma q \bar{q}$ vertex function is derived assuming that $m$ is the pole mass, so that the relation \eqref{MT_condition} corresponds to the on-shell mass renormalization condition. While this is different from the ``kinetic mass'' counterterm in our calculation in Sec.~\ref{selfenergycontribution}, this is a known feature of our regularization scheme~\cite{Mustaki:1990im,Zhang:1993dd,Zhang:1993is,Harindranath:1993de,Brodsky:1997de} with a cutoff in $k^+$ and dimensional regularization for transverse momenta. It is also known that at least the divergent pieces of the mass counterterms agree if one regularizes with a cutoff in $k^-$, which is however technically difficult to do for the finite terms.  In a future paper we will show how it is possible to develop a regularization scheme where the kinetic and vertex mass counterterms agree exactly.

\subsection{Additional check for $\mathcal{V}^{T},$ $\mathcal{V}^{L}$}

We have now verified that our LCWF's satisfy the constraints for the scalar functions $\mathcal{S}^T$ and $\mathcal{N}^T$, and used the condition on $\mathcal{M}^T$ for mass renormalization. There remains the condition \eqref{FD_VT_correspondence} for $\mathcal{V}^T$. This is, however, a 
a bit more problematic. Indeed, the Dirac form factor and $\mathcal{V}^T$ involve quantities which have soft divergences and UV divergences (related to quark wave function renormalization, not mass). Hence, there should be some regularization scheme dependence, making it more difficult to use these relations to compare our results with earlier ones in the literature. Nevertheless, such soft and UV divergences have some degree of universality. This allows us to obtain a constraint between two finite quantities by taking the difference between the constraints for the longitudinal and transverse photons. By subtracting \eq(H7) in Ref.~\cite{Beuf:2021qqa} for the longitudinal photon from its analogue \eqref{FD_VT_correspondence} for the transverse one, we find that
\begin{align}
\left[\frac{\alpha_s\, C_F}{2\pi}\right]\;\left.\bigg[{\cal V}^{T}-{\cal V}^{L}\bigg]\right|_{\Pt^2= -\overline{Q}^2 -m^2} 
& = - \frac{(2z\!-\!1)^2}{4z(1\!-\!z)}\; F_P(q^2/m^2)
\label{eq:FP_VT_minus_VL_correspondence} 
\, .
\end{align}
 
Note that the quark self-energy diagrams cancel in the difference. Then, one has, using the notations of Ref.~\cite{Beuf:2021qqa}
\begin{align}
{\cal V}^{T}-{\cal V}^{L} & = {\cal V}^{T}_{\ref{diag:vertexqbaremT}}-{\cal V}^{L}_{(c)}-{\cal V}^{L}_{(e)_1} +\Big(z \leftrightarrow (1\!-\!z) \Big)
\, .
\label{eq:VT_minus_VL_sym}
\end{align}
We will be comparing to results from our earlier paper~\cite{Beuf:2021qqa}, so we must start from intermediate results that are in a slightly less integrated form. Using \eqs (73) and~(78) from~\cite{Beuf:2021qqa} one finds
\begin{multline}
{\cal V}^{L}_{(c)}+{\cal V}^{L}_{(e)_1} =   
-\frac{(D_s-4)}{(D-4)}\frac{z}{2}
 +\int_{\frac{ \alpha}{z}}^{1} d\xi \Big[\frac{1}{\xi}-(1\!-\!z) - z\xi\Big]\, \mathcal{A}_0(\Delta_2)
-\frac{z^2 m^2}{(1\!-\!z)} \int_{0}^{1} d\xi\, (1\!-\!\xi)\xi^2\, \mathcal{B}_0 
\\ 
 +\int_{\alpha/z}^{1} d\xi\, (1\!-\!\xi) \Big[\frac{2}{\xi}+\frac{(2z\!-\!1)}{(1\!-\!z)}-\frac{z\xi}{(1\!-\!z)}\Big]\, 
      \left\{ (1\!-\!\xi)\left(\Pt^2 \!+\!\overline{Q}^2 \!+\!m^2\right)\, \mathcal{B}_0 -\Pt^j\, \mathcal{B}^j   
			+  (1\!-\!z)\xi m^2 \mathcal{B}_0 \right\}
+ O(D\!-\!4)
\, .
\label{eq:VL13a_1}
\end{multline}
and, from \eq\eqref{eq:VT} in this paper, 
\begin{multline}
{\cal V}^{T}_{\ref{diag:vertexqbaremT}} =  
-\frac{(D_s-4)}{(D-4)}\frac{z}{2}-z
 +\int_{\alpha / z}^{1} d\xi \Big[\frac{1}{\xi} -2z +z\xi\Big]\, 
\left\{\mathcal{A}_0(\Delta_2)-(1\!-\!\xi)\left(\Pt^2 \!+\!\overline{Q}^2 \!+\!m^2\right)\, \left[\frac{\Pt^j\, \mathcal{B}^j}{\Pt^2} +\frac{\xi \mathcal{B}_0}{(1\!-\!z)}\right]
\right\}
\\ 
 +\int_{\alpha/z}^{1} d\xi\, (1\!-\!\xi) \Big[\frac{2}{\xi}+\frac{z\xi}{(1\!-\!z)}\Big]\, 
      \left\{ (1\!-\!\xi)\left(\Pt^2 \!+\!\overline{Q}^2 \!+\!m^2\right)\, \mathcal{B}_0 -\Pt^j\, \mathcal{B}^j   
			+  (1\!-\!z)\xi m^2 \mathcal{B}_0 \right\}
			\\  
+z m^2 \int_{0}^{1} d\xi\, (1\!-\!\xi)\xi\,  \left\{ 
     \frac{(1\!-\!2z)}{(1\!-\!z)}\,\frac{\Pt^j\, \mathcal{B}^j}{\Pt^2}    
    +\xi\, \mathcal{B}_0 \right\}
+ O(D\!-\!4)
\, .
\label{eq:VT1_1}
\end{multline}
The scheme dependent rational  term $-\frac{(D_s-4)}{(D-4)} \frac{z}{2}$ in Eqs.~\eqref{eq:VL13a_1} and \eqref{eq:VT1_1} cancels in the difference between the two, whereas the term $-z$ in Eq.~\eqref{eq:VT1_1} is a scheme independent rational term induced by the tensor reduction.

Setting $\Pt^2= -\overline{Q}^2 -m^2$ and following the same method as previously, one arrives at the end of the calculation at
\begin{multline}
\left.\bigg[{\cal V}^{T}_{\ref{diag:vertexqbaremT}}-{\cal V}^{L}_{(c)}-{\cal V}^{L}_{(e)_1}\bigg]\right|_{\Pt^2= -\overline{Q}^2 -m^2} 
=  (1\!-\!2z) \left\{\Gamma\left(2\!-\!\frac{D}{2}\right)\,  \left[\frac{m^2}{4\pi\, \mu^2}\right]^{\frac{D}{2}\!-\!2} +2\right\}
\\ 
+\left[q^2 -\frac{3m^2}{4z(1\!-\!z)}\right]\int_{0}^{z} d\chi\, \frac{(1\!-\!2\chi)(1\!-\!2z)}{\left\{m^2 -\chi (1\!-\!\chi) q^2\right\}}
-\frac{(1\!-\!2z)^2}{4z(1\!-\!z)}\int_{0}^{z} d\chi\, \frac{m^2}{\left\{m^2 -\chi (1\!-\!\chi) q^2\right\}}
\label{eq:VT_minus_VL_1} 
\, .
\end{multline}
After symmetrizing in $\chi \leftrightarrow (1\!-\!\chi)$ and adding the other graphs according to Eq.~\eqref{eq:VT_minus_VL_sym}, one gets
\begin{align}
\left.\bigg[{\cal V}^{T}-{\cal V}^{L}\bigg]\right|_{\Pt^2= -\overline{Q}^2 -m^2} 
= & 
-\frac{(1\!-\!2z)^2}{4z(1\!-\!z)}\int_{0}^{1} d\chi\, \frac{m^2}{\left\{m^2 -\chi (1\!-\!\chi) q^2\right\}}
\label{eq:VT_minus_VL_2} 
\, ,
\end{align}
which indeed reproduces the constraint \eqref{eq:FP_VT_minus_VL_correspondence}.

\bibliography{spires}
\bibliographystyle{JHEP-2modlong}

\end{document}